\documentclass[a4paper,12pt,twoside,openright]{book}

\usepackage{color,lmodern,bm,url,amsmath,amssymb,epsfig,graphicx,verbatim,xspace,multirow,mathtools,booktabs,epigraph,fancyhdr,indentfirst,comment,lipsum,cite,appendix,titlesec,ragged2e}
\usepackage[hang,flushmargin]{footmisc} 
\hypersetup{
colorlinks=true,
citecolor=ultramarine,
linkcolor=cadmiumgreen,
urlcolor=indigo(dye)}
\usepackage[T1]{fontenc} 
\usepackage[skip=10pt,font=footnotesize,bf]{caption}
\definecolor{ultramarine}{rgb}{0.07, 0.04, 0.56}
\definecolor{cadmiumgreen}{rgb}{0.0, 0.42, 0.24}
\definecolor{indigo(dye)}{rgb}{0.0, 0.25, 0.42}
\definecolor{orangered}{RGB}{255,69,0}
\definecolor{rosso}{cmyk}{0,1,1,0.1}
\usepackage[Sonny]{fncychap}
\usepackage[english]{babel}
\usepackage[final]{pdfpages} 

\newcommand{\f}[2]{\frac{#1}{#2}}  
\newcommand{\mk}[1]{\left( #1 \right)}  
\newcommand{\kk}[1]{\left[ #1 \right]} 
\newcommand{\ck}[1]{\left\{ #1 \right\}}  
\newcommand{\be}{\begin{equation}}  
\newcommand{\ee}{\end{equation}}
\newcommand{\ba}{\begin{aligned}}  
\newcommand{\ea}{\end{aligned}}
\newcommand{\bali}{\begin{alignat}}
\newcommand{\eali}{\end{alignat}}
\newcommand{\bc}{}
\newcommand{\Mpl}{M_{\rm Pl}}
\newcommand{\pa}{\partial}

\newcommand{\diff}[2]{\ensuremath{\frac{\text{d}#1}{\text{d}#2}}}
\newcommand{\difs}[2]{\ensuremath{\frac{\text{d}^2 #1}{\text{d}#2^2}}}
\newcommand{\dif}{\text{d}}
\newcommand{\logn}{\text{ln\,}}
\newcommand{\ie}{\emph{i.e.}~}

\usepackage[hyperpageref]{backref}

\renewcommand*{\backref}[1]{}  
\renewcommand*{\backrefalt}[4]{
\ifcase #1 
No cited.
\or
{Cited on page} #2.
\else
{Cited on page} #2.
\fi} 

\makeatletter
\def \cleardoublepage {\clearpage \if@twoside
\ifodd \c@page
\else
\null\thispagestyle{empty}\clearpage
\fi
\fi}
\makeatother

\renewcommand{\headrulewidth}{0.0pt}

\fancypagestyle{plain}{%
\fancyhead{} 
\fancyfoot[C]{\thepage}
\renewcommand{\headrulewidth}{0pt} 
}

\begin{document}

\setlength{\unitlength}{1cm} 
\thispagestyle{empty}
\begin{center}

\begin{flushleft}
\includegraphics[scale=0.06]{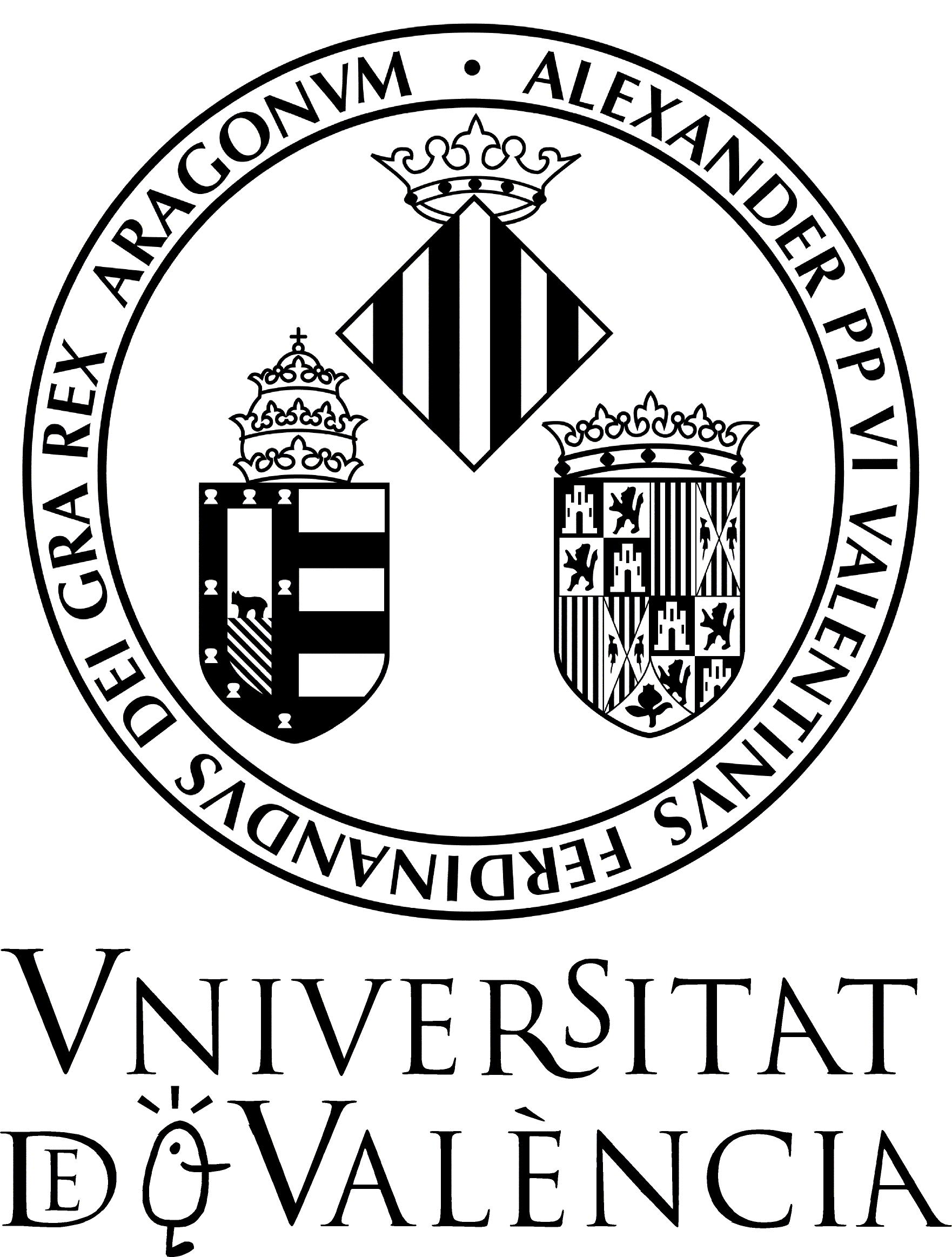}
\vspace{0.5cm}

{\large Universitat de Val\`{e}ncia -- IFIC/CSIC}\\[5pt]
{\large Departament de Física Teòrica}\\[5pt]
{\large Programa de Doctorado en F\'isica}
\vspace{0.5cm}

\noindent{\color{rosso} \rule{\textwidth}{0.05cm}}\vspace{0.5cm}
\textbf{\LARGE{NONCANONICAL APPROACHES\\ TO INFLATION
			}\\ \vspace{1cm}}
{\Large Ph.D. Dissertation}\\ \vspace{1cm}
{\Large \emph{H\'ector Ariel Ram\'irez Rodr\'iguez}}\vspace{0.5cm}
\noindent{\color{indigo(dye)} \rule{\textwidth}{0.05cm}}\vspace{1cm}
\end{flushleft}

\begin{flushright}
{\large Under the supervision of}\\[5pt]
{\Large {Dr. Olga Mena Requejo}}\\[4ex]
\end{flushright}

\vspace{2cm}

{\large \bf{Valencia, March 2019}}

\newpage

$$ $$\\[19cm]

\end{center}

\begin{titlepage}
\thispagestyle{empty}

\newpage
\cleardoublepage
\thispagestyle{empty}

\noindent {\large \textbf{Thesis Commitee Members}}\\[2cm]

\begin{flushleft}
\begin{tabular}{ l l }
{\bf Examiners} & \\[0.5cm]
Prof. Gabriela Barenboim & \qquad Universitat de València \\[0.5cm]
Prof. Nick E. Mavromatos & \qquad King's College London \\[0.5cm]
Prof. David Wands & \qquad University of Portsmouth\\[2cm]

{\bf Rapporteurs} & \\[0.5cm]
Prof. Mar Bastero Gil & \qquad Universidad de Granada \\[0.5cm]
Dr. Daniel G. Figueroa & \qquad École Polytechnique Fédérale de Lausanne \\[0.5cm]
Dr. Gonzalo Olmo & \qquad Universitat de València \\[1cm]
\end{tabular}
\end{flushleft}

\newpage
\thispagestyle{empty}

\frontmatter
\fancyfoot[C]{\thepage}
\chapter*{Agradecimientos / Acknowledgments}
\addcontentsline{toc}{chapter}{Acknowledgments}

\emph{First and foremost,} a Olga, porque solo gracias a ella esta tesis fue posible. Por haberme dado la oportunidad de trabajar con ella y haber confiado en mí desde el primer día. Han pasado más de cinco años desde entonces y nunca ha dejado de alentarme, aconsejarme y motivarme a hacer las cosas lo mejor posible. Nunca dejaré de agradecer la fortuna que fue haberla conocido y haber trabajado a su lado. 
\vspace{2pt}

Many thanks to Lotfi Boubekeur for introducing me to Inflation, for teaching me the basics of it and helping me grow as a researcher. Also for all the help, patience and friendship throughout these years. Many thanks to Hayato Motohashi as well, for his teaching and his enormous patience during the hours I spent in his office. Thanks both for the advice in work, physics and life, and for the reading and comments on the manuscript of this thesis.
\vspace{2pt}

I owe a great debt of gratitude to Prof. Wayne Hu for his time and teaching during my two visits to U. Chicago, and also for his help and patience regarding our project. I would also like to greatly thank Prof. David Wands for hosting me at ICG and making me feel at home every second since the beginning, but also for his time and endless help and support. Finally, to Prof. Shinji Tsujikawa for his interest, great support and help during the realization of our project and towards my academic life.
\vspace{2pt}

I'm obviously in debt to the rest of my collaborators for their help and contribution: Elena Giusarma, Stefano Gariazzo and Lavinia Heisenberg. But in particular, to Sam Passaglia for his friendship, help, motivation, advice and the good times in Japan. Además, muchísimas gracias también a todos los miembros de mi grupo en el IFIC por su ayuda y amistad a lo largo de estos años.
\vspace{2pt}

Agradecimientos especiales a los amigos más cercanos que me acompañaron a lo largo de estos años. A Miguel Escudero por absolutamente todo: amistad, compañía, ayuda, consejos, colaboración; desde el primer momento en el que empezó el máster y hasta el último minuto de doctorado, pasando por todos los group meetings, congresos, viajes y estancias. A José Ángel, por ser mi primer gran amigo desde mi primera etapa en Valencia, y por continuar siéndolo desde entonces, por los infinitos coffee breaks, charlas de la vida y su sana (e innecesaria) competitividad en todo. A Quique y a Gomis por su compañía todos estos años, ayuda, consejos y su (aún más innecesaria) competitividad dentro y fuera del despacho. También, al resto de mis amigos de Valencia que de una u otra forma se han distanciado pero de los que siempre tendré grandes recuerdos y agradecimientos por haberme incluido en el grupo cuando llegué. Last but not least, special thanks to Sam Witte for all his advice, help, grammar corrections and friendship inside and outside work.
\vspace{2pt}

I would also like to thank every single friend I made during these years not only in IFIC but also in Trieste, in Chicago, in Tokyo and in Portsmouth, who somehow contributed to this thesis, either through their help, advice, company or friendship. Many special thanks to every single friend I made in Portsmouth, I wish all the research institutes were like the ICG. Also, thanks a lot to every person I met in my visits to other institutes, people who hosted me or spent some time discussing my work. Finally, thanks to all the people I played football with in all these cities, I hope you enjoyed the games as much as I did.
\vspace{2pt}

Finalmente, a mi mamá, mi papá y mis hermanos. Porque, aunque hemos estado alejados, esta tesis no hubiera sido posible sin ellos, sin su apoyo incondicional.
\vspace{2cm}

\begin{flushright}
\emph{\Large HR}
\end{flushright}

\newpage
\cleardoublepage
\thispagestyle{empty}

$$ $$\\[4cm]

\begin{flushright}
\emph{``And the end of all our exploring\\ Will be to arrive where we started\\ And know the place for the first time.''} \\[1cm]
--- \raggedleft Frank Wilczek \& Betsy Devine,\\ Longing for the Harmonies\par
\end{flushright}

\newpage
\thispagestyle{empty}

\end{titlepage}

\fancyfoot[C]{\thepage}

\chapter*{List of Publications}
\addcontentsline{toc}{chapter}{List of Publications}  

This PhD thesis is based on the following publications: \\
\begin{itemize}

\item \emph{Phenomenological approaches of inflation and their equivalence}~\cite{Boubekeur:2014xva}\\
L.~Boubekeur, E.~Giusarma, O.~Mena and H.~Ram\'irez.\\
\href{https://journals.aps.org/prd/abstract/10.1103/PhysRevD.91.083006}{\emph{Phys.\ Rev.}\ D {\bf 91} (2015) no.8,  083006}, [\href{https://arxiv.org/abs/1411.7237}{\tt 1411.7237}].

\item \emph{Do current data prefer a nonminimally coupled inflaton?}~\cite{Boubekeur:2015xza}\\
L.~Boubekeur, E.~Giusarma, O.~Mena and H.~Ram\'irez.\\
\href{https://journals.aps.org/prd/abstract/10.1103/PhysRevD.91.103004}{\emph{Phys.\ Rev.\ D} {\bf 91} (2015) 103004}, [\href{https://arxiv.org/abs/1502.05193}{\tt 1502.05193}].

\item \emph{The present and future of the most favoured inflationary models after $Planck$ 2015}~\cite{Escudero:2015wba}\\
M.~Escudero, H.~Ram\'irez, L.~Boubekeur, E.~Giusarma and O.~Mena.\\  
 \href{http://dx.doi.org/10.1088/1475-7516/2016/02/020}{\emph{JCAP}  {\bf 1602} (2016) 020}, [\href{http://arxiv.org/abs/1509.05419}{\tt 1509.05419}]. 
 
\item \emph{Reconciling tensor and scalar observables in G-inflation}~\cite{Ramirez:2018dxe}\\
H.~Ram\'irez, S.~Passaglia, H.~Motohashi, W.~Hu and O.~Mena.\\  
 \href{http://iopscience.iop.org/article/10.1088/1475-7516/2018/04/039/meta}{\emph{JCAP}  {\bf 1804} (2018) no.04,  039}, [\href{https://arxiv.org/abs/1802.04290}{\tt 1802.04290}].
 
\item \emph{Inflation with mixed helicities and its observational imprint on CMB}~\cite{Heisenberg:2018erb}\\
L.~Heisenberg, H.~Ram\'irez and S.~Tsujikawa.\\
\href{https://journals.aps.org/prd/abstract/10.1103/PhysRevD.99.023505}{\emph{Phys.\ Rev.\ D} {\bf 99} (2019) no.2,  023505},
[\href{https://arxiv.org/abs/1812.03340}{\tt 1812.03340}].
  
\end{itemize}

\newpage

Other works not included in this thesis are:
\begin{itemize}

\item \emph{Primordial power spectrum features in phenomenological descriptions of inflation}~\cite{Gariazzo:2016blm}\\
S.~Gariazzo, O.~Mena, H.~Ram\'irez and L.~Boubekeur.\\  
 \href{https://www.sciencedirect.com/science/article/pii/S2212686417300390?via\%3Dihub}{\emph{Phys.\ Dark Univ.} {\bf 17} (2017) 38}, [\href{https://arxiv.org/abs/1606.00842}{\tt 1606.00842}]. 
 
\item \emph{Running of featureful primordial power spectra}~\cite{Gariazzo:2017akm}\\
S.~Gariazzo, O.~Mena, V.~Miralles, H.~Ram\'irez and L.~Boubekeur.\\
\href{https://journals.aps.org/prd/abstract/10.1103/PhysRevD.95.123534}{\emph{Phys.\ Rev.\ D} {\bf 95} (2017) no.12,  123534},  [\href{https://arxiv.org/abs/1701.08977}{\tt 1701.08977}].

\end{itemize}


\chapter*{Preface} 
\addcontentsline{toc}{chapter}{Preface} 

{\LARGE\bf T}he Universe is \emph{some} 13,800,000,000 years old. From its first moments to its current stage, the Universe has undergone several different thermodynamical processes that transformed and deformed it, going between hot and cold, and from dark to bright. Despite the immense complexity of these transformations, when we look at the sky, we can gain access to its \emph{story}. It is a story written in the stars and in the galaxies that we observe, and in the light coming from distant corners of the Universe. Thus, when we look at the sky, we are able to look into the past, and even into the \emph{far} past, back in time to what cosmologists call the \emph{early universe}. 
There, the \emph{tale} of Inflation is drafted---as an incomplete account of events that theoretical physicists are eager to fulfill. These events took place at \emph{some instant} during the first\footnote{There are 35 zeros there to the right of the decimal mark.} 0.000000000000000000000000000000000001 seconds after the \emph{big bang}. At that moment, the Universe is believed to have undergone an accelerated expansion and \emph{how was that so?} is the question we, theoretical physicists, wish to answer.


In this thesis we attempt to address a part of this fundamental question. We cover both phenomenological and theoretical approaches to the study of inflation: from model-independent parametrizations to modifications of the laws of gravity. It is divided in four parts. The first one, containing five chapters, consists of an introduction to the research carried out during the PhD: Chapter \S\ref{sec:Introduction} provides a short introduction to the standard cosmological model, in particular focusing on the epochs and the observables which motivate the need for the inflationary paradigm. 
In Chapter \S\ref{sec:inflation}, we review the dynamics of the canonical single-field inflationary scenario, showing that a scalar quantum field can produce an accelerated expansion of the Universe which effectively solves the problems identified in \S\ref{sec:Introduction}. 
Furthermore, we review the dynamics and the evolution of the primordial quantum fluctuations and their signatures on current observations. In Chapter \S\ref{sec:MIApp}, we discuss the Mukhanov parametrization, a model-independent approach to study the allowed parameter space of the canonical inflationary scenario. 
An alternative approach, using modified gravity, is proposed in Chapter \S\ref{sec:STtheories}. There, 
we review the construction of the most general scalar-tensor and scalar-vector-tensor theories of gravity yielding second-order equations of motion. Additionally, we discuss the main models of inflation developed within these frameworks. Finally, in Chapter \S\ref{sec:GSR}, we demonstrate new techniques that move beyond the slow-roll approximation to compute the inflationary observables more accurately, in both canonical and noncanonical scenarios. These chapters are complemented with detailed appendices on the cosmological perturbation theory and useful expressions for the main chapters.

The second Part is based on the most relevant peer-reviewed publications for this thesis. There, the reader can find the main results obtained during the Ph.D. Finally, in Part \ref{sec:summary} we summarize these results and draw our conclusions.



\tableofcontents
\hypersetup{linkcolor=cadmiumgreen}

\setlength{\parskip}{0.2em}

\mainmatter 
\renewcommand{\headrulewidth}{0.5pt} 
\cfoot{}

\part*{Part I\\[1cm] The Inflationary Universe\\[2cm] \normalfont \normalsize{ \emph{``The reader may well be surprised that scientists dare to study processes that took place so early in the history of the universe. On the basis of present observations, in a universe that is some 10 to 20 billion years old, cosmologists are claiming that they can extrapolate backward in time to learn the conditions in the universe just one second after the beginning! If cosmologists are so smart, you might ask, why can't they predict the weather? The answer, I would argue, is not that cosmologists are so smart, but that the early universe is much simpler than the weather!''}\\[1cm]
--- \raggedleft Alan H. Guth, The Inflationary Universe}
\addcontentsline{toc}{part}{{I~~ The Inflationary Universe}}
}\label{sec:intropart}\thispagestyle{empty}
\lhead[{\bfseries \thepage}]{ \rightmark}
\rhead[ Chapter \thechapter. \leftmark]{\bfseries \thepage}
\chapter{An introduction to $\Lambda$CDM}
\label{sec:Introduction}

At the beginning of the past century, the common belief was that the Universe we live in was static in nature, a space-time with infinite volume which would neither expand nor contract. When Albert Einstein was formulating the General Theory of Relativity (GR), during the second decade of the century, the equations he obtained would predict a scenario in which the Universe would collapse due to the gravitational force pulling on galaxies and clusters of galaxies. In order to counteract this effect, in 1917, Einstein introduced a \emph{cosmological constant}, $\Lambda$, into his equations, a term that induces a repulsive force, counterbalancing the attractive force of gravity, leading to a static universe. 

\begin{figure}[t]
\begin{center}
\includegraphics[keepaspectratio, width=11cm]{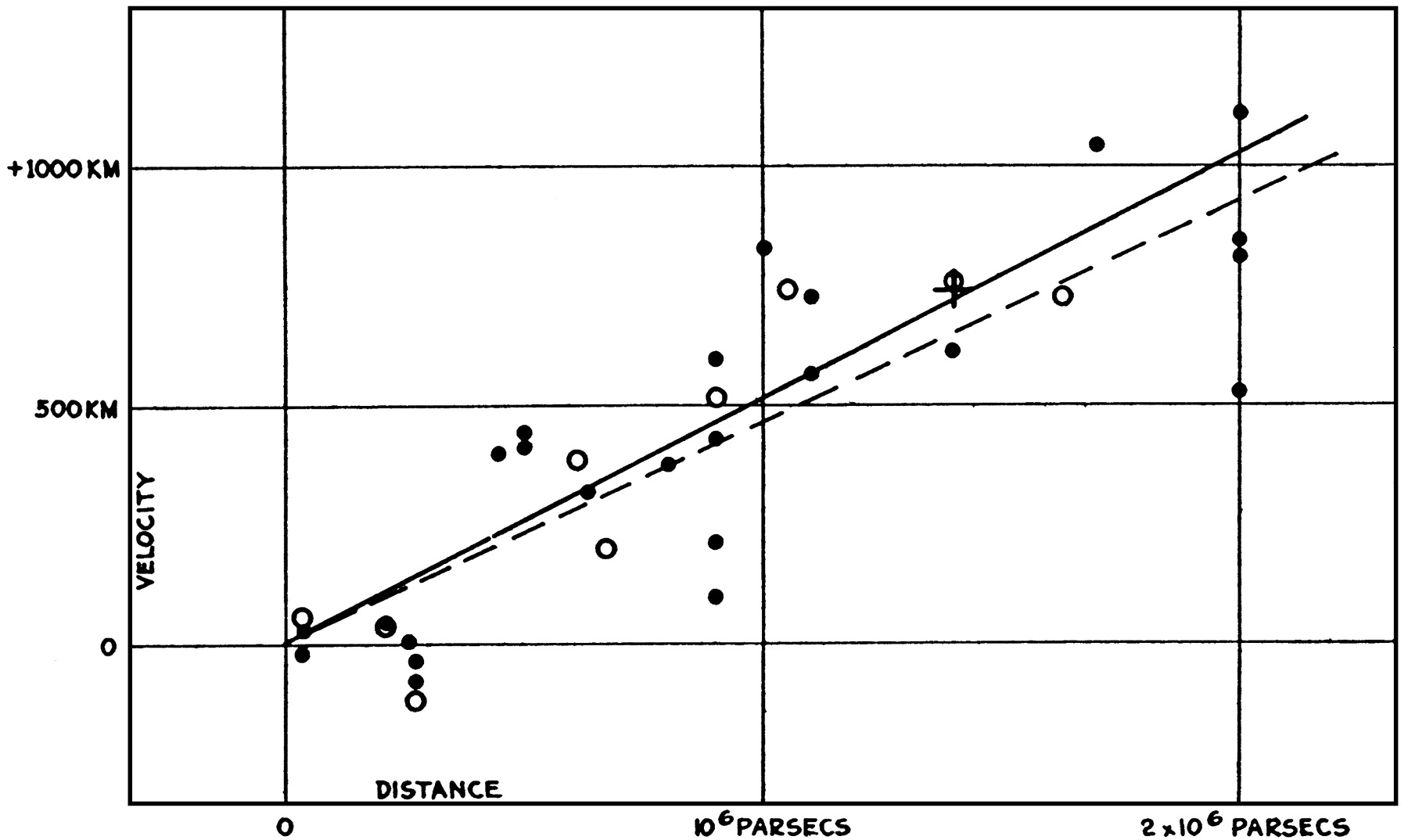}
\end{center}
\caption{\label{fig:Hubble} Hubble diagram: Velocity-distance relation among galaxies as observed by Edwin Hubble in 1929. The black circles and the solid line give the estimation for individual galaxies whereas white circles and the broken line give the estimation for combined galaxies into groups. The vertical axis is given in units of km/s whereas the horizontal axis is shown in parsecs (1 pc=$3.08\times10^{16}$ m). This plot is the original from Ref.~\cite{Hubble168}.}
\end{figure}

Soon after, and during the course of the last and current centuries, astronomers obtained an enormous amount of information about the origin and evolution of the Universe. First, in 1929, Edwin Hubble observed that galaxies were receding from us at a rate proportional to their distances~\cite{Hubble168} (see Fig.~\ref{fig:Hubble}). The \emph{Hubble law}---as it is now called---was then a clear evidence that the Universe was not only evolving but that it was dynamical! Einstein was forced to remove the cosmological constant from his equations in what he called his ``biggest blunder''.\footnote{I strongly suggest the reader Ref.~\cite{Ferreira2014} for an amazing exposition of the history of the General Theory of Relativity, from its developments to its latests cosmological consequences through the contributions of some of the greatest minds from the last century.}

After the groundbreaking observations made by E. Hubble on the expanding state of the Universe, the equations of GR still suggested that the Universe could come to a halt and eventually start to contract due to the effects of gravity; the question was \emph{when?} or, relatedly, how fast the most distant galaxies are receding from us. Unexpectedly however, further observations during the last decade of the past century made by the High-Z Supernova Search Team~\cite{Riess:1998cb} and, independently, by the Supernova Cosmology Project~\cite{Perlmutter:1998np}, revealed that the Universe was not decelerating, but all the contrary, galaxies are actually receeding from one another at an accelerated rate. Both teams looked at distant Supernovae whose (apparent) luminosity is well-known (this type of supernovae are called Type Ia). These supernovae are \emph{standard candles}: by measuring their flux and knowing their luminosity, we can determine the luminosity distance to these objects and compare to what we expect from the theory. Indeed, the luminosity distance is directly related to the expansion rate of the Universe and its energy content~\cite{Dodelson:2003ft,Amendola:2015ksp}. The two aforementioned independent groups observed that the Type Ia Supernovae were much fainter than what one would expect in a universe with only matter. Consequently, an additional ingredient was mandatory to make our Universe to expand in an accelerated way.

The accelerated nature of the expansion of the Universe has been confirmed by several experiments during the following years, however, its nature remains a mystery. The simplest explanation relies on an intrinsic source of energy of space itself 
which would act in the same way as the cosmological constant Einstein introduced 100 years ago. Even though the observed value for this vacuum energy density and the value computed from quantum field theory (QFT) calculations differ in many orders of magnitude, the cosmological constant $\Lambda$ is now a fundamental part of the standard model of Cosmology and it is referred to as the \emph{dark energy}.

This Chapter provides a brief introduction to the standard model of Cosmology---the so-called $\Lambda$CDM---by accounting for the evolution of the Universe from the big bang to the current observations of the late-time accelerated expansion. We shall then review the basic equations for the dynamics of an expanding universe and the main problems of the $\Lambda$CDM model, which indicate the strong need for an explanation of the initial conditions of the early universe.

\section{The expanding universe}

In an expanding universe, where each galaxy is receding from one another, one could perform the thought experiment of reversing the time flow. An expanding universe would become a collapsing one where all galaxies get closer and closer to each other. When we then look further back in time, we can see that all the matter and energy content fuse together in a very small and, hence, highly dense and energetic patch of space and time. At this point---dubbed as the hot \emph{big bang}---the equations of GR break down and a new formulation of gravity which includes the laws of quantum mechanics needs to be found. As we currently do not know the principles of such a theory, a given cosmological model must assume some initial conditions which otherwise should come up from a good quantum gravity candidate. As we shall see, these initial conditions need to account for the right amount of initial density perturbations as well as for the observed homogeneity and isotropy of the largest structures of the Universe. 

\begin{figure}[t]
\begin{center}
\includegraphics[keepaspectratio, width=11cm]{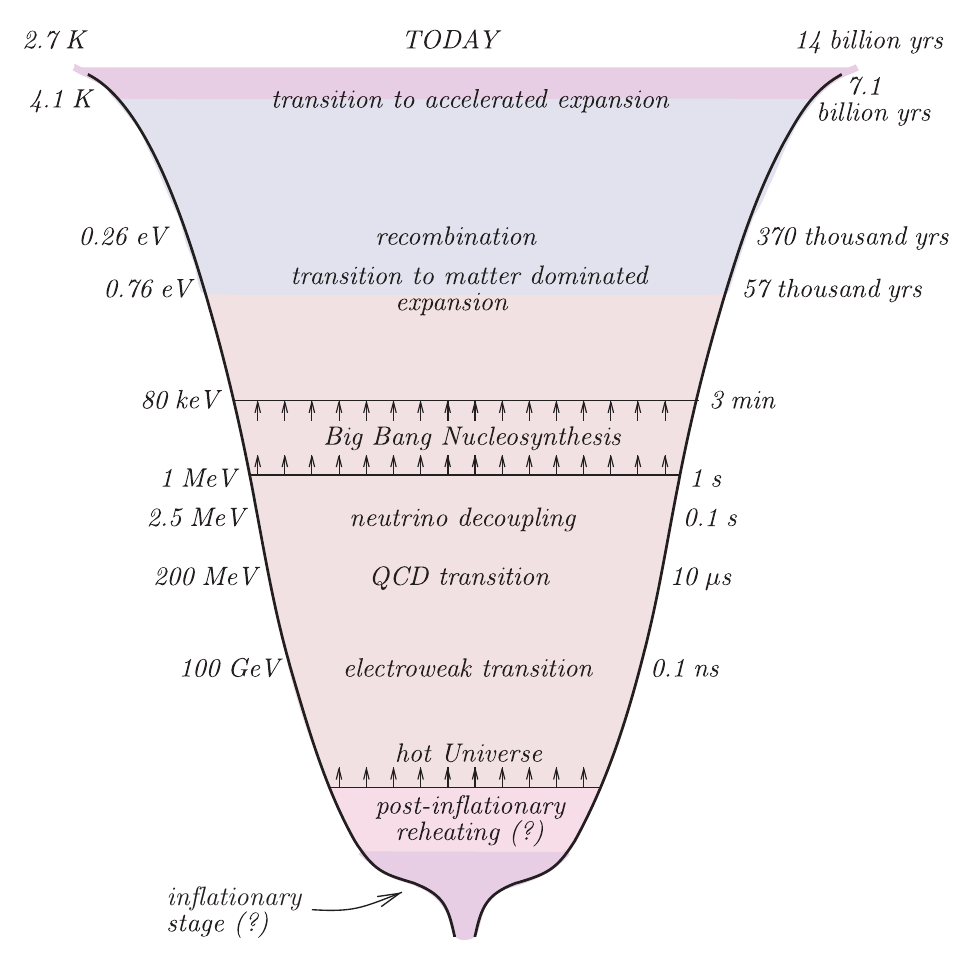}
\end{center}
\caption{\label{fig:Stages} Stages of the evolution of the Universe. Adapted from \emph{Introduction to the Theory of the Early Universe}~\cite{Gorbunov:2011zz} (page 20).}
\end{figure}

The Universe started to expand soon after the Big Bang, cooling down and following several proceses for a period of approximately 14 billion years\footnote{As in English: 1 billion = 1 thousand million.}---the current age of the Universe. During each of these proceses, the matter and energy content of the Universe went through different phases, each of which left imprints in different direct and indirect cosmological observations we measure nowadays. These indeed have helped us to uncover the history of the Universe we are about to briefly summarize~\cite{Kolb:1990vq,Dodelson:2003ft,Mukhanov:2005sc,Weinberg:2008zzc,Gorbunov:2011zz}. 
Figure~\ref{fig:Stages} shows a schematic summary of the different stages the Universe has gone through.

\subsection{Cosmological phase transitions}

As already pointed out, our starting point is the \emph{hot} big bang---we will see that the event previously described as the big bang is not the expected beginning of the Universe, but the residual of the \emph{inflationary} epoch. We call the hot `Big Bang' to the epoch where all the elementary particles, described in the Standard Model of Particle Physics~\cite{Weinberg:1995mt,Weinberg:1996kr,Patrignani:2016xqp}, were in thermal equilibrium---they were moving freely in the \emph{primordial plasma}---at energies of a few hundreds of GeV, approximately 10$^{15}$ degrees Kelvin.\footnote{1 GeV=$1.16\times10^{13}$ K. Given this equivalence, we shall sometimes refer to a given temperature in eV units.}

As the Universe started to cool down, it experimented phase transitions characterized by the change in the nature of the cosmic fluid. The first one resulted in the spontaneous breaking of the \emph{electroweak} (EW) symmetry~\cite{Kirzhnits:1972ut,Dolan:1973qd,Weinberg:1974hy}:\footnote{Let us emphasize that there is a reasonable expectation for a \emph{Grand Unification} epoch, where the QCD and the EW interactions are unified into a single force. Therefore the first phase transition would be at the energy-scale of the Grand Unified Theories (GUT) corresponding to temperatures of around $T\sim10^{16}~K$. However, even though the idea was proposed in 1974~\cite{Georgi:1974aa}, there are no experimental hints yet that confirm the theory and, furthermore, we will see that inflation is expected to take place at slightly lower energies. Therefore we will ignore the hypothesis of the GUT epoch in this thesis.} at energies above approximately 100 GeV---the energy-scale of the EW interaction---the EW $SU(2)\otimes U(1)$ gauge symmetry remained unbroken and, consequently, particles in the primordial fluid were massless. Once the temperature dropped, the Higgs field acquired a nonzero vacuum expectation value (vev) which, in turn, breaks the EW symmetry down to the $U(1)$ gauge electromagnetic group. The interaction of particles with the Higgs field provides them with mass (except for the photon which belongs to the unbroken $U(1)$ group)~\cite{Englert:1964et, Higgs:1964pj}. As a result, the new massive particles, as the $W^\pm$ and $Z$ gauge bosons, mediate only short-distance interactions.

Another phase transition, the QCD---\emph{Quantum Chromodynamics}---tran-\\sition, occurred at energies around $\Lambda_\text{QCD}\sim200$ MeV. The QCD theory describes the strong force between quarks and gluons, which are subject to an internal charge called \emph{colour}~\cite{Fritzsch:1973pi,Quigg:1983gw,Pich:1999yz}. The strong force has the peculiar characteristic of being weaker at shorter (rather than at larger) distances as opposed to the well-known electromagnetic force. This distinctive feature, called \emph{asymptotic freedom}~\cite{Gross:1973id,Politzer:1973fx}, allows the fluid of quarks and gluons to interact only weakly above this energy scale. Once the energy drops below $\Lambda_\text{QCD}$, quarks and gluons get confined into colourless states, called `hadrons', of regions with size of $\Lambda^{-1}_\text{QCD}\simeq10^{-15}$ m. Consequently, isolated quarks cannot exist below the confinement energy scale.

\subsection{Neutrino decoupling}

Neutrinos are weakly interacting particles. As such, they stopped interacting soon in the early universe, exactly when their interaction rate falls below the rate of the expansion of the Universe,
at an approximate temperature of 2-3 MeV~\cite{Kolb:1990vq,Gorbunov:2011zz,Lesgourgues:2006nd}.
Below this temperature, these relic neutrinos can travel freely through the Universe as they do today. Their temperature and number density are indeed of the same order as the measured relic photons that we shall describe later. However, although direct detection of the relic neutrinos is an extremely difficult task given their feeble interaction with matter~\cite{Betts:2013uya}, their energy density plays an important role on the Universe's evolution \cite{Bashinsky:2004aa,Follin:2015aa,Baumann_2019} and thus we are confident of their existence.

\subsection{Big Bang Nucleosynthesis}

Light elements form when freely streaming neutrons bind together with protons into nuclei. These processes happened at energies of a few MeV, corresponding to the binding energy of nuclei and, as a consequence, there was a production of hydrogen and helium-4, in large amounts, and deuterium, helium-3 and lithium-7 in smaller abundances.\footnote{Heavier elements need higher densities to form. Carbon and other elements synthesized from it, are the result of thermonuclear reactions in stars once after they have burned out their concentrations of hydrogen and helium.}

The calculation of the amount of light elements produced during this epoch requires the physics of the previous phase transitions---namely nuclear physics and weak interactions---as well as the use of the equations of GR~\cite{Alpher:1948ve,Wagoner:1966pv}. Consequently, the measurement of the primordial abundances of such elements and its agreement with the Big Bang Nucleosynthesis (BBN) theory is one of the greatest achievements of the $\Lambda$CDM model. This, furthermore, makes the BBN epoch the earliest epoch probed with observations~\cite{Patrignani:2016xqp} (see however \cite{Fields:2011zzb,Cyburt:2015mya} for a discussion on the controversial observed amount of Lithium and the theoretical expectations).

\subsection{Recombination}

We have reached an epoch where the constituents of the primordial fluid were nuclei, electrons and photons. During BBN, the photons were still energetic enough to excite electrons out of atoms. However, once the temperature of the Universe drops at energies around 0.26 eV ($\sim$3000 K), electrons are finally trapped by the nuclei, forming the first stable atoms. This made the remnant of the primordial fluid to become a neutral gas made mostly of hydrogen~\cite{Peebles:1968ja,Zeldovich:1969en}.


It is at this point where a crucial event takes place: 
photons stopped being actively scattered by the electrons and were able to propagate freely through the Universe, forming a relic radiation which has been freely propagating since then. This radiation is in fact the first light of the Universe and, furthermore, it can be measured today with antennas and satellites as some type of noise coming from all parts of the sky. This photon radiation is the so-called \emph{Cosmic Microwave Background} (CMB) and, as we will see, it plays a crucial role in the understanding of the inflationary epoch because it contains information about the primordial density perturbations and also about the degree of homogeneity and isotropy present during the recombination epoch.

\subsection{The Cosmic Microwave Background}
\label{sec:CMBintro}

The energy spectrum of the CMB, as measured today, is precisely that of a black body~\cite{Fixsen:1996nj} with a mean temperature of $T_0=2.726\pm0.001$ K~\cite{Fixsen:2009ug}. It was first detected in 1965 by Arno Penzias and Robert Wilson using their antenna from Bell Laboratories~\cite{Penzias:1965wn}. Once they ruled out any known source of noise, Dicke, Peebles, Roll and Wilkinson reported, in the same year, that the source of this radiation could be attributed to the relic photons that decoupled at the recombiation era~\cite{Dicke:1965zz}.

\begin{figure}[t]
\begin{center}
\includegraphics[keepaspectratio, width=11cm]{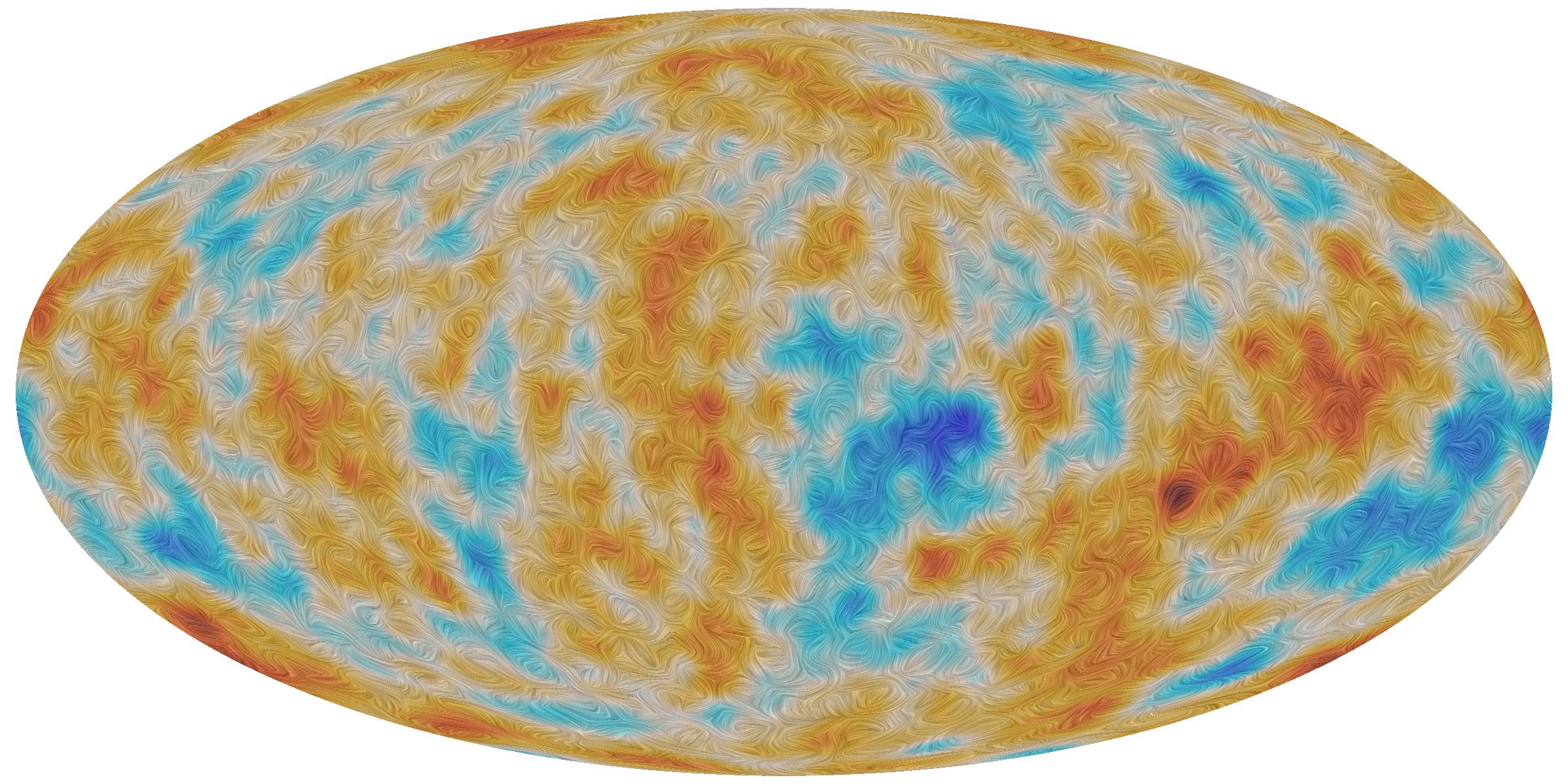}
\end{center}
\caption{\label{fig:CMB} Temperature anisotropies and polarization in the Cosmic Microwave Background. Variations in color indicate variations in temperature: the bluer (redder) regions correspond to colder (hotter) temperatures. On the other hand, the texture pattern represents the direction of polarized light. The illustration shows the anisotropies at an angular resolution of 5$^\circ$, however, the \emph{Planck} satellite has reached a resolution as accurate as $\sim0.16^\circ$~\cite{Aghanim:2018eyx}. \emph{Image Credit:} European Space Agency (ESA) and \emph{Planck} Collaboration.}
\end{figure}

The CMB spectrum with mean $T_0$ temperature is not, however, perfectly isotropic. There are small variations in temperature across the celestial sphere. A map of the CMB is shown in Fig.~\ref{fig:CMB} where the changes around the mean temperature---quantified by the differences in color---manifest as anisotropies across the angular scales observed in the sky. These anisotropies are of order of 10$^{-5}$ and are consequence of the slight difference in density across the particle fluid at the time of recombination. Therefore, the CMB is indeed a map of the Universe when it was about 380000 years old.

The differences in temperature across the sphere can be conveniently expanded in spherical harmonics as
\be \f{\delta T(\hat{n})}{T_0}=\sum_{\ell=1}^\infty\sum_{m=-\ell}^{m=\ell}a_{\ell m}Y_{\ell m}(\hat{n})~, \ee
where $\delta T(\hat{n})\equiv T(\hat{n})-T_0$ quantifies the deviation between the temperature $T(\hat{n})$ coming from the direction $\hat{n}$ and the mean temperature $T_0$. The coefficients $a_{lm}$ are themselves related to the amplitud of temperature fluctuations, whereas their ensemble average $\langle a_{\ell m}\rangle$ contains all the statistical information about an average of universes like ours.

One important measurement of the CMB is that the primordial density perturbations must have been close to Gaussian. Given that the $a_{lm}$ coefficients are linear functions of the primordial perturbations, then they are also Gaussian random variables. Hence, the spectrum $C_l$ of the two-point correlation function $\langle a_{\ell m}a_{\ell'm'}^*\rangle$ completely determines the CMB anisotropies.

Furthermore, as we have only one universe to experiment with, the ensemble average can be translated to an average over the single sky we can observe. For higher multipoles $\ell$, with a large number of different values for $m=-\ell,...,\ell$, this is a good approximation and indeed observations are consistent with the Gaussian hypothesis. For lower multipoles, however, the statistical analyses are limited by the cosmic variance. Specifically, the spectrum is defined as
\be C_\ell^{TT}\equiv\f{1}{2\ell+1}\sum_{m=-\ell}^{m=\ell}\langle a_{\ell m}a^{*}_{\ell m}\rangle~, \ee
where the statistical error is $1/\sqrt{\ell+1/2}$~, which is clearly larger for a smaller value of $\ell$.

Another important type of information contained in the CMB spectrum is its polarization. Figure~\ref{fig:CMB} also shows the pattern of polarized light measured in the CMB. The photons decoupled during the recombination era come with polarization states due to the Thompson scattering they experimented before decoupling~\cite{Rees_1968,Negroponte:1980aa,Bond_1984}; however, their polarization can be further affected during their subsequent travel by scattering with free electrons during the reionization era\footnote{At late times, star formation processes lead to a \emph{re}ionization period in the Universe. CMB photons can therefore interact with the new free electrons, changing their polarization.} or by lensing effects due to massive structures.\footnote{Massive structures bend the light that travels close to them. On one hand, stars, galaxies and galaxy clusters can act as enormous lenses for distant light passing through them, deforming it into \emph{Einstein rings}~\cite{Weinberg:1972kfs}. On the other hand, light rays traveling long distances during the early universe are also affected by mass sources surrounding their path but in a smaller amount. The statistical account for this effect is commonly known as \emph{weak lensing} and it can also modify the polarization state of the CMB photons.}

As for the temperature anisotropies, we can define two different scalar quantities of polarization in terms of the polarization factors $a^E_{lm}$ and $a^B_{lm}$ as
\be E(\hat{n})=\sum_{\ell=1}^\infty\sum_{m=-\ell}^{m=\ell}a^E_{\ell m}Y_{\ell m}(\hat{n})~,\qquad B(\hat{n})=\sum_{\ell=1}^\infty\sum_{m=-\ell}^{m=\ell}a^B_{\ell m}Y_{\ell m}(\hat{n})~. \ee
With these two different types of polarization, we can now define three different types of correlations---$TT$, $EE$ and $BB$---plus three cross-correlations---$TE$, $TB$ and $EB$,--- however, the last two vanish due to symmetry under parity~\cite{Dodelson:2003ft,Gorbunov:2011zz}. 

Measurements of the CMB can then determine the spectra $C_\ell^{TT}$, $C_\ell^{TE}$, $C_\ell^{EE}$ and $C_\ell^{BB}$. The shape shown in Fig.~\ref{fig:CMB} is characteristic of the $E$-mode polarization, the predominant type of polarization observed, whereas measurements of the $B$-mode polarization have only placed upper bounds on the $BB$ spectrum. The $B$-mode polarization on degree scales is produced by tensor modes present during inflation, thereby a measurement of this type of polarization would extremely help to understand the physics of inflation (see, \emph{e.g.}, Refs.~\cite{Kamionkowski:1996zd,Seljak:1996gy,Ade:2017uvt,Gorbunov:2011zzc}).

\subsection{Structure formation}
\label{sec:lss}

The starting point of structure formation is the assumption of initial regions of overdensities. During the epoch of radiation domination (before recombination), the amplitude of the density perturbations was small. However, at some point, the Universe becomes matter dominated and then matter starts to get trapped into overdensed regions due to the gravitational potentials.

The way galaxies and clusters of galaxies are currently distributed in space depends crucially on the primordial overdensity. The existence of these initial overdensities is indeed assumed, in the same way as the initial homogeneity and isotropy, as no mechanism within the $\Lambda$CDM model is able to produce it. We will see later that inflation, in fact, is exactly a mechanism that provides us with these initial perturbations, with predictions that are amazingly consistent with the data.

Furthermore, the theory of structure formation gives strong hints for the existence of an unknown type of matter which does not have electromagnetic interaction, \ie does not emit light. This \emph{dark matter} is indeed needed to understand the rotation curves of galaxies and to account for the rate of formation of the structures: without dark matter, structures would not have been formed yet! Consequently, the dark matter must be non-relativistic---it must cluster---and therefore it is said that dark matter is \emph{cold}. Current observations show that the dark matter accounts for the 85\% of the matter content in the Universe and therefore it is a key element in the development of the $\Lambda$CDM (\emph{lambda-cold dark matter}) model, together with the dark energy component.\footnote{We shall not further discuss the nature of dark matter as it is not the main topic of this work, see however Refs.~\cite{Bertone:2005aa,Bergstrom:2012aa,Kusenko:2013aa} for reviews on the subject.}

\begin{figure}[t]
\begin{center}
\includegraphics[keepaspectratio, width=11cm]{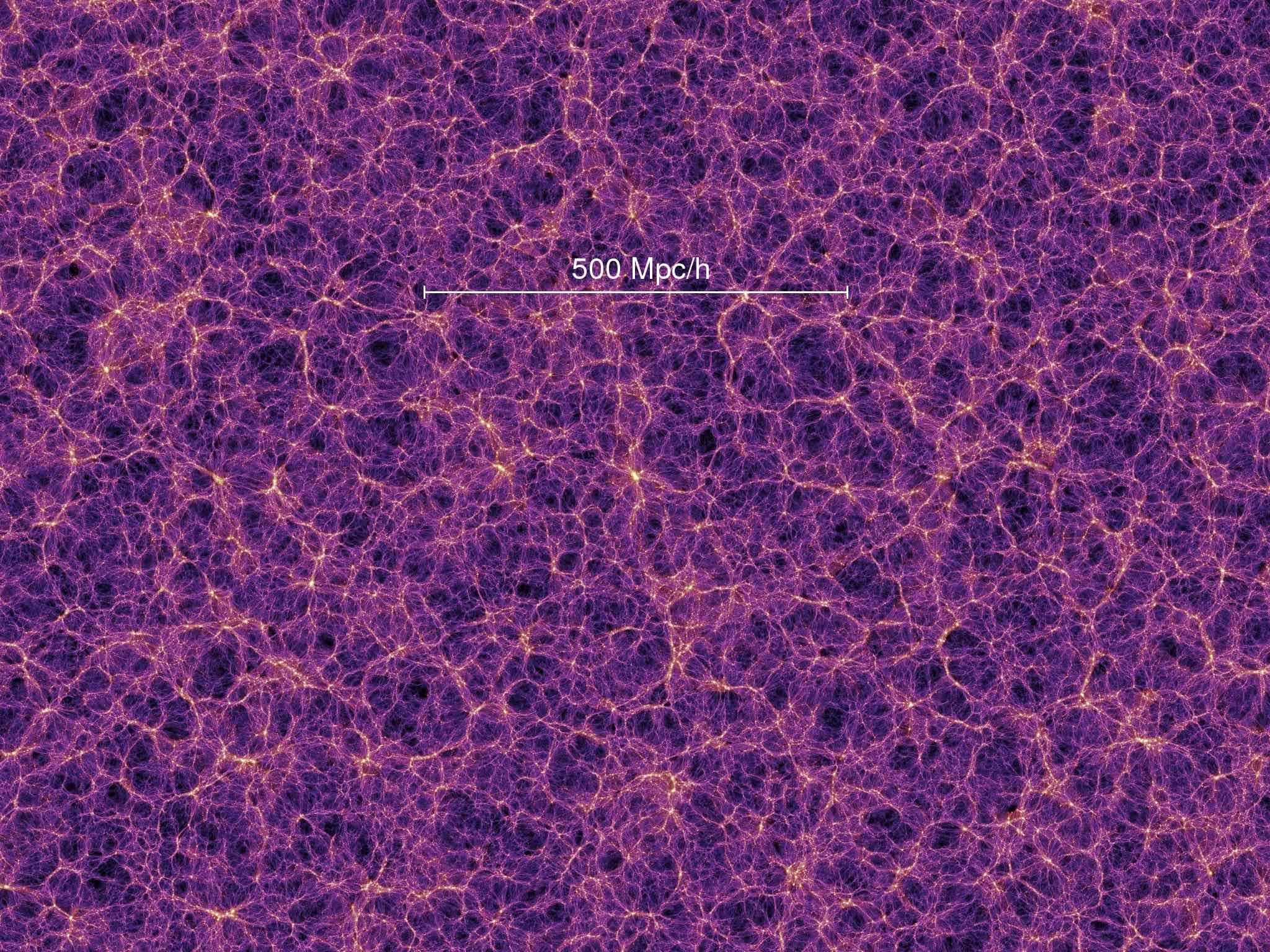}
\end{center}
\caption{N-body simulation of the dark matter density distribution at $t=13.6$ Gyr (today) using the $\Lambda$CDM model~\cite{Springel:2005nw}. It is shown the scale distance of 500 Mpc/h (see \S\ref{sec:dynofexpuni} for details) above which the distribution of matter is clearly homogeneous and isotropic as assumed by the $\Lambda$CDM model.}
\label{fig:lss} 
\end{figure}

As the evolution of the structure formation links the current state of the large structures with the initial conditions of the early universe, the observations of the \emph{Large Scale Structure} (LSS) and their statistical signatures have the power of constraining inflation apart from those from the CMB. The first important observation we note is that the Universe, as already stated, is highly homogeneous, \emph{i.e.}, at relatively large scales, it looks the same wherever we look. Figure~\ref{fig:lss} is an example of this fact: it shows the N-body simulation of 10$^{10}$ particles of a dark matter field evolved following the $\Lambda$CDM model~\cite{Springel:2005nw}.

\section{Dynamics of an expanding universe}
\label{sec:dynofexpuni}

So far we have briefly reviewed the evolution of the Universe which is consistent with observations. It can be summarized as a primordial fluid made by elementary particles filling the spacetime. Across this fluid, there must have existed density perturbations in order to lead to structure formation processes due to the gravitational potential wells. As the Universe expanded, this fluid cooled down experiencing several processes which left their imprint both indirectly and directly in the CMB photons and in the structures we measure today. From observations of these two, we can infer the required level of homogeneity and anisotropy the primordial fluid should have had. Let us now set the mathematical grounds upon which the theory is built (see Refs.~\cite{Dodelson:2003ft,Mukhanov:2005sc,Weinberg:2008zzc,Gorbunov:2011zz,Baumann:2009ds,Baumann:2018muz} for comprehensive studies in the literature).

\subsection{Geometry}

The geometry of an expanding homogenous and isotropic universe is simply described by the Friedmann-Lema\^itre-Robertson-Walker metric (FLRW)
\be \ba \dif s^2=&g_{\mu\nu}\dif x^\mu \dif x^\nu\\ =&-\dif t^2+a^2(t)g_{ij}\dif x^i\dif x^j~, \ea \label{eq:flrwm} \ee
where $g_{ij}$ is the metric of a unit 3-sphere given by
\be \dif l^2=\dif\chi^2+\Phi(\chi^2)\mk{\dif\theta^2+\sin^2\theta\dif\phi^2}~. \ee
Depending on the spatial curvature of the universe, the value of $\Phi(\chi^2)$ is given by
\be \label{eq:curv}
\Phi(\chi^2) \equiv  \left\{
\begin{array}{c} \sinh^2 \chi \\ \chi^2 \\ \sin^2 \chi \\
\end{array} \right. \quad \begin{array}{l} k=-1 \\ k=0 \\ k=+1 \end{array}~,
\ee
where the curvature parameter $k$ is +1, 0 and -1 for a positive-curvature, flat and negative-curvature universe respectively. 

The function $a(t)$, called \emph{scale factor}, grows with time and thus characterizes the distance between two distant objects in space at a given time. We can therefore define the rate of cosmological expansion characterized by the change of the scale factor in time as\footnote{Here and throughout this thesis, dots imply derivatives with respect to cosmic time $t$.}
\be H(t)=\f{\dot a(t)}{a(t)}~, \ee
which is another function of time, and is called the \emph{Hubble rate}. The present value of the Hubble parameter, denoted by $H_0$, is currently being constrained by the \emph{Planck} satellite. Its measured value is $H_0=(67.27\pm0.6)$ km s$^{-1}$Mpc$^{-1}=h\cdot100$ km s$^{-1}$ Mpc$^{-1}$.\footnote{A megaparsec (Mpc) is a standard cosmological unit of length given by $1~\text{Mpc}=3.1\times10^{24}~\text{cm}$. Also, $h\simeq0.66$ is a dimensionless parameter sometimes used to parametrize the value of $H_0$ (as in Fig.~\ref{fig:lss}).} However, local estimates from distance ladders find a value of $H_0=(73.8\pm2.4)$ km s$^{-1}$Mpc$^{-1}$, showing a discrepancy of around 3.5$\sigma$ level (see \cite{Aghanim:2018eyx} for details).

To understand the value of the intrinsic curvature, \ie the value of $k$ in Eq.~\eqref{eq:curv}, we again assume a homogeneous and isotropic universe filled with a perfect fluid (\ie with vanishing viscous shear and vanishing heat flux) characterized only by an energy density $\rho$ and an isotropic pressure $p$. With these ingredients we can define the ratio of energy density relative to the \emph{critical} one, $\rho_c$,\footnote{Where $\rho_c$, the energy density of an exactly flat spacetime, is to be carefully defined in \S\ref{sec:evolution}.} as $\Omega=\rho/\rho_c$, and the equation of state as $\omega\equiv p/\rho$. The curvature parameter is related to $\Omega$ as 
\be 1-\Omega=-\f{k}{\mk{aH}^2}~. \label{eq:flatness} \ee
Therefore the intrinsic curvature of the Universe today depends on its total energy density. Current CMB, LSS and BAO\footnote{Baryon acoustic oscillations (BAO) are pressure waves in the coupled baryon-photon fluid, similar to sound waves, which had visible effects on the CMB and LSS spectra~\cite{Peebles:1994xt,Aubourg:2014yra}.} 
combined observations~\cite{Aghanim:2018eyx} estimate a present value of $\Omega_k\equiv1-\Omega_0=0.0007\pm0.0019$ at 68\% confidence level, implying that, to a very good approximation, we are living in a flat universe ($k=0$).

\begin{figure}[t]
\begin{tabular}{cc}
\includegraphics[keepaspectratio, width=6.8cm]{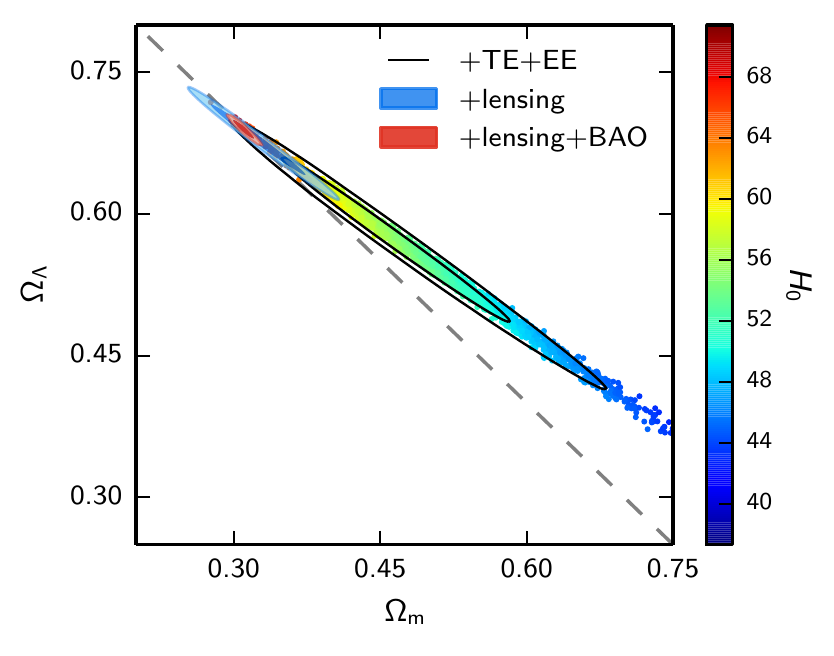} & \includegraphics[keepaspectratio, width=7cm]{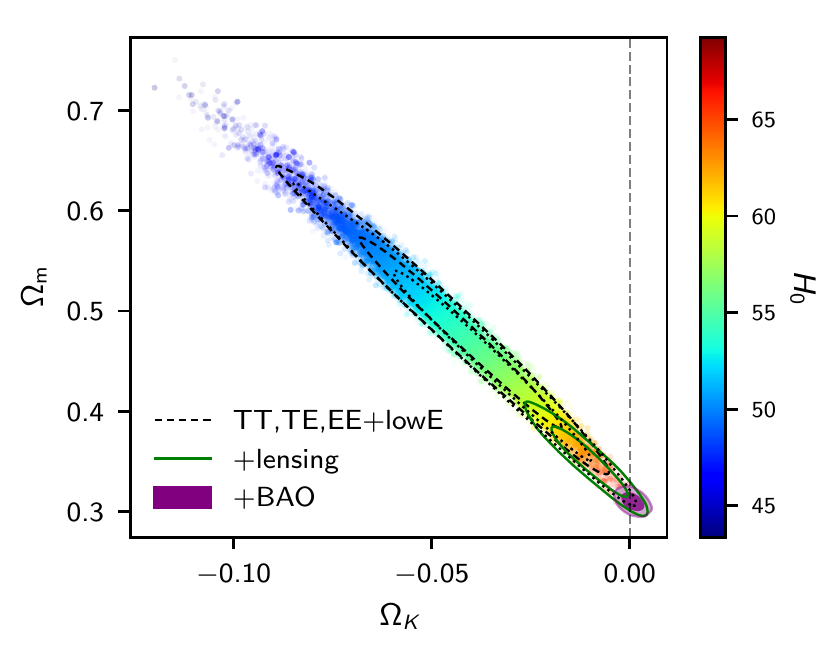}\\
\end{tabular}
\caption{\emph{Left: Planck} 2015 constraints in the $\Omega_m-\Omega_\Lambda$ plane~\cite{Ade:2015xua}. \emph{Right: Planck} 2018 constraints in the $\Omega_k-\Omega_m$ plane~\cite{Aghanim:2018eyx}. Both constraints are color-coded by the measurement of $H_0$ and are obtained by using CMB (TT, TE and EE), LSS (weak lensing) and BAO observations.}
\label{fig:omega}
\end{figure}

In the same way, we can define a ratio for both the total matter content, $\Omega_m$, and the contribution due to the dark energy, $\Omega_\Lambda$, the sum of which equals the total energy content of the Universe. Figure~\ref{fig:omega} shows the current constraints on the three ratios ($\Omega_k=1-\Omega_m-\Omega_\Lambda$) using CMB, LSS and BAO observations. We see that around the 70\% of the Universe is filled with the mysterious dark energy. 

\subsection{Evolution}
\label{sec:evolution}

The evolution of the Universe is governed by the Einstein's field equations of General Relativity written as
\be
R_{\mu\nu}-\frac12g_{\mu\nu}R=8\pi GT_{\mu\nu}~,
\label{eq:EGR}
\ee
where $G$ is Newton's gravitational constant, $R=g^{\mu\nu}R_{\mu\nu}$ is the scalar curvature, and the Ricci tensor $R_{\mu\nu}$ is defined in terms of the Christoffel symbols as
\be
R_{\mu\nu}=\Gamma^\lambda_{\mu\nu,\lambda}-\Gamma^\lambda_{\nu\lambda,\mu}+\Gamma^\lambda_{\mu\nu}\Gamma^\sigma_{\lambda\sigma}-\Gamma^\lambda_{\mu\sigma}\Gamma^\sigma_{\lambda\nu}~.
\ee
Here and throughout this thesis, commas denote partial derivatives $_{,\alpha}\equiv\partial/\partial x^\alpha$. The symbols themselves are affine connections defined in GR as
\be \Gamma_{\mu\nu}^\lambda=\frac12g^{\lambda\sigma}\mk{g_{\nu\sigma,\mu}+g_{\mu\sigma,\nu}-g_{\mu\nu,\sigma}}~. \ee

The energy-momentum tensor $T_{\mu\nu}$, in Eq.~\eqref{eq:EGR}, reads as
\be T^\mu_\nu=(\rho+p)u^\mu u_\nu-p\delta^\mu_\nu~, \label{eq:tmunudef} \ee
for a perfect, homogeneous and isotropic, and in a local reference frame fluid, where $u^\mu$ is its 4-velocity satisfying the condition $g_{\mu\nu} u^\mu u^\nu=-1$. In cosmology one usually chooses a reference frame which is comoving with the fluid. In this case, $u^\mu=(1,0,0,0)$ and then the energy-momentum tensor can be written as a diagonal matrix $T_\nu^\mu=$\emph{diag}$\mk{\rho,-p,-p,-p}$. Furthermore, the energy-momentum tensor is conserved, \ie
\be T^\mu_{\nu;\mu}=0~, \label{eq:cont1} \ee
where semicolons denote covariant derivatives $_{;\mu}\equiv\nabla_\mu T^\mu_\nu=T^\mu_{\nu,\mu}+\Gamma^\mu_{\mu\sigma}T^\sigma_\nu-\Gamma_{\mu\nu}^\sigma T^\mu_\sigma$. Equation~\eqref{eq:cont1} 
leads to the continuity equation
\be \diff{\rho}{t}+3\f{\dot a}{a}\mk{\rho+p}=0~, \qquad \leftrightarrow \qquad \diff{\ln\rho}{\ln a}=-3\mk{1+\omega}~, \label{eq:cont2} \ee
where we used the definition of the equation of state $\omega\equiv p/\rho$.

One needs to compute all the components of Eq.~\eqref{eq:EGR} considering the FLRW spacetime by means of the metric given by Eq.~\eqref{eq:flrwm}. The 00-component of the Einstein equations relates the rate of cosmological expansion given by $H$ to the total energy density as\footnote{Here and from now on, we will work in units given by $\Mpl=(8\pi G)^{-1/2}=1$, where $\Mpl$ is the Planck mass scale.}
\be \label{eq:fried}
\mk{\f{\dot a}{a}}^2=\f{1}{3}\rho-\f{k}{a^2}~.
\ee
This is called the first Friedmann equation. Notice that for a flat ($k=0$) universe, the energy density reads as $\rho_c=3H^2$ which we had defined before as the critical density, and therefore Eq.~\eqref{eq:fried} can be written as Eq.~\eqref{eq:flatness}.

Taking the derivative of Eq.~\eqref{eq:fried} and using the continuity equation~\eqref{eq:cont2}, one obtains the second Friedmann equation:
\be \label{eq:fried2}
\f{\ddot a}{a}=-\f{1}{6}\mk{\rho+3p}~,
\ee
which gives the acceleration of the scale factor in terms of $\rho$ and $p$.

The continuity equations \eqref{eq:cont2} can also be integrated for $\omega=const.$ to find the behavior of the total energy density as 
\be \rho\propto a^{-3\mk{1+\omega}}~, \ee
and thus, by plugging it into Eq.~\eqref{eq:fried2}, we could find the behavior of the scale factor for a universe dominated for different components (depending on the value of the equation of state $\omega$):
\be \label{eq:aastau}
a(t)\propto\left\{
\begin{array}{c} t^{2/3\mk{1+\omega}}\\ e^{Ht}\\\end{array} \right. 
\quad \begin{array}{l} \omega\neq-1\\ \omega=-1 \end{array}~.
\ee
Notice that an equation of state given by $\omega=p/\rho=-1$ implies that the universe is filled with a fluid with negative pressure. This is exactly the case of a universe dominated by a cosmological constant or by a scalar field driving an accelerated expansion.

\subsection{Horizons}
\begin{figure}[t]
\begin{center}
\includegraphics[keepaspectratio, width=13cm]{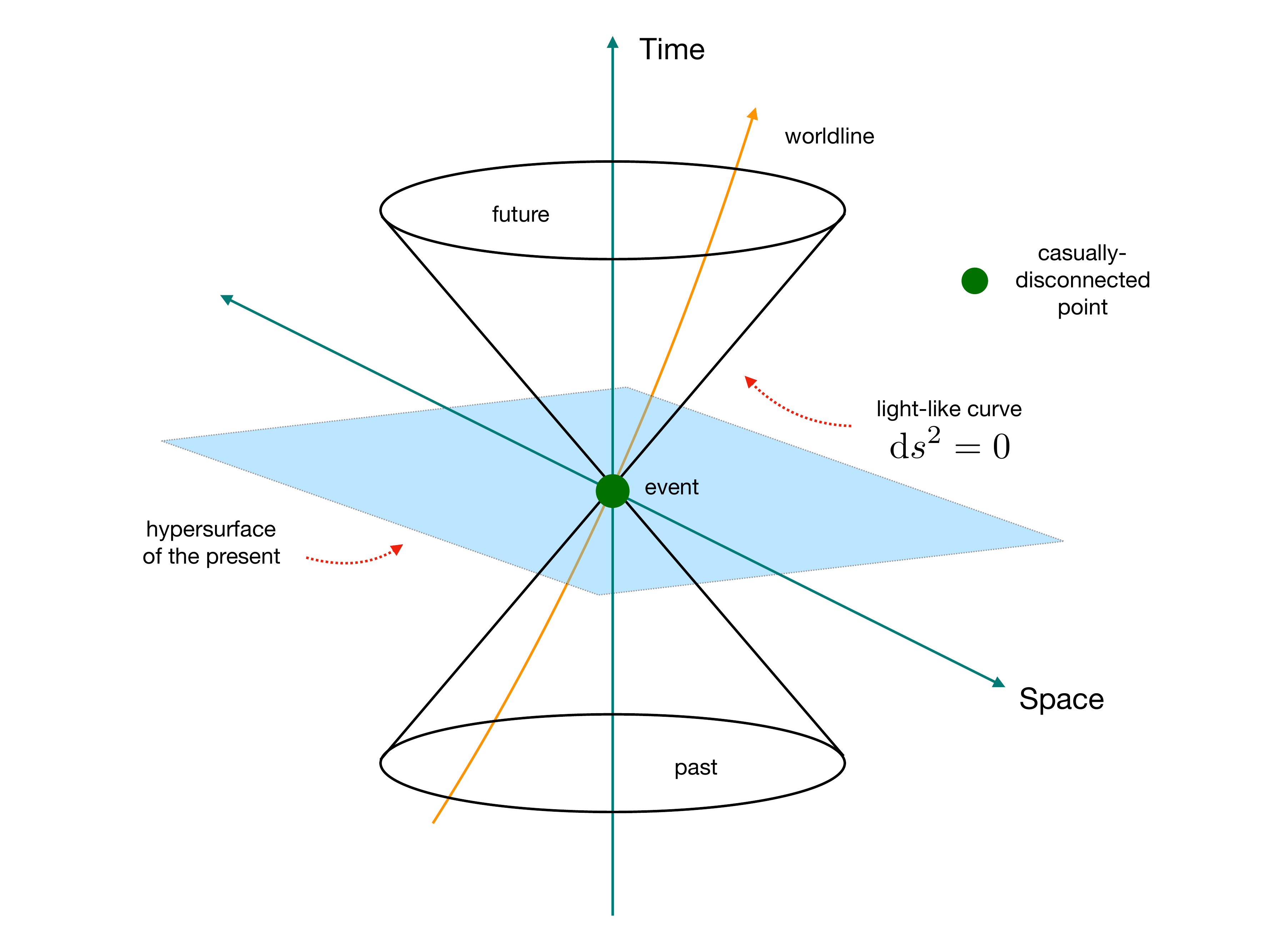}
\end{center}
\caption{\label{fig:lightcone} Light cone. The information coming from an event produced at a given point in spacetime can only travel with finite speed in time-like worldlines. Light-like curves then enclose all the regions that are and will be causally-connected to that event.} 
\end{figure}

Information across space can only travel with finite speed, as stated by the Special Theory of Relativity. This defines the causal structure of the Universe: an event originated at some point in spacetime will propagate with a speed which cannot surpass the speed of light. Photons, for instance, ---traveling at the speed of light---follow null (light-like) geodesics obeying $\dif s^2=0$. To better understand the consequences of this simple fact, we define a standard function of time, called \emph{conformal time} $\tau$, given by
\be \dif\tau=\f{\dif t}{a(t)}~. \label{eq:conftime} \ee
In terms of $\tau$, the FLRW line element, Eq.~\eqref{eq:flrwm}, with the spatially flat metric $g_{ij}=\delta_{ij}$, can be written as
\be \dif s^2=a^2(\tau)\mk{-\dif\tau^2+\delta_{ij}\dif x^i\dif x^j}~, \ee
\ie a static Minkowski metric ($g_{\mu\nu}^\text{Mink}=diag\mk{-1,1,1,1}$) rescaled by $a(\tau)$. It is simple to see then that null geodesics are described by straight lines of 45$^\circ$:
\be |\dif \vec{x}|=\dif\tau~. \label{eq:nullgeo} \ee
Figure \ref{fig:lightcone} sketches causally connected and disconnected regions of spacetime: null geodesics given by $\dif s^2=0$ enclose regions causally connected to a given event in a \emph{light cone}; regions outside the light cone do not have access to the event. The light cone grows with time, \ie causally disconnected regions will be reached by the cone at some future time.

Imagine then a photon emitted during the Big Bang; there is a finite physical distance this photon has traveled since then given by
\be d_H(t)=a(t)\tau(t)~. \ee
This distance, in fact, defines the radius of a sphere called \emph{cosmological horizon} or \emph{comoving particle horizon} which, for an observer at present time, represents the size of the observable universe.

Now imagine an observer lying at some position $\vec{x}=0$. For this observer, there will be a future event which will never reach her. For an arbitrary future time, Eq.~\eqref{eq:nullgeo} reads
\be \tau(t\to\infty)-\tau(t)=\int_t^\infty\f{\dif t'}{a(t')}~, \ee
which allows us to define the physical size
\be d_e=a(t)\int_t^\infty\f{\dif t'}{a(t')}~=\frac1{H_e}. \ee
This result implies that such an observer will never know about an event that happens at a distance larger than $d_e$. This distance is called the \emph{event horizon}.

As we shall see, the event horizon allows us to understand how, without an accelerated expansion during the early universe, most of the observable cosmological scales would have never been in causal contact, which is the core of the $\Lambda$CDM problems we are about to discuss.

\section{Problems of the standard cosmological model}

The $\Lambda$CDM model just described, consisting on different phases, each driven by very different physical processes, is able to explain with incredible accuracy a large amount of direct and indirect observations. However, as we have already stated several times, it does not provide neither the initial conditions for the primordial fluid in the very early universe---its assumed homogeneity and isotropy---nor the required density perturbations which are the seeds for the structures we observe today in our Universe; these ingredients are just assumed to be there.

On the one hand, it is indeed a puzzle the homogeneity observed in the Universe. Take for instance the CMB anisotropies. The differences in temperature are of order of 10$^{-5}$, however, the CMB at the time of decoupling consisted of $10^4$ causally disconnected patches which should have never been in thermal equilibrium. How is it that they have the same temperature then? (This is the so-called Horizon problem). On the other hand, for our universe to be flat now, it must have been flat to an incredibly degree in the far past, a value uncomfortably small to take as an initial condition. (This is the so-called flatness problem). These two issues are among the main problems of the standard model of Cosmology.

\subsection{Horizon problem}

The particle horizon presented in Eq.~\eqref{eq:conftime} can be rewritten as
\be \label{eq:horprob} \tau=\int_0^{a'}\f{\dif\ln a}{aH}~. \ee
Furthermore, from Eq.~\eqref{eq:aastau} one can use the definition $\dif t=a\dif\tau$ and find that the combination $(aH)^{-1}$ grows, for a matter (with $\omega=0$)- or radiation (with $\omega=-1/3$)-dominated  universe, as
\be \mk{aH}^{-1}\propto a^{\frac12\mk{1+3\omega}}~, \ee
and therefore the particle horizon \eqref{eq:horprob} grows in a similar way.


The quantity defined as $(aH)^{-1}$ is called the \emph{comoving Hubble radius}, and its implications are quite important: as the comoving Hubble radius has been growing monotonically with time during the evolution of the Universe, observable scales are now entering the particle horizon and, therefore, they were outside causal contact in the far past, at the CMB decoupling for instance. Consequently, the homogeneity problem is manifest: two points with an angular separation exceeding 2 degrees over the observable sky should have never been in thermal equilibrium and yet they have almost exactly the same temperature!

\subsection{Flatness problem}

We have now defined the comoving Hubble radius, which clearly is a function of time that monotonically grows during the evolution of the Universe. Evidently, Eq.~\eqref{eq:flatness} is therefore a function of time too. It can be then explicitly written as
\be |\Omega(a)-1|=\biggr|\frac{k}{(aH)^2}\biggr|~, \label{eq:flatness2} \ee
where we recall that $\Omega(a)\equiv\rho(a)/\rho_c(a)$. Because $(aH)^{-1}$ grows with time, $|\Omega(a)-1|$ must diverge and therefore the value $\Omega(a)=1$ is an unstable fixed point, as seen from the differential equation~\cite{Baumann:2009ds}
\be \diff{\ln \Omega}{\ln a}=\mk{1+3\omega}\mk{\Omega-1}~. \label{eq:fixpoint} \ee 
For the observed value $\Omega(a)\sim1$, the initial conditions for $\Omega$ then require an extreme fine tuning. For instance, to account for the flatness level observed today, $|\Omega(a_\text{BBN})-1|\leq\mathcal{O}\mk{10^{-16}}$ or $|\Omega(a_\text{GUT})-1|\leq\mathcal{O}\mk{10^{-61}}$. Setting these orders of magnitude as initial conditions imply a huge fine-tuning problem.
 
\subsection{Initial perturbations problem}

Finally, as we have already stated, even though the homogeneity and isotropy are evident, they are not perfect. There exist structures like galaxies, cluster of galaxies and cosmic voids which back in time were seeded by small density perturbations which differed in amplitude by $\delta\rho/\rho\sim10^{-5}$, according to the level of anisotropy observed in the CMB. These perturbations are, again, assumed and put \emph{by hand}, as the $\Lambda$CDM model has no mechanism which can produce them. To that end, a theory providing a mechanism for the generation of these primordial seeds is very appealing.

In the following, we shall see that both the horizon and flatness problems are trivially solved if we account for an epoch in which the comoving Hubble radius decreases before starting to increase again, and that this epoch must consist in an accelerated expansion of the Universe. Furthermore, in the quantum regime, vacuum fluctuations subject to this accelerated expansion could be stretched to classical scales, becoming into the primordial seeds we are looking for. Such a mechanism is now conceived as \emph{inflation} (for reasons we are about to discuss) and it is not only an artifact to solve the horizon and flatness problems, but a theory where the laws of GR and those of quantum mechanics are put to work together, converting inflation in the theory of the primordial quantum fluctuations.

\lhead[{\bfseries \thepage}]{ \rightmark}
\rhead[ Chapter \thechapter. \leftmark]{\bfseries \thepage}
\chapter{The Physics of Inflation}
\label{sec:inflation}

The \emph{inflationary paradigm} provides the Standard Model of Cosmology with a mechanism which easily solves the horizon and flatness problems and, at the same time, produces the primordial seeds that became the structures we see today in the sky. Independently of the precise nature of the mechanism, it consists on an accelerating stage during the early universe (similar to the current one driven by the dark energy component) which happened only for a \emph{brief} period, 
soon after the big bang. During this time, the Universe should have exponentially increased---\emph{inflated}---by a factor of $10^{24}$ in order to fit the current observational constraints. As we shall see, the comoving Hubble radius decreases during this stage and, therefore, observable scales were inside the horizon at the beginning, \ie in causally-connected regions. Hence, this solves the horizon problem. A similar analysis shows that the flatness problem is solved too.

Different mechanisms to inflate the universe have been proposed---the standard picture being that of a new field driving the accelerated expansion. The original one,\footnote{Alan Guth was the first one who proposed a scalar field for the inflationary mechanism and who coined the term `inflation.' However, historically, the first successful model of inflation is due to Alexei Starobinsky (1979). See \S\ref{sec:STintro} for a discussion on this model.} due to Alan Guth~\cite{Guth:1980zm}, consisted in a new scalar field trapped in a false vacuum state which energy density drives the accelerated expansion. The false vacuum is unstable and decays into a true vacuum by means of a process called \emph{quantum bubble nucleation}. The hot big bang was then generated by bubble collisions whose kinetic energy is obtained from the energy of the false vacuum. A deep analysis of this mechanism, however, showed that this method does not work for our Universe: for sufficiently long inflation to solve the horizon problem, the bubble collision rate is not even small but it does not happen at all as the bubbles get pushed to causally disconnected regions due to the expansion~\cite{Guth:1981uk,Cook:1981fz,Barrow:1981pa}. Even though Guth's mechanism did not work, he showed that an accelerated expanding universe could be able to solve the horizon and flatness problems. 

Soon after, Andrei Linde~\cite{LINDE1982389} and, independently, Andreas Albrecht and Paul Steinhardt~\cite{Albrecht:1982aa} introduced a new mechanism in which the new scalar field, instead of being trapped in a false vacuum, is rolling down a smooth potential. Inflation then takes place while the field rolls slowly compared to the expansion rate of the Universe. Once the potential becomes steeper, the field rolls towards the vacuum state, oscillates around the minimum and reheats the Universe. This new mechanism has prevailed up to now and it is the so-called \emph{Slow-Roll inflation}.

In 1981, Viatcheslav Mukhanov and Gennady Chibisov showed an amazing consequence of an accelerated stage of the primordial universe~\cite{Mukhanov:1981xt}: quantum fluctuations present during this epoch are able to generate the primordial density perturbations and their spectra amplitude are consistent with observations. Later, during the 1982 \emph{Nuffield Workshop on the Very Early Universe}, four different working groups, led by Stephen Hawking~\cite{HAWKING1982295}, Alexei Starobinsky~\cite{STAROBINSKY1982175}, Alan Guth and So-Young Pi~\cite{Guth:1982aa}, and James Bardeen, Paul Steinhardt and Michael Turner~\cite{Bardeen:1983aa}, computed the primordial density perturbations generated due to quantum fluctuations by the slow-roll mechanism. These calculations made inflation not only an artifact to solve the horizon and flatness problems, but a fully testable theory able to generate the initial conditions of the $\Lambda$CDM model.\footnote{Alan Guth himself is the author of a book on the history of inflation---\emph{The Inflationary Universe: The quest for a new theory of cosmic origins}~\cite{Guth:1997wk}. I suggest the interested reader to take a look at the book for an extraordinary account of the development of the Inflationary Theory.}

The simplified picture of inflation consists then in an accelerated epoch driven by the energy density of a new scalar field, dubbed the \emph{inflaton}, which slowly rolls down its potential. Once the inflaton acquires a large velocity, inflation ends and the inflaton oscillates around the minimum of the potential, reheating the Universe \ie giving birth to the hot big bang universe we described in the previous chapter. During the inflaton's evolution, vacuum fluctuations of  the inflaton field are continuously created everywhere in space. These fluctuations, which were in causal contact, get stretched to classical levels, exiting the horizon and originating overdensity fluctuations that seeded the structure formation of the Universe.

Along this Chapter, we firstly focus on the classical dynamics of slow-roll inflation: the solution to the $\Lambda$CDM problems and the dynamics of a scalar field coupled to Einstein's gravity (GR). Secondly, we shall introduce the theory of cosmological perturbations and follow the quantization prescription for a scalar field in order to compute the predictions for the primordial perturbations. Finally, we shall describe the cosmological observations able to test and discern between different realizations of inflation.


\section{The horizon and flatness problems revisited}

As already pointed out, the core of the $\Lambda$CDM problems is the growing nature of the comoving Hubble radius $\mk{aH}^{-1}$---a region enclosing events that are causally-connected at a given time---during the evolution of the Universe. As a consequence, most of the observable scales must have been disconnected in the past. The intuitive solution is then a mechanism which makes the comoving Hubble radius decrease during the early times. This would imply that observable scales were causally-connected at some initial time and then exited the horizon when it decreased. The horizon problem would then be solved as currently disconnected regions across space would have been allowed to be causally-connected in the past.

\begin{figure}[t]
\begin{center}
\includegraphics[keepaspectratio, width=13cm]{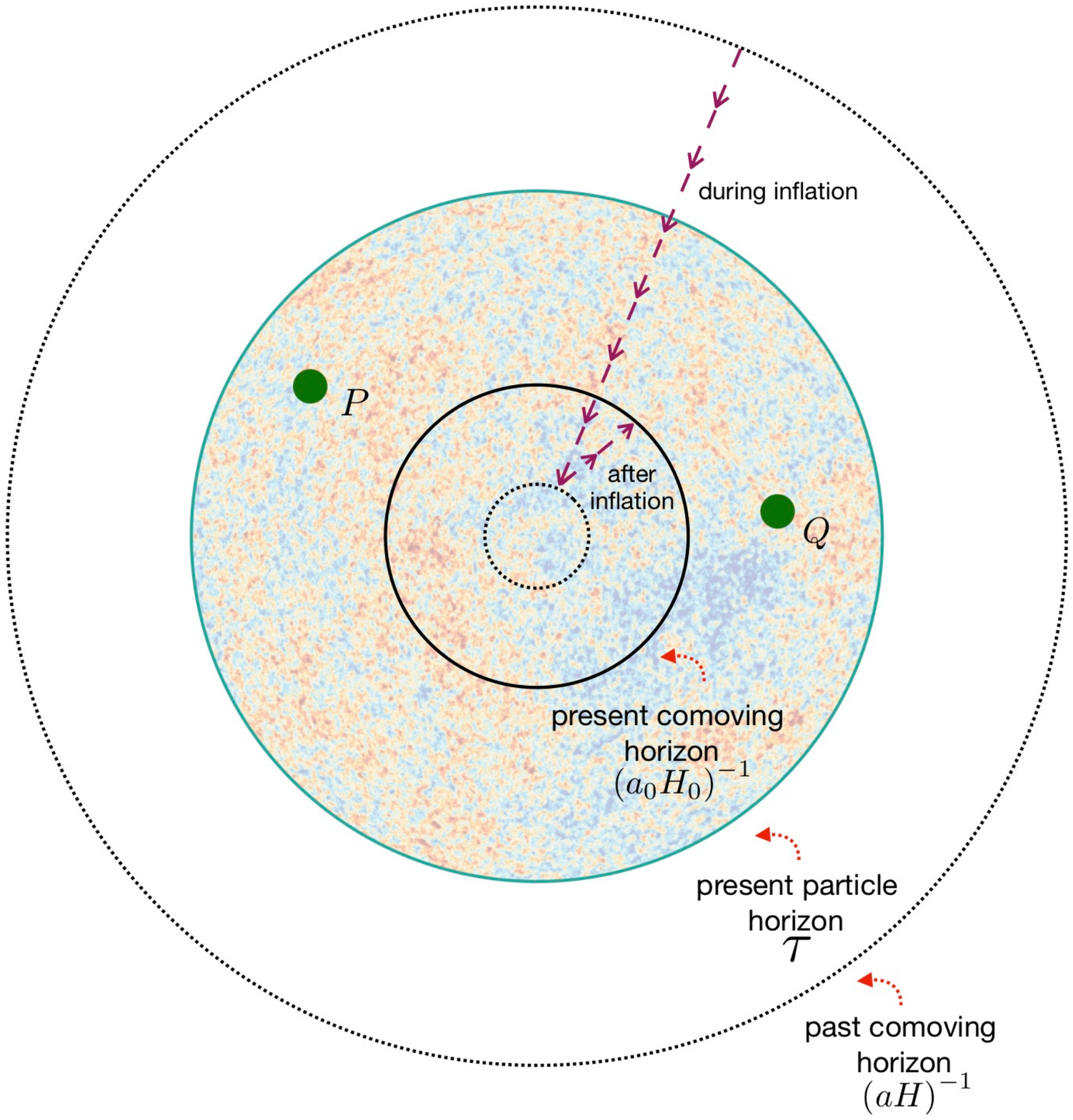}
\end{center}
\caption{\label{fig:comhor} Evolution of the comoving Hubble radius $\mk{aH}^{-1}$. At early times, the horizon was large enough so that observable scales were in causal contact. As inflation took place, the horizon shrank and scales came out to disconnected regions. Inflation then finished and the horizon started to grow to the present size. Two casually-disconnected regions, $P$ and $Q$, were then in causal contact at some point in the past, thus resolving the horizon problem.}
\end{figure}

As we shall see in \S\ref{sec:condforinf}, during inflation, the Hubble parameter $H$ is approximately constant. Therefore, the particle horizon $\tau$, given by Eq.~\eqref{eq:horprob}, can be integrated explicitly as
\be \tau\simeq-\f{1}{aH}~. \label{eq:parthorsin}\ee
So one can see that a large past Hubble horizon $\mk{aH}^{-1}$ would make $\tau$ fairly large today, larger than the present Hubble horizon $\mk{a_0H_0}^{-1}$, \ie two largely-separated points in the CMB would not communicate today but would have done so in the past if they were inside the particle horizon $\tau$. Figure~\ref{fig:comhor} sketches this reasoning.

Furthermore, it is evident from Eq.~\eqref{eq:flatness} that a decreasing Hubble radius drives the Universe towards flatness, and just deviating from it at present times. Thereby $\Omega=1$, which previously was an unstable fix point (see Eq.~\eqref{eq:fixpoint}), became an attractor solution thanks to inflation, thus also solving the flatness problem.

\subsection{Conditions for inflation}
\label{sec:condforinf}

The shrinking Hubble radius entails important consequences for the evolution of the scale factor $a$, \ie for the evolution of the Universe. First, lets note that the change of the decreasing $\mk{aH}^{-1}$ over time is
\be \diff{}{t}\mk{\f{1}{aH}}=-\f{\ddot a}{\mk{aH}^2}<0~, \ee
and therefore, from the inequality,
\be \ddot a>0~, \label{eq:adotdot} \ee
is a necessary condition for the shrinking of the Hubble radius. It is evident then that we require an accelerated expansion to solve the horizon and flatness problems.

Furthermore, Eq.~\eqref{eq:adotdot} has implications on the evolution of the Hubble parameter due to the relation $\dot H=(\ddot a/a)-H^2$, and hence
\be \f{\ddot a}{a}=H^2\mk{1-\epsilon_H}~, \ee
where we have implicitly defined the first \emph{slow-roll parameter} as
\be \epsilon_H\equiv-\f{\dot H}{H^2}<1~. \label{eq:epsilonh} \ee
As we shall see, $\epsilon_H$ is one of the most important parameters in inflation, as it quantifies its duration and, equivalently, determines when it ends. 

Furthermore, from the second Friedmann equation \eqref{eq:fried2},
\be \ba \f{\ddot a}{a}=H^2\mk{1-\epsilon_H}&=-\frac16\mk{\rho+3p}\\ &=-\frac\rho6\mk{1+3\omega}~, \label{eq:epome} \ea \ee
where, for $\epsilon_H\to0$ and a flat Universe with $\rho=\rho_c=3H^2$, we find that Eq.~\eqref{eq:epome} leads to
\be \omega\to-1~, \qquad \leftrightarrow \qquad a\propto e^{Ht}~, \ee
as already obtained from Eq.~\eqref{eq:aastau}. This means that the expansion increases exponentially or, in other words, the universe inflates! In a general case, Eq.~\eqref{eq:epome} suggests a more general condition for an accelerated expansion:
\be p<-\frac13\rho~, \ee
which, as discussed in \S\ref{sec:evolution}, implies that the accelerated expansion is driven by a fluid with negative pressure. 

\begin{figure}[t]
\begin{center}
\includegraphics[keepaspectratio, width=14cm]{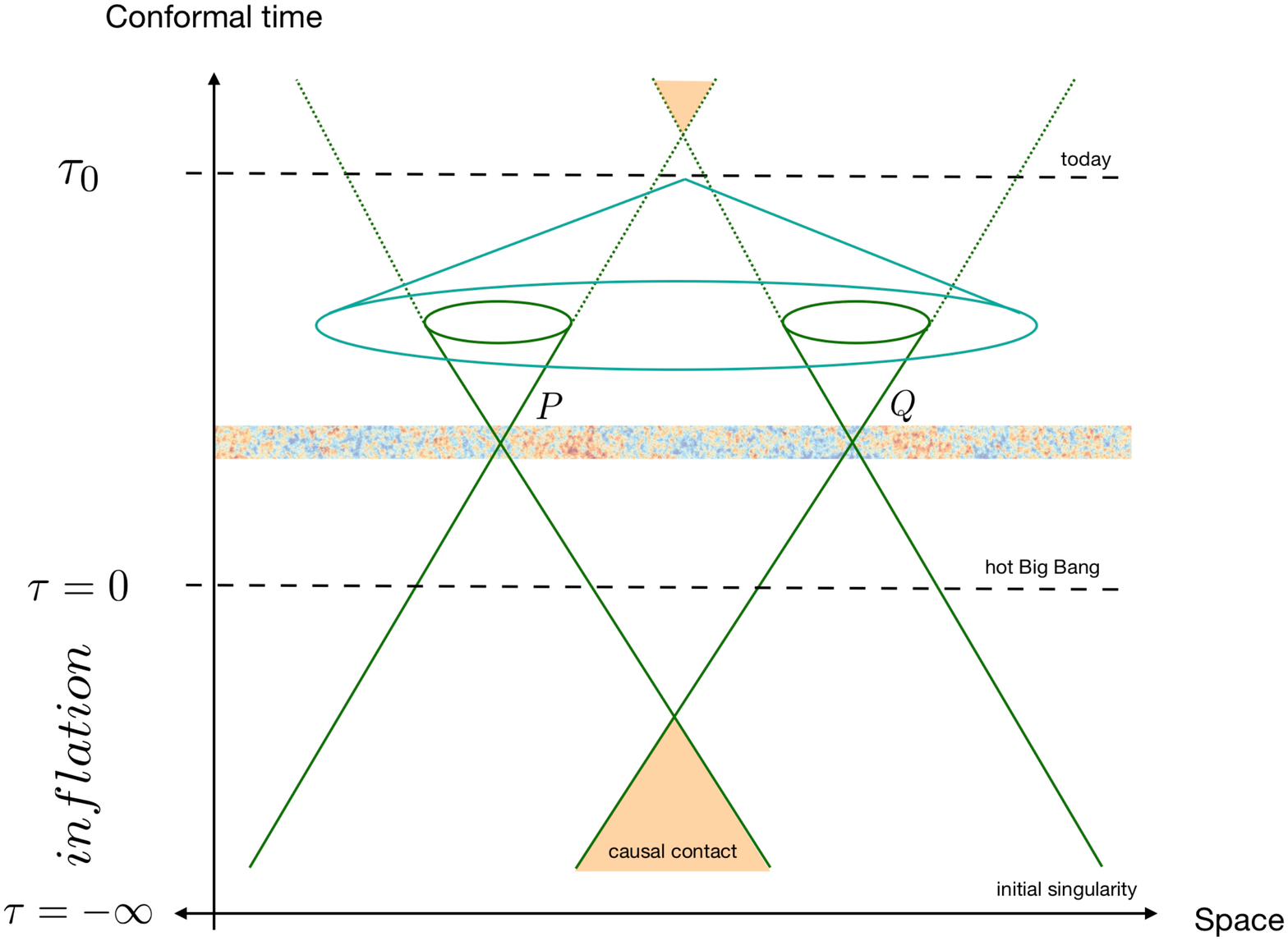}
\end{center}
\caption{\label{fig:conftime} Conformal diagram including the inflationary epoch. Inflation shifts the initial singularity to $\tau=-\infty$ (see Eqs.~\eqref{eq:dSdefi}) allowing the light cones of two CMB points, $P$ and $Q$, which are causally disconnected now, to be causally connected at some point in the past, thus solving the horizon problem.}
\end{figure}

Finally, notice that Eq.~\eqref{eq:epsilonh} shows that during this accelerated expansion, the rate of change of the Hubble parameter is required to be small, meaning that $H$ is approximately constant during inflation. This has important consequences on the conformal time, namely (see Eq.~\eqref{eq:parthorsin})
\be \tau=-\frac{1}{aH}~, \qquad \leftrightarrow \qquad a=-\f{1}{H\tau}~, \label{eq:dSdefi} \ee
and therefore a singularity $a=0$ corresponds to $\tau\to-\infty$ . Consequently, at $\tau=0$ the scale factor is not well defined and inflation must end before reaching this epoch (that is, $H\simeq const.$ stops being a good approximation). The spacetime defined with these characteristics is called \emph{de Sitter space} and it is exactly the spacetime of inflation. To see the consequences of this in the evolution of two CMB points, let us take Fig.~\ref{fig:comhor} and put it in perspective as a function of the conformal time $\tau$, shown in Fig.~\ref{fig:conftime}. If we take only the period containing the hot big bang (from $\tau=0$ to $\tau_0$), two CMB points could have never been in contact, whereas once we assume inflation took place, the light cones of these two points intersect in the far past, during inflation, allowing them to be causally connected.

Before continuing, and to summarize, let us emphasize that whatever the mechanism for inflation is, the simple fact that the comoving Hubble radius shrinks implies that the following conditions must be (mutually) satisfied:
\be \ddot a>0~, \qquad \epsilon_H\equiv-\f{\dot H}{H^2}<1~, \qquad p<-\frac13\rho~. \label{eq:condforinf} \ee
Now, let us discuss how the energy density of a scalar field driving inflation, subject to the \emph{slow-roll approximation}, effectively satisfies these conditions. 

\section{Canonical single-field inflation}

At the background level, we consider a single scalar and homogeneous field $\phi(t,x^i)=\phi(t)$, which we shall name the `inflaton', minimally coupled to Einstein's gravity. The action is then given by the sum of the Einstein-Hilbert action and the action for the scalar field. It reads as
\be \mathcal{S}=\mathcal{S}_\text{EH}+\mathcal{S}_\phi=\int\dif^4x\sqrt{-g}\kk{\frac12R+\frac12g^{\mu\nu}\phi_{,\mu}\phi_{,\nu}-V(\phi)}~, \label{eq:EinHil} \ee
where $g=$det$\mk{g_{\mu\nu}}$, and $V(\phi)$ is the potential energy of the inflaton $\phi$. As we shall see, the predictions for a given inflationary model are, in general, highly dependent on the form of $V(\phi)$.

The variation of the Einstein-Hilbert action leads to the Einstein equations in the vacuum $R_{\mu\nu}-\frac12g_{\mu\nu}R=0$. On the other hand, the variation of $\mathcal{S}_\phi$ defines the energy-momentum tensor for the scalar field:
\be \delta\mathcal{S}_\phi=\frac12\int\dif^4x\sqrt{-g}T_{\mu\nu}\delta g^{\mu\nu}~, \ee
which can be solved for $T_{\mu\nu}$ as
\be T_{\mu\nu}=\phi_{,\mu}\phi_{,\nu}-g_{\mu\nu}\kk{\frac12g^{\rho\sigma}\phi_{,\rho}\phi_{,\sigma}-V(\phi)}~. \label{eq:sctmunu} \ee
Using the FLRW metric \eqref{eq:flrwm}, the 00- and $ii$-components of Eq.~\eqref{eq:sctmunu} can be related to those in Eq.~\eqref{eq:tmunudef} for a perfect fluid. Consequently, the energy density and pressure for a homogeneous minimally coupled scalar field are given by:
\begin{alignat}{2} \rho=&\frac12\dot\phi^2+V(\phi)~,\label{eq:denpre} \\ p=&\frac12\dot\phi^2-V(\phi)~. \label{eq:denpre2} \end{alignat}

If we now take the continuity equation~\eqref{eq:cont2} and substitute Eqs.~\eqref{eq:denpre}-\eqref{eq:denpre2} into it, we obtain the Klein-Gordon equation for a scalar field in the gravitational background:
\be \ddot\phi+3H\dot\phi+V'(\phi)=0~. \label{eq:KG} \ee
Here primes denote derivatives with respect to the field, as $'\equiv\dif/\dif\phi$. Furthermore, it is possible to do the same for the Friedmann equations~\eqref{eq:fried} and \eqref{eq:fried2} to obtain the evolution equation for the Hubble parameter and the constraint equation respectively as
\begin{alignat}{2} H^2&=\frac13\kk{\frac12\dot\phi^2+V(\phi)}~, \label{eq:conseq} \\ 0&=\dot\phi^2-V(\phi)+3\mk{H^2+\dot H}~. \label{eq:conseq2} \end{alignat}
Together with the Klein-Gordon equation~\eqref{eq:KG}, Eqs.~\eqref{eq:conseq}-\eqref{eq:conseq2} completely determine the dynamics of the scalar field in the gravitational backgr-ound---and hence are the so-called \emph{background equations of motion}. Now, we shall discuss how this set of equations behaves under the conditions for inflation obtained in \S\ref{sec:condforinf}.

\subsection{Conditions for inflation revisited}
\label{sec:conrev}

Recall the conditions for inflation in Eqs.~\eqref{eq:condforinf}. The third equation, for the energy density and pressure of $\phi$, can be written as
\be \omega_\phi=\f{p_\phi}{\rho_\phi}=\f{\frac12\dot\phi^2-V(\phi)}{\frac12\dot\phi^2+V(\phi)}<-\frac13~, \ee
which the last inequality can be recast as $\dot\phi^2<V(\phi)$~. The same can be noticed from the second equation in~\eqref{eq:condforinf}, where the slow-roll parameter can be written, using Eqs.~\eqref{eq:conseq}-\eqref{eq:conseq2}, as
\be \epsilon_H=\frac12\f{\dot\phi^2}{H^2}<1~. \label{eq:epscond} \ee
In this case, the inflationary limit $\epsilon_H\to0$ places the even stronger condition
\be \dot\phi\ll V(\phi)~. \ee
In addition, the second derivative, \ie the acceleration of $\phi$, must be negligible compared to the rate of expansion. This places the second condition
\be |\ddot\phi|\ll |3H\dot\phi|,|V'(\phi)|~. \ee
This inequality allows us to introduce the second slow-roll parameter $\eta_H$, defined as
\be \ba \eta_H=&\epsilon_H-\f{\dif\ln\epsilon_H}{2H\dif t}\\ =&-\f{\ddot\phi}{\dot\phi H}~. \label{eq:etaHdef} \ea \ee
Then, the condition 
\be |\eta_H|<1~, \label{eq:etacond} \ee
ensures that the fractional change of $\epsilon_H$ is small. We shall sometimes use the slow-roll parameter $\delta_1=-\eta_H$ which will help us to better define a hierarchy of slow-roll parameters $\delta_i$ (see \S\ref{sec:GSR}).

\begin{figure}[t]
\begin{center}
\includegraphics[keepaspectratio, width=13cm]{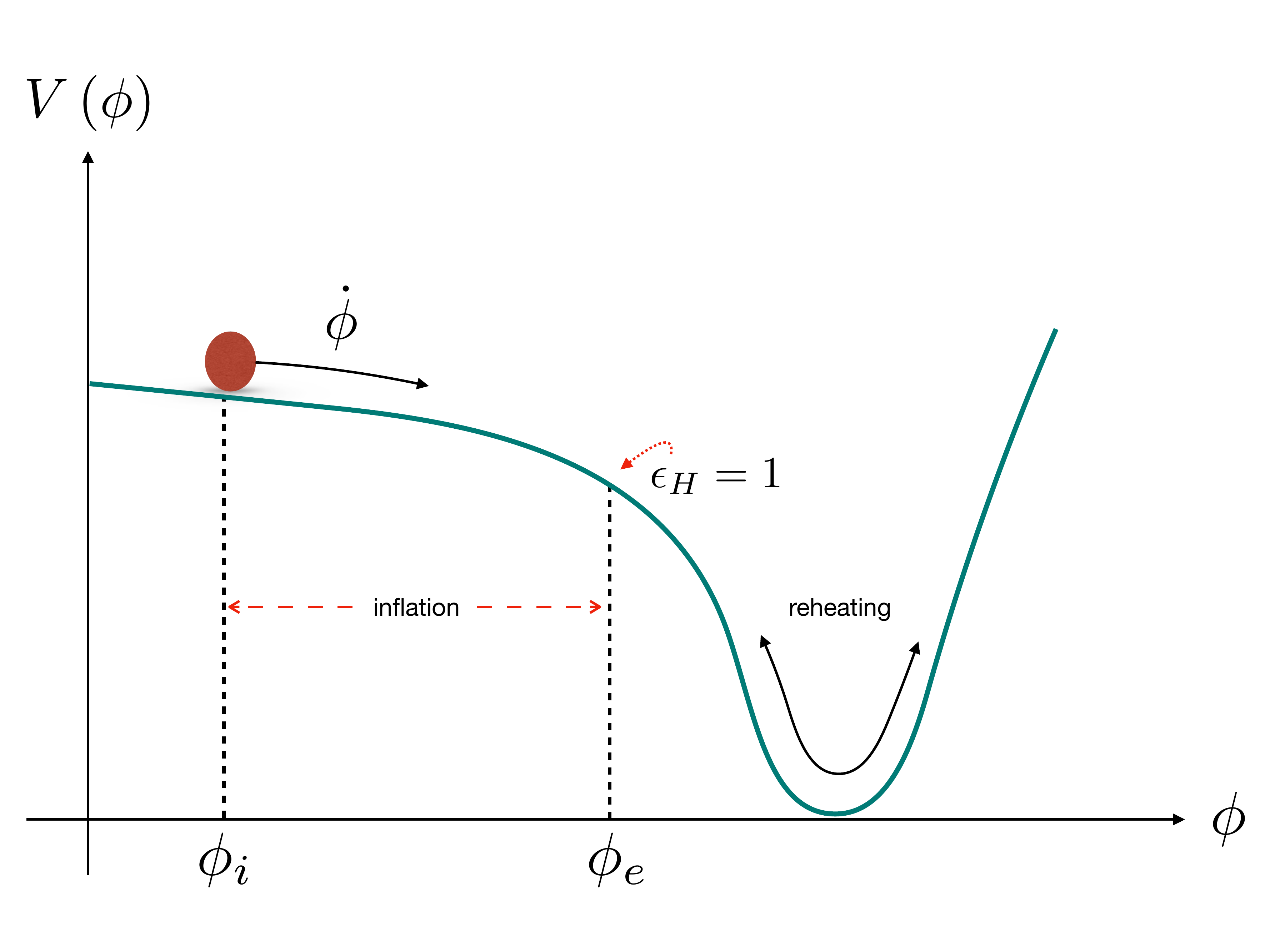}
\end{center}
\caption{\label{fig:potential} Evolution of the inflaton. The inflaton rolls down the potential, inflating the Universe. Once it acquires a large velocity, the slow-roll conditions break and inflation finishes. Afterwards, the inflaton oscillates around the potential's minimum and reheats the Universe. Note that, in general, $\phi_i>\phi_e$, so the field decreases towards the right in this sketch.}
\end{figure}

\begin{figure}[t]
\begin{center}
\includegraphics[keepaspectratio, width=13cm]{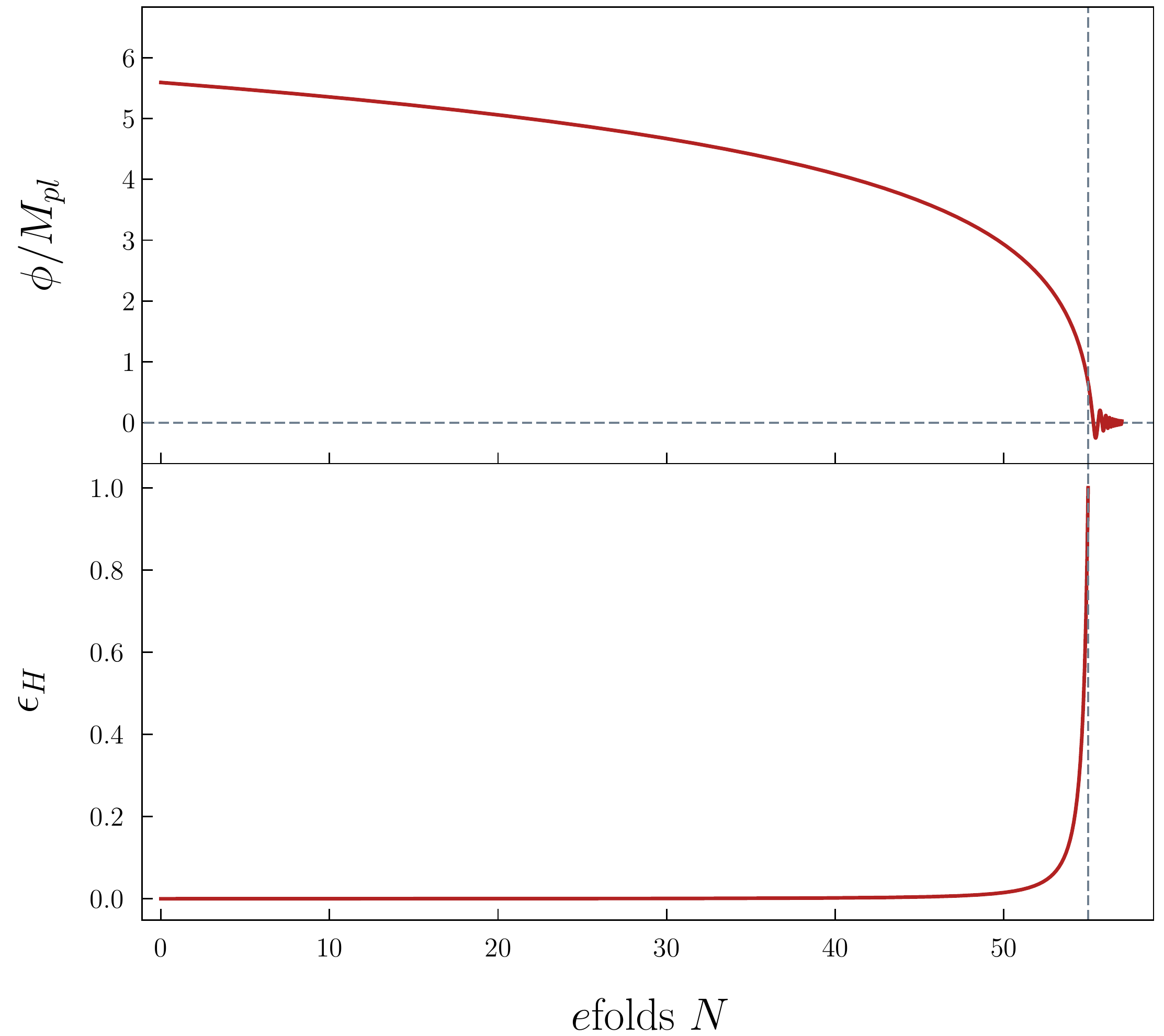}
\end{center}
\caption{\label{fig:alphaplots} Evolution for the field $\phi$ and the first slow-roll parameters $\epsilon_H$ for the model given in Eq.~\eqref{eq:alphattrac} with $\alpha_c=1$, by solving the background equations \eqref{eq:KG}-\eqref{eq:conseq2} numerically. The plot is normalized such as the end of inflation $\epsilon_H=1$ coincides with $N=55$ (gray vertical dashed line).}
\end{figure}

Therefore, the conditions for inflation Eqs.~\eqref{eq:condforinf} were recast as the slow-roll conditions $\{\epsilon_H,|\eta_H|\}<1$ which place constraints for the velocity of the field $\phi$. Namely, the potential energy $V(\phi)$ should dominate over the kinetic energy $\dot\phi^2/2$ or, in other words, the field should \emph{roll slowly} down its potential. This is sketched in Fig.~\ref{fig:potential}, where a sufficiently flat potential would make the field roll slowly towards the minimum: once the potential gets steeper, the field acquires a large velocity, breaking the condition \eqref{eq:epscond}; finally, the field oscillates around the minimum and reheats the Universe. In addition, we illustrate, in Fig.~\ref{fig:alphaplots}, the solution for the field $\phi$ and the first slow-roll parameter $\epsilon_H$ computed by solving numerically the background equations \eqref{eq:KG}-\eqref{eq:conseq2} for the $\alpha$-attractor potential given in Eq.~\eqref{eq:alphattrac} with $\alpha_c=1$. Notice that $\phi$ and $\epsilon_H$ evolve slowly during most of the evolution, parametrized by the number of $e$folds $N=\int H\dif t$ (quantity that we shall carefully describe in \S\ref{sec:efolds}) and that the field enters the oscillatory stage when inflation finishes at $\epsilon_H=1$, as expected. 

\subsection{Slow-roll approximation}

The conditions obtained in \S\ref{sec:conrev} allow us to simplify the Einstein equations for the inflaton, Eqs.~\eqref{eq:KG}-\eqref{eq:conseq2}. In particular,
\begin{alignat}{2} \dot\phi&\simeq-\f{V'(\phi)}{3H}~, \label{eq:srapp} \\ H^2&\simeq\frac13V(\phi)~, \label{eq:srapp2} \end{alignat}
which is the so-called \emph{slow-roll approximation} (SR).\footnote{Along this thesis, `SR' shall refer to the (slow-roll) approximation only, which helps to differentiate it from other approximations discussed in \S\ref{sec:GSR}.} Notice that the second equation also implies that $H^2$ is approximately constant as expected. Also, from Eqs.~\eqref{eq:srapp}-\eqref{eq:srapp2}, one can see that the conditions for inflation in terms of the field velocity can be once more recast as conditions for the shape of the potential $V(\phi)$. This allows us to define the potential slow-roll parameters as
\be \epsilon_V\equiv\frac12\mk{\f{V'(\phi)}{V(\phi)}}^2~, \qquad \eta_V\equiv\f{V''(\phi)}{V(\phi)}~, \label{eq:srpotparams} \ee
which are related to the \emph{Hubble} slow-roll parameters as $\epsilon_H\simeq\epsilon_V$ and $\eta_H\simeq\eta_V-\epsilon_V$, respectively, as long as the SR approximation (Eqs.~\eqref{eq:srapp}-\eqref{eq:srapp2}) holds. They are also subject to the slow-roll conditions, \ie inflation finishes when $\epsilon_V$, $\eta_V\sim1$.

\subsection{Reheating}
\label{sec:reheating}

After inflation has finished, the inflaton rolls to the global minimum of the potential where it oscillates. There will be energy losses due to oscillations, corresponding to the decay of individual $\phi$-particles. 
The equation of motion of $\phi$ then becomes
\be \ddot\phi+3H\dot\phi+\Gamma\dot\phi+V'(\phi)=0~,\ee
after having expanded the potential around the minimum value and where $\Gamma$ is the decay rate of $\phi$, which acts as an additional friction term and depends on how the inflaton couples to the Standard Model particles. One important feature is that reheating occurs at $t~\sim H^{-1}\sim\Gamma^{-1}$, \ie the reheating temperature is given by $T_\text{reh}\sim\sqrt{\Mpl\Gamma}$. 

As we shall discuss, an important and surprising feature of inflation is that the primordial perturbations freeze after inflation has finished, \ie their subsequent evolution is not affected by the physics of reheating (see Refs.~\cite{Kofman:1997yn,Mukhanov:2005sc,Bassett:2006aa,Lyth:2009zz,Gorbunov:2011zz} for more details on the reheating processes in the early universe). 

\subsection{Duration of inflation}
\label{sec:efolds}

As the expansion is exponentially accelerated, the duration of inflation is parametrized by means of the number of $e$folds $\Delta N\equiv\ln\mk{a_\text{end}/a_\text{initial}}$. Therefore, the number of $e$folds elapsed from a particular epoch to the end of inflation is given by
\be \ba N=&\int^{t_\text{end}}_{t}H\dif t'~,\\ \simeq&\int^\phi_{\phi_\text{end}}\f{V(\phi')}{V'(\phi')}\dif\phi'=\int^\phi_{\phi_\text{end}}\f{\dif\phi'}{\sqrt{2\epsilon_V}}~, \ea \label{eq:efoldint} \ee
where in the second line we assumed the SR approximation, and thus we can approximate the duration of inflation by means of the field excursion $\Delta\phi$. 

The precise value of $N$, needed to solve the horizon and flatness problems, depends then on the energy scale of inflation and also on the physics of reheating. The latter in fact provides the following relation~\cite{Lyth:2009zz,Kolb:1990vq}:
\be N=56-\frac23\ln\f{10^{16}~\text{GeV}}{\rho_*^{1/4}}-\frac13\ln\f{10^9~\text{GeV}}{T_\text{reh}}~, \label{eq:efoldsreh} \ee
where $\rho_*$ is the energy density at the end of inflation. Thus, we can estimate $N$ for some well-motivated values of $T_\text{reh}$. In particular, to solve the aforementioned problems, it is found that $\Delta N\geq60$~\cite{Lyth:2009zz,Kolb:1990vq}. Furthermore, CMB scales should have exited the horizon around 55 $e$folds before inflation ended (see references in \S\ref{sec:reheating}):
\be N_\text{CMB}=\int_{\phi_\text{end}}^{\phi_\text{CMB}}\f{\dif\phi}{\sqrt{2\epsilon_H}}\simeq55~. \label{eq:ncmb} \ee

Before moving on, a comment is in order. Introducing units back, the first slow-roll condition tells us that $|V'(\phi)/V(\phi)|\ll \sqrt{2}/\Mpl$, for which $N\gg(\phi-\phi_\textrm{end})/\sqrt{2}\Mpl$ in Eq.~\eqref{eq:efoldint}. This means that we will get a sufficient amount of inflation as long as the excursion $\Delta\phi$ changes at least as large as $\sqrt{2}\Mpl$. These \emph{super-Planckian} values (encountered in many inflationary models as the one used in Fig.~\ref{fig:alphaplots}) do not represent a breakdown of the classical theory. In fact, the condition for neglecting quantum gravitational effects is that the field energy density is much smaller than the Planck energy density: $|V(\phi)|\ll2\Mpl^2$ \cite{Weinberg:2008zzc,Gorbunov:2011zzc}. This condition can be simply satisfied by supposing that $V(\phi)$ is proportional to a small coupling constant which, in turn, does not affect the slow-roll conditions nor the value of $N$.

\section{Models of inflation}
\label{sec:modelsofinfl}

So far we have not made any prediction but just found that, under the assumption that there exists a single field minimally coupled to Einstein's gravity, the conditions for inflation require that the potential energy dominates over the kinetic one. Then, in order to exploit the theory, we need to choose a particular function for $V(\phi)$ and solve the background equations. Their computation is often performed analytically given the simplifications one can do using the SR approximation. However, there exist numerous potentials proposed in the literature which break per se the slow-roll conditions and hence the background equations must be solved numerically. In the following we discuss the usual approximations to choose a model in which we include noncanonical models, which are a central part of this thesis. We do not attempt to give a full list of models but only a taste of the most popular and phenomenologically well-behaved ones. For a well-known and exhaustive classification see Ref.~\cite{Martin:2013tda}.

\subsection*{Single-field canonical models}

A general and historical classification of single-field models relies on whether the field in a particular model takes super- or sub-Planckian values. The former class is dubbed \emph{large-field inflation} whereas \emph{small-field inflation} the latter. The requirement of the flatness of the potential $V(\phi)$ is the same for both and therefore we do not discuss their further conceptual differences but the interested reader is referred to Refs.~\cite{Lyth:2009zz,Gorbunov:2011zzc}.

\subsubsection*{Chaotic inflation}

Unarguably, the simplest model is given by the potential energy $V(\phi)=m^2\phi^2/2$ which belongs to the class of models called \emph{chaotic inflation}~\cite{LINDE1983177}, generally written as 
\be V(\phi)=\lambda_p\phi^p~. \label{eq:monpot} \ee 
In the next section we shall see that this class of models, in the canonical framework, are in tension with CMB observations~\cite{Collaboration:2018aa}, however we will often use it as a working example given its simplicity. For instance, the potential slow-roll parameters for this model are simply given by $\epsilon_V=\eta_V=2/\phi^2~.$ Furthermore, the end of inflation---~$\epsilon_V=1$~---sets the final value for the field as $\phi_e=\sqrt{2}\Mpl$, where we recovered the units for illustration. Then the field value at which CMB fluctuations must have been created can be computed by solving Eq.~\eqref{eq:ncmb}. This gives us $\phi_\text{CMB}=2\sqrt{N_\text{CMB}-1/2}\simeq15\Mpl$ for $N_\text{CMB}\simeq55$. Notice that this model takes super-Planckian values, \ie $\Delta\phi>\Mpl$; models with this characteristic produce in general a large amplitud of primordial tensor modes and thus they are in tension with observations~\cite{Collaboration:2018aa}.

\subsubsection*{Small-field inflation}

A model of inflation with super-Planckian values might be subject to quantum effects which affect the evolution of $\phi$ in a way we currently do not know. Therefore, models with short excursions $\Delta\phi$ are attractive. Among the most popular ones, \emph{Hiltop inflation}---similar to that sketched in Fig.~\ref{fig:potential}---given by the potential~\cite{Boubekeur:2005zm}
\be V(\phi)=V_0\kk{1-\mk{\f{\phi}{\mu}}^p}~, \ee
is able to fit observations for $p=4$~\cite{Collaboration:2018aa}.

\subsubsection*{High-energy physics models}

Other class of models are inspired from high-energy theories. Historically, from GUT, the Coleman-Weinberg potential~\cite{LINDE1982389,Albrecht:1982aa}
\be V(\phi)=V_0\ck{\mk{\f{\phi}{\mu}}^4\kk{\ln\mk{\f{\phi}{\mu}}-\frac14}+\frac14}~, \ee
was used when inflation was first being studied. However, calculations of the primordial perturbations were incompatible with the phenomenological values of $V_0$ and $\mu$ coming from particle physics. The same problem arises from the widely studied Higgs potential~\cite{Bezrukov:2007ep,LINDE1983177}. 

Along of the lines of GUT theories, supersymmetric realizations provide the potential
\be V(\phi)=\Lambda^4\kk{1+\alpha_h{\rm log}\mk{\phi/\Mpl}}~, \ee
where $\alpha_h>0$. In this scenario, inflation is driven by loop corrections in spontaneously broken supersymmetric (SB SUSY) GUT theories~\cite{Dvali:1994aa}.

Another widely studied model comes from axion physics, called \emph{natural inflation}~\cite{Freese:1993bc,Kim:2004rp,ArkaniHamed:2003mz,ArkaniHamed:2003wu}, and is given by a periodic potential of the form
\be V(\phi)=V_0\kk{\cos\mk{\f{\phi}{f}}+1}~. \ee
However, this model is becoming disfavored by the latest measurements~\cite{Collaboration:2018aa}.

From string theory, \emph{brane inflation}---driven by a D-brane---is characterized by the effective potential 
\be V(\phi)=M^2 M_{\rm pl}^2 \left[ 1-\left(\frac{\mu}{\phi} \right)^p+\cdots \right]~, \ee
where $p$ and $\mu$ are positive constants. In general, one assumes that inflation ends around $\phi\sim\mu$, before the additional terms denoted by the ellipsis contribute to the potential. The models arising from the setup of D-brane and anti D-brane configuration have the power $p=2$ \cite{GarciaBellido:2001ky} or $p=4$ \cite{Dvali:2001fw,Kachru:2003sx}. 

More recently, from supergravity theories, the $\alpha$-attractors with the potential energy~\cite{Kallosh:2013yoa,Ferrara:2013rsa,Collaboration:2018aa}
\be V(\phi)=\frac34\alpha_cM^2\Mpl^2\kk{1-\text{exp}\mk{{-\sqrt{\f{2}{3\alpha_c}}\f{\phi}{\Mpl}}}}^2~, \label{eq:alphattrac} \ee
have been used mainly due to their flexibility to fit observational predictions, depending on the value of $\alpha_c$ (which, interestingly, coincides with Starobinsky inflation, Eq.~\eqref{eq:Staromodel}, in the limit $\alpha_c=1$ and with the $\phi^2$ model of chaotic inflation in the limit $\alpha_c\to\infty$).

\subsection*{Multifield models}

It would be very natural that different species of particles were present during inflation. They may have not played any role in the evolution of the Universe, but any interaction between the inflaton field and other particles will inevitably lead to new phenomenology and to different mechanisms for the production of perturbations. The study of \emph{multifield inflation} deserves a thesis of its own, but the interested reader is encouraged to look at the comprehensive review by D. Wands~\cite{Wands:2007bd} or in~\cite{Lyth:2009zz,Gorbunov:2011zz}.

\subsection*{Noncanonical models}

Here we consider cases in which we do not only choose a potential energy $V(\phi)$ but also modify either the kinetic energy of the field, the gravitational interaction, or both.

\subsubsection*{$k$-inflation}
\label{sec:kinflation}

Instead of taking $\mathcal{L}_\phi=\frac12g^{\mu\nu}\phi_{,\mu}\phi_{,\nu}-V(\phi)$, one can consider more general kinetic terms proportional to $\phi$ and its velocity $\dot\phi$ as
\be \mathcal{L}_\phi=K(\phi,X)-V(\phi)~, \ee
where $X\equiv\frac12g^{\mu\nu}\phi_{,\mu}\phi_{,\nu}$. These kind of models are called $k$-inflation and it can be shown that inflation can indeed be driven by the kinetic term and take place even for a steep potential~\cite{ArmendarizPicon:1999rj,Garriga:1999vw}.

\subsubsection*{Nonminimal couplings}

Equation \eqref{eq:EinHil} assumes a minimal coupling between $\phi$ and $R$, however, a term like $\xi\phi R$, where $\xi$ is a coupling constant, is also allowed and introduces new phenomenology for different values of the coupling. In this configuration, simple potentials can be reconciled with observations for a range of values of $\xi$. Furthermore, it can be shown that the theory can be recast as one with a minimal coupling with an effective potential if one performs a conformal transformation of the metric as~\cite{Boubekeur:2015xza,Salopek:1988qh,Futamase:1987ua,Fakir:1990eg,Kaiser:1994vs,Komatsu:1999mt,Bezrukov:2007ep,Barvinsky:2008ia,Barvinsky:2008cya,Bezrukov:2009db,Hertzberg:2010dc,Okada:2010jf,Linde:2011nh}
\be g_{\mu\nu}\to\Omega^2(\phi)g_{\mu\nu}~. \label{eq:conftrans} \ee

\subsubsection*{Scalar-tensor theories}
\label{sec:STintro}

The two approaches described above can be extended to general theories of modified gravity. In general, any modification of GR will introduce new degrees of freedom, from which a scalar field can be identified as the inflaton. Currently, the most general scalar-tensor theories are the so-called \emph{Horndeski}~\cite{Horndeski:1974wa,Nicolis:2008in,Deffayet:2009mn,Deffayet:2009wt} and \emph{beyond Horndeski}~\cite{Gleyzes:2014dya,Langlois:2015cwa,Ezquiaga:2016nqo,Motohashi:2016ftl,Crisostomi:2016czh,BenAchour:2016fzp,Achour:2016rkg,Motohashi:2017eya} theories of gravity. These are fully characterized by a few functions, $G_i(\phi,X)$, coupled to the Ricci and Einstein tensors and to derivatives of the field. Therefore, any choice of these functions will inevitably introduce new phenomenology to the inflationary evolution.

Historically, the first successful model of inflation was due to Starobinsky~\cite{STAROBINSKY198099}. He realized that an early exponential acceleration comes as a solution of the Einstein equations with quantum corrections, due to the conformal anomaly of free scalar fields interacting with the classical gravitational background.\footnote{In the classical theory, a conformally-invariant free scalar field ($m=0$), \ie respecting the symmetry given in Eq.~\eqref{eq:conftrans}, satisfies $T^\mu_{~\mu}=0$. However, the quantum expectation value $\langle0|T^\mu_{~\mu}|0\rangle$ differs from 0, contributing with linear combinations of the scalar curvature $R$. This is called in the literature a conformal (or \emph{trace}) anomaly (see, \emph{e.g.}, Ref.~\cite{Parker:2009uva} for details).} This conformal anomaly contributes with higher-order terms, in the scalar curvature $R$, to the Einstein-Hilbert action. The action then reads
\be S=\int\dif^4x\sqrt{-g}\f{\Mpl}{2}\mk{R+\f{R^2}{6M^2}}~, \label{eq:Staromodel} \ee
where, in the absence of a quantum-gravity description of the theory, $M$ is a phenomenological parameter with dimensions of mass. This model belongs to the class of theories called $f(R)$, where suitable functions of $R$ can be written. Furthermore, these classes allow the same conformal transformation, Eq.~\eqref{eq:conftrans}, as the nonminimal-coupling models and, in particular, Eq.~\eqref{eq:Staromodel} can be recast as a canonical action of a scalar field with the potential given in Eq.~\eqref{eq:alphattrac} (with $\alpha_c=1$), \ie the Starobinsky model is a limit case of the $\alpha$-attractors~\cite{Whitt:1984pd,Coule:1987wt,Barrow:1988xh,Maeda:1988ab}.

The first models of inflation in the framework of general \emph{Horndeski-like} theories were called $G$-inflation and have been studied for very different potentials. In particular, one can show that simple potentials as those of chaotic inflation can be reconciled with observations for simple choices of $G_i(\phi,X)$~\cite{Ramirez:2018dxe,Burrage:2010cu,Kamada:2010qe,Ohashi:2012wf,Kamada:2013bia}.

The study of this class of theories for inflation is one of the main goals of this thesis. Consequently, they are fully discussed in \S\ref{sec:STtheories}.


\section{The theory of primordial quantum fluctuations}
\label{sec:theoryofquantum}

We have thus far discussed the classical physics of the inflationary theory: a mechanism able to drive the expansion of the early universe in an accelerated way, solving the horizon and flatness problems. Furthermore, we showed that a scalar field, evolving slowly compared to the expansion rate, satisfies the requirements for the inflationary mechanism.

Yet we are halfway into the story inflation has to tell. As already stated, inflation is also able to provide with the initial conditions for the hot big bang model, \ie with the primordial density perturbations that led to the CMB anisotropies and the large scale structure. The origin of these lies on the vacuum fluctuations of the inflaton field itself, which is subject to quantum effects. 

The inflaton fluctuations backreact on the spacetime geometry, leading to metric perturbations. The full set of quantum perturbations then get stretched to cosmological scales due to the accelerated expansion. As we shall see, these fluctuations in the inflaton field lead to time differences in the evolution of different patches of the Universe, \ie inflation finishes at different times in different places across space. Each of these patches will then evolve as independent causally-disconnected \emph{universes}, each one with different energy density, and it is once these patches come back inside the horizon, during recent times, when they become causally-connected again.

In Appendix~\ref{app:pert} we review the Cosmological Perturbation Theory, useful for this chapter. There we compute the primordial curvature perturbation which power spectrum is related to current CMB measurements. One important feature of this perturbation is that it freezes when it comes out the horizon during inflation. Consequently, its evolution is not modified by reheating processes and, in this way, we can connect the physics at the end of inflation with the density perturbations during the latter epochs, including the CMB anisotropies. We shall study the statistical properties of the primordial curvature and tensor perturbations that inflation creates and, in the next section, compare them to current observations.

We will start by finding the second-order action for scalar and tensor perturbations. Then we will quantize the field perturbations and find their equations of motion. Their solutions are not trivial in general so we will explain different approaches to solve them. Finally we shall give the exact formula for the power spectra of these primordial perturbations.

\subsection{Scalar and tensor perturbations}

To compute the second-order action for perturbations, we first adopt the Arnowitt-Deser-Misner (ADM) formalism which allows us to split the metric in such a useful way that the constraint equations clearly manifest~\cite{Arnowitt:1962hi}. The line element, following this splitting, then reads
\be \dif s^2=-N^2\dif t^2+g_{ij}\mk{\dif x^i+N^i\dif t}\mk{\dif x^j+N^j\dif t}~, \label{eq:ADM} \ee
where $g_{ij}$ is the three-dimensional metric on slices of constant $t$, $N(x^i)$ is called the \emph{lapse function} and $N_i(x_i)$ is called the \emph{shift function}. As we shall see, both $N$ and $N_i$ are Lagrange multipliers and, furthermore, they contain the same information as the metric perturbations $\Phi$ and $B$ introduced in Appendix~\ref{app:pert}.

By inserting Eq.~\eqref{eq:ADM} into Eq.~\eqref{eq:EinHil}, the action becomes
\be \ba \mathcal{S}=\frac12\int\dif^4x\sqrt{-g}&\kk{N^{(3)}R-2NV+N^{-1}\mk{E_{ij}E^{ij}-E^2}\right. \\&\left.+N^{-1}\mk{\dot\phi-N^i\phi_{,i}}^2-Ng^{ij}\phi_{,i}\phi_{,j}-2V}~, \label{eq:admact} \ea \ee
where $^{(3)}R$ is the three-dimensional curvature and
\be E_{ij}\equiv\frac12\mk{\dot g_{ij}-N_{i;j}-N_{j;i}}~, \qquad E=E^i_i=g^{ij}E_{ij}~. \ee
One can see that neither $N$ nor $N_i$ have temporal derivatives and therefore they are subject to dynamical constraints (the only dynamical variables are then $\phi$ and $g_{ij}$). Consequently, by varying the action \eqref{eq:admact} with respect to $N$ and $N^i$, we get the following constraint equations
\begin{alignat}{2} ^{(3)}R-2V-g^{ij}\phi_{,i}\phi_{,j}-N^{-2}\kk{E_{ij}E^{ij}-E^2+\mk{\dot\phi-N^i\phi_{,i}}^2}&=0~, \label{eq:constN1} \\ \notag \\\kk{N^{-1}\mk{E_j^i-E\delta^i_j}}_{;i}&=0~. \label{eq:constN2} \end{alignat}

Now that the splitting, \ie the foliation of the spacetime is evident, we introduce the metric and inflaton perturbations defined in Appendix \ref{app:pert}. For this, it is customary to choose the \emph{comoving} gauge to fix time and spatial reparametrizations.\footnote{See, \emph{e.g.}, Refs. \cite{Mukhanov:2005sc,Baumann:2009ds,Malik:2009aa} for relations in this and other gauges.} In this gauge, the inflaton perturbation $\delta\phi$ and $E$ vanish, and thus we adopt a coordinate system which \emph{moves} with the cosmic fluid; furthermore, most of the energy density is driven by the inflaton field during inflation, \ie $\delta\rho\sim\delta\phi$. A consequence of this is that the curvature perturbation on density hypersurfaces, $\zeta_\phi$, and the spatial curvature $\Psi$ relate as $\zeta_\phi\simeq-\Psi$ (see Eq.~\eqref{eq:zetaphi}) and, therefore, the perturbed spatial metric $g_{ij}$ in the comoving gauge reads as (see Eq.~\eqref{eq:A2})\footnote{We drop the subscript `$\phi$' as the distinction between $\zeta$ and $\zeta_\phi$ is not further neccesary.}
\be g_{ij}=a^2\kk{\mk{1+2\zeta}\delta_{ij}+h_{ij}}~, \ee
where we assumed that the vector perturbation $F_i$ is subdominant. Also, $h_{ij}$ is the only tensor perturbation and obeys the equation of a gravitational wave (see Eq.~\eqref{eq:eqprimgravwave}), \ie the generation of a background of primordial tensor modes $h_{ij}$ is equivalent to the generation of a background of primordial gravitational waves (primordial GW). This waves could polarize the CMB, as discussed in \S\ref{sec:CMBintro}.

We then expand the lapse and shift into background and perturbed quantities. Furthermore, the shift admits a helicity decomposition (see Appendix \ref{app:pert} for details) in such a way that we can write $N$ and $N^i$ as
\be N=\overline{N}+N_{(1)}~, \qquad \qquad N_i=\overline{N_i}+\chi_{(1),i}+\omega_{i(1)}~, \label{eq:lapseshift} \ee
to first order in perturbations.

Plugging Eqs.~\eqref{eq:lapseshift} into the constraint equations \eqref{eq:constN1}-\eqref{eq:constN2} we find to zero order the Friedmann equation \eqref{eq:conseq}, which means that it is a constraint equation and not an equation of motion. On the other hand, to first order in perturbations, we find that~\cite{Maldacena:2002vr}
\be N_{(1)}=\f{\dot\zeta}{H}~, \qquad \text{and} \qquad \chi_{(1)}=-\f{\zeta}{H}+a^2\f{\dot\phi^2}{2H^2}\partial^{-2}\dot\zeta~, \label{eq:lapseshift2} \ee
where $\partial^{-2}$ is defined through the relation $\partial^{-2}\mk{\partial^2\phi}=\phi$.

Finally, by expanding the action Eq.~\eqref{eq:admact} to first order in scalar perturbations and substituting Eqs.~\eqref{eq:lapseshift2} into it, we arrive to the quadratic action for scalar perturbations\footnote{This equation is popularly identified with the `(2)' superscript and called `quadratic', although it is composed with \emph{first}-order perturbations identified in this thesis with the `(1)' subscript.}
\be \mathcal{S}^{(2)}_\zeta=\frac12\int\dif^4xa^3\frac{\dot\phi}{H^2}\kk{\dot\zeta^2-a^{-2}\mk{\pa\zeta}^2}~. \ee

For tensor perturbations, the computation of the quadratic action is much simpler, given that we only have $h_{ij}$. The tensor perturbation can be decomposed into its polarization states as
\be h_{ij}=\gamma_+e^+_{ij}+\gamma_\times e^\times_{ij}~, \ee
and thus we only study the evolution of the scalar components $\gamma_+$ and $\gamma_\times$. The quadratic action for tensor perturbations then reads as
\be \mathcal{S}_\gamma^{(2)}=\sum_{\lambda=+,\times}\frac18\int\dif^4xa^3\kk{\dot\gamma^2_\lambda-a^{-2}(\partial\gamma_{\lambda})^2}~, \ee
where the sum is over the two polarization states.

\subsection{Quantization}

We define the scalar and tensor \emph{Mukhanov variables}, $u_s\equiv z_s\zeta$ and $u_t\equiv z_t\gamma$ with
\be z_s^2\equiv a^2\f{\dot\phi^2}{H^2}=2a^2\epsilon_H~, \qquad \qquad z_t^2\equiv\f{a^2}{2}~. \label{eq:zetas} \ee
In terms of these variables, the quadratic actions become
\be \mathcal{S}^{(2)}_p=\frac12\int\dif\tau\dif^3x\kk{(u'_p)^2-\mk{\partial u_{p}}^2+\f{z''_p}{z_p}u_p^2}~, \label{eq:quadraup} \\ \ee
where $p=s,t$ stands for either scalars or tensors. Also, we changed to conformal time and, therefore, from now on primes refer to derivatives with respect to $\tau$, unless otherwise stated.

In order to quantize the field $u_p$, we define its Fourier expansion as 
\be u(\tau,x^i)=\int\f{\dif^3k}{(2\pi)^3}u_{k_i}(\tau)e^{ik_ix^i}~, \label{Fourierexpu} \ee
where we omit here the subscript `$p$' in both $u(\tau,x^i)$ and $u_{k_i}(\tau)$ to simplify the notation. By varying the quadratic action Eq.~\eqref{eq:quadraup} with respect to $u_p$ one obtains the \emph{Mukhanov-Sasaki} equation in Fourier space as
\be u''_{p}+\mk{k^2-\f{z''_p}{z_p}}u_{p}=0~, \label{eq:MSeqdef} \ee
where here $u_p=u_k(\tau)$, from Eq.~\eqref{Fourierexpu}, after removing the vector subscript $i$ for the wavenumbers $k$, given that equation \eqref{eq:MSeqdef} depends only on their magnitude.

To specify the solutions of the evolution equation~\eqref{eq:MSeqdef} we first need to promote $u_p$ to a quantum operator in the standard way as
\be \hat{u}=\int\f{\dif k_i^3}{(2\pi)^3}\kk{u_k(\tau)\hat{a}_{k_i}e^{ik_ix^i}+u_k^*(\tau)\hat{a}^\dagger_{k_i}e^{-ik_ix^i}}~, \ee
where the creation and annihilation operators satisfy the usual commutation relation
\be \kk{\hat{a}_{k_i},\hat{a}_{k'_i}^\dagger}=(2\pi)^3\delta\mk{k_i-k'_i}~, \ee
only if the following normalization condition of $u_k$ and its conjugate momenta $\pi=u'_k$ is satisfied:
\be u'_ku_k^{*}-u_ku'^{*}_k=i~. \label{eq:bouneq1} \ee

Secondly, we need to choose a vacuum state. In the far past, \ie for $\tau\to-\infty$ (or, equivalently $k\gg aH$), Eq.~\eqref{eq:MSeqdef} becomes
\be u''_p+k^2u_p=0~, \ee
which is the equation of a (quantum) simple harmonic oscillator with time-independent frequency. It can thus be shown that the requirement of the vacuum state to be the state with minimum energy implies that~\cite{Mukhanov:2005sc}
\be u_p\mk{\tau\to-\infty}=\frac1{\sqrt{2k}}e^{-ik\tau}~, \label{eq:BDavies} \ee
which defines a unique physical vacuum---the \emph{Bunch-Davies vacuum}---and, along with Eq.~\eqref{eq:bouneq1}, completely fixes the mode functions.

\subsection{Solutions to the Mukhanov-Sasaki equation}
\label{sec:soluMukhSas}

The Mukhanov-Sasaki equation \eqref{eq:MSeqdef} is not simple to solve in general, as it depends on the specific inflationary background, encoded in $z_p$. For canonical inflation, \ie a background with a smooth inflaton evolution, one can simplify the $z_p''/z_p$ factor by assuming that the evolution is close to a \emph{de Sitter} phase and find analytic solutions by means of the SR approximation. Conversely, the background could not be smooth---features in the inflaton potential can be present---or can be given by a scalar-tensor theory different than GR. In these cases, different techniques must be used or numerical integration must be performed.

\subsubsection*{Quasi-de Sitter solution}

In de Sitter space where de Hubble parameter $H$ is constant, Eq.~\eqref{eq:MSeqdef} reduces to
\be u''_p+\mk{k^2-\f{2}{\tau^2}}u_p=0~, \label{eq:MSdeSitter}\ee
which, using the initial condition Eq.~\eqref{eq:BDavies}, has as solution
\be u_k(\tau)=\f{e^{-ik\tau}}{\sqrt{2k}}\mk{1-\f{i}{k\tau}}~, \label{eq:quasidS} \ee
which is the same solution for either scalars or tensors, so we dropped the subscript $p$.

Observations are to be compared with the spectrum of the primordial quantum fluctuations. In this case, the spectrum of $u_p$ is defined as
\be \langle\hat u_{k_i}(\tau),\hat u_{k'_i}\rangle=\mk{2\pi}^3\delta\mk{k_i+k'_i}\mathcal{P}_{u_p}(k)~, \ee
where $\mathcal{P}_{u_p}(k)\equiv|u_{k}(\tau)|^2$ is the \emph{power spectrum} of the variable $u_p$, while the dimensionless power spectrum, $\Delta^2(k)$, reads as
\be \Delta_{u_p}^2(k)\equiv\f{k^3}{2\pi^2}\mathcal{P}_{u_p}(k)~. \label{eq:genPS} \ee

Notice that on superhorizon scales, $|k\tau|\ll1$,
\be |u_k(\tau)|^2=\f{1}{2k^3\tau^2}\mk{1+k^2\tau^2}\simeq\f{a^2H^2}{2k^3}~, \ee
where, in the approximation, we took the de Sitter condition on the conformal time Eq.~\eqref{eq:dSdefi}. Furthermore, using the relations $\zeta=u_s/z_s$ and $\gamma=u_t/z_t$, we can compute the dimensionless power spectrum for the primordial scalar and tensor perturbations in quasi-de Sitter space, using therefore the solution given in Eq.~\eqref{eq:quasidS}, as
\begin{alignat}{2} \Delta_\zeta^2(k)&=\f{k^3}{2\pi^2}\biggr|\f{u_p(\tau)}{z_s}\biggr|^2=\f{H^2}{8\pi^2\epsilon_H}\biggr|_{k=aH}~, \label{eq:psquasidS} \\ \notag \\ \Delta_\gamma^2(k)&=\f{k^3}{2\pi^2}\biggr|\f{u_p(\tau)}{z_t}\biggr|^2=\f{H^2}{2\pi^2}\biggr|_{k=aH}~, \label{eq:psquasidSt} \end{alignat}
where it has been explicitly stated that they must be evaluated at horizon crossing $k=aH$. 

\subsubsection*{First-order in slow-roll solution}

We can take weaker restrictions for the $z''_p/z_p$ factor in Eq.~\eqref{eq:MSeqdef} if we expand it in slow-roll parameters. On the one hand, the tensor sector is not modified as $z_t^2=a/2$ does not contain any slow-roll parameter. On the other hand, the scalar factor $z''_s/z_s$ can be expanded as
\be \f{z''}{z}=a^2H^2\mk{2+2\epsilon_H+\epsilon_H^2+3\delta_1+4\delta_1\epsilon_H+\delta_2}~, \label{eq:zprim} \ee
where we dropped the subscript $s$ to make the notation simpler, and we employed the Hubble slow-roll parameter convention:
\allowdisplaybreaks
\begin{eqnarray} \label{osrsrparam}
\delta_1&\equiv&\frac12\diff{\,\logn\epsilon_H}{N}-\epsilon_H~, \qquad 
\delta_{2}\equiv\diff{\delta_1}{N}+\delta_1\mk{\delta_1-\epsilon_H}~.
\end{eqnarray}
Equation \eqref{eq:zprim} is exact, \ie no slow-roll hierarchy approximation has been used at that point (namely, we kept $\mathcal O(\epsilon_H^2)$ terms).\footnote{See \S\ref{sec:GSR} for details on the hierarchy of slow-roll parameters.}

To first order in SR approximation, where the quasi-de Sitter condition reads as
\be aH=-\frac1{\tau}(1+\epsilon_H)~, \qquad\qquad (\text{first order in SR}) \label{eq:qdS} \ee
Eq.~\eqref{eq:zprim} is reduced to
\be
\f{z''}{z}\simeq\frac1{\tau^2}\mk{2+6\epsilon_H+3\delta_1}\equiv\frac{\nu^2-\frac14}{\tau^2}~, \qquad (\text{first order in SR}) \label{eq:nu} \ee
where
\be \nu^2\equiv\frac94+6\epsilon_H+3\delta_1~, \quad \leftrightarrow \quad \nu\simeq\frac32+2\epsilon_H+\delta_1~. \ee
Hence, to first order in slow-roll parameters, the scalar Mukhanov-Sasaki equation \eqref{eq:MSeqdef},
\be u''_k+\mk{k^2-\f{\nu^2-\frac14}{\tau^2}}u_k=0~, \ee
can be recast as a Bessel equation and thus it has an exact solution given by
\be u_k(\tau)=\sqrt{-\tau}\kk{\alpha H_\nu^{(1)}\mk{-k\tau}+\beta H_\nu^{(2)}\mk{-k\tau}}~, \label{eq:uasHank} \ee
where $H_\nu^{(1)}$ and $H_\nu^{(2)}$ are the Hankel functions of the first and second kind, respectively. These functions are equal, $H_\nu^{(1)}(x)=H_\nu^{(2)}(x)$, for a real argument $x$ and have the following asymptotic limits:
\begin{alignat}{2} H_\nu^{(1)}(x\to\infty)&\simeq\sqrt{\f{2}{\pi x}}e^{i\kk{x-\mk{\nu+\frac12}\frac\pi2}}~,\\ 
H_\nu^{(1)}(x\to0)&\simeq-i\f{\mk{\nu-1}!}{\pi}\mk{\frac2x}^\nu=\sqrt{\frac2\pi}\mk{2^{\nu-\frac32}}\f{\Gamma(\nu)}{\Gamma\mk{\frac32}}x^{-\nu}e^{-i\frac\pi2}~. \end{alignat}
Therefore, in the far past $|k\tau|\to-\infty$, Eq.~\eqref{eq:uasHank} is written as
\be \ba u_k(\tau\to-\infty)&=\sqrt{\frac2\pi}\kk{\alpha\f{e^{-ik\tau}}{\sqrt{k}}+\beta\f{e^{ik\tau}}{\sqrt{k}}}~,\\ &=\sqrt{\frac\pi2}\sqrt{-\tau}H_v^{(1)}(-k\tau)~, \label{eq:BDalaHank} \ea \ee
where we dropped the factors $e^{\pm i\frac\pi2\mk{\nu+\frac12}}$ and, in the second line, we took $\alpha=\sqrt{\pi/2}$ and $\beta=0$ by comparison with Eq.~\eqref{eq:BDavies}, \ie Eq.~\eqref{eq:BDalaHank} fixes the Bunch-Davies mode functions to first order in the slow-roll parameters.

Finally, the dimensionless power spectrum, computed in the limit $k\tau\ll1$, reads as
\be \ba \Delta_\zeta^2(k)&=\f{k^3}{2\pi^2}\biggr|\f{u_k(\tau\to0)}{z}\biggr|^2\\ &=\f{2^{2\nu-3}}{(2\pi)^2}\kk{\f{\Gamma(\nu)}{\Gamma\mk{\frac32}}}^2\mk{\f{H}{-a\tau\dot\phi}}^2\mk{-k\tau}^{3-2\nu}\biggr|_{k=aH}~, \label{eq:dimpshank} \ea \ee
where one can notice that in the limit $\epsilon_H=\delta_1\to0$ (or, equivalently, $\nu\to3/2$), Eq.~\eqref{eq:dimpshank} reduces to Eq.~\eqref{eq:psquasidS} as expected.

\subsubsection*{Integral solutions}

In the previous approximate solutions, the validity of the slow-roll conditions, Eqs.~\eqref{eq:epscond} and \eqref{eq:etacond}, was assumed. The first condition, $\epsilon_H\ll1$, is required to not terminate inflation prematurely, whereas $|\delta_1|\ll1$~\footnote{Recall that $\eta_H=-\delta_1$.} enforces that $\epsilon_H$ evolves slowly and thus the only deviation from quasi-de Sitter is due to the end of inflation, which ensures that the SR approximation remains valid. However, in canonical inflation, the evolution of these parameters depends 
on the shape of the potential, meaning that an irregular potential---with features of some sort---would make one of the parameters increase before the end of inflation, violating the slow-roll conditions. This does not necessarily mean that inflation is terminated, but that the SR approximation cannot be trusted. This has become an issue as more models with features in the potential have become popular due to their particular signatures in the power spectrum. For such cases, numerical integration of Eq.~\eqref{eq:MSeqdef} has been usually performed.

Alternatively, new techniques to solve the Mukhanov-Sasaki equation in a semi-analytical way have been developed to overcome the deficiencies of the SR approximation. In \S\ref{sec:GSR} we will carefully review two powerful methods: the \emph{Generilized slow-roll} (GSR)~\cite{Dvorkin:2009ne,Adshead:2011bw,Adshead:2011jq,Hu:2011vr,Miranda:2012rm,Miranda:2013wxa,Adshead:2013aa,Hu:2014hoa,Miranda:2014wga,Obied:2018qdr}, and the \emph{Optimized slow-roll} (OSR)~\cite{Motohashi:2015hpa,Motohashi:2017gqb} approximations. The former relies on an integral, iterative solution of Eq.~\eqref{eq:MSeqdef}, whereas the latter relies on analytical formulas in terms of the slow-roll parameters as in the standard SR approximation, but with a different and more accurate order counting of slow-roll parameters. In both cases, $\epsilon_H$ is still required to be small in amplitude, so inflation is not terminated, but its evolution can be as fast as the $e$folding scale. As we shall see in \S\ref{sec:GSR}, both techniques were recently extended to include the full Horndeski background described in \S\ref{sec:STintro} and to be detailed in \S\ref{sec:Hornsec}, making these methods even more powerful.

In the case in which neither the conditions for the SR approximation nor those for the GSR/OSR techniques are satisfied, direct numerical integration of Eq.~\eqref{eq:MSeqdef} is required, for each wavenumber $k$ and with the initial condition given by Eq.~\eqref{eq:quasidS}.

\subsection{Scale dependence, the amplitude of gravitational waves and current observational bounds}
\label{sec:bounds}

The scale dependence of the primordial spectra of scalar and tensor perturbations is quantified through the spectral indices
\be n_s-1\equiv\diff{\ln \Delta_\zeta^2}{\ln k}~, \qquad \qquad n_t\equiv\diff{\ln \Delta_\gamma^2}{\ln k}~. \ee
Equations \eqref{eq:psquasidS} and \eqref{eq:psquasidSt} allow to relate the spectral indices (sometimes called `primordial tilts') with the slow-roll parameters and thus, working to linear order, we can write them as
\begin{alignat}{2} n_s-1&=-4\epsilon_H-2\delta_1~, \qquad\qquad (\text{first order in SR}) \label{eq:ns}\\ n_t&=-2\epsilon_H~, \qquad\qquad\qquad~~ (\text{first order in SR}) \label{eq:nt} \end{alignat}
where the slow-roll parameters should be evaluated at some fixed scale $k_*$---usually being when CMB scales left the horizon, in order to compare the spectral indices with CMB observations.

Moreover, another parameters obtained from a further quantification of the scale-dependence of the scalar spectral index have been proved to be useful while testing models of inflation against observations~\cite{Escudero:2015wba,Cabass:2016aa,Bruck:2016aa}. In particular, the running of the scalar spectral index and its own running can be written, respectively, as
\be \alpha_s\equiv\diff{n_s}{\ln k}~, \qquad\qquad \beta_s\equiv\diff{\alpha_s}{\ln k}~, \label{eq:inflaparamalphabeta} \ee
and analogously for tensors. Notice that, by taking Eq.~\eqref{eq:ns}, $\alpha_s$ and $\beta_s$ can also be written in terms of the slow-roll parameters, and that the hierarchy of these parameters implies that $\alpha_s=\mathcal O(\epsilon_H^2)$ and $\beta_s=\mathcal O(\epsilon_H^3)$. Therefore, it is customary to parametrize the scalar spectrum in a power-law form, in terms of the scalar parameters, as
\be \Delta_\zeta^2(k)=A_s\mk{\f{k}{k_*}}^{n_s-1+\frac12\alpha_s\ln\mk{k/k_*}+\frac1{3!}\beta_s\ln^2\mk{k/k_*}}~, \label{eq:powerlawform}\ee
where $A_s=\Delta_\zeta^2\mk{k_*}$. \emph{Planck} 2018 measurements take a pivot scale of $k_*=0.05$ Mpc$^{-1}$, for which the scalar power spectrum amplitude $A_s$ is measured as
\be A_s=(2.0989\pm0.014)\times10^{-9}~, \ee
at 68\% confidence level (CL), using the \emph{Planck} TT,TE,EE+lowE+lensing\footnote{Here, `TT', `TE' and `EE' stand for the combined likelihood using TT, TE, and EE spectra, introduced in \S\ref{sec:CMBintro}, at $\ell\geq30$; `lowE' stands for the low-$\ell$ temperature-only likelihood plus the low-$\ell$ EE-only likelihood in the range 2$\leq\ell\leq29$; and `lensing' stands for the \emph{Planck} 2018 lensing likelihood which uses the lensing trispectrum to estimate the power spectrum of the lensing potential~\cite{Collaboration:2018aa}.} constraints~\cite{Collaboration:2018aa} (we shall take the same constraints throughout this thesis unless otherwise stated). At this $k_*$, the measurements on the scalar parameters then read
\begin{alignat}{3} n_s&=0.9625\pm0.0048~ \label{eq:ns obs} ,\\ \alpha_s&=0.002\pm0.010~,\\ \beta_s&=0.010\pm0.013~, \end{alignat}
at 68\% CL, which is consistent with the expectations of the slow-roll hierarchy. 

Additionally, the amplitude of tensor perturbations is parametrized through the \emph{tensor-to-scalar ratio} as
\be r\equiv\f{4\Delta_\gamma^2}{\Delta_\zeta^2}~, \label{eq:ttsr} \ee
where the factor 4 comes after taking into account the two different polarizations of tensor modes. Using Eqs.~\eqref{eq:psquasidS} and \eqref{eq:psquasidSt}, one can see that the tensor-to-scalar ratio can be related to the slow-roll parameters, in which case, using the quasi-de Sitter approximation, it reads as $r=16\epsilon_H$. By using Eq.~\eqref{eq:nt}, it is straightforward to see that the tensor-to-scalar ratio is related to the tensor tilt as
\be r=-8n_t~, \label{eq:rnt} \ee
which is called the \emph{consistency relation}. Any deviation from Eq.~\eqref{eq:rnt} would be a signature of noncanonical inflationary physics.

\begin{figure}[t]
\begin{center}
\includegraphics[keepaspectratio, width=14cm]{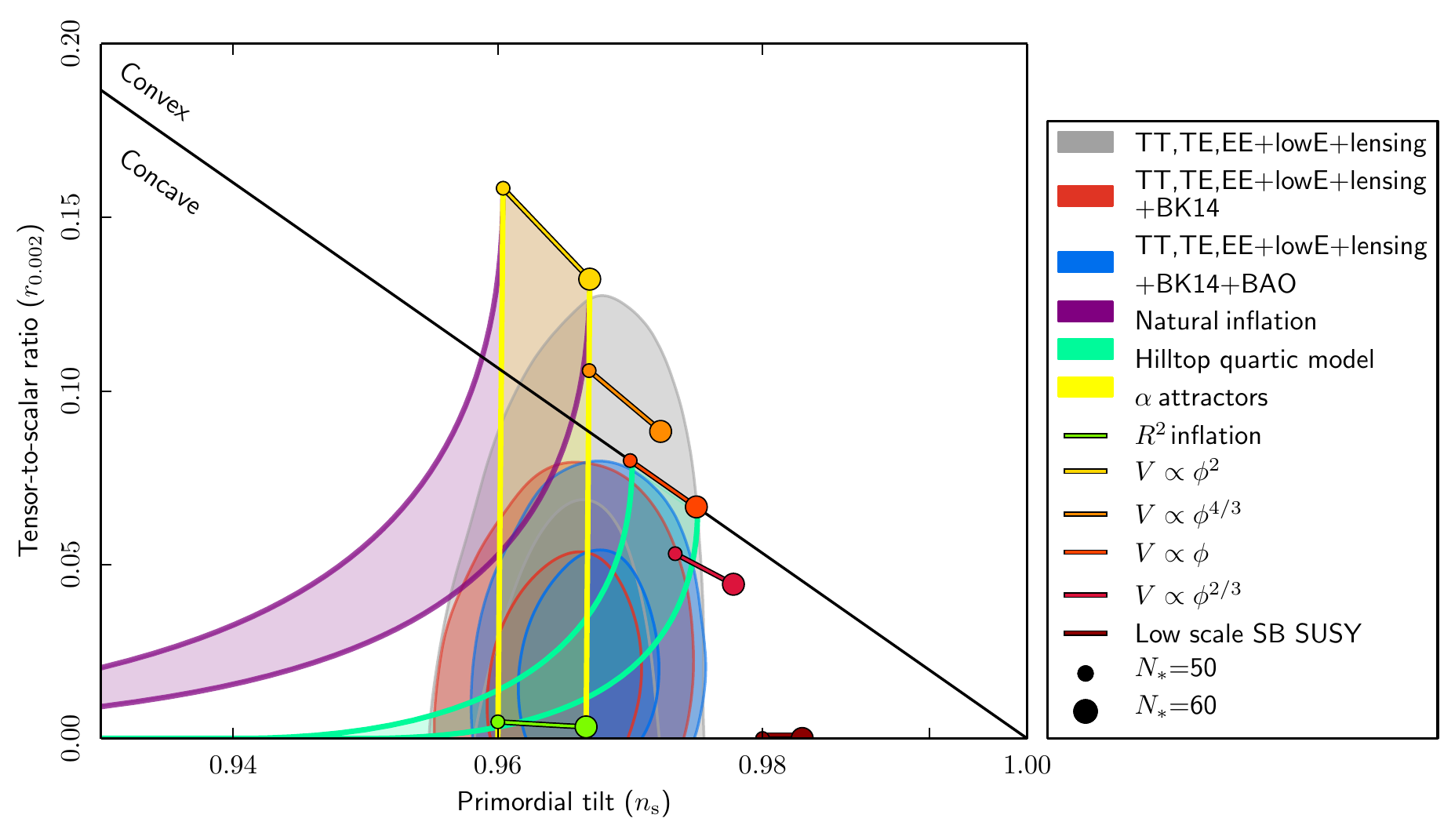}
\end{center}
\caption{\label{fig:planckmodels} \emph{Planck} 2018 constraints on the scalar spectral index $n_s$ and the tensor-to-scalar ratio $r$ at $k_*=0.002$ Mpc$^{-1}$ from \emph{Planck} measurements alone and in combination with BK14 or BK14+BAO data. The 68\% and 95\% CL regions are shown and compared to the theoretical predictions of selected inflationary models. Adapted from \cite{Collaboration:2018aa}.}
\end{figure}

Figure \ref{fig:planckmodels}, adapted from \cite{Collaboration:2018aa}, shows the 68\% and 95\% CL constraints coming from \emph{Planck} TT,TE,EE+lowE+lensing measurements alone and also from the combined BICEP2/Keck Array 2014 polarization data~\cite{Array:2015xqh}. Superimposed are several inflationary models, all of them reviewed in \S\ref{sec:modelsofinfl},\footnote{Notice that the monomial potentials $\phi^p$ fall into the class of chaotic inflation given by Eq.~\eqref{eq:monpot}.}  and the theoretical line separating concave- and convex-shaped potentials where one can see that the latter are in tension with observations. Notice that all the predictions for the theoretical models are shown for a window of values of the number of $e$folds $N_{\rm CMB}=50-60$, this given the uncertainty in Eq.~\eqref{eq:efoldsreh}. Furthermore, notice that the tensor-to-scalar ratio is consistent with a negligible amplitude of primordial GW, being the current upper bound
\be r<0.064~, \label{eq:rbound} \ee
at $k_*=0.002$ Mpc$^{-1}$, using \emph{Planck} TT,TE,EE+lowE+lensing+BK14 constraints. This comes from the fact that no signal of $B$-mode polarization generated by the primordial GW has yet been detected. Future experiments as, \emph{e.g.}, CORE (a CMB space satellite~\cite{Finelli:2016cyd,Remazeilles:2017szm,Delabrouille:2017rct}), could be able to improve the current sensitivity to $r$.


\begin{figure}[t]
\begin{center}
\includegraphics[keepaspectratio, width=13cm]{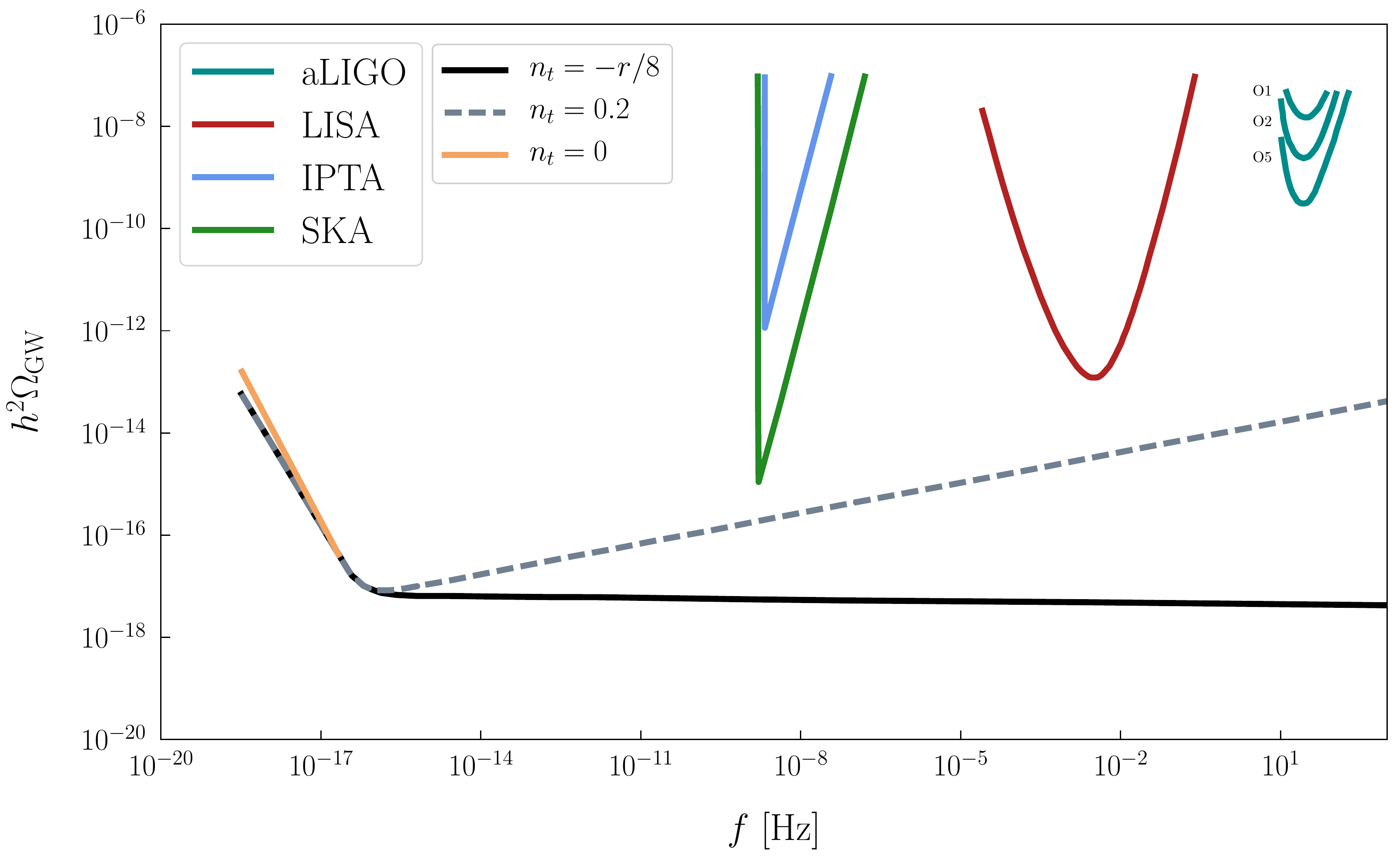}
\end{center}
\caption{\label{fig:GWspectrum} Gravitational-wave sensitivity curves for different detectors: Advanced LIGO~\cite{2015,:2016ab}---showing the first (O1) and second (O2) runs and at designed sensitivity (O5)---, LISA~\cite{Amaro-Seoane:2017aa}, IPTA~\cite{Hobbs_2010}, and SKA~\cite{KRAMER2004993}; along with the GW energy density, given by Eq.~\eqref{eq:omegaGW}, of three different inflationary scenarios: canonical inflation given by the consistency relation (solid black), a blue-tilted model (dashed gray) and a flat spectrum (orange), all of them by taking the saturated bound $r=0.064$.}
\end{figure}


Additionally, current GW experiments, as the ground-based interferometer aLIGO~\cite{2015,:2016ab}, have proved their efficacy in measuring GW coming from astrophysical processes~\cite{Abbott:2016blz,Abbott:2016nmj,TheLIGOScientific:2016pea,GBM:2017lvd,TheLIGOScientific:2017qsa,Abbott:2017oio,Abbott:2017gyy,Abbott:2017vtc,Monitor:2017mdv,LIGOScientific:2018mvr}. Even though the main goal of these interferometers is to measure astrophysically-sourced GW, a detection of primordial GW is potentially viable. Figure \ref{fig:GWspectrum} shows the gravitational wave-sensitivity curves of several experiments able to detect GW~\cite{Moore_2014,Caprini:2018mtu}. Theoretically, for primordial GW, the GW spectrum is given in terms of the primordial scalar spectrum and the tensor-to-scalar ratio as~\cite{Caprini:2018mtu}
\be \Omega_{\rm GW}(f)=\mk{\frac3{128}}\Omega_{\rm rad}\Delta_\zeta^2\mk{\f{f}{f_*}}^{n_t}\kk{\frac12\mk{\f{f_{\rm eq}}{f}}^2+\frac49\mk{\sqrt{2}-1}}r~, \label{eq:omegaGW} \ee
where $f_*=k_*/(2\pi\,a_0)$ is the pivot frequency related to the pivot scale $k_*$ and $f_{\rm eq}=H_0\Omega_m/\mk{\pi\sqrt{2\Omega_{\rm rad}}}$ is the frequency entering the horizon at matter-radiation equality. Using Eq.~\eqref{eq:omegaGW}, one can compute the predictions of a given inflationary model in terms of the tensor tilt $n_t$ as shown in Fig.~\ref{fig:GWspectrum}. By taking the bound in Eq.~\eqref{eq:rbound} as a true value for $r$, it is shown the theoretical predictions for three different scenarios: the consistency relation in canonical inflation, Eq.~\eqref{eq:rnt}, a blue-tilted scenario with $n_t=0.2$ and a flat spectrum ($n_t=0$). In particular, some inflationary scenarios, belonging to the classes discussed in \S\ref{sec:STintro}, predict a large-enough blue tilt of tensor perturbations which could reach future interferometer sensitivities~\cite{Cook:2011hg,Senatore:2011sp,Mylova:2018yap}.

To finish this section, let us notice that although we do not yet know the true nature of inflation, we can still estimate the energy scale at which it took place given the bound \eqref{eq:rbound}. Recall that $\Delta_t^2\propto H^2$ and, because of the SR approximation, $H^2\propto V(\phi)$.  Hence, from Eq.~\eqref{eq:ttsr}, the energy at CMB scales was approximately
\be E\equiv V^{1/4}\simeq\mk{\f{3\pi^2}{2}r\Delta_\zeta^2}^{1/4}\Mpl<7\times10^{-3}\Mpl~, \ee
\ie the final stages of inflation occurred at sub-Planckian energy densities (if Nature chose canonical slow-roll inflation).


\lhead[{\bfseries \thepage}]{ \rightmark}
\rhead[ Chapter \thechapter. \leftmark]{\bfseries \thepage}
\chapter{Model-independent approaches}
\label{sec:MIApp}

Whenever a new well-motivated model of inflation is found, its dynamics must be tested against observations, as discussed in \S\ref{sec:bounds}. While this is the standard approach in \emph{testing} inflation, not much information (if any) about the inflationary period is gained. For instance, on the one side, it may be argued that canonical chaotic inflation is in tension with the data and that this particular class of models could be ruled-out in the near future (see Fig.~\ref{fig:planckmodels}). On the other side, it would be a strong claim to say that inflation is noncanonical.

Furthermore, from the observational side, the consistency relation of canonical inflation, Eq.~\eqref{eq:rnt}, may be more challenging to test than expected. With the final results of the \emph{Planck} satellite already released~\cite{Collaboration:2018aa}, a new generation of experiments is required for an improvement on the measurement (or, in the absence of a signal, on the upper limit) on the amplitude of primordial gravitational waves through the tensor-to-scalar ratio $r$. Moreover, the subsequent measurement of its scale dependence, $n_t$, entails an additional experimental challenge.

With all these considerations, it is desirable to look for more robust ways to formulate inflation by capturing its generic features without committing to a specific model. Such features may be simply given by the conditions for inflation, Eqs.~\eqref{eq:condforinf}, and the required amount of inflation to solve the inflationary problems, Eq.~\eqref{eq:ncmb}, consistent with CMB and LSS data. Indeed, a model independent approach, developed by V. Mukhanov~\cite{Mukhanov:2013tua}, relies on these conditions by parametrizing the equation of state $w$ during inflation.

\section{Mukhanov parametrization}
\label{sec:mukpar}

As discussed in \S\ref{sec:condforinf}, the equation of state during inflation is given by (see Eq.~\eqref{eq:epome})
\be \omega\equiv\frac p\rho=-1+\frac23\epsilon_H~, \label{eq:omegainfMP} \ee
\ie well before the end of inflation, $\epsilon_H\to0$, the Universe is driven by a fluid with negative pressure, $p=-\rho$. On the other hand, $\omega$ approaches to $-1/3$ when $\epsilon_H\to1$. 

This behavior can be instead parametrized in terms of the number of (remaining) $e$folds of inflation $\mathcal N$.\footnote{\ie $\mathcal N$ approaches to 0 as $\epsilon_H\to1$.} One can thus propose the ansatz~\cite{Mukhanov:2013tua,Barranco:2014ira,Boubekeur:2014xva}
\be \omega=-1+\f{\beta}{\mk{\mathcal N+1}^\alpha}~, \label{eq:MPansatz} \ee
which reproduces the same aforementioned behavior 
for the positive and order-unity parameters $\alpha$ and $\beta$. However, more interesting is the fact that we can further parametrize the scalar tilt $n_s$ and the tensor-to-scalar ratio $r$ with the same parameters. Indeed, recall that $n_s-1=-4\epsilon_H-2\delta_1=-2\epsilon_H-\dif\ln\epsilon_H/\dif N$ to first order in slow-roll parameters, and in single-field canonical inflation (see \S\ref{sec:bounds}). Then, substituting Eq.~\eqref{eq:omegainfMP}, the scalar tilt can be written as
\be \ba \label{eq:nsMP}
n_s-1=&-3\mk{\omega+1}-\diff{}{N}\kk{\ln\mk{\omega+1}}\\
=&-\f{3\beta}{\mk{\mathcal N +1}^\alpha}-\f{\alpha}{\mathcal N+1}~, 
\ea \ee
where the ansatz~\eqref{eq:MPansatz} was applied in the second line. In the same manner, the canonical-inflation consistency relation, Eq.~\eqref{eq:rnt}, written as $r=16\epsilon_H$ to first order in slow-roll parameters, can be parametrized as
\be r=\f{24\beta}{\mk{\mathcal N+1}^\alpha}~. \label{eq:MPr} \ee
Notice then that Eqs.~\eqref{eq:nsMP} and \eqref{eq:MPr} provide generic, model-independent predictions for canonical single-field inflation. 
For instance, notice that for $\alpha>1$, the second term in Eq.~\eqref{eq:nsMP} dominates and $n_s$ approximates to $n_s-1\approx-\alpha/\mathcal N_*$ at CMB scales. Then, taking the central value of $n_s$ in Eq.~\eqref{eq:ns obs} and $\mathcal N_*\sim55$, it follows that $\alpha\lesssim2$. This in turn places a lower bound on $r$:
\be r\gtrsim\mk{6\times10^{-3}}\beta~, \label{eq:lboundMP} \ee
where, again, $\beta=\mathcal O(1)$. Consequently, we were allowed to predict a lower bound on the tensor-to-scalar ratio---assuming that inflation is driven by a canonical single field and for a given measured value of $n_s$---just by following the conditions for inflation provided the ansatz \eqref{eq:MPansatz}. Notice that the bound in Eq.~\eqref{eq:lboundMP} is one order of magnitude smaller than the current observational upper bound \eqref{eq:rbound} and thus it could be reached with future CMB experiments~\cite{Collaboration:2011aa}.

In addition, the behavior of the majority of single-field scenarios can be recovered for specific values of $\alpha$ and $\beta$ as discussed in Ref.~\cite{Mukhanov:2013tua}. To mention a few examples, chaotic inflation corresponds to $\alpha=1$ for which the observables read $n_s-1=-(3\beta+1)/(\mathcal N+1)$ and $r=24\beta/(\mathcal N+1)$; for $\alpha=2$ and $\beta=1/2$, $n_s-1\approx-2/\mathcal N$ and $r\approx12/\mathcal N^2$, which corresponds to the Starobisnky model, Eq.~\eqref{eq:Staromodel}. For further examples and details, see Ref.~\cite{Mukhanov:2013tua}.

Finally, let us mention that this hydrodynamical approach can be easily extended to $k$-inflation (see \S\ref{sec:kinflation}) where another two phenomenological parameters are required in order to parametrize the nontrivial sound speed of primordial perturbations as
\be c_s=\f{\gamma}{\mk{\mathcal N+1}^\delta}~. \ee
Here, $\delta\geq0$ but $\gamma$ is an arbitrary positive number \ie the sound speed grows towards the end of inflation. With this addition, the scalar tilt and the tensor-to-scalar ratio change to
\be n_s-1=-\f{3\beta}{\mk{\mathcal N +1}^\alpha}-\f{\alpha+\delta}{\mathcal N+1}~, \qquad\qquad\quad
r=\f{24\beta\gamma}{\mk{\mathcal N+1}^{\alpha+\delta}}~. \ee
Notice then that the lower bound \eqref{eq:lboundMP} can be further suppressed. On the other hand, it is well known that a too small sound speed induces large non-Gaussianities, for which $c_s$ cannot be much lower than 0.1~\cite{Mukhanov:2013tua,Gariazzo:2016blm}. In this case, the lower bound on $r$ would be further pushed one order of magnitude at maximum.

\lhead[{\bfseries \thepage}]{ \rightmark}
\rhead[ Chapter \thechapter. \leftmark]{\bfseries \thepage}
\chapter{Inflation beyond General Relativity}
\label{sec:STtheories}

We have seen that the simplest scenarios of inflation, \ie monomial potentials, are in tension with current CMB observations (see \S\ref{sec:bounds}). In this regard, the straightforward model-building approach is to consider more complicated potential functions $V(\phi)$ for the inflaton field $\phi$ which fit observations. Currently, our corresponding  approach relies on finding inflationary models that come from well-motivated high-energy theories of particle physics. However, one drawback of considering these models is the lack of simpler (and perhaps more natural) predictions of such a high-energy theory or, even worst, their failure on experimental confirmation.\footnote{Take for instance supersymmetry (SUSY), a very elegant solution to many problems in the Standard Model. 
It was developed in the 1970's and it is actively searched for mainly in accelerators. Hints of a minimal supersymmetric theory were already expected at the current working energies of the Large Hadron Collider (LHC) (see, \emph{e.g.}, Ref.~\cite{Draper:2011aa}); yet, many inflationary models based on SUSY are still being considered.}

Canonical inflation, being driven by a (new) quantum field, is also a theory of gravity based on General Relativity. Einstein's theory of GR is perhaps one of the most successful theories in physics. It has passed the most stringent solar-system tests and predicted several observations over the course of the last century, being one of its most amazing confirmations just achieved in 2015 with the measurement of the gravitational waves produced by a two black-hole merger~\cite{Abbott:2016blz,Abbott:2016nmj,TheLIGOScientific:2016pea,GBM:2017lvd,TheLIGOScientific:2017qsa,Abbott:2017oio,Abbott:2017gyy,Abbott:2017vtc,Monitor:2017mdv,LIGOScientific:2018mvr}. 

Yet, there are huge hints on the incompleteness of GR, the most intriguing one being posed by the dark energy (DE) issue. As discussed in \S\ref{sec:Introduction}, there is no natural explanation for the DE, although several proposals have been studied. In a similar way that for inflation, a new scalar field could be able to drive the current expansion, however, due to the low energy scale of the current acceleration, new difficulties arise when one tries to construct consistent particle physics models for this new field~\cite{Amendola:2015ksp}. A second approach relies on modifying GR at large scales such that these modifications are able to explain the accelerated expansion without modifying the local dynamics, where GR has been tested to be highly accurate. In fact, this approach is one of the most active research lines in Cosmology and has provided numerous kinds of modifications of GR 
to explain the nature of the DE.\footnote{The interested reader is referred to Ref.~\cite{Amendola:2015ksp} for a pedagogical review of the different approaches to DE.}

The same modifications of GR proposed to explain the DE, if realized during the early universe, are able to produce different predictions for the same inflationary potentials $V(\phi)$ previously studied. Indeed, there exist simple modifications of GR which lead to a better agreement of the monomial potentials, previously ruled-out, with the current CMB observations. This is in fact one common approach: instead of proposing complicated potential functions $V(\phi)$, we could keep the simplest realizations and just find well-motivated modifications of GR which predict observables satisfying current constraints. While this well-motivated modifications were originally kept simple too---as in the noncanonical models of inflation reviewed in \S\ref{sec:modelsofinfl}---in the recent years several classes and, as we shall see, frameworks of generalizations of GR have been worked out. Their study started with a simple question: \emph{what is the most general modification of GR, respecting its symmetries and principles, and which propagates only physical degrees of freedom?}

Indeed, the construction of such frameworks became itself a research line in the fields of gravitation and also of cosmology, where new terms and interactions between the inflaton and the gravity sector have been considered. 
In this chapter, we shall review the most popular frameworks of general modifications of GR involving new fields (scalars and vectors) coupled to the gravity sector. Indeed, by keeping its symmetries and constraints---namely Lorentz invariance, unitarity, locality and a (pseudo-)Riemannian spacetime---any modification of GR will inevitably introduce new degrees of freedom which could either be in the form of new scalar, vector or tensor fields~\cite{Heisenberg:2018aa}.\footnote{In other words, GR is the only Lorentz-invariant theory of gravity for a massless spin-2 particle.}

We shall follow a bottom-up approach: we start by reviewing the motivations for the construction of the aforementioned frameworks and the considerations one should take. Then, first, we shall review the scalar-tensor interactions which lead to the so-called \emph{Horndeski theory}; second, we move into the discussion of vector-tensor interactions which, in turn, lead to the \emph{generalized Proca theories}; third, we shall discuss the more general framework which aims to join the first two into the most general class of scalar-vector-tensor (SVT) interactions. In a second part, we shall review some of the most popular models of cosmological inflation that are developed within these general frameworks.\footnote{We would like to emphasize that these are mathematical frameworks rather than physical theories (as the name may suggests), \ie they only provide us with a full set of modifications of GR allowed by physical symmetries and other constraints, and not with a fixed set of fundamental laws of gravity.}

\section{Towards the most general SVT framework}
\label{sec:mostgenSVT}

General Relativity describes a theory of a massless spin-2 particle which propagates only two degrees of freedom as a result of constraints coming from the invariance under differentiable coordinate transformations---\emph{diffeomorphism invariance}. 
As already stated, any modification of GR will introduce new degrees of freedom in the form of scalar, vector or tensor fields. Take for instance the addition of a mass term for a gravitational field $h_{\mu\nu}$: Lorentz invariance restricts the metric combinations allowed for the mass term to be proportional to\footnote{Only scalar combinations of the metric are allowed. In this regard, $h_{\mu\nu}^2=h_{\mu\nu}h^{\mu\nu}$ and $h^2=h_\mu^\mu h_\nu^\nu$.}
\be
m^2\mk{h_{\mu\nu}^2-h^2}~,
\ee
where $h_{\mu\nu}$ is a symmetric tensor field. The presence of this mass term makes the theory no longer diffeomorphism invariant and thus more degrees of freedom, apart of the two tensor polarizations, must propagate. This symmetry, however, can be restored by means of the \emph{Stueckelberg trick}, a field redefinition $h_{\mu\nu}\to h_{\mu\nu}+2\chi_{\mk{\mu,\nu}}$ which introduces additional Stueckelberg fields $\chi^\alpha$, and which transforms the mass term into\footnote{$\chi_{\mk{\mu,\nu}}\equiv\mk{\pa_\mu\chi_\nu+\pa_\nu\chi_\mu}/2$~.}
\be
m^2\kk{\mk{h_{\mu\nu}+2\chi_{\mk{\mu,\nu}}}^2-\mk{h+2\chi^\alpha_{~,\alpha}}^2}~.
\ee
Furthermore, the Stueckelberg field can be split into its transverse and longitudinal parts $\chi^\alpha\to A^\alpha+\partial^\alpha\pi$ in order to make the degrees of freedom explicitly shown. In this way, the mass term becomes
\be \ba
m^2\mk{h_{\mu\nu}^2-h^2}-m^2F_{\mu\nu}^2-2m^2\mk{h_{\mu\nu}A^{\mu,\nu}-h\pa_\mu A^\mu}-2m^2\mk{h_{\mu\nu}\pi^{,\mu\nu}-h\pa_\mu\pa^\mu\pi}~,
\ea \ee
\ie the Stueckelberg trick produced the kinetic terms proportional to $h_{\mu\nu}\pi^{,\mu\nu}$ and $F_{\mu\nu}^2=\mk{A_{\mu,\nu}-A_{\nu,\mu}}^2$ for the scalar field $\pi$ and the vector field $A_\mu$, respectively.\footnote{This happens after performing a canonical normalization $A_\mu\to A_\mu/m$ and $\pi\to\pi/m^2$. See \cite{deRham:2014zqa,Heisenberg:2018aa} for details.} Therefore, a fully consistent theory of \emph{massive gravity}---where diffeomorphism invariance is broken---propagates five degrees of freedom (compared to the two of GR): two tensor (helicity-2) modes from $h_{\mu\nu}$, two vector (helicity-1) modes from $A_\mu$~\footnote{The other two degrees of freedom from $A_\mu$ are removed by means of the gauge invariance.} and the scalar (helicity-0) mode $\pi$, the last two coming from the Stueckelberg field $\chi^\alpha$.

Also interesting, apart from the kinetic terms for the Stueckelberg fields, the mass term further produces interaction terms of the form
\be m^2\mk{h_{\mu\nu}\pi^{,\mu\nu}-h\pa_\mu\pa^\mu\pi}~, \qquad m^2\mk{h_{\mu\nu}A^{\nu,\mu}-h\pa_\mu A^\mu}~, 
\ee
\ie mixing terms between different helicity modes. Phenomenologically, it has been shown that the helicity-0 mode present in a consistent theory of massive gravity provides a self-accelerating solution and thus it could potentially explain DE~\cite{deRham:2010tw}; therefore, one would expect interesting cosmological implications from several different mixing combinations coming from more general modifications of GR.

These mixings can be classified as scalar-tensor, vector-tensor or scalar-vector interactions. In this sense, one could follow the theory-independent approach of constructing all the different possible combinations allowed by Lorentz invariance and further restrictions as locality and unitarity, and write down all the possible combinations between scalar and tensor modes, vector and tensor modes, and scalar and vector modes coupled to gravity. In doing so, one would notice that combinations of arbitrarily high-order derivatives are allowed; however, it is well known that a nondegenerate Lagrangian, with temporal derivatives higher than order one, yields equations of motion (EoM) higher than order two. This fact incorporates new pathologies: in the Hamiltonian picture, a Lagrangian of this kind yields a Hamiltonian which is unbounded from below and thus the energy of the system in consideration can take either positive or negative values, \ie it can excite either positive or negative degrees of freedom. A negative degree of freedom of this type is known as \emph{Ostrogradsky instability} or, colloquially, \emph{Ostrogradsky's ghost}.\footnote{See Footnote 2 in Ref. \cite{Motohashi:2015aa} for a discussion on the terminology of the instability and the associated theorem.} We can therefore state \emph{Ostrogradsky's theorem} as: \emph{`Higher-derivative theories contain extra degrees of freedom, and are usually plagued by negative energies and related instabilities}'. Consequently, and in order to maintain a pathologically-free theory of gravity, any modification of GR, involving higher derivatives, must still yield second-order EoM.\footnote{This condition is in fact respected in Nature, as no higher order EoM describe physical phenomena---for instance, Newton's, Maxwell's and, again, Einstein's equations are all of them of second-order.} We shall see that this is achieved by imposing constraints which allow us to remove the \emph{ghostly} terms from the EoM.

\subsection{Scalar-tensor interactions}
\label{sec:STinst}

The simplest terms allowed in a scalar-tensor theory are given in the Lagrangian for a scalar field minimally coupled to GR (shown in Eq.~\eqref{eq:EinHil}), namely those of a canonical kinetic energy, $X\equiv-\frac12g^{\mu\nu}\phi_{,\mu}\phi_{,\nu}$, and a potential energy, $V(\phi)$---this term being already a generalization of the even simpler $m^2\phi^2$ mass term. This theory is of first order in derivatives and thus it propagates only real (positive-energy) fields.

In order to construct more general terms, the first natural step relies in combining the canonical kinetic and potential terms into a general function of the field, $f(X,\phi)$. Theories of this type, known as \emph{k-essence}, have been widely considered in the context of both DE and inflation (named as \emph{k-inflation} in the latter context). In particular, terms such as $f(\phi)X+G(\phi)X^2$ or $G(\phi)\sqrt{1-f(\phi)X}$ show up naturally in models inspired from string theory and supersymmetry realizations (see Ref.~\cite{Amendola:2015ksp} and references therein for details). Indeed, several new interactions have been discovered in the context of higher-dimensional theories. Another example comes from the Dvali-Gabadadze-Porrati (DGP) model of brane cosmology where in a 3+1 spacetime, embedded in a 4+1 dimensional Minkowski space, the graviton helicity-0 mode appears with a self-interaction term, $\Box\phi\mk{\partial\phi}^2$~\footnote{We define the d'Alambertian operator in the standard way: $\Box\phi\equiv\nabla^\mu\nabla_\mu\phi:=\phi^{;\mu}_{~;\mu}$. Notice however that the interaction present in the DGP model comes from a theory in flat spacetime and, moreover, a covariant derivative acting on a scalar quantity is just a partial differentiation, \ie we are still writing partial derivatives.}, able to drive an accelerated expansion~\cite{Dvali:2000hr}. Notice that this term contains two derivatives acting on the scalar field $\phi$, however, its EoM, ${\Box\phi}^2-\mk{\phi_{,\mu\nu}}^2=0$, are still second-order and thus the model avoids the Ostrogradsky instability. Following this spirit, one could carefully construct higher-order derivative terms which, by means of some particular constraints which remove the higher-order terms, yield second-order equations of motion. This task led to the development of the Galileon theories---a general scalar-tensor theory in flat spacetime, with second-order EoM, which is invariant under the Galilean transformations $\phi\to\phi+x_\mu b^\mu+c$~\cite{Nicolis:2008in}. The generalization to a nonflat spacetime was named \emph{covariant Galileons}~\cite{Deffayet:2009wt}, now known as \emph{Horndeski theory}~\cite{Horndeski:1974wa}.\footnote{In 1974, Gregory Horndeski precisely studied the most general four-dimensional scalar-tensor theory of gravity which yield second-order EoM. His work was rather unnoticed until its rediscovery as the covariant Galileons.}

\subsubsection{Horndeski theory}
\label{sec:Hornsec}

It is possible to construct a theory order by order in derivatives following the generalization procedure mentioned above. We can write the Lagrangians $\mathcal{L}_2=G_2(\phi,X)$ and $\mathcal{L}_3=G_3(\phi,X)\Box\phi$ where the subscript makes reference to the number of times the field $\phi$ appears. The fourth Lagrangian allows two types of interactions: $\alpha_1G_4(\phi,X)\mk{\Box\phi}^2$ and $\alpha_2G_4(\phi,X)\mk{\phi_{,\mu\nu}}^2$; however, by inspection of the EoM, one notices that in order to maintain only second-order terms, the constraint $\alpha_1=-\alpha_2$ must be satisfied---this then fixes the form of $\mathcal{L}_4$. Following this procedure, one finds that as long as we restrict ourselves to second-order EoM in four dimensions, only four Lagrangians can be written down, \ie up to $\mathcal{L}_5$. Next, we shall promote the partial derivatives to covariant derivatives and thus \emph{covariantize} the theory. In doing so, the number of derivatives increases and therefore the order of the EoM changes accordingly. The correct order is recovered by introducing nonminimal couplings to gravity into the Lagrangians at the desired order (see \cite{Heisenberg:2018aa} for details).

The application of the previous algorithm leads to the full Lagrangian of the Horndeski theory which is then given by four Lagrangians, $\mathcal{L_H}=\sum_{i=2}^5\mathcal{L}_i,$ each of them proportional to an arbitrary function $G_i(\phi,X)$:
\begin{align} \label{eq:Horn}
\mathcal{L_H}=&~G_2 \notag\\ &+G_3\Box\phi \notag\\ &+G_4R+G_{4,X} [ (\Box\phi)^2 -\phi^{;\mu\nu}\phi_{;\mu\nu}] \\
&+G_5 G^{\mu\nu} \phi_{;\mu\nu}- \f{G_{5,X}}{6} [ (\Box\phi)^3 - 3 (\Box\phi) \phi_{;\mu\nu}\phi^{;\mu\nu} + 2 \phi_{;\mu\nu}\phi^{;\mu\sigma} {\phi^{;\nu}}_{\!;\sigma}]~, \notag \end{align} 
where $G_{i,X}\equiv \pa G_i/\pa X$, $R$ is the scalar curvature and $G_{\mu\nu}=R_{\mu\nu}-\frac12g_{\mu\nu}R$ the Einstein tensor. Notice that the canonical Lagrangian, Eq.~\eqref{eq:EinHil}, is recovered for the choice $G_2=X-V(\phi)$, $G_4=1/2$, and $G_3=G_5=0$. Equation~\eqref{eq:Horn} represents therefore the most general theory of gravity involving scalar and tensor fields which yields second-order EoM and is free of Ostrogradsky ghosts.

Horndeski gravity is not however the most general theory of gravity free from instabilities. It is now known that having second-order EoM as a condition for the avoidance of Ostrogradsky instabilities is actually not necessary as long as there exists an additional constraint equation which helps to remove the higher-order terms. This inspired the construction of the \emph{Degenerate Higher-Order Scalar-Tensor} (DHOST) theories of gravity which are now the most general theories of gravity, at cubic order in second-order derivatives, with additional primary constraints ensuring the propagation of only three physical degrees of freedom~\cite{Langlois:2015cwa,Langlois:2015skt,BenAchour:2016fzp,Achour:2016rkg}. In the rest of this thesis, we shall however restrict ourselves to the phenomenology of the Horndeski theory for simplicity.

\subsection{Vector-tensor interactions}

We are now in the pursuit of the most general theory of a spin-1 field $A_\mu$, coupled to gravity, yielding second-order EoM, \ie propagating only real vector and tensor modes. As we shall see, the total number or physical degrees of freedom will depend on whether we restrict ourselves to maintain the \emph{gauge} symmetry or not---equivalently, whether we allow the field to be massive. Both theories provide new interesting phenomenology and thus one has the freedom to choose either one. Nevertheless, each case is constructed in the same spirit as the covariant Galileons were obtained: we need to write down all possible combinations order by order by respecting the second-order EoM condition, then covariantize the theory by promoting the partial derivatives to covariant ones and reduce to the correct order by introducing nonminimal couplings to the gravity sector.

\subsubsection{Maxwell theory}

For a massless U(1) field $A_\mu$, the allowed interactions linear in derivatives have the form $\alpha_1\partial_\mu A_\nu\pa^\mu A^v+\alpha_2\pa_\mu A_\nu\pa^\nu A^\mu$. The number of propagating degrees of freedom is fixed by the existence of a primary constraint that imposes $\alpha_1=-\alpha_2$ which makes the temporal mode $A_0$ nondynamical. Furthermore, $\alpha_1<0$ must be satisfied in order to ensure that the Hamiltonian is bounded from below (see~\cite{Heisenberg:2018aa} for details). These conditions generate a gauge symmetry which further removes the longitudinal mode. Consequently, we obtain a Lorentz invariant theory of a massless spin-1 field invariant under gauge transformations, $A_\mu\to A_\mu+\pa_\mu\theta$ (where $\theta$ is a real arbitrary constant) which guarantees that only two vector degrees of freedom propagate. This theory is nothing but the Maxwell's theory of electromagnetism (in absence of external currents)
\be \mathcal{L}_{A_\mu}^{\rm Maxwell}=-\frac14F_{\mu\nu}^2~, \qquad\qquad\qquad F_{\mu\nu}\equiv\pa_\mu A_\nu-\pa_\nu A_\mu~, \ee
after canonically normalizing the vector field by setting $\alpha_1=-1/2$. Similar to the Galileons case, one might look for higher-order self-interactions; however, a no-go theorem states that it is the Maxwell kinetic term the only possible combination yielding second-order EoM for an Abelian vector field as long as we restrict ourselves to keep the gauge symmetry~\cite{Deffayet:2010zh,Deffayet:2013tca,Deffayet:2016von,Deffayet:2017eqq}.

By promoting the partial derivatives to covariant ones, nonminimal couplings are required as in the scalar-tensor case. Additionally, in order to preserve gauge invariance, only couplings of the field strength $F_{\mu\nu}$, and not direct couplings of the vector field, must be considered. In this case, it can be shown that $F_{\mu\nu}$ can only couple to the double dual Riemann tensor $L^{\mu\nu\alpha\beta}=\frac{1}{4}\mathcal{E}^{\mu\nu\rho\sigma}\mathcal{E}^{\alpha\beta\gamma\delta} R_{\rho\sigma\gamma\delta}$, where $\mathcal{E}^{\mu\nu\alpha\beta}$ is the antisymmetric Levi-Civita tensor satisfying the normalization $\mathcal{E}^{\mu\nu\alpha\beta}\mathcal{E}_{\mu\nu\alpha\beta}=-4!$. Consequently, the most general Lagrangian for a massless vector field on curved spacetime yielding second-order EoM is given by~\cite{doi:10.1063/1.522837,Barrow:2012ay,Jimenez:2013qsa}
\be \label{eq:genmax} \mathcal{L}_{\rm Maxwell}=\frac12R-\frac14F_{\mu\nu}^2+\frac1{4M}L^{\alpha\beta\gamma\delta}F_{\alpha\beta}F_{\gamma\delta}~, \ee
where $M$ is the relevant mass scale.

\subsubsection{Proca theory}

The Proca theory describes a massive U(1) vector field. The mass term proportional to $A^\mu A_\mu$ breaks the gauge symmetry and therefore one degree of freedom more is allowed to propagate---three in total. Nevertheless, as in the massive gravity case, the gauge symmetry can be restored using the Stueckelberg trick by means of the change of variables $A_\mu\to A_\mu+\pa\phi$, where $\phi$ is a scalar Stueckelberg field. Under this change, the Proca theory becomes
\be \label{eq:proca} \mathcal{L}_{A_\mu}^{\rm Proca}=-\frac14F_{\mu\nu}^2-\frac12m_AA_\mu^2-\frac12\mk{\pa\phi}^2-m_AA_\mu\pa^\mu\phi~, \ee
where we have canonically normalized the scalar field as $\phi\to\phi/m_A$. Now, Eq.~\eqref{eq:proca} in invariant under simultaneous gauge, $A_\mu\to A_\mu+\pa_\mu\phi$, and shift, $\phi\to\phi-\theta$, symmetries and, more interestingly, the Stueckelberg trick produced an interaction term between the vector field and the scalar Stueckelberg field, where the latter comes with a kinetic term. Therefore, associating $\phi$ to the longitudinal vector mode, the third degree of freedom is explicitly shown---indeed, the change of variables $A_\mu\to A_\mu+\pa\phi$ can be seen as a helicity decomposition of the vector field.

Unlike the massless case, the Proca theory allows for more general interactions made by higher-order derivatives and thus avoiding the aforementioned no-go theorem. Then, to construct Galileon vector theories, called \emph{generalized Proca} in the Literature, we keep the second-order EoM restriction and add a second restriction: the temporal mode $A_0$ most remain nondynamical, otherwise it would unavoidably be a ghost mode. The algorithm is similar to the one previously discussed for scalar Galileons and therefore we shall focus on the covariantized version (see~\cite{Heisenberg:2014rta,Allys:2015sht,Jimenez:2016isa,Heisenberg:2018aa} for details). By replacing partial derivatives with covariant ones and introducing the corresponding nonminimal couplings, the generalized Proca theories in curved space become:
\begin{align} \label{eq:genProca}
\mathcal{L}_{\rm Proca}=&~G_2(X,F,Y)\notag\\
& +G_3(X)A^\mu_{~;\mu}\notag\\
& +G_4(X)R+G_{4,X}(X)\kk{(A^\mu_{~;\mu})^2-A_{\rho;\sigma}A^{\sigma;\rho}}\notag\\
& +G_5(X)G_{\mu\nu}A^{\mu;\nu}-\frac16G_{5,X}(X)\bigl[(A^\mu_{~;\mu})^3-3A^\mu_{~;\mu}A_{\rho;\sigma}A^{\sigma;\rho}\notag\\
&~~~+2A_{\rho;\sigma}A^{\sigma;\gamma}A_\gamma^{~;\rho}\bigr]-\tilde{G}_5(X)\tilde{F}^{\alpha \mu}\tilde{F}^\beta_{~\mu}A_{\beta;\alpha} \notag \\
& +G_6(X)L^{\mu\nu\alpha\beta}A_{\nu;\mu}A_{\beta;\alpha}+\frac12 G_{6,X}(X)\tilde{F}^{\alpha\beta}\tilde{F}^{\mu\nu}A_{\mu;\alpha}A_{\nu;\beta}~,~
\end{align}
where $\tilde{F}^{\mu\nu}=\mathcal{E}^{\mu\nu\alpha\beta}F_{\alpha\beta}/2$ is the dual of the strength tensor, and we explicitly showed the dependence of the $G_i$ functions in terms of the quantities
\be \label{eq:vectcomb} X=-\frac12 A_\mu A^\mu~,\qquad F=-\frac14F_{\mu\nu}F^{\mu\nu}~,\qquad Y=A^\mu A^\nu F_\mu^{~\alpha}F_{\nu\alpha}~. \ee
Consequently, Eq.~\eqref{eq:genProca} is the most general theory of gravity with a vector field $A_\mu$ yielding second-order EoM, \ie propagating only real fields---two tensor modes, two transverse vector modes and the longitudinal mode. These theories have brought important new phenomenology in the study of DE~\cite{Tasinato:2014eka,Tasinato:2014mia,DeFelice:2016uil,DeFelice:2016yws,deFelice:2017paw} and compact objects as black holes and neutron stars~\cite{Cisterna:2016nwq,Minamitsuji:2016ydr,DeFelice:2016cri,Fan:2016jnz,Chagoya:2016aar,Heisenberg:2017xda,Heisenberg:2017hwb,Chagoya:2017fyl}.

Beyond Generalized Proca theories have been constructed in the same spirit as beyond Horndeski theories. We shall not discuss them in this thesis but the interested reader is referred to Refs.~\cite{Heisenberg:2016eld,Heisenberg:2018aa}.

\subsection{Scalar-vector-tensor interactions}
\label{sec:SVTtheories}

Recall that the Stueckelberg trick performed to the Proca theory, Eq.~\eqref{eq:proca}, produced a kinetic term for the scalar field $\phi$ and a genuine new interaction between $\phi$ and $A_\mu$. Nothing has been said about this scalar field, however it can lead to interesting dynamics while being coupled to the Proca vector field in a gravitational background. Bearing this in mind, it is interesting to consider different kind of combinations between these two helicity modes and to construct a general theory of gravity with both scalar and tensor fields. As in the case of the Proca theories, one can construct interactions depending on whether the gauge symmetry is kept or not. As we shall see, both theories contain new interesting phenomenology that can be applied to different physical scenarios.

\subsubsection{Gauge-invariant theory}

On the one hand, it is possible to allow independent self-interactions of the scalar field via derivative terms---such as the third term in Eq.~\eqref{eq:proca}---which, restricting to second-order EoM, lead to shift-symmetric Horndeski interactions $\mathcal{L_H}^{\rm shift}$. On the other hand, it is also possible to construct order by order interactions between the vectorial combinations in Eqs.~\eqref{eq:vectcomb} and the $\nabla\phi$ term and its derivatives. By restricting to gauge-invariant combinations and explicitly breaking the shift symmetry to allow more general ones, one obtains the most general gauge-invariant scalar-vector-tensor theory yielding second-order EoM~\cite{Heisenberg:2018acv}:
\be \ba \label{eq:svt1}
\mathcal{L}=&~\mathcal{L_H}\\
&+f_2(F,\tilde{F},Y) \\
&+\mathcal{M}_3^{\mu\nu}\phi_{;\mu\nu}\\
&+\mathcal{M}_4^{\mu\nu\alpha\beta}\phi_{;\mu\alpha}\phi_{;\nu\beta}+f_4(\phi,X)L^{\mu\nu\alpha\beta}F_{\mu\nu}F_{\alpha\beta}~,\\ \ea \ee
where $\mathcal{L_H}$ is given in Eq.~\eqref{eq:Horn} and here $Y=\nabla_\mu\phi\nabla_\nu\phi F^{\mu\alpha}F^\nu_{~\alpha}$. We also defined the rank-2 and rank-4 tensors, $\mathcal{M}_3^{\mu\nu}$ and $\mathcal{M}_4^{\mu\nu\alpha\beta}$, respectively, as
\begin{alignat}{2} \mathcal{M}^{\mu\nu}_3&=\kk{f_3(\phi,X)g_{\rho\sigma}+\tilde{f}_3(\phi,X)\phi_\rho\phi_\sigma}\tilde{F}^{\mu\rho}\tilde{F}^{\nu\sigma}~, \label{eq:M3}\\
\mathcal{M}^{\mu\nu\alpha\beta}_4&=\kk{\frac12f_{4,X}(\phi,X)+\tilde{f}_4(\phi)}\tilde{F}^{\mu\nu}\tilde{F}^{\alpha\beta}~, \label{eq:M4} \end{alignat}
where we note that the function $\tilde{f}_4$ depends on $\phi$ alone. Notice that in the limit of constant $\phi$ and $f_4$ one recovers the Maxwell theory in Eq.~\eqref{eq:genmax}.

\subsubsection{Broken gauge-invariant theory}

Abandoning the gauge invariance, the vector field cannot only enter via the terms in Eqs.~\eqref{eq:vectcomb} but also via $S_{\mu\nu}=\nabla_\mu A_\nu+\nabla_\nu A_\mu$. In this regard, we can introduce an effective metric tensor constructed from possible combinations of $g_{\mu\nu}$, $A_{\mu}$, and $\nabla_{\mu}\phi$, given as~\cite{Heisenberg:2018acv}
\be \mathcal{G}_{\mu\nu}^{h_n}=h_{n1}(\phi,X_i)g_{\mu\nu}+h_{n2}(\phi,X_i)\phi_{;\mu}\phi_{;\nu}+h_{n3}(\phi,X_i)A_\mu A_\nu+h_{n4}(\phi,X_i)A_\mu\phi_{;\nu}~ \notag \ee
where the $X_i$ are defined below. Then, following the same procedure as before, the most general broken gauge-invariant scalar-vector-tensor theories yielding second-order EoM are written as
\begin{alignat}{7} \label{eq:svt2} \mathcal{L}_{\rm SVT}=&~f_2(\phi,X_1,X_2,X_3,F,\tilde{F},Y_1,Y_2,Y_3)\notag\\
&+f_{3}(\phi,X_3)g^{\mu\nu}S_{\mu\nu}+\tilde{f}_{3}(\phi,X_3)A^{\mu}A^{\nu} S_{\mu\nu}\notag\\
&+f_{4}(\phi,X_3)R+f_{4,X_3}(\phi,X_3)\kk{(A^\mu_{~;\mu})^2-A_{\mu;\nu}A^{\mu;\mu}}\\
&+f_5(\phi,X_3)G^{\mu\nu}A_{\mu;\nu}-\frac{1}{6}f_{5,X_3}(\phi,X_3)\bigl[(A^\mu_{~;\mu})^3-3A^\mu_{~;\mu}A_{\rho;\sigma}A^{\sigma;\rho} \notag\\
&~~~+2A_{\rho;\sigma}A^{\sigma;\gamma}A_\gamma^{~;\rho}\bigr]+\mathcal{M}_5^{\mu\nu}\phi_{;\mu\nu}+\mathcal{N}_5^{\mu\nu}S_{\mu\nu}\notag\\
&+f_6(\phi,X_1)L^{\mu\nu\alpha\beta}F_{\mu\nu}F_{\alpha\beta}+\mathcal{M}_6^{\mu\nu\alpha\beta}\phi_{\mu\alpha}\phi_{\nu\beta}+\tilde{f}_6(\phi,X_3)L^{\mu\nu\alpha\beta}F_{\mu\nu}F_{\alpha\beta}\notag\\
&~~~+\mathcal{N}_6^{\mu\nu\alpha\beta}S_{\mu\alpha}S_{\nu\beta}~,\notag \end{alignat}
where we now use the notation
\be X_1=-\frac12\phi_{;\mu}\phi^{;\mu}~, \qquad X_2=-\frac12A^\mu\phi_{;\mu}~, \qquad X_3=-\frac12A_\mu A^\mu~,\ee
and
\be Y_1=\phi_{;\mu}\phi_{;\nu}F^{\mu\alpha}F^\nu_{~\alpha}~, \qquad Y_2=\phi_{;\mu}A_\nu F^{\mu\alpha}F^\nu_{~\alpha}~,\qquad Y_3=A_\mu A_\nu F^{\mu\alpha}F^\nu_{~\alpha}~, \ee
the latter of which corresponds to the interactions arising from pure vector modes. Furthermore, the rank-2 tensors $\mathcal{M}^{\mu\nu}_5$ and $\mathcal{N}^{\mu\nu}_5$, which encode intrinsic vector interactions, are given by
\be \mathcal{M}^{\mu\nu}_5=\mathcal{G}_{\rho\sigma}^{h_5}\tilde{F}^{\mu\rho}\tilde{F}^{\nu\sigma}~, \qquad \mathcal{N}^{\mu\nu}_5=\mathcal{G}_{\rho\sigma}^{\tilde{h}_5}\tilde{F}^{\mu\rho}\tilde{F}^{\nu\sigma}~,\ee
where the functions $h_{5j}$ and $\tilde{h}_{5j}$ ($j=1,2,3,4$) appearing in ${\cal G}_{\rho\sigma}^{h_5}$ and ${\cal G}_{\rho\sigma}^{\tilde{h}_5}$ are functions of $\phi$ and $X_1, X_2, X_3$. On the other hand, the rank-4 tensors $\mathcal{M}_6^{\mu\nu\alpha\beta}$ and $\mathcal{N}^{\mu\nu\alpha\beta}_6$ are defined as 
\be \mathcal{M}^{\mu\nu\alpha\beta}_6=2f_{6,X_1}(\phi, X_1)\tilde{F}^{\mu\nu}\tilde{F}^{\alpha\beta}~, \qquad \mathcal{N}^{\mu\nu\alpha\beta}_6=\frac12\tilde{f}_{6,X_3}(\phi, X_3)\tilde{F}^{\mu\nu}\tilde{F}^{\alpha\beta}~.\ee

Notice that the functions $f_3, \tilde{f}_3, f_4, f_5, \tilde{f}_6$ depend on $\phi$ and $X_3$, whereas $f_6$ has dependence on $\phi$ and $X_1$. Furthermore, the Generalized Proca theories, Eq.~\eqref{eq:genProca}, are recovered by using the correspondence 
\begin{alignat}{4} 
& \phi\to0~, \qquad X_{1,2}\to0~, \qquad X_3\to X~, \qquad Y_{1,2}\to0~, \qquad Y_3\to Y~, \notag\\
& f_2\to G_2(X,F,Y)~, \qquad 2f_3\to G_3(X)~, \qquad \tilde{f}_3\to0~, \qquad f_4\to G_4(X)~, \notag\\
& f_5\to G_5(X)~, \qquad h_{5j}\to0~, \qquad \tilde{h}_{51}\to-\frac12\tilde{G}_5(X), \qquad \tilde{h}_{52}, \tilde{h}_{53}, \tilde{h}_{54}\to0~, \notag\\ 
&f_6\to0~, \qquad 4\tilde{f}_6\to G_6(X)~. \notag \end{alignat}

Finally, we note that the full scalar-vector-tensor theory with second-order EoM is completed by adding the Horndeski interactions $\mathcal{L_H}$, in Eq.~\eqref{eq:Horn}, to $\mathcal{L}_{\rm SVT}$, Eq.~\eqref{eq:svt2}. Therefore, we end up with a theory with six propagating degrees of freedom: two tensor modes, two vector modes and two scalar modes. This general theory has been developed just recently, but applications for DE~\cite{Kase:2018nwt}, black holes~\cite{Heisenberg:2018vti} and inflation~\cite{Heisenberg:2018erb} (to be discussed in \S\ref{sec:mixhel}) have already been performed.

\section{Inflation in scalar-tensor theories}

Our goal here is to apply the Horndeski theory and the SVT theories to the physics of inflation. To that end, one needs to consider a background FLRW spacetime and compute the EoM for the background and for the primordial perturbations, the latter of which will lead us to compute the power spectra of these perturbations and to make predictions from the theory (see \S\ref{sec:bounds}).

The background EoM and the quadratic actions of primordial scalar and tensor perturbations for the Horndeski theory, Eq.~\eqref{eq:Horn}, were computed in Ref. \cite{Kobayashi:2011nu} and are shown in Appendix \S\ref{app:eqsofmo}. 
Here, 
we shall focus on the novel phenomenology coming from specific models of inflation already tested, some of which constitute a part of the original results presented in this thesis. 
We shall firstly discuss the addition of nonminimal couplings between the scalar field and the scalar curvature $R$ to the canonical action, mediated by some coupling $\xi$ which alleviates the tension between the canonical model and the data. Secondly, we shall discuss the class of models named as \emph{G-inflation}, derived from taking into account a nonvanishing function $G_3$ in Eq.~\eqref{eq:Horn}---this class of models has been studied due to its ability to reconciling simple inflationary potentials $V(\phi)$ with the data.

\subsection{Nonminimal coupling to gravity}
\label{sec:nonmin}

The Horndeski theory has become a rich framework to construct phenomenological models of both early- and late-time cosmology. The most common modification of the canonical action, Eq.~\eqref{eq:EinHil}, comes from accounting for a nonminimal coupling between the scalar field and the gravity sector via the term $f(\phi)R$. From Eq.~\eqref{eq:Horn}, notice that this term can be obtained by setting $G_4(\phi,X)=f(\phi)$. 
However, nonminimal couplings of this form have been considered long before the Galileon theories were formulated~\cite{Salopek:1988qh,Futamase:1987ua,Fakir:1990eg,Kaiser:1994vs,Komatsu:1999mt,Hertzberg:2010dc}, and reconsidered when such coupling was found in the framework of supergravity theories~\cite{Einhorn:2009bh,Kallosh:2010ug,Ferrara:2010yw,Lee:2010hj,Ferrara:2010in,Kallosh:2010xz}. In particular, the simple function $f(\phi)=(1+\xi\phi^2)/2$ has been extensively studied, where $\xi$ is a dimensionless coupling expected to be small in this model in order for $\phi$ to 
successfully reheat the Universe. Indeed, \emph{Planck} places the lower bound on this parameter to be ${\rm log}_{10}\xi>-1.6$ at 95\% CL for the quartic potential $\phi^4$ which is highly disfavored in the canonical picture~\cite{Collaboration:2018aa}.

By introducing the nonminimal coupling, the action of such a theory is given by
\be \mathcal{S}_{\rm NM}=\int\dif^4x\sqrt{-g}\kk{\frac12\mk{1+\xi\phi^2}R-\frac12g^{\mu\nu}\phi_{,\mu}\phi_{,\nu}-U(\phi)}~, \label{eq:nonminact} \ee
where $U(\phi)$ is the potential function in the \emph{Jordan frame}. Indeed, it can be shown that the theory in Eq.~\eqref{eq:nonminact} can be recast as a canonical action (the \emph{Einstein frame}) by means of a \emph{conformal transformation} of the form
\be g_{\mu\nu}^{\rm E}=\Omega(\phi)g_{\mu\nu}~, \ee
where, in this case,
\be \Omega(\phi)\equiv1+\xi\phi^2~. \ee
Under this transformation, 
the action \eqref{eq:nonminact} becomes
\be \mathcal{S}_{\rm NM}^{\rm E}=\int\dif^4x\sqrt{-g_{\rm E}}\mk{\frac12R_{\rm E}-\frac12g_{\rm E}^{\mu\nu}\varphi_{,\mu}\varphi_{,\nu}-V\kk{\varphi\mk\phi}}~, \label{eq:nonminact2} \ee
where the index `E' emphasizes that the action is written in the Einstein frame, \ie in canonical form, with an effective potential function 
\be \label{eq:effpot} V\kk{\varphi(\phi)}=\frac{U(\phi)}{\Omega^2(\phi)}~, \ee
of the rescaled field
\be \mk{\diff{\varphi}{\phi}}^2=\frac1\Omega+\frac32\mk{\frac{\Omega_{,\phi}}\Omega}^2~. \ee

\begin{figure}[t]
\begin{center}
\includegraphics[keepaspectratio, width=13cm]{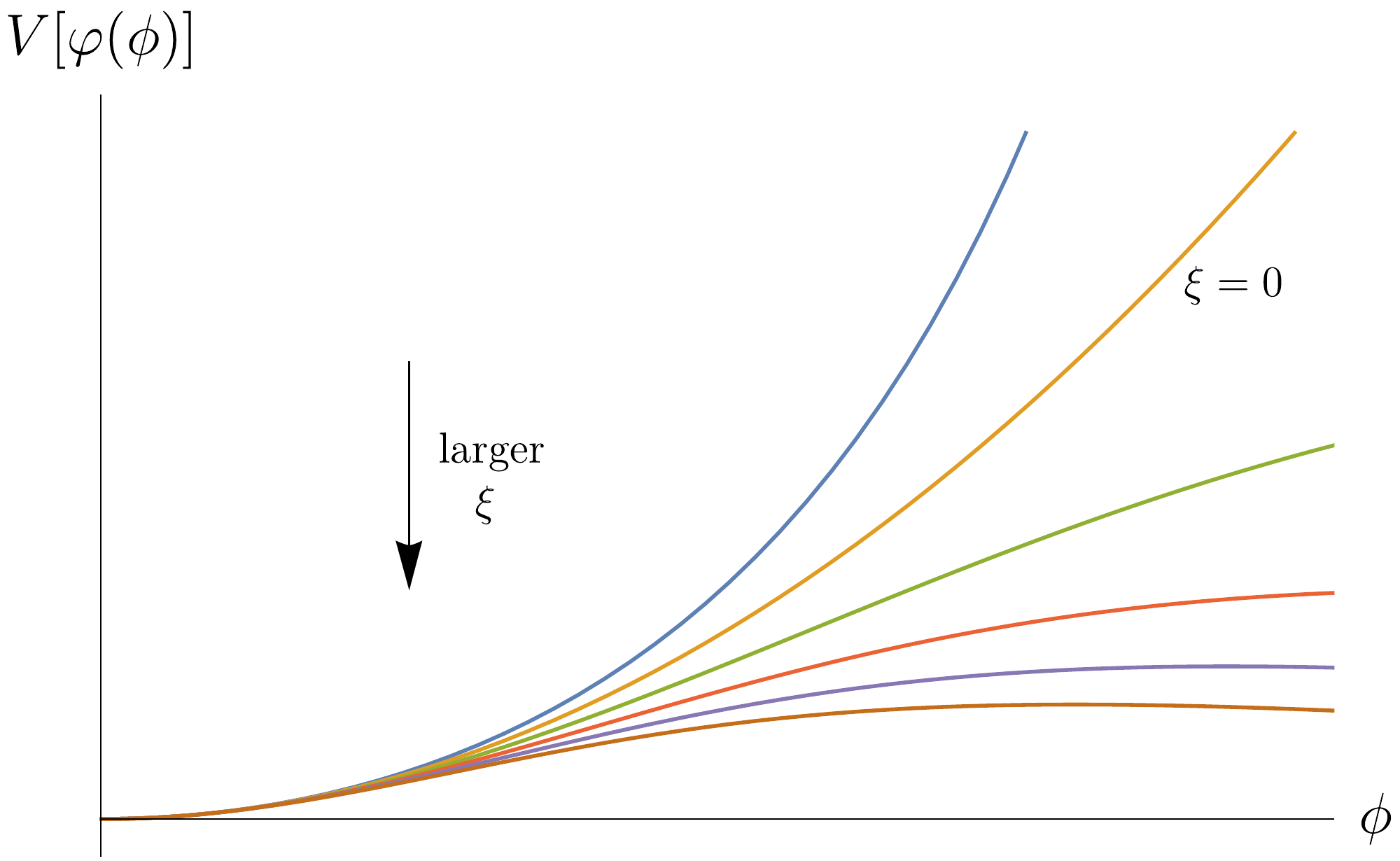}
\end{center}
\caption{\label{fig:nonminimal} Schematic representation of the effective potential \eqref{eq:effpot} for $U(\phi)\propto\phi^2$. The different lines represent different values of the coupling constant $\xi$ in the function $\Omega=1+\xi\phi^2$, where a larger $\xi$ corresponds to a flatter potential (a less concave one). From the observationally point of view, a flatter potential gives rise to a suppression of the tensor-to-scalar ratio $r$ and, consequently, a larger value of $\xi$ drives the canonical model to be in a better agreement with observations (see \S\ref{sec:bounds}).}
\end{figure}

As already stated, for a range of values of $\xi$, several canonical models of inflation can be reconciled with CMB observations, among which the chaotic model $m^2\phi^2$ and the quartic potential $\lambda\phi^4$ have been exhaustively studied~\cite{Okada:2010jf,Linde:2011nh,Kaiser:2013sna,Chiba:2014sva,Pallis:2014cda,Boubekeur:2015xza,Tenkanen:2017jih}. The explanation for this is quite simple, as seen in the Einstein frame: any different value of $\xi$ changes the shape of the effective potential and consequently its inflationary predictions; namely, the ability of $\xi$ to make the potential $V(\varphi)$ flatter will induce a suppression in the tensor-to-scalar ratio $r$ and thus make the potentials $U(\phi)$ more favored with respect to CMB observations (see Fig.~\ref{fig:nonminimal}).

\subsection{G-inflation}
\label{sec:ginfl}

Notice from Eq.~\eqref{eq:Horn} that the simplest nontrivial modification of the canonical action, Eq.~\eqref{eq:EinHil}, beyond linear order, comes from the third-order Lagrangian proportional to $G_3$. With this term, the action becomes
\be \mathcal{S}_G=\int\dif^4x\sqrt{-g}\kk{\frac12R+G_2(\phi,X)+G_3(\phi,X)\Box\phi}~, \label{eq:ginfl} \ee
where we have set $G_4(\phi,X)=1/2$ in order to account for the Einstein-Hilbert term. 
This class of models is called `G-inflation' in the literature, and its cosmological implications have been extensively explored---it was first studied in Ref. \cite{Kobayashi:2010cm} as a kinetically-driven model of inflation, \ie $G_2(\phi,X)=G_2(X)$ and thus no potential term was introduced. However, potential-driven versions, considered in subsequent works, realized that simple potentials as the ones of chaotic inflation, Eq.~\eqref{eq:monpot}, could be reconciled with CMB observations in the same spirit as in the presence of a nonminimal coupling (see Ref. \cite{Ohashi:2012wf}). Further extensions as, for instance, a Higgs boson driving inflation in this framework~\cite{Kamada:2010qe,Kamada:2013bia}, and studies on potential signatures on higher correlation functions \cite{DeFelice:2011zh} or reheating \cite{BazrafshanMoghaddam:2016tdk}, have also been carried out.

The equations of motion, for the full theory in Eq.~\eqref{eq:Horn} were computed in Ref. \cite{Kobayashi:2011nu} (also shown in Appendix \S\ref{app:eqsofmo}) assuming a homogeneous field $\phi=\phi(t)$ and the flat FLRW spacetime metric $\dif s^2=-N^2(t)\dif t^2+a^2(t)g_{ij}\dif x^i\dif x^j$ (where the lapse function $N(t)$ is introduced for convenience and later set to one). Particularly, for the G-inflation model in Eq.~\eqref{eq:ginfl}, the variation of the action with respect to $N(t)$ yields the Friedmann equation
\be 3H^2+G_2-2XG_{2,X}-2XG_{3,\phi}+6X\dot{\phi}HG_{3,X}=0~. \label{eq:gmov1} \ee
On the other hand, variation with respect to the scale factor $a(t)$ gives the evolution equation
\be 3H^2+2\dot{H}+G_2+2X\mk{G_{3,\phi}+\ddot{\phi}G_{3,X}}=0~. \label{eq:gmov2} \ee
Finally, the variation with respect to $\phi(t)$ gives the scalar-field equation of motion
\be 3H\dot{\phi}-G_2-6\mk{3H^2+\dot{H}}G_3+\mk{1-2G_{3,\phi\phi}}X+\mk{1+4G_{3\phi}-6H\dot{\phi}G_{3X}}\ddot{\phi}=0~. \label{eq:gmov3} \ee

Furthermore, the quadratic actions for scalar and tensor perturbations, from the full Horndeski background, were also computed in Ref. \cite{Kobayashi:2011nu} and are given by
\begin{alignat}{2}
\mathcal{S}_\zeta^{(2)}&=\int\dif{}^4x\frac{a^3b_s\epsilon_H}{c_s^2}\mk{\dot\zeta^2-\frac{c_s^2k^2}{a^2}\zeta^2}~, \label{eq:acthorn}\\
\mathcal{S}_\gamma^{(2)}&=\sum_{\lambda=+,\times}\int\dif{}^4x\frac{a^3b_t}{4c_t^2}\mk{\dot\gamma_\lambda^2-\frac{c_t^2k^2}{a^2}\gamma_\lambda^2}~, \label{eq:acthorn2}
\end{alignat}
where $c_{s,t}^2$ and $b_{s,t}$ are normalization factors which depend on the background, \ie on the $G_i(\phi,X)$ functions, as it is shown in \S\ref{sec:norfactapp}. Particularly, for the model in Eq.~\eqref{eq:ginfl}, they read as
\begin{align} \label{eq:cbs}
b_s&=\f{2 \mu_1H-2\dot\mu_1- \mu_1^2}{\epsilon_H }~, \quad &b_t=1~, \notag\\ 
c_s^2&=\f{3\mk{2 \mu_1H-2\dot\mu_1-\mu_1^2}}{4 \mu_2+9\mu_1^2}~, \quad &c_t^2=1~,\\
\intertext{where} 
\mu_1&=2H+2\dot\phi G_3~, \notag\\
\mu_2&=-9H^2+6\dot\phi^2G_{3,\phi}+\f{3}{2}\mk{\dot\phi-24HG_3}\dot\phi~.
\end{align}
Notice that the tensor normalization factors correspond to those of the canonical tensor quadratic action, meaning that the choice $G_4=1/2$ and $G_5=0$ does not modify the tensor sector.\footnote{This statement holds for any choice of $G_2$ and $G_3$, see Refs. \cite{Kobayashi:2011nu,Ohashi:2012wf}.}

In order to compute the Mukhanov-Sasaki equations for the quadratic actions \eqref{eq:acthorn}-\eqref{eq:acthorn2}, in terms of the Mukhanov variables $u_s=z_s\zeta$ and $u_t=z_t\gamma$, we need to redefine the $z_p$ variables, Eqs.~\eqref{eq:zetas}, as
\be z_s=\frac{a\sqrt{2b_s\epsilon_H}}{c_s}~, \qquad\qquad z_t=\frac a{c_s}\sqrt{\frac{b_t}2}~, \ee
for which the evolution equations read
\be u''_{p}+\mk{c_p^2k^2-\f{z''_p}{z_p}}u_{p}=0~. \label{eq:MSeqHorn} \ee
As already stated, for the model in Eq.~\eqref{eq:ginfl}, the tensor sector is not modified and thus the evolution equations and their solutions remain as in canonical inflation. On the other hand, for scalar perturbations, the solution of Eq.~\eqref{eq:MSeqHorn} is not trivial---one should study carefully the background evolution for a given choice of the $G_i(\phi,X)$ functions and, from there, determine whether the SR approximation is suitable or numerical integration must be performed. Furthermore, closer attention needs to be devoted to the evolution of the normalization factors $b_p$ and $c_p^2$ as they may develop instabilities; namely, $c_p^2$ represent the sound speeds of primordial perturbations that need to be positive defined in order to avoid for \emph{Laplacian} instabilities, whereas the factors $b_p$ are required with the same condition so they do not contribute with a wrong sign to the kinetic term, otherwise they would represent a ghost instability.

\begin{figure}[t]
\begin{center}
\includegraphics[keepaspectratio, width=13cm]{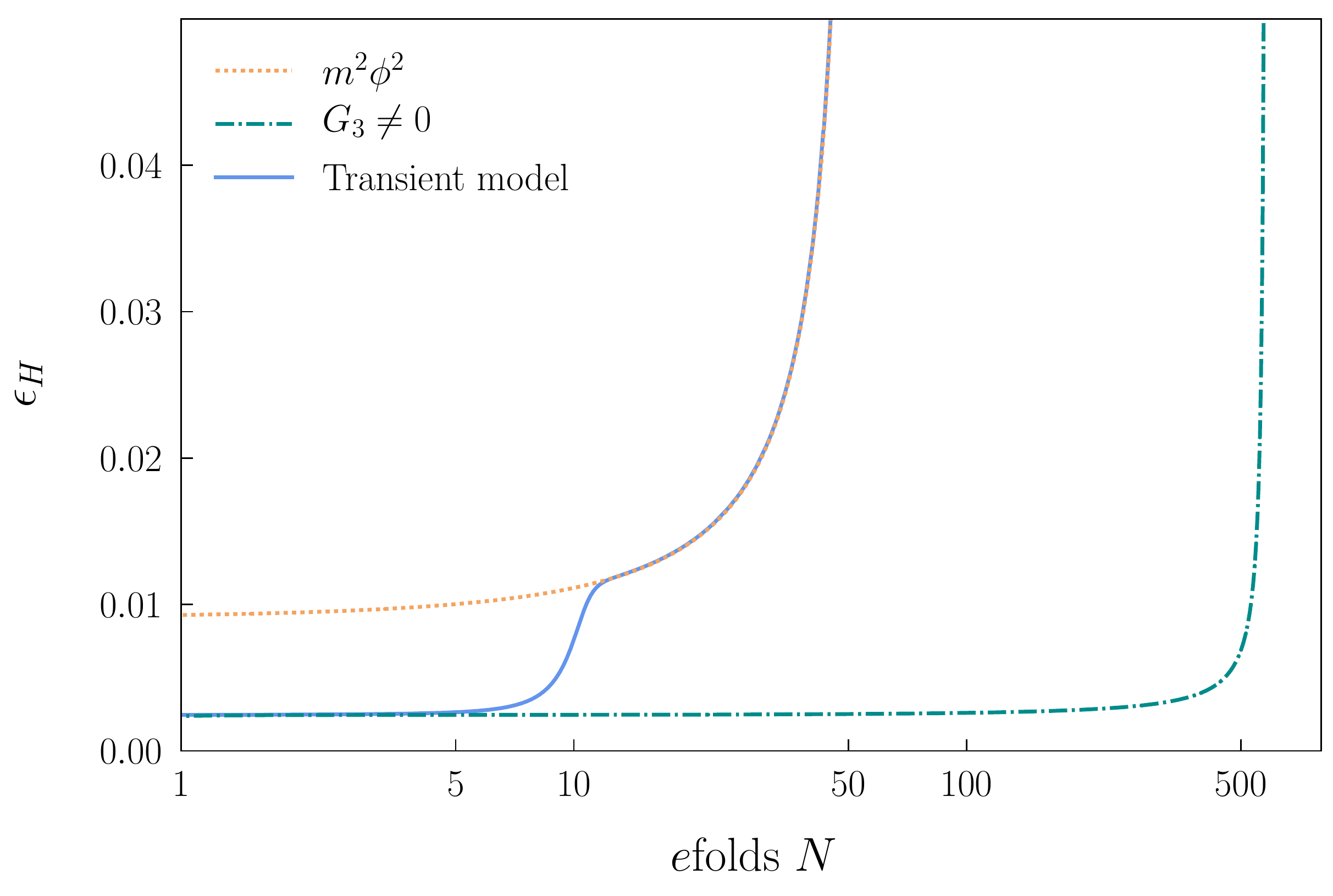}
\end{center}
\caption{\label{fig:transientmodel} Slow-roll parameter $\epsilon_H=-\dot H/H^2$ for three different models: the canonical quadratic potential $V(\phi)=m^2\phi^2/2$ (dotted, orange), the quadratic potential plus $G_3(\phi,X)=-X/(2M^3)$ (dash-dotted, green), and a transient model given by $G_3(\phi,X)=-X\kk{1+\tanh\mk{(\phi-\phi_r)/d}}/(2M^3)$ with $\ck{M,m,\phi_r,d}=\ck{1.303\times10^{-4},2.58\times10^{-6},13.87,0.086}$. The hyperbolic tangent provides a mechanism to switch off the contribution of the $G_3$ term and thus to transition to the canonical regime. As a consequence, the transient model safely reheats the Universe and suppresses the tensor to scalar ratio, $r=16\epsilon_H$, at CMB scales $(N=0)$---the plot is normalized such as in both canonical and transient models inflation ends at $N=55$. However, CMB scales for the model in green lie at $N\sim500$, \ie the suppression of $r$ in such a model is small. See Ref.~\cite{Ramirez:2018dxe} for details.}
\end{figure}

Until recently, the avoidance of instabilities at the perturbations level represented a severe problem on the construction of G-inflation models. Reference \cite{Ohashi:2012wf}, for instance, studied a potential driven-version based on the function $G_3(\phi,X)=-X/(2M^3)$ and found that they could reconcile the quadratic potential $m^2\phi^2$ (among others) with CMB observations for small values of the mass scale $M$ compared to $\Mpl$, however with a lower bound of $M=4.2\times10^{-4}\Mpl$. Although the tension between the model and the data is recovered when we consider the most recent \emph{Planck} data (see Ref.~\cite{Ramirez:2018dxe}), the issue with smaller values of $M$ 
remained interesting as it was due to the appearance of Laplacian instabilities during reheating. Indeed, the $G_3$ term still affects the dynamics of the inflaton field after the end of inflation, which translates into the lack of coherent oscillations during the reheating epoch. 
Nevertheless, it has been shown that these instabilities can be avoided by terminating the influence of $G_3$ before the end of inflation; this mechanism can be simply achieved by a transition from a G-inflation domain to a canonical inflationary era able to properly reheat the Universe. Furthermore, this transition should be carefully introduced after CMB scales in order to contrast with the canonical predictions of a given potential $V(\phi)$. This can be seen from Fig.~\ref{fig:transientmodel}, where the slow-roll parameter $\epsilon_H$, Eq.~\eqref{eq:epsilonh}, is shown for a transient model which was carefully constructed in order to be placed after CMB scales $(N=0)$ and before the end of inflation $(N=55)$. In addition, recall that under the SR approximation, the tensor-to-scalar-ratio at CMB scales reads as $r=16\epsilon_H$, \ie a suppression of $r$ is expected for the \emph{Transient} model in comparison with the canonical quadratic scenario for the same value of $n_s$. Consequently, such a transient model is able to reconcile chaotic inflation with observations and avoid Laplacian instabilities (see~\cite{Ramirez:2018dxe} for details).

\section{Inflation in scalar-vector-tensor theories}

The background EoM and the quadratic actions of primordial scalar, vector and tensor perturbations for the SVT theories, Eq.~\eqref{eq:svt2}, were fully computed in Ref. \cite{Heisenberg:2018mxx} and are shown in Appendix \S\ref{app:eqsofmo}. Here, we shall review their consequences on inflation by constructing a simple model, yet with phenomenological implications, with a scalar-vector coupling of the form $A^\mu\nabla_\mu\phi$. As we shall see, the longitudinal vector is able to affect the cosmic expansion during inflation which will be translated into a suppression of the tensor-to-scalar ratio for large-field models~\cite{Heisenberg:2018erb}.

\subsection{Inflation with mixed helicities}
\label{sec:mixhel}

The scalar-vector-tensor theories allow for extra interactions in the form of scalar-vector mixings. In the context of inflation, the vector field is able to modify the dynamics of the expansion driven by the scalar field and, consequently, the predictions for a given potential function $V(\phi)$. As it can be noticed from Eq.~\eqref{eq:svt2}, a scalar-vector mixing can be included in several different forms. Among these possible forms, the simplest one is given by $X_2=-A^\mu\nabla_\mu\phi/2$, already present in $f_2$. This term is genuine, coming from the helicity decomposition provided by the Stueckelberg trick and therefore it is interesting to study the dynamics it offers when added to a canonical model of inflation.

\begin{figure}[t]
\begin{center}
\includegraphics[keepaspectratio, width=13cm]{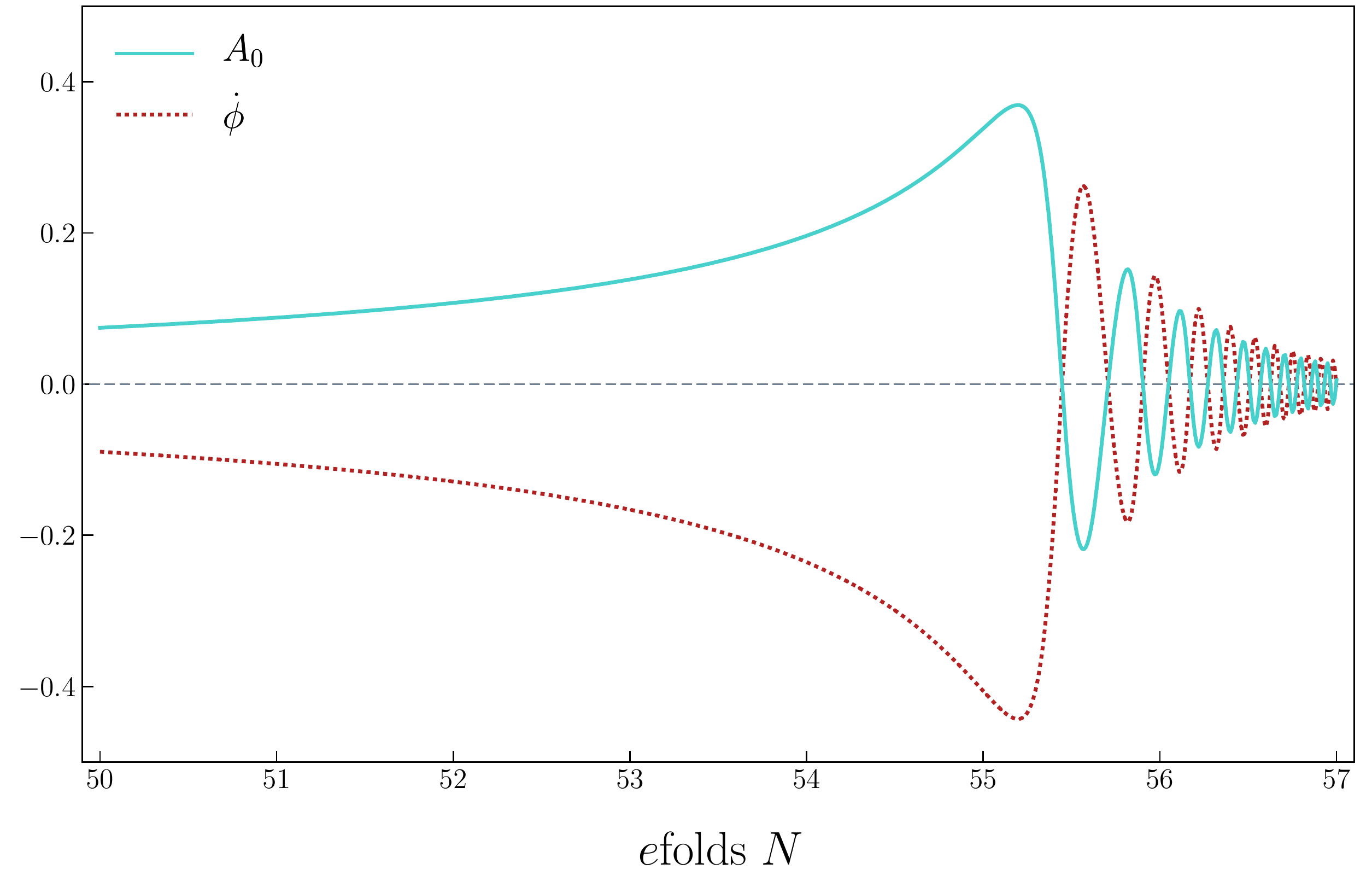}
\end{center}
\caption{\label{fig:azeroconst} Background evolution of the temporal mode $A_0$ and the scalar-field velocity $\dot\phi$, by the end of inflation $(N=55)$ and during reheating
, computed for the model given in Eq.~\eqref{eq:mixmodel} and for the potential given in Eq.~\eqref{eq:alphattrac} with $\alpha_c=\sqrt{6}/3$. Notice that, as expected from Eq.~\eqref{eq:Adotphi}, the ratio $A_0/\dot\phi$ remains constant during the whole evolution.}
\end{figure}

We then focus here on a model of inflation driven by a helicity-0 mode, $\phi$, mixed with a helicity-1 mode, $A_\mu$, where both fields are allowed to propagate, \ie the vector kinetic and self-interaction terms are included.~\footnote{Recall, however, that $F=-F_{\mu\nu}F^{\mu\nu}/4$ does not contribute to the dynamics on a FLRW spacetime due to the conformal invariance of the Maxwell Lagrangian.} The action then reads as
\be \mathcal{S}_{\rm mix}=\int\dif^4x\sqrt{-g}\kk{\frac12R+F+X_1-V(\phi)+\beta_mMX_2+\beta_AM^2X_3}~, \label{eq:mixmodel} \ee
where we recall that
\be X_1=-\frac12\nabla_\mu\phi\nabla^\mu\phi~, \qquad X_2=-\frac12A^\mu\nabla_\mu\phi~, \qquad X_3=-\frac12A_\mu A^\mu~, \ee
and where $M$ is the positive, constant vector mass, and $\beta_m$ and $\beta_A$ are dimensionless constants. The equations of motion, computed on a FLRW spacetime metric \eqref{eq:flrwm}, with a compatible vector profile $A_\mu=(A_0(t),0,0,0)$ and a homogeneous scalar field $\phi(t)$, read as
\begin{alignat}{4}
3H^2-\frac12\dot\phi^2-V(\phi)+\frac12\beta_AM^2A_0^2&=0~, \label{eq:svtHm1} \\
2\dot H+\dot\phi^2+\frac12\beta_mM\dot\phi A_0&=0~, \label{eq:svtHm2}\\
\ddot\phi+3H\dot\phi+V_{,\phi}+\frac12M\beta_m\mk{\dot A_0+3HA_0}&=0~, \label{eq:svtHm3}\\
2\beta_AMA_0+\beta_m\dot\phi&=0~. \label{eq:Adotphi}
\end{alignat}
Notice that we now have a fourth EoM corresponding to the variation of the action with respect to $A_0$. Interestingly enough, Eq.~\eqref{eq:Adotphi} tells us that the ratio $A_0/\dot\phi$ remains constant during the evolution, as depicted in Fig.~\ref{fig:azeroconst}. This fact allows us to substitute $\dot\phi\propto A_0$ into Eqs.~\eqref{eq:svtHm1}-\eqref{eq:svtHm3} and to introduce the parameter
\be \beta\equiv1-\frac{\beta_m^2}{4\beta_A}~, \label{eq:betadef}\ee
for convenience, as we shall see. Furthermore, we can define a rescaled field $\varphi$ in terms of $\beta$ as
\be \dif\varphi=\sqrt{\beta}\dif\phi~, \label{eq:varphi} \ee
and rewrite the EoM as
\begin{alignat}{3}
3H^2-\frac12\dot\varphi^2-V(\varphi)&=0~, \\
2\dot H+\dot\varphi^2&=0~, \\
\ddot\varphi+3H\dot\varphi+V_{,\varphi}&=0~,
\end{alignat}
\ie the proportionality between $A_0$ and $\dot\phi$ leads to an effective single-field dynamics driven by the $\varphi$ field---therefore the computation of the power spectra can be easily performed using the standard SR approximation.

\begin{figure}[t]
\begin{center}
\includegraphics[keepaspectratio, width=13cm]{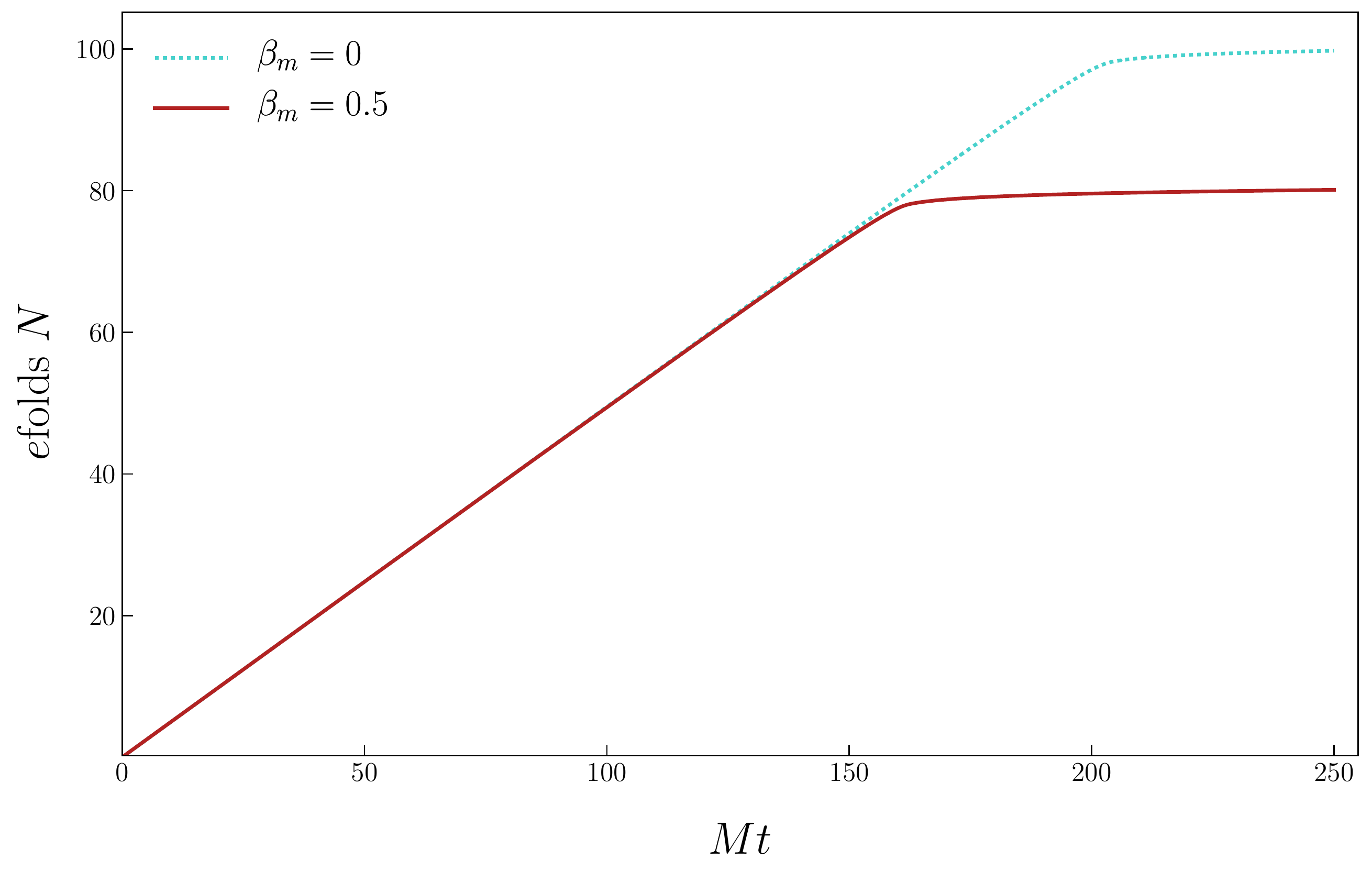}
\end{center}
\caption{\label{fig:redefolds} Number of $e$folds of inflation $\dif N=H\dif t$, as a function of time rescaled by the vector mass $M$, for the same specifications than those in Fig.~\ref{fig:azeroconst}. Notice that for a nonvanishing scalar-vector mixing, mediated by $\beta_m$, the expansion is shorter; regarding the inflationary observables, a shorter expansion would require to start the inflaton's evolution from a flatter part of the potential in order to have enough inflation which, in turn, translates into a suppression of the tensor-to-scalar ratio $r$ (see~\cite{Heisenberg:2018erb}).}
\end{figure}

The conditions for the avoidance of scalar ghosts, worked out in Ref.~\cite{Heisenberg:2018mxx} for the full Lagrangian, trivially provide the constraint $4\beta_A>\beta_m^2\geq0$ for this model and, consequently, $\beta$ lies in the range $0<\beta\leq1$ (see Eq.~\eqref{eq:betadef}). The deviation of $\beta$ from unity, induced by a nonvanishing scalar-vector mixing, makes the rescaled field $\varphi$ to evolve slower compared to the inflaton field $\phi$, which in turns makes the expansion shorter---there are fewer $e$folds $N$ for the same field excursion---as seen from Fig.~\ref{fig:redefolds}. This has important consequences on the inflationary observables. Namely, in order to have enough inflation, the field $\varphi$ needs to evolve from a flatter part of the potential $V(\varphi)$ which will produce a suppression on the tensor-to-scalar ratio $r$ specially noticeable for small-field models---while small-field models currently satisfy the CMB bounds on $r$, they could be in tension in the near future and thus a scalar-vector-mixing model may reconcile such potential with observations (see Ref.~\cite{Heisenberg:2018erb} for details).

The complete, general quadratic actions for scalar, vector and tensor perturbations, along with the conditions for the avoidance of ghosts and Laplacian instabilities for the theory in Eq.~\eqref{eq:svt2}, were computed in Ref.~\cite{Heisenberg:2018mxx} (which we omit to show here for brevity), whereas their particularizations for the model in Eq.~\eqref{eq:mixmodel} were computed in Ref.~\cite{Heisenberg:2018erb}, as well as the power spectra for the three helicity modes and the predictions for several inflationary potentials; and thus we refer the interested reader to these works for details. Further studies concerning the epoch of reheating, imprints on higher correlation functions and more complicated mixings with potential new phenomenology as, for instance, a non-negligible amplitude of vector perturbations, are expected to be carried out in the near future.

\lhead[{\bfseries \thepage}]{ \rightmark}
\rhead[ Chapter \thechapter. \leftmark]{\bfseries \thepage}
\chapter{Generalized Slow-Roll Approximation}
\label{sec:GSR}

In \S\ref{sec:condforinf} we discussed the conditions required for a successful period of inflation---in doing so, we defined the slow-roll parameters $\epsilon_H$ and $\eta_H$, Eqs.~\eqref{eq:epsilonh} and \eqref{eq:etaHdef}. The condition $\epsilon_H\ll1$ is required so the evolution remains close to de Sitter and inflation does not end earlier than expected; whereas $|\eta_H(=-\delta_1)|\ll1$ ensures that the evolution of $\epsilon_H$ is slow, which is usually understood as a requirement for the inflaton's slow evolution, needed for a sufficient amount of inflation. 

Satisfying the slow-roll conditions unwittingly defines a hierarchy of the so-called \emph{Hubble slow-roll} parameters
\be \delta_p\equiv\frac1{H^p\dot\phi}\mk{\diff{}{t}}^{p+1}\phi~, \ee
where a given parameter $\delta_{p}$ is of order $\mathcal O(\epsilon_H^p)$. This hierarchy is helpful to obtain approximative solutions of the Mukhanov-Sasaki equation \eqref{eq:MSeqdef}, as discussed in \S\ref{sec:soluMukhSas}. Furthermore, a consequential hierarchy of inflationary observables, $n_s-1=\mathcal O(\epsilon_H)$, $\alpha_s=\mathcal O(\epsilon_H^2)$, $\beta_s=\mathcal O(\epsilon_H^3)$, etc., is implicitly defined (see Eqs.~\eqref{eq:inflaparamalphabeta}). As discussed in \S\ref{sec:bounds}, this hierarchy of observational parameters is compatible with the current data (given the large uncertainties on both $\alpha_s$ and $\beta_s$), however, it is not really required by the observations. Furthermore, it is not a consequence of the slow-roll (SR) approximation either. Interpretations of the aforementioned CMB constraints in terms of the slow-roll parameters could then provide misleading results, even so at second-order in the SR approximation which is usually assumed to be more accurate.

Features in the inflationary potential $V(\phi)$ translate into oscillations or other type of glitches in the primordial power spectra. These features leave the spectra nearly scale-invariant although no longer scale-free~\cite{Adams:2001vc,Covi:2006ci,Hamann:2007pa,Pahud:2008ae,Joy:2008qd,Mortonson:2009qv}. Consequently, treating models of this sort with the standard SR approximation is known to fail, even in canonical inflation, due to large local\footnote{At a specific scale $k$.} tilt and running of the tilt (being equivalently to a large $|\delta_1|$), \ie numerical integration of the mode-function equation is usually performed.

In this chapter, we review the \emph{Generalized Slow-Roll} (GSR) approximation, which was developed to overcome the deficiencies of the standard SR approximation. Here, the evolution of the first slow-roll parameter $\epsilon_H$, sourced by features, is only assumed to be small in amplitude, \ie nothing is assumed for its frequency. Conversely, if their frequency is of order $\Delta N\geq1$, a Taylor expansion of the sources around an optimized horizon exit epoch leads to analytical expressions for the power spectra observables with a correct order counting of the slow-roll parameters. This approach is named \emph{Optimized Slow-Roll} (OSR). In addition, we shall assume a general scalar-tensor background given by the Horndeski framework, Eq.~\eqref{eq:Horn}, for which the mode-function evolution equations are given by Eq.~\eqref{eq:MSeqHorn}.\footnote{The original GSR approximation was developed by E. Stewart~\cite{Stewart:2001cd} in the framework of canonical inflation to remove the assumptions yielding to the hierarchy of the inflationary observables. However, this approximation still required $|\eta_H|\ll1$ and thus only applied for small deviations from scale-invariance. The techniques reviewed in this chapter were developed to improve the original GSR and later extended to noncanonical models.}

For convenience, we define several new variables: a rescaling of the mode functions
\be y=\sqrt{2c_{s,t}k}u_{s,t}~, \label{eq:relosr1}\ee
a horizon epoch $x$, in terms of the sound horizons $s_{s,t}$,
\be x \equiv ks_{s,t}~, \qquad\qquad s_{s,t}(N)\equiv\int_{N}^{N_f} \frac{c_{s,t}}{aH}\dif{N}~, \ee 
and the source functions
\begin{align} \label{eq:relosr2}
f_s&\equiv2\pi z_s\sqrt{c_s}s_s=\sqrt{8\pi^2\f{b_s\epsilon_Hc_s}{H^2}}\f{aHs_s}{c_s}~,\notag\\
f_t&\equiv2\pi z_t\sqrt{c_t}s_t=\sqrt{2\pi^2\f{b_tc_t}{H^2}}\f{aHs_t}{c_t}~,
\end{align}
for scalars and tensors, respectively. In terms of these variables, the Mukhanov-Sasaki equation \eqref{eq:MSeqHorn}, can be written as
\be  \label{eq:msgsr}
\difs{y}{x}+\mk{1-\f{2}{x^2}}y=\f{f''-3f'}{f}\f{y}{x^2}~,
\ee
where primes will represent derivatives with respect to $\ln x$ along this chapter. As it can be noticed, Eq.~\eqref{eq:msgsr} resembles the mode-function equation in de Sitter space, Eq.~\eqref{eq:MSdeSitter}, with an extra term sourced by the function
\be g(x)=\f{f''-3f'}{f}~. \ee
Therefore, $g(x)$ encodes all the deviations from the de Sitter solution due to excitations of the source functions $f$. Bear in mind that so far we have not made any assumption for the evolution of $\epsilon_H$ or the other slow-roll parameters and therefore, in these variables, the dimensionless power spectra, Eqs.~\eqref{eq:genPS}, are given by
\be
\Delta^2_{\zeta,\gamma}=\lim_{x\to0}\bigg{\rvert}\mk{\f{xy}{f}}_{s,t}\bigg{\rvert}^2~.
\ee
Notice now that in the case where the source function $f$ remains nearly constant, the scalar and tensor power spectra, to the lowest order in excitations, approximate to~\cite{Motohashi:2017gqb}
\be \Delta_\zeta^2\approx\frac1{f^2_s}\approx\f{H^2}{8\pi^2\epsilon_Hc_sb_s}~, \qquad\qquad
\Delta_\gamma^2\approx\frac1{f^2_t}\approx\f{H^2}{2\pi^2c_tb_t}~, \label{eq:srhorn}\ee
which correspond to the de Sitter results for the spectra in the Horndeski background~\cite{Kase:2014cwa}.

\section{Generalized Slow-Roll}

Equation \eqref{eq:msgsr} can be solved using Green function techniques provided that the amplitude of $f$ remains small, \ie the solution does not deviate considerably from the de Sitter solution, the Bunch-Davies vacuum,
\be y_0(x)=\mk{1+\frac ix}e^{ix}~. \ee
Again, this requirement only assumes small deviations of scale invariance over an average of time, but nothing on the local tilt. The formal solution to Eq.~\eqref{eq:msgsr} then reads as
\be y(x)=y_0(x)-\int^\infty_x\f{\dif u}{u^2}\f{f''-3f'}{f}y(u){\rm Im}\kk{y_0^*(u)y_0(x)}~, \ee
from which we can replace $y\to y_0$ on the right-hand side and iteratively improve the solution order by order in deviations from de Sitter. To first order we obtain,\footnote{See, {\it e.g.}, Refs. \cite{Dvorkin:2009ne,Motohashi:2015hpa} for details and the formulas to second order in deviations from the de Sitter background.}
\be \ba \label{eq:gsrfor}
\ln \Delta_{\zeta,\gamma}^2(k)&\approx G(\ln x_*)+\int^\infty_{x_*}\f{\dif x}{x}W(x)G'(\ln x)~,\\
&\approx-\int_0^\infty\f{\dif{x}}{x}W'(x)G\mk{\ln x}~, \ea \ee
where $x_*\ll1$ and integration by parts was performed in the second line. Furthermore, $G(\ln x)$ is a source function that now encodes all the deviations from the de Sitter solution, written as 
\be \label{eq:gsource} G\equiv-2\ln f+\frac23\mk{\ln f}'~, \ee
and $W(x)$ is a window function given by
\be
W(x)=\frac{3\sin(2x)}{2x^3}-\frac{3\cos(2x)}{x^2}-\frac{3\sin(2x)}{2x}~,
\ee
which determines the freezout of the mode functions~\cite{Dvorkin:2009ne}.

Equation \eqref{eq:gsrfor} is known as the \emph{Generalized Slow-Roll} formula. It still requires numerical integration, though it is more computationally efficient than solving Eq.~\eqref{eq:MSeqHorn}. Moreover, the source function $G$ can be used as a model-independent mean to connect observational constraints with any inflationary model that belongs to the \emph{effective field theory} class \cite{Dvorkin:2011ui,Obied:2017tpd}. In addition, the tilts $n_{s,t}$ and the higher-order running parameters can also be efficiently computed by taking derivatives of Eq.~\eqref{eq:gsrfor} with respect to the scale $k$, whereas the tensor-to-scalar ratio is computed in the standard way, using Eq.~\eqref{eq:ttsr}.

\section{Optimized Slow-Roll}
\label{sec:osr}

In the GSR expansion, Eq.~\eqref{eq:gsrfor}, local scale-dependence of the power spectra is encoded in a nonvanishing $G'(\ln x)$. The condition for small departures from the de Sitter solution then implies that the average of $G'$, over several $e$folds, is of order $\mathcal O(1/N)$, which is consistent with CMB and LSS observations where $N\sim55$. On the other hand, as previously stated, the sources are allowed to vary on a shorter scale $\Delta N$ and, consequently, $G''=\mathcal O(1/\Delta N)G'$. In general, there exists a hierarchy of $G^{(p)}$ functions given by
\be G^{(p)}\equiv\f{\dif^pG}{\dif\ln x^p}=\mathcal O\mk{\f{1}{N\Delta N^{p-1}}}~, \label{eq:gorders} \ee
which therefore can be distinguished from the standard $\mathcal O(1/N^p)$ slow-roll hierarchy.

It can be shown that the source in the GSR formula can be expanded in Taylor series around the horizon exit epoch, provided that $1\lesssim\Delta N\leq N$ \cite{Motohashi:2015hpa}. Compared with the usual SR approximation, this expansion creates a hierarchy of parameters separated by $1/\Delta N$ rather than $1/N$.\footnote{For $\Delta N\sim1$ numerical integration is needed, either by exactly solving the Mukhanov-Sasaki equation \eqref{eq:MSeqHorn} or by performing the GSR approximation by means of Eq.~\eqref{eq:gsrfor}. On the other hand, if $\Delta N\ll1$, the hierarchy is inverted and different techniques can be performed (see \cite{Miranda:2015cea}).} For the first-order GSR formula, the expansion reads as \cite{Motohashi:2015hpa,Motohashi:2017gqb}
\be \ln\Delta_{\zeta,\gamma}^2\approx G(\ln x_f)+\sum_{p=1}^\infty q_p(\ln x_f)G^{(p)}(\ln x_f)~, \label{eq:osrtaylor}\ee
where the $q_p(\ln x_f)$ coefficients are given by
\be q_1(\ln x_f)=\ln x_1- \ln x_f~, \qquad\qquad \ln x_1\equiv\frac73-\ln 2-\gamma_{\rm E}~, \label{eq:q1} \ee
and
\begin{alignat}{2} 
q_p(\ln x_f)&=\sum_{n=0}^p\f{c_{p-n}}{n!}q_1^n(\ln x_f)~,\\
c_p&=\frac1{p!}\lim_{z\to0}\f{\dif^p}{\dif z^p}\kk{e^{-z\mk{\frac73-\gamma_{\rm E}}}\cos\mk{\f{\pi z}{2}}\f{3\Gamma\mk{2+z}}{(1-z)(3-z)}}~.
\end{alignat}
Here, $\gamma_{\rm E}$ is the Euler-Mascheroni constant. The coefficients $q_p$ depend only on the evaluation epoch $x_f$ and thus they do not depend on the inflationary model and are equal for scalars and tensors.

\subsection{Optimization}

The sound horizon exit epoch corresponds to $\ln x_f=0$, for which the standard slow-roll results are recovered by truncating Eq.~\eqref{eq:osrtaylor} to leading order, \ie $\ln \Delta^2\approx G(0)$. In this case, the next-to-leading (NLO) order slow-roll (SR) correction ($p=1$) is suppressed by $q_1(0)/\Delta N$.

However, we can improve the truncation of Eq.~\eqref{eq:osrtaylor} by optimizing the evaluation point $x_f$. For instance, notice that $q_1(\ln x_f)$ vanishes for $x_f=x_1$ and therefore the NLO order correction vanishes as well. The first correction would then come from the next-to-next-to-leading (NNLO) order optimized (OSR) correction $q_2(\ln x_1)/\Delta N^2$.

For large features, $\Delta N\sim N$, both NLO SR and NNLO OSR corrections are small and thus the leading-order solutions are accurate enough, as expected. On the other hand, if the sources vary, for instance, as $\Delta N\sim3$, the first SR correction (NLO) is expected to be of 35\%, as usual, whereas the first OSR correction (NNLO) is just about 4\%. Consequently, more accurate approximations for the observables are obtained by optimizing the evaluation point $x_f$. Since $\ln x_1\approx1.06$, notice that the optimization corresponds to evaluate the observables at around $\sim1$ $e$fold before the sound horizon exit.

We can therefore establish the $p$-th order optimization by fixing the evaluation epoch to be $\ln x_f=\ln x_{p+1}$, so that the next-order correction identically vanishes as a consequence of the $q_{p+1}(\ln x_{p+1})=0$ solution.

The tilts and the higher order running parameters can be obtained by differentiating Eq.~\eqref{eq:osrtaylor} and using the fact that \cite{Motohashi:2015hpa,Motohashi:2017gqb}
\be \diff{G^{(p)}(\ln x_f)}{\ln k}=-G^{(p+1)}(\ln x_f)~. \ee
Therefore the first observables read, to leading order, as $\dif\ln\Delta^2/\dif\ln k\approx-G'(\ln x_f)$ and $\alpha\approx G''(\ln x_f)$. As previously stated, this implies a hierarchy of the $G^{(p)}$ functions defined by Eq.~\eqref{eq:gorders}. However, it is more convenient to relate the observables to the standard Hubble slow-roll parameters.

\subsection{Correspondence to the Hubble slow-roll parameters}

In the context of a general scalar tensor theory, parametrized by $\epsilon_H=-\dif\ln H/\dif N$ and the normalization factors $c_{s,t}^2$ and $b_{s,t}$, the Hubble slow-roll parameter convention is given by the hierarchies\footnote{Consistent with the Horndeski theory parametrization (\S\ref{sec:Hornsec}), for which the normalization factor are given in Eqs.~\eqref{eq:norfactapp}. However, the OSR approximation holds for more general theories belonging to the effective field theory class, see Ref.~\cite{Motohashi:2017gqb} for details.}
\allowdisplaybreaks
\begin{eqnarray} \label{osrsrparam}
\delta_1&\equiv&\frac12\diff{\,\logn\epsilon_H}{N}-\epsilon_H~, \,\,\,\,\,\,\,\,\,
\delta_{p+1}\equiv\diff{\delta_p}{N}+\delta_p\mk{\delta_1-p\epsilon_H}~,\notag\\ 
\sigma_{i,1}&\equiv&\diff{\, \logn c_i}{N}~, \,\,\,\,\,\,\,\,\,\,\,\,\,\,\,\,\,\,\,\,\,\,\,\,\,\,\,
\sigma_{i,p+1}\equiv\diff{\sigma_{i,p}}{N}~, \\
\xi_{i,1}&\equiv&\diff{\logn b_i}{N}~, \,\,\,\,\,\,\,\,\,\,\,\,\,\,\,\,\,\,\,\,\,\,\,\,\,\,\,\,\,
\xi_{i,p+1}\equiv\diff{\xi_{i,p}}{N}~\notag,
\end{eqnarray}
where here $i=s,t$ and $p\geq1$.

The previously stated assumptions, $G'=\mathcal O(1/N)$ and $G^{(p+1)}\sim\mathcal O(1/\Delta N)G^{(p)}$, then fix the expectations for the slow-roll parameters as
\be \ba
\ck{G',\epsilon_H,\delta_1,\sigma_{i,1},\xi_{i,1}}&=\mathcal O\mk{\frac 1N}~,\\
\ck{G^{(p+1)},\delta_{p+1},\sigma_{i,p+1},\xi_{i,p+1}}&=\mathcal O\mk{\frac 1{N\Delta N^p}}~.
\ea \ee
Therefore, a relation between the $G^{(p)}$ functions and the slow-roll parameters can be established by means of Eq.~\eqref{eq:gsource}, using Eqs.~\eqref{eq:relosr2}. In doing so, a convention regarding the expansion in inverse powers of $N$ and $\Delta N$ is adopted, namely, expressions are expanded up to $\mathcal O(1/N^2)$, \ie terms of order $\mathcal O(1/N\Delta N^p)$ are kept but not $\mathcal O(1/N^2\Delta N^p)$ terms (see Refs.~\cite{Motohashi:2015hpa,Motohashi:2017gqb} for details).

The first order optimized slow-roll formulas, in terms of the slow-roll parameters, then read as~\cite{Motohashi:2017gqb}
\begin{alignat}{6} \label{osrscalars}
\logn\Delta_\zeta^2&\simeq\logn\mk{\f{H^2}{8\pi^2 b_sc_s\epsilon_H}}-\f{10}{3}\epsilon_H-\f{2}{3}\delta_1-\f{7}{3}\sigma_{s1}-\f{1}{3}\xi_{s1}\Big|_{x=x_1}~,\notag\\
n_s-1&\simeq-4\epsilon_H-2\delta_1-\sigma_{s1}-\xi_{s1}-\f{2}{3}\delta_2-\f{7}{3}\sigma_{s2}-\f{1}{3}\xi_{s2}\Big|_{x=x_1}~, \\
\alpha_s&\simeq-2\delta_2-\sigma_{s2}-\xi_{s2}-\f{2}{3}\delta_3-\f{7}{3}\sigma_{s3}-\f{1}{3}\xi_{s3}-8\epsilon_H^2-10\epsilon_H\delta_1+2\delta_1^2\Big|_{x=x_1}~\notag,\\
\intertext{for scalar, and}
\logn\Delta_{\gamma}^2&\simeq\logn\mk{\f{H^2}{2\pi^2b_tc_t}}-\f{8}{3}\epsilon_H-\f{7}{3}\sigma_{t1}-\f{1}{3}\xi_{t1}\Big|_{x=x_1}~,\notag\\ 
n_t&\simeq-2\epsilon_H-\sigma_{t1}-\xi_{t1}-\f{7}{3}\sigma_{t2}-\f{1}{3}\xi_{t2}\Big|_{x=x_1}~, \label{osrtensors}\\
\alpha_t&\simeq-\sigma_{t2}-\xi_{t2}-\f{7}{3}\sigma_{t3}-\f{1}{3}\xi_{t3}-4\epsilon_H^2-4\epsilon_H\delta_1\Big|_{x=x_1}~\notag,
\end{alignat}
for tensor perturbations. Notice then that the OSR approximation introduces corrections to the standard slow-roll results, Eqs.~\eqref{eq:srhorn}, even at leading order, aided by the different and optimized evaluation point $x=x_1$. Furthermore, the OSR expressions \eqref{osrscalars} can accurately relate inflationary models to the standard power-law, Eq.~\eqref{eq:powerlawform}, in cases when $|\alpha_s|$ is of order $|n_s-1|$, unlike the traditional second-order SR approximation \cite{Motohashi:2015hpa,Ramirez:2018dxe}.

Finally, the tensor-to-scalar ratio can be computed in the standard way through Eq.~\eqref{eq:ttsr}. Note however that it is taken at fixed scale $k$ which in general corresponds to an evaluation point $x=x_1$ at two different epochs $N$ due to the different sound speeds $c_s^2$ and $c_t^2$ for scalars and tensors, respectively. 

The efficiencies of the GSR and OSR approximations have been tested and compared to the standard leading and NLO SR approximation for models with features in the potential as well as for noncanonical models as G-inflation (see, {\it e.g.}, Refs.~\cite{Miranda:2012rm,Miranda:2013wxa,Miranda:2014wga,Motohashi:2015hpa,Motohashi:2017kbs,Ramirez:2018dxe}) and have been further extended for the computation of the bispectrum \cite{Adshead:2011jq,Adshead:2012aa,Adshead:2013ab,Passaglia:2018aa}.\\

\newpage

In the following, we present two appendices which complement some of the topics discussed in this Part I: the \emph{Cosmological perturbation theory} (Appendix \ref{app:pert}) sets the basis for the calculations performed in \S\ref{sec:theoryofquantum}, whereas Appendix \ref{app:eqsofmo} provides the complete set of equations of motion for the theories discussed in \S\ref{sec:STtheories}.\\

After these appendices, Part \ref{sec:papers} contains the publications where the main original research developed during the realization of this thesis is presented.

\lhead[{\bfseries \thepage}]{ \rightmark}
\rhead[ Appendix \thechapter. \leftmark]{\bfseries \thepage}
\appendix
\chapter{Cosmological perturbation theory}
\label{app:pert}

In this appendix we review the cosmological perturbation theory for a FLRW spacetime. We start by defining the group of gauge transformations---coordinate changes---that a given perturbation is subject to. Then, starting from the most general perturbed FLRW metric, we explicitly show the scalar, vector and tensor perturbations composing the perturbed line element, as well as the perturbations composing the energy-momentum tensor of an ideal fluid, as the one described in \S\ref{sec:Introduction}. We later describe how these perturbations transform under the gauge transformations and thus we compute the gauge-invariant variables used in \S\ref{sec:inflation}, relevant for the inflationary theory. Here we mainly follow Refs.~\cite{Mukhanov:2005sc,Baumann:2009ds,Malik:2009aa} and, for the sake of simplicity, we will work only to first-order in perturbations---which suffices for the computation of the power spectrum of primordial perturbations (the computation of the bispectrum requires going to second order, however we do not discuss it in this thesis).

\section{Gauge transformations}
\label{sec:gaugetra}

Now that we want to study perturbations living in a spacetime, the choice of a coordinate system is not as straightforward as in an homogeneous universe. In the latter, we were used to define the threading---curves of constant spatial coordinates $x^i$---as curves corresponding to the motion of free-falling observers with zero momentum density, and the slicing---hypersurfaces of constant time $t$---corresponding to a homogeneous universe.

When perturbations are present, there is no preferred coordinate system anymore and, furthermore, the threading and slicing choice is not unique. This implies that we would be defining the perturbations by specifying the coordinates. It is then important for Cosmology to find how perturbations transform under a change of coordinates---a \emph{gauge} transformation---and to study the evolution of gauge-invariant variables in order to avoid ambiguities due to a given gauge choice.

In general, any quantity can be split in its background component and its perturbations as
\be \ba T(t,x^i)&=\overline T(t)+\delta T(t,x^i)\\ &=\overline T(t)+\sum_{n=1}^\infty\mk{\frac{\epsilon^n}{n!}}\delta T_{(n)}(t,x^i)~, \label{eq:genpertqu} \ea \ee
where overlines represent unperturbed background quantities and $n$ represents the order of the perturbation. Furthermore, $T(t,x^i)$ transforms under a gauge transformation as
\be \widetilde T=e^{\pounds_\xi}T~, \ee
where $\pounds$ denotes a Lie derivative with respect to an auxiliary vector field $\xi$ generating the transformation.\footnote{The gauge transformations form a Lie group with an associated Lie algebra of group generators~\cite{Malik:2009aa}.} Under such a transformation, the right-hand side of Eq.~\eqref{eq:genpertqu} transforms as
\be \ba \widetilde T=&\mk{1+\epsilon^1\pounds_{\xi_1}+\frac12\epsilon^2\pounds^2_{\xi_1}+\frac12\epsilon^2\pounds^2_{\xi_2}+\mathcal{O}(\epsilon^3)}\kk{\overline T+\epsilon^1\delta T_{(1)}+\frac12\epsilon^2\delta T_{(2)}+\mathcal{O}(\epsilon^3)}\\ =&\overline T+\epsilon^1\mk{\delta T_{(1)}+\pounds_{\xi_1}\overline T}+\epsilon^2\mk{\frac12\delta T_{(2)}+\pounds_{\xi_1}\delta T_{(1)}+\frac12\pounds^2_{\xi_1}\overline T+\frac12\pounds_{\xi_2}\overline T}+\mathcal{O}(\epsilon^3)~. \ea \ee
Then, it is evident that background quantities are gauge invariant, whereas first- and second-order perturbations transform as
\begin{alignat}{2} \widetilde{\delta T}_{(1)}&=\delta T_{(1)}+\pounds_{\xi_1}\overline T~, \label{eq:fogtra}\\ \widetilde{\delta T}_{(2)}&=\delta T_{(2)}+\pounds_{\xi_2}\overline T+\pounds^2_{\xi_1}\overline T+2\pounds_{\xi_1}\delta T_{(1)}~. \end{alignat}
Notice that the specific form of the Lie-derivative terms depends on whether the perturbation is a scalar, a vector or a tensor.

\subsection*{Lie derivatives}

The Lie derivatives with respect to the vector field $\xi^\mu$ applied to a scalar $\varphi$, a covariant vector $v_\mu$ and a covariant tensor $t_{\mu\nu}$ are given, respectively, by~\cite{Malik:2009aa}
\begin{alignat}{3} \pounds_\xi\varphi&=\varphi_{,\alpha}\xi^\alpha~,\\ \pounds_\xi v_\mu&=v_{\mu,\alpha}\xi^\alpha+v_\alpha\xi^\alpha_{~,\mu}~,\\ \pounds_\xi t_{\mu\nu}&=t_{\mu\nu,\alpha}\xi^\alpha+t_{\mu\alpha}\xi^\alpha_{~,\nu}+t_{\alpha\nu}\xi^\alpha_{~,\mu}~, \end{alignat}
where we recall that the notation $_{,\alpha}\equiv\partial/\partial x^\alpha$ is used.

In the following, we shall define the cosmological perturbations and apply the transformation rules obtained here to them, where we will keep the analyses to first-order in perturbations.

\section{Metric perturbations}

We start by reviewing the metric perturbations of a FLRW line element given by
\be \dif s^2=\mk{\overline g_{\mu\nu}+\delta g_{\mu\nu}}\dif x^\mu\dif x^\nu~, \ee
where $\overline g_{\mu\nu}(t)$ is the homogeneous FLRW metric given in Eq.~\eqref{eq:flrwm} and $\delta g_{\mu\nu}(t,x_i)$ is composed by the perturbations. Therefore, the most general first-order perturbed FLRW metric  can be written as
\be \dif s^2=-\mk{1+2\Phi}\dif t^2+2aB_i\dif x^i\dif t+a^2\kk{\mk{1-2\Psi}\delta_{ij}+E_{ij}}\dif x^i\dif x^j~. \label{eq:A2} \ee
Here $\Phi$---the \emph{lapse}, which specifies the relation between $t$ and the proper time along the threading---and $\Psi$---the spatial \emph{curvature} perturbation---are 3-scalars, whereas the vector and tensor perturbations $B_i$---the \emph{shift}, which specifies the velocity between the threading and the worldlines orthogonal to the slicing---and $E_{ij}$---the \emph{shear}---can be further decomposed as\footnote{This is called the scalar-vector-tensor decomposition of perturbed quantities into different helicity modes: scalar, vector and tensor perturbations have helicity 0, $\pm1$ and $\pm2$, respectively. Perturbations of different helicity evolve independently and thus they can be studied separately.}
\be B_i\equiv B_{,i}-S_{,i}~, \qquad \text{where} \quad  S_i^{~,i}=0~, \label{eq:svtdecom1} \ee
and
\be E_{ij}\equiv2E_{,ij}+F_{i,j}+F_{j,i}+h_{ij}~, \qquad \text{where} \quad F_{i}^{~,i}=0~, \quad h_i^i=0~, \quad h_{ij}^{~,i}=0~.\ee
Consequently, we have defined two more scalar perturbations, $B$ and $E$, two vector perturbations, $S_i$ and $F_i$, with zero divergence, and a 3-tensor perturbation $h_{ij}$ that is traceless and transverse. 

\subsection*{Gauge transformations of metric perturbations}

Using the gauge transformation properties obtained in \S\ref{sec:gaugetra}, we now explicitly show how the scalar metric perturbations $\Phi$, $B$, $\Psi$ and $E$ transform to first order. Conversely, one can show that vector perturbations $S_i$ and $F_i$ decay very quickly during the expansion and they are actually not produced during inflation~\cite{Weinberg:2008zzc}. Furthermore, the tensor perturbation $h_{ij}$ is gauge invariant--- it does not change under coordinate transformations~\cite{Mukhanov:2005sc}.

For scalar perturbations then, the perturbed metric components are given as
\begin{alignat}{3} \delta g_{00}&=-2\Phi~,\\ \delta g_{0i}&=aB_{,i}\Phi~,\\ \delta g_{ij}&=-2a^2\mk{\Psi\delta_{ij}-E_{,ij}}~, \end{alignat}
which transform to first order, according to Eq.~\eqref{eq:fogtra}, as
\begin{alignat}{3} \widetilde{\delta g}_{00}&=\delta g_{00}-2\dot\alpha~, \label{eq:mettrans15}\\ 
\widetilde{\delta g}_{0i}&=\delta g_{0i}-\alpha_{,i}+a^2\dot\beta_{,i}~, \label{eq:mettrans16}\\ 
\widetilde{\delta g}_{ij}&=\delta g_{ij}+a^2\kk{2H\alpha\delta_{ij}+2\beta_{,ij}}~, \label{eq:mettrans17} \end{alignat}
where we have decomposed the generating vector as $\xi^\mu=(\xi^0,\xi^i)\equiv(\alpha,\beta^{,i}+\gamma^i)$, and set $\gamma^i=0$. Equations~\eqref{eq:mettrans15}-\eqref{eq:mettrans17} give the transformation of each of the scalar perturbations respectively as
\begin{alignat}{4} \widetilde\Phi&=\Phi+\dot\alpha~,\\ \widetilde B&=B-a^{-1}\alpha+a\dot\beta~,\\ \widetilde\Psi&=\Psi-H\alpha~,\\ \widetilde E&=E+\beta~.  \end{alignat}

\section{Matter perturbations}

We consider perturbations present in an ideal fluid characterized by its energy density $\rho$, pressure $p$, 4-velocity $u^\mu$ and anisotropic stress $\Sigma^{\mu\nu}$. Recall that the 4-velocity obeys $g_{\mu\nu}u^\mu u^\nu=-1$ and its only nonvanishing background components are $\overline u^0=-\overline u_0=1$. Therefore we write the perturbed 4-velocity to first order in perturbations, using Eq.~\eqref{eq:A2}, as
\be u^0=\overline u^0+\delta u^0=1-\Phi~, \qquad u_0=\overline u_0+\delta u_0=-1-\Phi~, \ee
and
\be u^i=\delta u^i=\frac1a\mk{v^i-B^i}~, \qquad u_i=\delta u_i=av_i~, \ee
where the linear perturbation $v^i$ is the physical velocity of the fluid (defined with respect to its proper time).

Furthermore, the energy density and pressure can be split in the standard way as
\be \rho(t,x^i)=\overline\rho(t)+\delta\rho(t,x^i)~, \qquad p(t,x^i)=\overline p(t)+\delta p(t,x^i)~. \ee

With these definitions, we can construct the perturbed energy momentum tensor, $T^\mu_\nu=\mk{\rho+p}u^\mu u_\nu+p\delta^\mu_\nu+\Sigma^\mu_\nu$, to first order, as
\begin{alignat}{4} T^0_0&=-\mk{\overline\rho+\delta\rho}~,\\ T^0_i&=\mk{\overline\rho+\overline p}av_i~,\\ T^i_0&=-\frac1a\mk{\overline\rho+\overline p}\mk{v^i-B^i}~,\\ T^i_j&=\mk{\overline p+\delta p}\delta^i_j+\Sigma^i_j~. \end{alignat}
The anisotropic stress tensor $\Sigma^\mu_\nu$ vanishes for the homogeneous FLRW Universe and, furthermore, it is constrained by $\Sigma^{\mu\nu}u_\nu=0$ and $\Sigma^\mu_\mu=0$, \ie only its spatial components are nonzero and define a perturbation.

\subsection*{Gauge transformations of matter perturbations}

In a very similar way as for the metric scalar perturbations, the energy density, pressure and momentum density perturbations transform as
\begin{alignat}{3} \widetilde\delta\rho&=\delta\rho+\dot{\overline\rho}\alpha~,\\ \widetilde\delta p&=\delta p+\dot{\overline p}\alpha~,\\ \widetilde\delta q&=\delta q-\mk{\overline\rho+\overline p}\alpha~, \end{alignat}
where the momentum density perturbation was defined as $(\delta q)_{,i}\equiv\mk{\overline\rho+\overline p}v_i$. Furthermore, the anisotropic stress $\Sigma^i_j$ is gauge invariant.

Analogously, a scalar particle field decomposed as $\phi=\overline\phi+\delta\phi$ transforms to first order as
\be \widetilde\phi=\overline\phi+\delta\phi_{(0)}+\dot{\overline\phi}\alpha~. \ee

Finally, we consider a vector field $A^\mu=(A^0,A^i)$. This field could be present during inflation and play some role in the evolution. We split its temporal and spatial components as
\be A^0=-\overline A+\delta A~, \qquad\qquad A_i=\psi_{,i}~, \ee
in which case, the new scalar perturbations transform according to Eq.~\eqref{eq:fogtra} as
\be \widetilde{\delta A}=\delta A-\dot A_0\alpha+A_0\dot\alpha~, \qquad\qquad \widetilde\psi=\psi+A_0\alpha~. \ee

\section{The primordial curvature perturbation}

We have defined the transformation rules for metric and matter perturbations. However, it is desirable to study the evolution of gauge-invariant variables instead of keeping track of the full set of perturbations plus the generators $\alpha$ and $\beta$, once a particular threading and slicing is defined. By studying only gauge-invariant combinations of these perturbations, we can avoid \emph{fictitious} perturbations or avoid to remove real ones---as James Bardeen stated: \emph{`only quantities that are explicitly invariant under gauge transformations should be considered.'}

\subsection{Gauge invariant variables}


The first two gauge invariant combinations are called \emph{Bardeen potentials} and are written as~\cite{Bardeen:1980kt}
\bali{2} \Phi_\text{B}&\equiv\Phi-\diff{}{t}\kk{a^2\mk{\dot{E}-\frac{B}{a}}}~,\\ \Psi_\text{B}&\equiv\Psi+a^2H\mk{\dot{E}-\f{B}{a}}~. \end{alignat}
One can see that $\Phi_\text{B}$ and $\Psi_\text{B}$ are invariant under gauge transformations, \ie a change of coordinates. Furthermore, if both are equal to zero, then metric perturbations, if present, must be fictitious.

Regarding matter perturbations, we define the following gauge-invariant combinations:
\bali{2} -\zeta&\equiv\Psi+\frac{H}{\dot{\overline\rho}}\delta\rho~, \label{eq:zetapert}\\ \mathcal{R}&\equiv\Psi-\f{H}{\overline\rho+\overline p}\delta q~, \label{eq:Rpert} \end{alignat}
where $\zeta$ is the curvature perturbation on uniform density hypersurfaces, whereas $\mathcal{R}$ is the comoving curvature perturbation. In the following, we shall see that $\zeta$ is conserved after inflation and, therefore, its power spectrum $\mathcal{P}_\zeta$ directly relates the CMB statistical properties with the physics of inflation. $\zeta$ is therefore called \emph{the primordial curvature perturbation}.

\subsection{Einstein equations}

Matter perturbations in a curved spacetime backreact creating geometric perturbations. Consequently, the Einstein equations \eqref{eq:EGR}, written as
\be \delta R_{\mu\nu}-\frac12\delta g_{\mu\nu}\delta R=\delta T_{\mu\nu}~, \label{eq:EGRpert} \ee
determine the evolution of the perturbations previously defined.

The evolution of a given perturbation is usually described in Fourier space, where each perturbed quantity can be decomposed as
\be \delta T(t,k_i)=\int\dif^3x^i\delta T(t,x^i)e^{-ik^ix^i}~, \label{eq:Fourierexp} \ee
where, due to translation invariance, different wavenumbers $k$ evolve independently at linear order~\cite{Baumann:2009ds}.

\subsubsection*{Scalars}

In Fourier space, the Einstein equations can then be written as~\cite{Baumann:2009ds,Gorbunov:2011zzc}
\bali{4} 3H\mk{\dot\Psi+H\Phi}+\f{k^2}{a^2}\kk{\Psi+H\mk{a^2\dot E-aB}}&=-\frac12\delta\rho~,\\ \dot\Psi+H\Phi&=-\frac12\delta q~, \\ \ddot\Psi+3H\dot\Psi+H\dot\Phi+\mk{3H^2+2\dot H}\Phi&=\frac12\mk{\delta\rho-\frac23k^2\delta\Sigma}~,\\ \mk{\Psi_\text{B}-\Phi_\text{B}}&=a^2\delta\Sigma~. \end{alignat}
In addition, the energy-momentum conservation gives the continuity equation and the Euler equation as
\bali{2} \dot{\delta\rho}+3H\mk{\delta\rho+\delta p}&=\frac{k^2}{a^2}\delta q+\mk{\overline\rho+\overline p}\kk{3\dot\Psi+k^2\mk{\dot E-\frac Ba}}~, \label{eq:conteqpert}\\ \dot{\delta q}+3H\delta q&=-\delta p+\frac23k^2\delta\Sigma-\mk{\overline\rho+\overline p}\Phi~. \end{alignat}

Using Eq.~\eqref{eq:zetapert}, Eq.~\eqref{eq:conteqpert} can be written as
\be \dot\zeta=-H\f{\delta p_\text{en}}{\overline\rho+\overline p}+\f{k^2}{3H}\kk{\dot E-\frac Ba+\f{\delta q}{a^2\mk{\overline\rho+\overline p}}}~, \ee
where we have introduced
\be \delta p_\text{en}\equiv\delta p-\f{\dot{\overline p}}{\dot{\overline\rho}}\delta\rho~, \ee
which measures the non-adiabatic part of the pressure perturbation. In inflation, perturbations are adiabatic in general, \ie $\delta p_\text{en}$ vanishes; furthermore, on superhorizon scales where $k/(aH)\ll1$, the second term vanishes as well, \ie the curvature perturbation $\zeta$ remains constant after inflation until scales enter the again the horizon. Consequently, and because the energy density during inflation is $\delta\rho\sim\delta\phi$, we are interested in computing the primordial power spectrum of
\be -\zeta_\phi\simeq\Psi+\f{H}{\dot{\overline\phi}}\delta\phi~, \label{eq:zetaphi} \ee
at horizon exit $k\sim aH$, and ignore the subsequent physics.

In the same way, one can define the curvature perturbation $\zeta_{\psi}$ for the scalar component of a vector field $A_i$ as
\be -\zeta_{\psi}\simeq\Psi+\f{H}{A_0}\psi~, \ee
and, furthermore, define a total curvature perturbation in the case in which both fields, $\phi$ and $A^\mu$, are playing a role in the inflationary dynamics, as
\be -\zeta=\Psi-\f{H\mk{\dot\phi\delta\phi+A_0\psi}}{\dot\phi^2+A_0^2}~, \ee
which is analogous to a two-field model of inflation~\cite{Malik:2009aa}.

\subsubsection*{Vectors}

The evolution equations for vector perturbations are sourced by an anisotropic stress perturbation $\delta\Sigma_i$ and are given by
\begin{alignat}{2} \dot{\delta q}_i+3H\delta q_i=k^2\delta\Sigma_i~,\\ k^2\mk{\dot F_i+\f{S_i}{a}}=2\delta q_i~. \end{alignat}
However, $\delta\Sigma_i$ is not created by inflation and, in its absence, $\delta q_i$ decays with the expansion, \ie the perturbation $\dot F_i+S_i/a$ vanishes. Therefore, vector perturbations are, in general, subdominant.

\subsubsection*{Tensors}

The evolution equation for the tensor perturbation $h_{ij}$ is given by
\be \ddot h_{ij}+3H\dot h_{ij}+\f{k^2}{a^2}h=0~, \label{eq:eqprimgravwave} \ee
which is the equation for a gravitational wave. They are produced by inflation and, in the same way as vectors, they decay with the expansion; however some models of inflation predict an observable amount of gravitational waves during the recombination epoch, \ie they can be distinguished in the CMB polarization spectrum.


\chapter{Equations of motion of general theories of gravity}
\label{app:eqsofmo}

In this appendix we show the equations of motion for the full Horndeski and SVT theories in a FLRW spacetime. The former were first computed in Ref. \cite{Kobayashi:2011nu} whereas the latter can be found in Ref. \cite{Heisenberg:2018mxx}.

\section{Horndeski theory}

We take a homogeneous scalar field $\phi=\phi(t)$ and assume a flat FLRW background with the line element given as
\be \label{eq:flrw2} \dif s^2=-N^2(t) dt^2+a^2(t)\delta_{ij}\dif x^i\dif x^j~, \ee
to the action
\be \mathcal{S_H}=\int\dif^4x\sqrt{-g}\mathcal{L_H}~, \label{eq:hornact} \ee
where $\mathcal{L_H}$ is given by Eq.~\eqref{eq:Horn}. The variation of Eq.~\eqref{eq:hornact} with respect to $N(t)$ gives the constraint equation
\be \sum_{i=2}^5\mathcal{E}_i=0~, \ee
where
\begin{alignat}{4}
\mathcal{E}_2=&2XG_{2,X}-G_2~, \\
\mathcal{E}_3=&2XG_{3,\phi}-6X\dot\phi HG_{3,X}~, \\
\mathcal{E}_4=&-6H^2G_4+24H^2X\mk{G_{4,X}+XG_{4,XX}}-12HX\dot\phi G_{4,\phi X} \notag\\
&-6H\dot\phi G_{4,\phi }~, \\
\mathcal{E}_5=&2H^3X\dot\phi\mk{5G_{5,X}+2XG_{5,XX}}-6H^2X\mk{3G_{5,\phi}+2XG_{5,\phi X}}~.
\end{alignat}

The variation with respect to $a(t)$ yields the evolution equation
\be \sum_{i=2}^5\mathcal{P}_i=0~, \ee
where
\begin{alignat}{6}
\mathcal{P}_2=&G_2~, \\
\mathcal{P}_3=&2X\mk{G_{3,\phi}+\ddot\phi G_{3,X}}~, \\
\mathcal{P}_4=&2\mk{3H^2+2\dot H}G_4-4\mk{3H^2X+H\dot X+2\dot HX}G_{4,X}-8HX\dot XG_{4,XX} \notag\\
&+2\mk{\ddot\phi+2H\dot\phi}G_{4,\phi}+4XG_{4,\phi\phi} +4X\mk{\ddot\phi-2H\dot\phi}G_{4,\phi X}~, \\
\mathcal{P}_5=&-2X\mk{2H^3\dot\phi+2H\dot H\dot\phi+3H^2\ddot\phi}G_{5X}-4H^2X^2\ddot\phi G_{5,XX}+4HX\dot\phi G_{5,\phi\phi} \notag\\
&+4HX\mk{\dot X-HX}G_{5,\phi X}+2\kk{2\mk{\dot HX+H\dot X}+3H^2X}G_{5,\phi}~.
\end{alignat}

Finally, the variation with respect to $\phi(t)$ gives the scalar-field equation of motion
\be \frac1{a^3}\diff{}{t}\mk{a^3 J}=P_\phi~, \ee
where
\begin{alignat}{4}
J=&\dot\phi G_{2,X}-6HXG_{3,X}+2\dot\phi G_{3,\phi}+6H^2\dot\phi\mk{G_{4,X}+2XG_{4,XX}}-12HXG_{4,\phi X} \notag\\
&+2H^3X\mk{3G_{5,X}+2XG_{5,XX}}-6H^2\dot\phi\mk{G_{5,\phi}+XG_{5,\phi X}}~, \\
P_\phi=&G_{2,\phi}+2X\mk{G_{3,\phi\phi}+\ddot\phi G_{3,\phi X}}+6\mk{2H^2+\dot H}G_{4,\phi}+6H\mk{\dot X+2HX}G_{4,\phi X} \notag\\
&-6H^2XG_{5,\phi\phi}+2H^3X\dot\phi G_{5,\phi X}~. 
\end{alignat}

For the particular choice of $G_4=1/2$ and $G_5=0$, the above equations reduce to the set of equations \eqref{eq:gmov1}-\eqref{eq:gmov3} corresponding to the G-inflation model discussed in \S\ref{sec:ginfl}.

\subsection{Normalization factors}
\label{sec:norfactapp}

Additionally, let us show the dependence on the $G_i(\phi,X)$ functions of the normalization factors $c_{s,t}^2$ and $b_{s,t}$ appearing in the quadratic actions of primordial perturbations, Eqs.~\eqref{eq:acthorn} and \eqref{eq:acthorn2}. As given in Ref.~\cite{Kobayashi:2011nu}, they read as
\be \label{eq:norfactapp} c_{s,t}^2=\f{\mathcal F_{s,t}}{\mathcal G_{s,t}}~, \qquad\qquad b_s=\f{\mathcal F_s}{\epsilon_H}~, \qquad\qquad b_t=4\mathcal F_t~, \ee
where
\be \ba
\mathcal{F}_s=\frac1{a}\diff{}{t}\mk{\frac a\Theta\mathcal{G}_t^2}-\mathcal{F}_t~, \qquad\qquad
\mathcal{G}_s=\frac\Sigma{\Theta^2}\mathcal{G}_t^2+3\mathcal{G}_t~,
\ea \ee
and
\be \ba
\mathcal{F}_t= & 2\kk{G_4-X\mk{\ddot\phi G_{5,X}+G_{5,\phi}}}~,\\ 
\mathcal{G}_t= & 2\kk{G_4-2XG_{4,X}-X\mk{H\dot\phi G_{5,X} -G_{5,\phi}}}~, \\
\Sigma= &XG_{2,X}+2X^2G_{2,XX}+12H\dot\phi XG_{3,X}+6H\dot\phi X^2G_{3,XX}-2XG_{3,\phi} \\
&-2X^2G_{3,\phi X}-6H^2G_4+6\biggl[H^2\mk{7XG_{4,X}+16X^2G_{4,XX}+4X^3G_{4,XXX}} \\
&-H\dot\phi\mk{G_{4,\phi}+5XG_{4,\phi X}+2X^2G_{4,\phi XX}}\biggr]+30H^3\dot\phi XG_{5,X}+26H^3\dot\phi X^2G_{5,XX} \\ &+4H^3\dot\phi X^3G_{5,XXX}-6H^2X\mk{6G_{5,\phi}+9XG_{5,\phi X}+2 X^2G_{5,\phi XX}}~,\\
\Theta= & -\dot\phi XG_{3,X}+2HG_4-8HXG_{4,X}-8HX^2G_{4,XX}+\dot\phi G_{4,\phi}+2X\dot\phi G_{4,\phi X}\\
&-H^2\dot\phi\mk{5XG_{5,X}+2X^2G_{5,XX}}+2HX\mk{3G_{5,\phi}+2XG_{5,\phi X}}~.
\ea \ee

\section{Scalar-vector-tensor theories}

We assume the line element in Eq.~\eqref{eq:flrw2} and consider homogeneous scalar and vector field configurations, $\phi(t)$ and $A_\mu(t)$, the latter of which is given by 
\be A_{\mu}(t)=\mk{A_0(t)N(t),0,0,0}~, \ee
where $A_0(t)$ is a time-dependent temporal vector component. Furthermore, the quantities $\ck{F, Y_1, Y_2, Y_3}$, the last row of Eq.~\eqref{eq:svt2}, corresponding to the sixth-order Lagrangian ${\cal L}_6$, and the interactions proportional to $\mathcal{M}^{\mu\nu}_5$ and $\mathcal{N}^{\mu\nu}_5$, do not affect the background cosmology.~\footnote{Furthermore, the parity-violating term $\tilde{F}$ in $f_2$, Eq.~\eqref{eq:svt2}, is not considered in this chapter as it was originally not considered in Ref.~\cite{Heisenberg:2018mxx} for simplicity.} Finally, the quantities $X_1$, $X_2$, $X_3$ are given, respectively, by 
\be X_1=\frac{\dot{\phi}^2}{2N^2}~,\qquad X_2=\frac{\dot{\phi}A_0}{2N}~,\qquad X_3=\frac{A_0^2}{2}~.\ee

With the above considerations, varying the action
\be \mathcal{S}_{\rm SVT}=\int\dif^4x\sqrt{-g}\mathcal{L}_{\rm SVT}~, \ee
on the spacetime metric \eqref{eq:flrw2}, with respect to $N$ gives the constraint equation
\begin{alignat}{11}
6f_4H^2+f_2-\dot{\phi}^2f_{2,X_1}-\frac12\dot{\phi}A_0f_{2,X_2}+6H\mk{\dot{\phi}f_{4,\phi}-HA_0^2 f_{4,X_3}}& \notag\\ +2A_0H^2\mk{3\dot{\phi}f_{5,\phi}-A_0^2Hf_{5,X_3}}&=0~. \label{eq:svtback1}
\end{alignat}
Varying the action \eqref{eq:svtback1} with respect to the scale factor $a(t)$ yields the evolution equation
\begin{alignat}{11}
2f_4\mk{2\dot{H}+3H^2}+f_2+2\dot{A}_0A_0^2\mk{f_{3,X_3}+\tilde{f}_3}+2\dot{\phi}A_0f_{3,\phi}+2\mk{\ddot{\phi}+2H\dot{\phi}}f_{4,\phi}& \notag\\
-2A_0\kk{A_0\mk{2\dot{H}+3H^2}+2\dot{A}_0H}f_{4,X_3}+2\dot{\phi}\dot{A}_0A_0f_{4,X_3 \phi}+2\dot{\phi}^2f_{4,\phi\phi}& \notag\\
-4HA_0^2\mk{\dot{A}_0A_0f_{4,X_3X_3}+\dot{\phi}f_{4,X_3 \phi}}+\bigl[2A_0\mk{H\ddot{\phi}+\dot{H}\dot{\phi}}& \notag\\
+\dot{\phi}\mk{2H\dot{A}_0+3H^2A_0}\bigr]f_{5,\phi}-HA_0^2\kk{2A_0\mk{\dot{H}+H^2}+3\dot{A}_0H}f_{5,X_3}& \notag\\ 
+H\dot{\phi}A_0^2\mk{2\dot{A}_0-HA_0}f_{5,X_3 \phi}+HA_0\mk{2\dot{\phi}^2f_{5,\phi \phi}-\dot{A}_0A_0^3 Hf_{5,X_3 X_3}}&=0~. \label{eq:svtback2}
\end{alignat}
The variation with respect to $\phi(t)$ gives the scalar-field equation of motion
\begin{alignat}{11}
\mk{f_{2,X_1}+\dot{\phi}^2f_{2,X_1X_1}+\dot{\phi}A_0f_{2,X_1X_2}+\frac14A_0^2f_{2,X_2 X_2}}\ddot{\phi}+3Hf_{2,X_1}\dot{\phi}-f_{2,\phi}& \notag\\
+\dot{\phi}^2f_{2,X_1\phi}-6\mk{\dot{H}+2H^2}f_{4,\phi}+\biggl[\frac12f_{2,X_2}+\frac12\dot{\phi}^2f_{2,X_1 X_2}+2f_{3,\phi}-3H^2 f_{5,\phi}& \notag \\
+A_0\mk{\dot{\phi}f_{2,X_1 X_3}+\frac14\dot{\phi}f_{2,X_2X_2}-6Hf_{4,X_3\phi}}+\frac{A_0^2}{2}\bigl(f_{2,X_2 X_3}-4\tilde{f}_{3,\phi}& \notag\\
-6H^2f_{5,X_3\phi}\bigr)\biggr]\dot{A}_0+\biggl[\frac12\dot{\phi}f_{2,X_2 \phi}+\frac32H f_{2,X_2}+6Hf_{3,\phi}-6A_0H^2f_{4,X_3\phi}& \notag\\
-3H\mk{2\dot{H}+3H^2}f_{5,\phi}-A_0^2H^3f_{5,X_3 \phi}\biggr]A_0&=0~, \label{eq:svtback3}
\end{alignat}
whereas the variation with respect to $A_0$ gives the temporal-vector equation of motion
\begin{alignat}{11}
2\mk{f_{2,X_3}+6H^2f_{4,X_3}-6H\dot{\phi}f_{4,X_3 \phi}}A_0+12H^2f_{4,X_3X_3}A_0^3+2H^3f_{5,X_3X_3}A_0^4& \notag\\
-2\mk{6Hf_{3,X_3}+6H\tilde{f}_3+2\dot{\phi}\tilde{f}_{3,\phi}-3H^3 f_{5,X_3}+3H^2\dot{\phi}f_{5,X_3\phi}}A_0^2& \notag\\
+\mk{f_{2,X2}+4f_{3,\phi}-6H^2 f_{5,\phi}}\dot{\phi}&=0~. \label{eq:svtback4}
\end{alignat}

Notice from Eqs.~\eqref{eq:svtback3} and \eqref{eq:svtback3} that the scalar field $\phi$ and the temporal vector component $A_0$ are coupled to each other in a non-trivial way (see Ref. \cite{Heisenberg:2018mxx} for an exhaustive discussion on the implications of this fact). For the particular model in Eq.~\eqref{eq:mixmodel}, Eqs.~\eqref{eq:svtback1}-\eqref{eq:svtback4} simplify to those given by Eqs.~\eqref{eq:svtHm1}-\eqref{eq:Adotphi}.


\setcounter{part}{2}

\part*{\addcontentsline{toc}{part}{{II~~ Scientific Research}} Part II\\[1cm] Scientific Research\\[2cm] \normalfont \normalsize{ 
This Part originally includes the publications as they are provided in the journals. For this version, however, only the front pages are kept. The interested reader may follow their references:~\cite{Boubekeur:2014xva,Boubekeur:2015xza,Escudero:2015wba,Ramirez:2018dxe,Heisenberg:2018erb}.}
}\label{sec:papers}\thispagestyle{empty}

\renewcommand{\headrulewidth}{0pt}

\phantomsection\addcontentsline{toc}{section}{{\bf1}~~Phenomenological approaches of inflation and their equivalence}
\includepdf[pages=1]{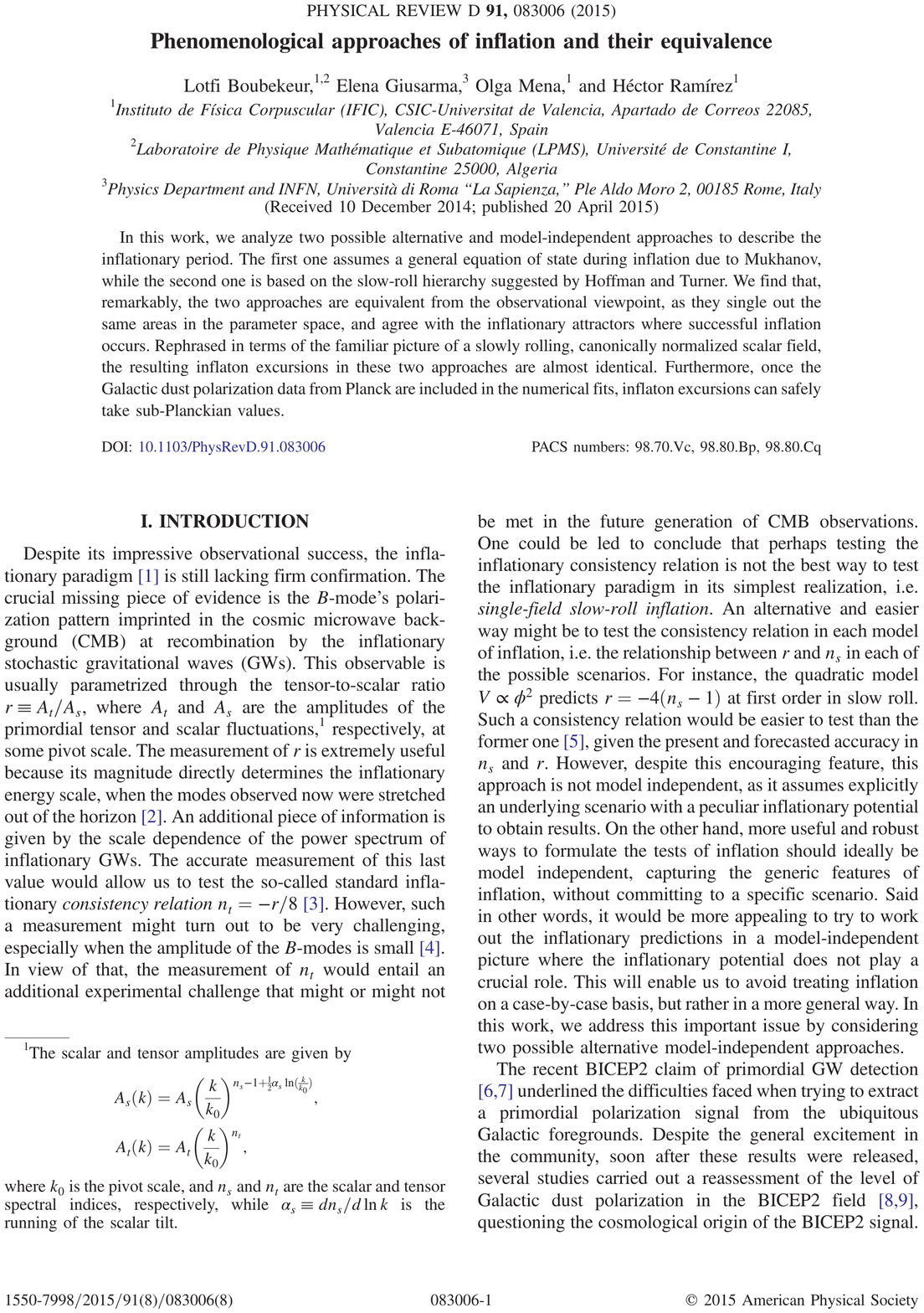}
\phantomsection\addcontentsline{toc}{section}{{\bf2}~~Do current data prefer a nonminimally coupled inflaton?}
\includepdf[pages=1]{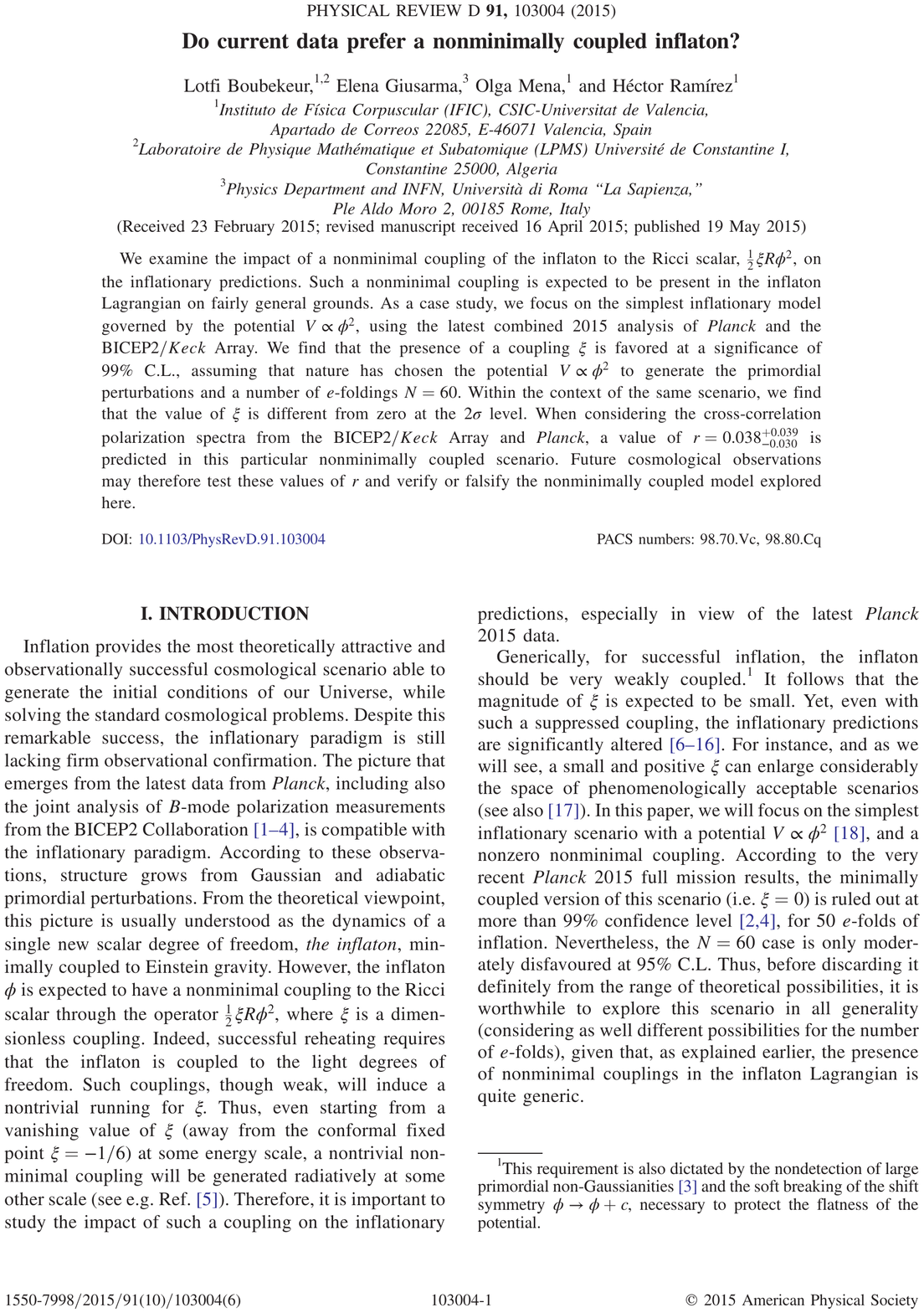}
\phantomsection\addcontentsline{toc}{section}{{\bf3}~~The present and future of the most favoured inflationary models after Planck 2015}
\includepdf[pages=1]{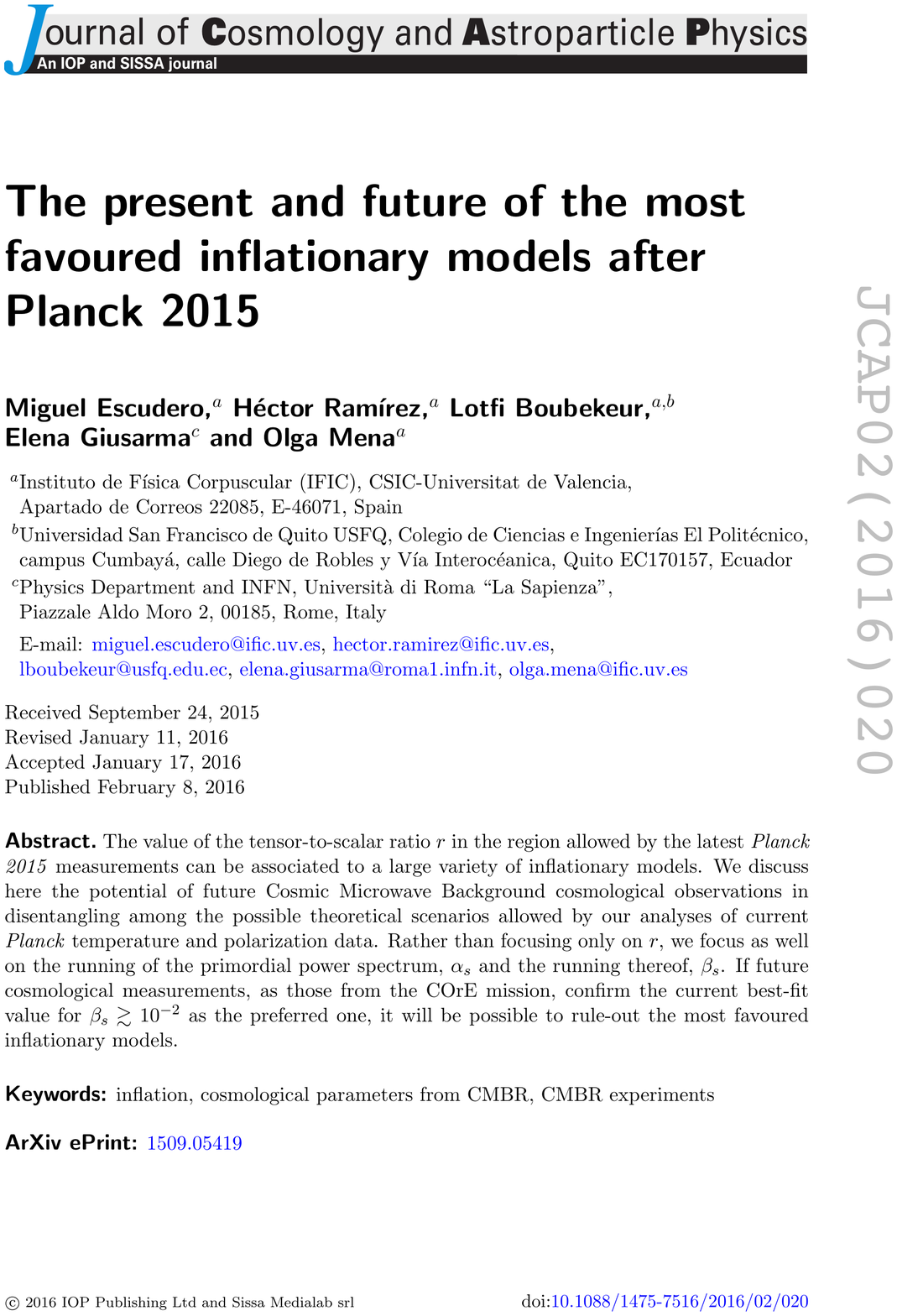}
\phantomsection\addcontentsline{toc}{section}{{\bf4}~~Reconciling tensor and scalar observables in G-inflation}
\includepdf[pages=1]{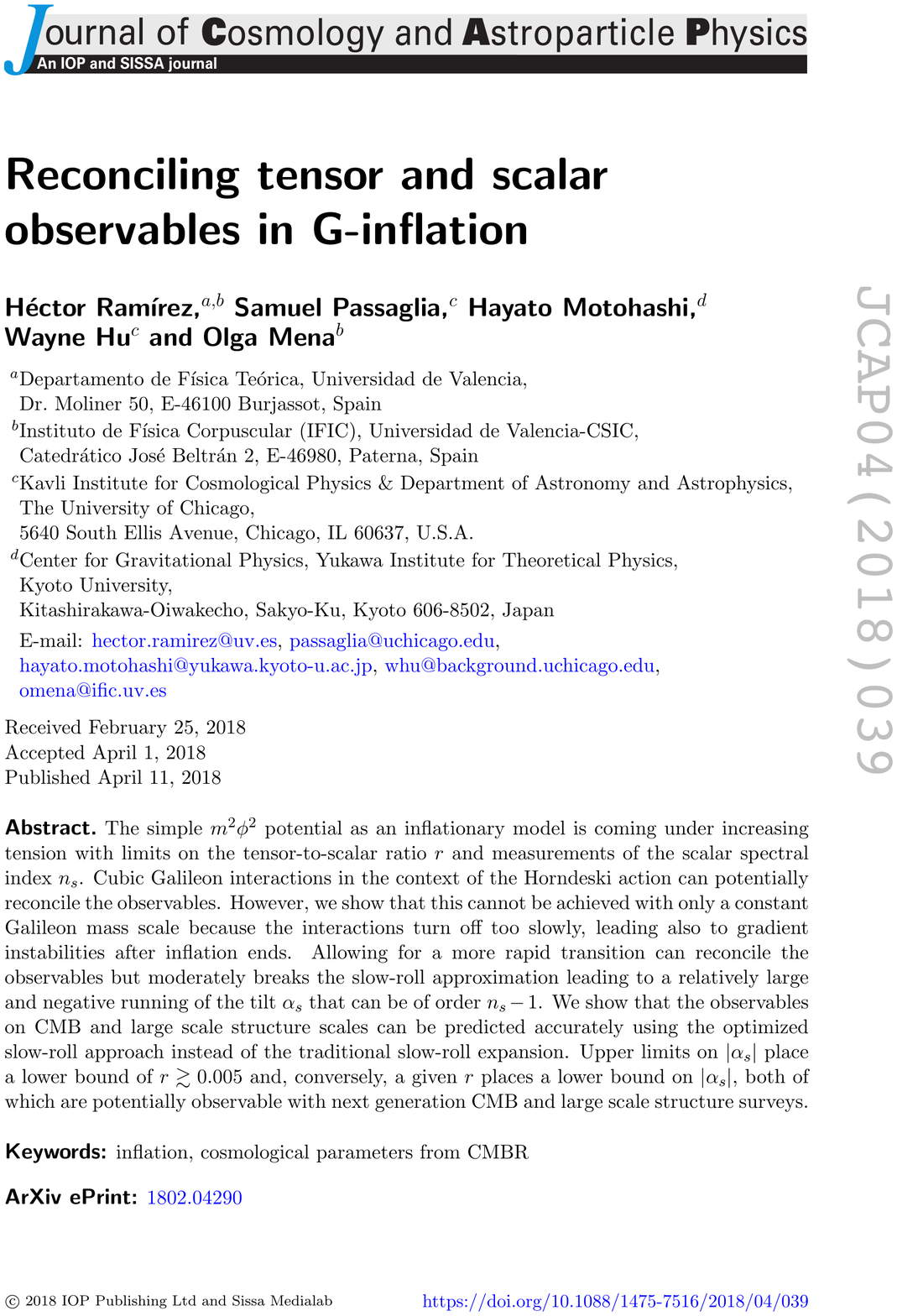}
\phantomsection\addcontentsline{toc}{section}{{\bf5}~~Inflation with mixed helicities and its observational imprint on CMB}
\includepdf[pages=1]{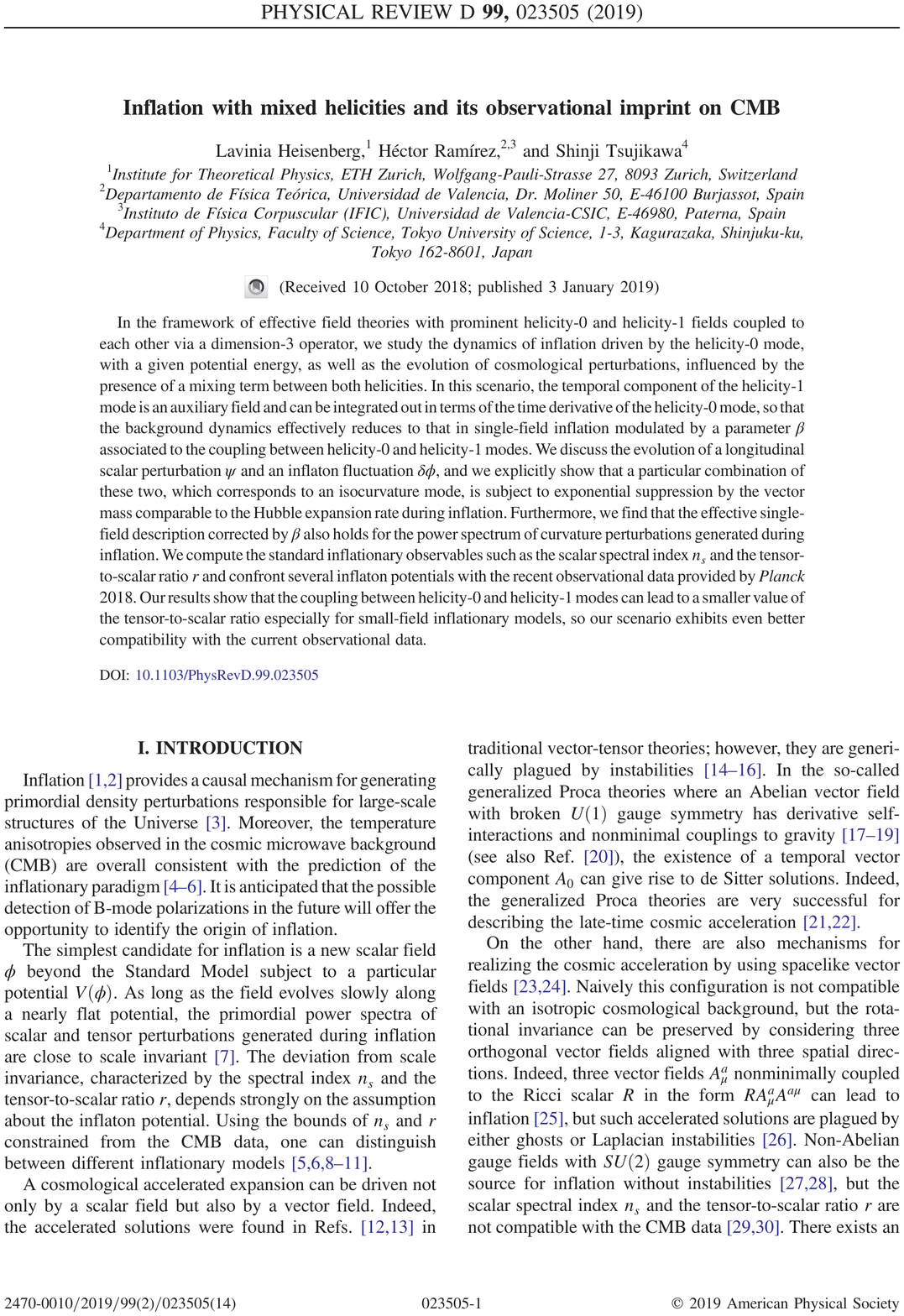}

\part{Summary \& Conclusions}\label{sec:summary}\thispagestyle{empty}
\renewcommand{\headrulewidth}{0.5pt}
\ChRuleWidth{0pt}

\lhead[{\bfseries \thepage}]{ \rightmark}
\rhead[ Summary \& Conclusions. \leftmark]{\bfseries \thepage}
\subsection*{}
\label{sec:summary}
\thispagestyle{plain}

\noindent {\LARGE\bf I}nflation elegantly solves the main problems of the standard cosmological model---the observed homogeneity, isotropy, and flatness of the Universe are simple outcomes of an early accelerated epoch. Furthermore, quantum fluctuations during this epoch are stretched by the expansion to classical scales, becoming the seeds for the structures. Yet, the canonical inflationary theory is becoming in worrisome tension with state-of-the-art cosmological observations. Assuming a single-field picture, \ie a scalar field slowly rolling down its sufficiently flat potential, the simplest monomial potential functions and similar constructions are in the edge of being ruled out---indeed, they predict large tensor power, meanwhile upper bounds on the tensor-to-scalar ratio $r$ have been significantly reduced by the latest \emph{Plank} satellite's measurements. In general, models embedded in high-energy particle-physics theories are in better shape, and therefore seeking new inflationary scenarios within the framework of particle physics became a natural approach. However, given the lack of evidence of these theories, different approaches seem more appealing: on the one hand, modeling-independent realizations have the potential of unveiling allowed parameter regions for general classes of inflationary models. On the other hand, a different model-building approach consists on keeping the simplest potentials but at the cost of modifying the underlying theory of gravity. These two alternative approaches constituted the main subject of this thesis.


In canonical single-field scenarios, the equation of state $w$ can be parame-trized with only two phenomenological parameters, $\alpha$ and $\beta$, in addition to the number of remaining $e$folds of inflation, as discussed in Chapter \S\ref{sec:MIApp}. Also, $w$ is related to the first slow-roll parameter $\epsilon_H$. Consequently, it is possible to relate the tilts $n_{s,t}$ and the tensor-to-scalar ratio $r$ to the parameters $\alpha$ and $\beta$. In other words, predictions on $n_{s,t}$ or $r$ can be obtained by constraining the Mukhanov phenomenological parameters without making any reference to a specific potential function $V(\phi)$. Indeed, in \S\ref{sec:mukpar} we showed that a lower bound on $r$ can be predicted only by taking the current constraints on $n_s$ on account of this parametrization, in the case that Nature has chosen canonical single-field inflation. Furthermore, in Ref. \cite{Boubekeur:2014xva} we have explicitly shown that this parametrization is in agreement with a more familiar one based on a hierarchy of the slow-roll parameters as they both single out the same parameter space when they are fitted to CMB temperature and polarization data.\\

The dark energy issue points to the possibility that the laws of gravity, \ie General Relativity, may need to be modified at large scales. This question brought up a large research area from which many theories of modified gravity have been proposed. Theories based on scalar fields, similar to inflation, led to the proposal of interesting couplings between a scalar field $\phi$ and the gravity sector. When studied for inflation, nonminimal couplings (expected indeed to be generated at some energy scale) are able to modify the predictions of a canonical potential $V(\phi)$. In particular, we showed in Ref. \cite{Boubekeur:2015xza} that a coupling $\xi\phi^2R$ is favored by present observations for small---$\mathcal O(10^{-3})$---and positive values of the coupling parameter $\xi$ at the 2$\sigma$ level, when a simple quadratic potential function, $m^2\phi^2$ is considered. Furthermore, in the presence of such a coupling, a nonzero value for the tensor-to-scalar ratio is also favored at the same confidence level.

These phenomenological outcomes obtained by simple nonminimal couplings of the inflaton field to the gravity sector lead to the search for phenomenological signatures derived from different types of terms allowed to be present in the theory. Keeping the symmetries and constraints of General Relativity (such as Lorentz invariance, unitarity and locality), along with the condition for second-order equations of motion (in order to avoid Ostrogradsky instabilities), only few combinations between self-derivative terms of $\phi$ and the gravity sector are allowed. This led to the constructions of general scalar-tensor theories of gravity, discussed in \S\ref{sec:STinst}, and from which the Horndeski framework stands out. This framework has become a playground for the construction of well-behaved models of inflation from which G-inflation, discussed in \S\ref{sec:ginfl}, is the simplest nontrivial one. Its action differentiates from the canonical due to the addition of the self-derivative term $G_3\Box\phi$, which introduces new phenomenology for a given choice of the free function $G_3(\phi, X)$, irrespectively of the potential function $V(\phi)$. Therefore, it allows inflation to be driven by monomial potentials and still satisfy observational constraints. A basic $G_3$ function with constant mass scale, however, is not able to fit the latest \emph{Planck} 2018 constraints due to their prediction of small scalar power relative to the tensor one as a consequence of a slow transition between the epoch when the mass scale is relevant to the canonical epoch. Furthermore, failure to complete the transition before the end of inflation leads to the appearance of instabilities during the reheating epoch. In Ref. \cite{Ramirez:2018dxe} we showed that a sufficiently fast, step-like transition is able to reconcile the monomial potentials (special attention was taken for the simplest $m^2\phi^2$) with observations and, at the same time, to solve the pathological issues presented in this class of models. Interestingly, by assuming a scalar tilt of $n_s\simeq0.966$, a lower bound on the tensor-to-scalar ratio of $r\gtrsim5\times10^{-3}$ (\ie potentially observable with next-generation satellites) is obtained for this transient model due to a nontrivial large running $\alpha_s$ originated by the sharp transition.\\

It is possible that vector fields were also present during the inflationary era. Regardless of their interaction with the inflaton field, they can affect and modify the dynamics of the expansion. However, couplings between the inflaton and the vector field, on top of the gravitational background, are of special interest and can be tested against CMB observations. In this regard, a general framework of scalar-vector-tensor theories of gravity was recently developed in the same spirit as the scalar-tensor Horndeski theories. Indeed, in \S\ref{sec:mostgenSVT} we showed that any modification of general relativity will introduce new degrees of freedom which can be in the form of new scalar, vector or even tensor fields. The simplest nontrivial combination of a scalar field $\phi$ and a $U(1)$ vector $A_\mu$, according to the scalar-vector-tensor framework, is given by $\beta_mA^\mu\partial_\mu\phi$. In Ref. \cite{Heisenberg:2018erb} we studied the inflationary signatures of several potential functions in the presence of this coupling. At the background level, inflation is still driven by the scalar field, whereas the temporal component of the vector field, $A_0$, is nondynamical. As a consequence, a single-field description of the background dynamics (modulated by a parameter $\beta=1-const.\times\beta_m^2$
) arises due to a nontrivial relation between $A_0$ and the scalar-field velocity $\dot\phi$. At the perturbations level, the longitudinal vector mode contributes to an isocurvature perturbation along with the standard inflaton fluctuation. This perturbation, however, is suppressed for a vector-field mass scale comparable to the Hubble parameter $H$, and, as a result, the power spectrum of the primordial curvature perturbation follows the same single-field description corrected by $\beta$. The spectral indices and the tensor-to-scalar ratio are further modified by the presence of $\beta$ but, interestingly, the canonical consistency relation $r=-8n_t$ is left unmodified. While confronting these results with CMB data, we showed that small-field models of inflation are considerably affected by the presence of the vector coupling. In particular, we found that for $\alpha$-attractors, $n_s\simeq1-2/N$ and $r\simeq12\alpha_c\beta/N^2$ for $\alpha_c<\mathcal O(10)$ (which includes Starobinsky inflation), \ie there exists a suppression of the tensor-to-scalar ratio for a small $\beta$ compared to the canonical models ($\beta=1$). Similar results were obtained for Brane inflation with $p=2$ and $p=4$ indices. These results follow from the fact that a nonvanishing $\beta$ increases $\dot\phi$ and, therefore, inflation needs to start from a flatter region of the potential, relevant for CMB scales, where a small value of the slow-roll parameter $\epsilon_H$ is maintained compared to the canonical cases.\\

The computation of the inflationary observables in noncanonical classes of inflation is, in general, far from being trivial. As discussed in \S\ref{sec:soluMukhSas}, different methods can be used to solve the mode-function equation, from where the slow-roll approximation (SR) usually stands out as the one which leads to analytical results. However, for noncanonical models, the slow-roll conditions are sometimes too restrictive and the use of the slow-roll approximation is not always allowed. This comes from the fact that new noncanonical terms affect the background dynamics of the inflaton field which usually depart from the standard smooth evolution. Furthermore, the slow-roll approximation is based on an assumed hierarchy of Hubble slow-roll parameters which, in turn, define a restrictive hierarchy of the primordial tilt and its running parameters ($\alpha_s$, $\beta_s$, etc.), as it was discussed at the beginning of \S\ref{sec:GSR}. To overcome these deficiencies, the generalized (GSR) and optimized (OSR) slow-roll approximations were developed and tested for several inflationary models with features in the potential. Additionally, these techniques were recently promoted to cover inflationary models belonging to the Horndeski and beyond Horndeski classes. Indeed, in Ref. \cite{Ramirez:2018dxe} we showed that the predictions of the aforementioned transient G-inflation model can be accurately computed using these techniques. On the one hand, GSR provides accurate results at the $\sim10\%$ level around the transition (of size $\Delta N\sim3$) to the canonical epoch; OSR, on the other hand, provides analytical results accurate at the $\sim20\%$ level. Both being compared with the standard SR approximation which deviates at the $\sim50\%$ level. However, due to the properties of the model, it is worth mentioning that both GSR and OSR give predictions at the percent level at CMB scales, whereas SR still deviates at the $\sim10\%$ level. We further showed that these results imply that the scalar power spectrum can still be described in its power-law form, around the relevant scales, as long as $n_s$ and $\alpha_s$ are computed using OSR. This is due to the fact that $\alpha_s$ can be of the same order of $n_s$ and thus the standard slow-roll hierarchy is not valid (in which case, OSR overcomes this wrong order-counting).

Indeed, a correct computation of the inflationary parameters is needed as further parameters, in particular the running of the running of the tilt, $\beta_s$, will play an important role to discern between inflationary models of inflation---this in addition to the possibility of a further unobservable amplitude of primordial gravitational waves---, as we showed in Ref. \cite{Escudero:2015wba}. In this work, a forecast for the CORE mission was carried out and confronted to the most favored 
models of inflation. We showed that there exists the possibility that the running $\alpha_s$ will not be as important as its own running $\beta_s$, as the latter may have the power to exclude all the models studied in the case that the best-fit value of \emph{Planck}, $\beta_s\simeq0.01$, prevails future observations.\\

To conclude, in this thesis we have developed a comprehensive novel exploration and a detailed study of the inflationary paradigm using different nonstandard approaches. Firstly, we covered model-independent parametrizations to clarify the allowed parameter space of canonical single-field inflation. Secondly, we demonstrated the potential of nonstandard inflationary parameters, the running of the running of the primordial tilt in particular, which may have the potential of ruling out the vast majority of the currently favored inflationary models.  And, finally, we explored the possibility that the inflaton field coupled differently as in the canonical version by introducing self-derivative terms belonging to general scalar-tensor theories; or the possibility that a gauge vector field, coupled to the inflaton, affected the dynamics in an observable way.

Future satellites, interferometers and different ground-based experiments will further guide us towards unveiling the true nature of the early universe. And, whether Nature chose a canonical model embedded in a more fundamental quantum field theory or nontrivial gravitational dynamics, model-building approaches along with a correct understanding of the observational parameters will keep helping in showing us the correct theoretical path.

\part{Bibliography}\label{sec:references}\thispagestyle{empty}
\renewcommand{\headrulewidth}{0.5pt}

\cleardoublepage
\phantomsection
\renewcommand{\headrulewidth}{0.5pt}
\lhead[{\bfseries \thepage}]{Bibliography}
\rhead[{Bibliography}]{\bfseries \thepage}
\begingroup
   \def\chapter*#1{}
\bibliographystyle{jhep}
\bibliography{Thesis}

\providecommand{\href}[2]{#2}\begingroup\raggedright\begin{thebibliography}{100}

\bibitem{Boubekeur:2014xva}
L.~Boubekeur, E.~Giusarma, O.~Mena and H.~Ram{\'\i}rez, \emph{{Phenomenological
  approaches of inflation and their equivalence}},
  \href{http://dx.doi.org/10.1103/PhysRevD.91.083006}{\emph{Phys. Rev.} {\bf
  D91} (2015) 083006}, [\href{http://arxiv.org/abs/1411.7237}{{\tt
  1411.7237}}].

\bibitem{Boubekeur:2015xza}
L.~Boubekeur, E.~Giusarma, O.~Mena and H.~Ram{\'\i}rez, \emph{{Do current data
  prefer a nonminimally coupled inflaton?}},
  \href{http://dx.doi.org/10.1103/PhysRevD.91.103004}{\emph{Phys. Rev.} {\bf
  D91} (2015) 103004}, [\href{http://arxiv.org/abs/1502.05193}{{\tt
  1502.05193}}].

\bibitem{Escudero:2015wba}
M.~Escudero, H.~Ram{\'\i}rez, L.~Boubekeur, E.~Giusarma and O.~Mena, \emph{{The
  present and future of the most favoured inflationary models after $Planck$
  2015}}, \href{http://dx.doi.org/10.1088/1475-7516/2016/02/020}{\emph{JCAP}
  {\bf 1602} (2016) 020}, [\href{http://arxiv.org/abs/1509.05419}{{\tt
  1509.05419}}].

\bibitem{Ramirez:2018dxe}
H.~Ram{\'\i}rez, S.~Passaglia, H.~Motohashi, W.~Hu and O.~Mena,
  \emph{{Reconciling tensor and scalar observables in G-inflation}},
  \href{http://dx.doi.org/10.1088/1475-7516/2018/04/039}{\emph{JCAP} {\bf 1804}
  (2018) 039}, [\href{http://arxiv.org/abs/1802.04290}{{\tt 1802.04290}}].

\bibitem{Heisenberg:2018erb}
L.~Heisenberg, H.~Ram{\'\i}rez and S.~Tsujikawa, \emph{{Inflation with mixed
  helicities and its observational imprint on CMB}},
  \href{http://dx.doi.org/10.1103/PhysRevD.99.023505}{\emph{Phys. Rev.} {\bf
  D99} (2019) 023505}, [\href{http://arxiv.org/abs/1812.03340}{{\tt
  1812.03340}}].

\bibitem{Gariazzo:2016blm}
S.~Gariazzo, O.~Mena, H.~Ram{\'\i}rez and L.~Boubekeur, \emph{{Primordial power
  spectrum features in phenomenological descriptions of inflation}},
  \href{http://dx.doi.org/10.1016/j.dark.2017.07.003}{\emph{Phys. Dark Univ.}
  {\bf 17} (2017) 38--45}, [\href{http://arxiv.org/abs/1606.00842}{{\tt
  1606.00842}}].

\bibitem{Gariazzo:2017akm}
S.~Gariazzo, O.~Mena, V.~Miralles, H.~Ram{\'\i}rez and L.~Boubekeur,
  \emph{{Running of featureful primordial power spectra}},
  \href{http://dx.doi.org/10.1103/PhysRevD.95.123534}{\emph{Phys. Rev.} {\bf
  D95} (2017) 123534}, [\href{http://arxiv.org/abs/1701.08977}{{\tt
  1701.08977}}].

\bibitem{Hubble168}
E.~Hubble, \emph{A relation between distance and radial velocity among
  extra-galactic nebulae},
  \href{http://dx.doi.org/10.1073/pnas.15.3.168}{\emph{Proceedings of the
  National Academy of Sciences} {\bf 15} (1929) 168--173}.

\bibitem{Ferreira2014}
P.~Ferreira, \emph{The Perfect Theory: A Century of Geniuses and the Battle
  over General Relativity}.
\newblock Little, Brown, 2014.

\bibitem{Riess:1998cb}
{\scshape Supernova Search Team} collaboration, A.~G. Riess et~al.,
  \emph{{Observational evidence from supernovae for an accelerating universe
  and a cosmological constant}},
  \href{http://dx.doi.org/10.1086/300499}{\emph{Astron. J.} {\bf 116} (1998)
  1009--1038}, [\href{http://arxiv.org/abs/astro-ph/9805201}{{\tt
  astro-ph/9805201}}].

\bibitem{Perlmutter:1998np}
{\scshape Supernova Cosmology Project} collaboration, S.~Perlmutter et~al.,
  \emph{{Measurements of Omega and Lambda from 42 high redshift supernovae}},
  \href{http://dx.doi.org/10.1086/307221}{\emph{Astrophys. J.} {\bf 517} (1999)
  565--586}, [\href{http://arxiv.org/abs/astro-ph/9812133}{{\tt
  astro-ph/9812133}}].

\bibitem{Dodelson:2003ft}
S.~Dodelson, \emph{{Modern Cosmology}}.
\newblock Academic Press, Amsterdam, 2003.

\bibitem{Amendola:2015ksp}
L.~Amendola and S.~Tsujikawa, \emph{{Dark Energy}}.
\newblock Cambridge University Press, 2015.

\bibitem{Kolb:1990vq}
E.~W. Kolb and M.~S. Turner, \emph{{The Early Universe}}, {\emph{Front. Phys.}
  {\bf 69} (1990) 1--547}.

\bibitem{Mukhanov:2005sc}
V.~Mukhanov, \emph{{Physical Foundations of Cosmology}}.
\newblock Cambridge University Press, Oxford, 2005.

\bibitem{Weinberg:2008zzc}
S.~Weinberg, \emph{{Cosmology}}.
\newblock Oxford, UK: Oxford Univ. Pr. (2008) 593 p, 2008.

\bibitem{Gorbunov:2011zz}
V.~A. Rubakov and D.~S. Gorbunov, \emph{{Introduction to the Theory of the
  Early Universe}}.
\newblock World Scientific, Singapore, 2017,
  \href{http://dx.doi.org/10.1142/10447}{10.1142/10447}.

\bibitem{Weinberg:1995mt}
S.~Weinberg, \emph{{The Quantum theory of fields. Vol. 1: Foundations}}.
\newblock Cambridge University Press, 2005.

\bibitem{Weinberg:1996kr}
S.~Weinberg, \emph{{The quantum theory of fields. Vol. 2: Modern
  applications}}.
\newblock Cambridge University Press, 2013.

\bibitem{Patrignani:2016xqp}
{\scshape Particle Data Group} collaboration, C.~Patrignani et~al.,
  \emph{{Review of Particle Physics}},
  \href{http://dx.doi.org/10.1088/1674-1137/40/10/100001}{\emph{Chin. Phys.}
  {\bf C40} (2016) 100001}.

\bibitem{Kirzhnits:1972ut}
D.~A. Kirzhnits and A.~D. Linde, \emph{{Macroscopic Consequences of the
  Weinberg Model}},
  \href{http://dx.doi.org/10.1016/0370-2693(72)90109-8}{\emph{Phys. Lett.} {\bf
  42B} (1972) 471--474}.

\bibitem{Dolan:1973qd}
L.~Dolan and R.~Jackiw, \emph{{Symmetry Behavior at Finite Temperature}},
  \href{http://dx.doi.org/10.1103/PhysRevD.9.3320}{\emph{Phys. Rev.} {\bf D9}
  (1974) 3320--3341}.

\bibitem{Weinberg:1974hy}
S.~Weinberg, \emph{{Gauge and Global Symmetries at High Temperature}},
  \href{http://dx.doi.org/10.1103/PhysRevD.9.3357}{\emph{Phys. Rev.} {\bf D9}
  (1974) 3357--3378}.

\bibitem{Georgi:1974aa}
H.~Georgi, \emph{Unity of all elementary-particle forces},
  \href{http://dx.doi.org/10.1103/PhysRevLett.32.438}{\emph{Physical Review
  Letters} {\bf 32} (1974) 438--441}.

\bibitem{Englert:1964et}
F.~Englert and R.~Brout, \emph{{Broken Symmetry and the Mass of Gauge Vector
  Mesons}}, \href{http://dx.doi.org/10.1103/PhysRevLett.13.321}{\emph{Phys.
  Rev. Lett.} {\bf 13} (1964) 321--323}.

\bibitem{Higgs:1964pj}
P.~W. Higgs, \emph{{Broken Symmetries and the Masses of Gauge Bosons}},
  \href{http://dx.doi.org/10.1103/PhysRevLett.13.508}{\emph{Phys. Rev. Lett.}
  {\bf 13} (1964) 508--509}.

\bibitem{Fritzsch:1973pi}
H.~Fritzsch, M.~Gell-Mann and H.~Leutwyler, \emph{{Advantages of the Color
  Octet Gluon Picture}},
  \href{http://dx.doi.org/10.1016/0370-2693(73)90625-4}{\emph{Phys. Lett.} {\bf
  47B} (1973) 365--368}.

\bibitem{Quigg:1983gw}
C.~Quigg, \emph{{Gauge Theories Of The Strong, Weak And Electromagnetic
  Interactions}}, {\emph{Front. Phys.} {\bf 56} (1983) 1--334}.

\bibitem{Pich:1999yz}
A.~Pich, \emph{{Aspects of quantum chromodynamics}},  in \emph{{Proceedings,
  Summer School in Particle Physics: Trieste, Italy, June 21-July 9, 1999}},
  pp.~53--102, 1999.
\newblock \href{http://arxiv.org/abs/hep-ph/0001118}{{\tt hep-ph/0001118}}.

\bibitem{Gross:1973id}
D.~J. Gross and F.~Wilczek, \emph{{Ultraviolet Behavior of Nonabelian Gauge
  Theories}}, \href{http://dx.doi.org/10.1103/PhysRevLett.30.1343}{\emph{Phys.
  Rev. Lett.} {\bf 30} (1973) 1343--1346}.

\bibitem{Politzer:1973fx}
H.~D. Politzer, \emph{{Reliable Perturbative Results for Strong
  Interactions?}},
  \href{http://dx.doi.org/10.1103/PhysRevLett.30.1346}{\emph{Phys. Rev. Lett.}
  {\bf 30} (1973) 1346--1349}.

\bibitem{Lesgourgues:2006nd}
J.~Lesgourgues and S.~Pastor, \emph{{Massive neutrinos and cosmology}},
  \href{http://dx.doi.org/10.1016/j.physrep.2006.04.001}{\emph{Phys. Rept.}
  {\bf 429} (2006) 307--379},
  [\href{http://arxiv.org/abs/astro-ph/0603494}{{\tt astro-ph/0603494}}].

\bibitem{Betts:2013uya}
S.~Betts, W.~R. Blanchard, R.~H. Carnevale, C.~Chang, C.~Chen, S.~Chidzik
  et~al., \emph{Development of a relic neutrino detection experiment at
  ptolemy: Princeton tritium observatory for light, early-universe,
  massive-neutrino yield},  \href{http://arxiv.org/abs/1307.4738}{{\tt
  1307.4738}}.

\bibitem{Bashinsky:2004aa}
S.~Bashinsky and U.~Seljak, \emph{Signatures of relativistic neutrinos in cmb
  anisotropy and matter clustering}, {\emph{Phys.Rev.D} {\bf 69} (2004)
  083002}, [\href{http://arxiv.org/abs/astro-ph/0310198}{{\tt
  astro-ph/0310198}}].

\bibitem{Follin:2015aa}
B.~Follin, L.~Knox, M.~Millea and Z.~Pan, \emph{A first detection of the
  acoustic oscillation phase shift expected from the cosmic neutrino
  background}, {\emph{Phys. Rev. Lett.} {\bf 115} (2015) 091301},
  [\href{http://arxiv.org/abs/1503.07863}{{\tt 1503.07863}}].

\bibitem{Baumann_2019}
D.~Baumann, F.~Beutler, R.~Flauger, D.~Green, A.~Slosar, M.~Vargas-Maga{\~n}a
  et~al., \emph{First constraint on the neutrino-induced phase shift in the
  spectrum of baryon acoustic oscillations},
  \href{http://dx.doi.org/10.1038/s41567-019-0435-6}{\emph{Nature Physics}
  (Feb, 2019) }.

\bibitem{Alpher:1948ve}
R.~A. Alpher, H.~Bethe and G.~Gamow, \emph{{The origin of chemical elements}},
  \href{http://dx.doi.org/10.1103/PhysRev.73.803}{\emph{Phys. Rev.} {\bf 73}
  (1948) 803--804}.

\bibitem{Wagoner:1966pv}
R.~V. Wagoner, W.~A. Fowler and F.~Hoyle, \emph{{On the Synthesis of elements
  at very high temperatures}},
  \href{http://dx.doi.org/10.1086/149126}{\emph{Astrophys. J.} {\bf 148} (1967)
  3--49}.

\bibitem{Fields:2011zzb}
B.~D. Fields, \emph{{The primordial lithium problem}},
  \href{http://dx.doi.org/10.1146/annurev-nucl-102010-130445}{\emph{Ann. Rev.
  Nucl. Part. Sci.} {\bf 61} (2011) 47--68},
  [\href{http://arxiv.org/abs/1203.3551}{{\tt 1203.3551}}].

\bibitem{Cyburt:2015mya}
R.~H. Cyburt, B.~D. Fields, K.~A. Olive and T.-H. Yeh, \emph{{Big Bang
  Nucleosynthesis: 2015}},
  \href{http://dx.doi.org/10.1103/RevModPhys.88.015004}{\emph{Rev. Mod. Phys.}
  {\bf 88} (2016) 015004}, [\href{http://arxiv.org/abs/1505.01076}{{\tt
  1505.01076}}].

\bibitem{Peebles:1968ja}
P.~J.~E. Peebles, \emph{{Recombination of the Primeval Plasma}},
  \href{http://dx.doi.org/10.1086/149628}{\emph{Astrophys. J.} {\bf 153} (1968)
  1}.

\bibitem{Zeldovich:1969en}
{\relax Ya}.~B. Zeldovich, V.~G. Kurt and R.~A. Sunyaev, \emph{{Recombination
  of hydrogen in the hot model of the universe}}, {\emph{Sov. Phys. JETP} {\bf
  28} (1969) 146}.

\bibitem{Fixsen:1996nj}
D.~J. Fixsen, E.~S. Cheng, J.~M. Gales, J.~C. Mather, R.~A. Shafer and E.~L.
  Wright, \emph{{The Cosmic Microwave Background spectrum from the full COBE
  FIRAS data set}}, \href{http://dx.doi.org/10.1086/178173}{\emph{Astrophys.
  J.} {\bf 473} (1996) 576}, [\href{http://arxiv.org/abs/astro-ph/9605054}{{\tt
  astro-ph/9605054}}].

\bibitem{Fixsen:2009ug}
D.~J. Fixsen, \emph{{The Temperature of the Cosmic Microwave Background}},
  \href{http://dx.doi.org/10.1088/0004-637X/707/2/916}{\emph{Astrophys. J.}
  {\bf 707} (2009) 916--920}, [\href{http://arxiv.org/abs/0911.1955}{{\tt
  0911.1955}}].

\bibitem{Penzias:1965wn}
A.~A. Penzias and R.~W. Wilson, \emph{{A Measurement of excess antenna
  temperature at 4080-Mc/s}},
  \href{http://dx.doi.org/10.1086/148307}{\emph{Astrophys. J.} {\bf 142} (1965)
  419--421}.

\bibitem{Dicke:1965zz}
R.~H. Dicke, P.~J.~E. Peebles, P.~G. Roll and D.~T. Wilkinson, \emph{{Cosmic
  Black-Body Radiation}},
  \href{http://dx.doi.org/10.1086/148306}{\emph{Astrophys. J.} {\bf 142} (1965)
  414--419}.

\bibitem{Aghanim:2018eyx}
{\scshape Planck} collaboration, N.~Aghanim et~al., \emph{{Planck 2018 results.
  VI. Cosmological parameters}},  \href{http://arxiv.org/abs/1807.06209}{{\tt
  1807.06209}}.

\bibitem{Rees_1968}
M.~J. Rees, \emph{Polarization and spectrum of the primeval radiation in an
  anisotropic universe}, \href{http://dx.doi.org/10.1086/180208}{\emph{The
  Astrophysical Journal} {\bf 153} (Jul, 1968) L1}.

\bibitem{Negroponte:1980aa}
J.~Negroponte, \emph{Polarization of the primeval radiation in an anisotropic
  universe},
  \href{http://dx.doi.org/10.1103/PhysRevLett.44.1433}{\emph{Physical Review
  Letters} {\bf 44} (1980) 1433--1437}.

\bibitem{Bond_1984}
J.~R. Bond and G.~Efstathiou, \emph{Cosmic background radiation anisotropies in
  universes dominated by nonbaryonic dark matter},
  \href{http://dx.doi.org/10.1086/184362}{\emph{The Astrophysical Journal} {\bf
  285} (Oct, 1984) L45}.

\bibitem{Weinberg:1972kfs}
S.~Weinberg, \emph{{Gravitation and Cosmology}}.
\newblock John Wiley and Sons, New York, 1972.

\bibitem{Kamionkowski:1996zd}
M.~Kamionkowski, A.~Kosowsky and A.~Stebbins, \emph{{A Probe of primordial
  gravity waves and vorticity}},
  \href{http://dx.doi.org/10.1103/PhysRevLett.78.2058}{\emph{Phys. Rev. Lett.}
  {\bf 78} (1997) 2058--2061},
  [\href{http://arxiv.org/abs/astro-ph/9609132}{{\tt astro-ph/9609132}}].

\bibitem{Seljak:1996gy}
U.~Seljak and M.~Zaldarriaga, \emph{{Signature of gravity waves in polarization
  of the microwave background}},
  \href{http://dx.doi.org/10.1103/PhysRevLett.78.2054}{\emph{Phys. Rev. Lett.}
  {\bf 78} (1997) 2054--2057},
  [\href{http://arxiv.org/abs/astro-ph/9609169}{{\tt astro-ph/9609169}}].

\bibitem{Ade:2017uvt}
{\scshape POLARBEAR} collaboration, P.~A.~R. Ade et~al., \emph{{A Measurement
  of the Cosmic Microwave Background $B$-Mode Polarization Power Spectrum at
  Sub-Degree Scales from 2 years of POLARBEAR Data}},
  \href{http://dx.doi.org/10.3847/1538-4357/aa8e9f}{\emph{Astrophys. J.} {\bf
  848} (2017) 121}, [\href{http://arxiv.org/abs/1705.02907}{{\tt 1705.02907}}].

\bibitem{Gorbunov:2011zzc}
D.~S. Gorbunov and V.~A. Rubakov, \emph{{Introduction to the theory of the
  early universe: Cosmological perturbations and inflationary theory}}.
\newblock 2011, \href{http://dx.doi.org/10.1142/7874}{10.1142/7874}.

\bibitem{Bertone:2005aa}
G.~Bertone, D.~Hooper and J.~Silk, \emph{Particle dark matter: Evidence,
  candidates and constraints}, {\emph{Phys.Rept.} {\bf 405} (2005) 279--390},
  [\href{http://arxiv.org/abs/hep-ph/0404175}{{\tt hep-ph/0404175}}].

\bibitem{Bergstrom:2012aa}
L.~Bergstr{\"o}m, \emph{Dark matter evidence, particle physics candidates and
  detection methods},  \href{http://arxiv.org/abs/1205.4882}{{\tt 1205.4882}}.

\bibitem{Kusenko:2013aa}
A.~Kusenko and L.~J. Rosenberg, \emph{Snowmass-2013 cosmic frontier 3 (cf3)
  working group summary: Non-wimp dark matter},
  \href{http://arxiv.org/abs/1310.8642}{{\tt 1310.8642}}.

\bibitem{Springel:2005nw}
V.~Springel et~al., \emph{{Simulating the joint evolution of quasars, galaxies
  and their large-scale distribution}},
  \href{http://dx.doi.org/10.1038/nature03597}{\emph{Nature} {\bf 435} (2005)
  629--636}, [\href{http://arxiv.org/abs/astro-ph/0504097}{{\tt
  astro-ph/0504097}}].

\bibitem{Baumann:2009ds}
D.~Baumann, \emph{{Inflation}},  in \emph{{Physics of the large and the small,
  TASI 09, proceedings of the Theoretical Advanced Study Institute in
  Elementary Particle Physics, Boulder, Colorado, USA, 1-26 June 2009}},
  pp.~523--686, 2011.
\newblock \href{http://arxiv.org/abs/0907.5424}{{\tt 0907.5424}}.
\newblock \href{http://dx.doi.org/10.1142/9789814327183_0010}{DOI}.

\bibitem{Baumann:2018muz}
D.~Baumann, \emph{{Primordial Cosmology}},
  \href{http://dx.doi.org/10.22323/1.305.0009}{\emph{PoS} {\bf TASI2017} (2018)
  009}, [\href{http://arxiv.org/abs/1807.03098}{{\tt 1807.03098}}].

\bibitem{Peebles:1994xt}
P.~J.~E. Peebles, \emph{{Principles of physical cosmology}}.
\newblock Princeton, USA: Univ. Pr. (1993) 718 p, 1994.

\bibitem{Aubourg:2014yra}
{\~A}.~Aubourg et~al., \emph{{Cosmological implications of baryon acoustic
  oscillation measurements}},
  \href{http://dx.doi.org/10.1103/PhysRevD.92.123516}{\emph{Phys. Rev.} {\bf
  D92} (2015) 123516}, [\href{http://arxiv.org/abs/1411.1074}{{\tt
  1411.1074}}].

\bibitem{Ade:2015xua}
{\scshape Planck} collaboration, P.~A.~R. Ade et~al., \emph{{Planck 2015
  results. XIII. Cosmological parameters}},
  \href{http://dx.doi.org/10.1051/0004-6361/201525830}{\emph{Astron.
  Astrophys.} {\bf 594} (2016) A13},
  [\href{http://arxiv.org/abs/1502.01589}{{\tt 1502.01589}}].

\bibitem{Guth:1980zm}
A.~H. Guth, \emph{{The Inflationary Universe: A Possible Solution to the
  Horizon and Flatness Problems}},
  \href{http://dx.doi.org/10.1103/PhysRevD.23.347}{\emph{Phys. Rev.} {\bf D23}
  (1981) 347--356}.

\bibitem{Guth:1981uk}
A.~H. Guth and E.~J. Weinberg, \emph{{Cosmological Consequences of a First
  Order Phase Transition in the SU(5) Grand Unified Model}},
  \href{http://dx.doi.org/10.1103/PhysRevD.23.876}{\emph{Phys. Rev.} {\bf D23}
  (1981) 876}.

\bibitem{Cook:1981fz}
G.~P. Cook and K.~T. Mahanthappa, \emph{{Supercooling in the SU(5) Phase
  Transitions and Magnetic Monopole Suppression}},
  \href{http://dx.doi.org/10.1103/PhysRevD.23.1321}{\emph{Phys. Rev.} {\bf D23}
  (1981) 1321}.

\bibitem{Barrow:1981pa}
J.~D. Barrow and M.~S. Turner, \emph{{Inflation in the Universe}},
  \href{http://dx.doi.org/10.1038/292035a0}{\emph{Nature} {\bf 292} (1981)
  35--38}.

\bibitem{LINDE1982389}
A.~Linde, \emph{A new inflationary universe scenario: A possible solution of
  the horizon, flatness, homogeneity, isotropy and primordial monopole
  problems},
  \href{http://dx.doi.org/https://doi.org/10.1016/0370-2693(82)91219-9}{\emph{Physics
  Letters B} {\bf 108} (1982) 389 -- 393}.

\bibitem{Albrecht:1982aa}
A.~Albrecht, \emph{Cosmology for grand unified theories with radiatively
  induced symmetry breaking},
  \href{http://dx.doi.org/10.1103/PhysRevLett.48.1220}{\emph{Physical Review
  Letters} {\bf 48} (1982) 1220--1223}.

\bibitem{Mukhanov:1981xt}
V.~F. Mukhanov and G.~V. Chibisov, \emph{{Quantum Fluctuations and a
  Nonsingular Universe}}, {\emph{JETP Lett.} {\bf 33} (1981) 532--535}.

\bibitem{HAWKING1982295}
S.~Hawking, \emph{The development of irregularities in a single bubble
  inflationary universe},
  \href{http://dx.doi.org/https://doi.org/10.1016/0370-2693(82)90373-2}{\emph{Physics
  Letters B} {\bf 115} (1982) 295 -- 297}.

\bibitem{STAROBINSKY1982175}
A.~Starobinsky, \emph{Dynamics of phase transition in the new inflationary
  universe scenario and generation of perturbations},
  \href{http://dx.doi.org/https://doi.org/10.1016/0370-2693(82)90541-X}{\emph{Physics
  Letters B} {\bf 117} (1982) 175 -- 178}.

\bibitem{Guth:1982aa}
A.~H. Guth, \emph{Fluctuations in the new inflationary universe},
  \href{http://dx.doi.org/10.1103/PhysRevLett.49.1110}{\emph{Physical Review
  Letters} {\bf 49} (1982) 1110--1113}.

\bibitem{Bardeen:1983aa}
J.~M. Bardeen, \emph{Spontaneous creation of almost scale-free density
  perturbations in an inflationary universe},
  \href{http://dx.doi.org/10.1103/PhysRevD.28.679}{\emph{Physical Review D}
  {\bf 28} (1983) 679--693}.

\bibitem{Guth:1997wk}
A.~H. Guth, \emph{{The inflationary universe: The quest for a new theory of
  cosmic origins}}.
\newblock 1997.

\bibitem{Kofman:1997yn}
L.~Kofman, A.~D. Linde and A.~A. Starobinsky, \emph{{Towards the theory of
  reheating after inflation}},
  \href{http://dx.doi.org/10.1103/PhysRevD.56.3258}{\emph{Phys. Rev.} {\bf D56}
  (1997) 3258--3295}, [\href{http://arxiv.org/abs/hep-ph/9704452}{{\tt
  hep-ph/9704452}}].

\bibitem{Bassett:2006aa}
B.~A. Bassett, S.~Tsujikawa and D.~Wands, \emph{Inflation dynamics and
  reheating}, {\emph{Rev.Mod.Phys.} {\bf 78} (2006) 537--589},
  [\href{http://arxiv.org/abs/astro-ph/0507632}{{\tt astro-ph/0507632}}].

\bibitem{Lyth:2009zz}
D.~H. Lyth and A.~R. Liddle, \emph{{The primordial density perturbation:
  Cosmology, inflation and the origin of structure}}.
\newblock 2009.

\bibitem{Martin:2013tda}
J.~Martin, C.~Ringeval and V.~Vennin, \emph{{Encyclopdia Inflationaris}},
  \href{http://dx.doi.org/10.1016/j.dark.2014.01.003}{\emph{Phys. Dark Univ.}
  {\bf 5-6} (2014) 75--235}, [\href{http://arxiv.org/abs/1303.3787}{{\tt
  1303.3787}}].

\bibitem{LINDE1983177}
A.~Linde, \emph{Chaotic inflation},
  \href{http://dx.doi.org/https://doi.org/10.1016/0370-2693(83)90837-7}{\emph{Physics
  Letters B} {\bf 129} (1983) 177 -- 181}.

\bibitem{Collaboration:2018aa}
P.~Collaboration, Y.~Akrami, F.~Arroja, M.~Ashdown, J.~Aumont, C.~Baccigalupi
  et~al., \emph{Planck 2018 results. x. constraints on inflation},
  \href{http://arxiv.org/abs/1807.06211}{{\tt 1807.06211}}.

\bibitem{Boubekeur:2005zm}
L.~Boubekeur and D.~H. Lyth, \emph{{Hilltop inflation}},
  \href{http://dx.doi.org/10.1088/1475-7516/2005/07/010}{\emph{JCAP} {\bf 0507}
  (2005) 010}, [\href{http://arxiv.org/abs/hep-ph/0502047}{{\tt
  hep-ph/0502047}}].

\bibitem{Bezrukov:2007ep}
F.~L. Bezrukov and M.~Shaposhnikov, \emph{{The Standard Model Higgs boson as
  the inflaton}},
  \href{http://dx.doi.org/10.1016/j.physletb.2007.11.072}{\emph{Phys. Lett.}
  {\bf B659} (2008) 703--706}, [\href{http://arxiv.org/abs/0710.3755}{{\tt
  0710.3755}}].

\bibitem{Dvali:1994aa}
G.~Dvali, Q.~Shafi and R.~Schaefer, \emph{Large scale structure and
  supersymmetric inflation without fine tuning}, {\emph{Phys.Rev.Lett.} {\bf
  73} (1994) 1886--1889}, [\href{http://arxiv.org/abs/hep-ph/9406319}{{\tt
  hep-ph/9406319}}].

\bibitem{Freese:1993bc}
K.~Freese, \emph{{Natural Inflation}},  in \emph{{Evolution of the universe and
  its observational quest. Proceedings, 37th Yamada Conference, Tokyo, Japan,
  June 8-12, 1993}}, pp.~49--58, 1993.
\newblock \href{http://arxiv.org/abs/astro-ph/9310012}{{\tt astro-ph/9310012}}.

\bibitem{Kim:2004rp}
J.~E. Kim, H.~P. Nilles and M.~Peloso, \emph{{Completing natural inflation}},
  \href{http://dx.doi.org/10.1088/1475-7516/2005/01/005}{\emph{JCAP} {\bf 0501}
  (2005) 005}, [\href{http://arxiv.org/abs/hep-ph/0409138}{{\tt
  hep-ph/0409138}}].

\bibitem{ArkaniHamed:2003mz}
N.~Arkani-Hamed, H.-C. Cheng, P.~Creminelli and L.~Randall,
  \emph{{Pseudonatural inflation}},
  \href{http://dx.doi.org/10.1088/1475-7516/2003/07/003}{\emph{JCAP} {\bf 0307}
  (2003) 003}, [\href{http://arxiv.org/abs/hep-th/0302034}{{\tt
  hep-th/0302034}}].

\bibitem{ArkaniHamed:2003wu}
N.~Arkani-Hamed, H.-C. Cheng, P.~Creminelli and L.~Randall, \emph{{Extra
  natural inflation}},
  \href{http://dx.doi.org/10.1103/PhysRevLett.90.221302}{\emph{Phys. Rev.
  Lett.} {\bf 90} (2003) 221302},
  [\href{http://arxiv.org/abs/hep-th/0301218}{{\tt hep-th/0301218}}].

\bibitem{GarciaBellido:2001ky}
J.~Garcia-Bellido, R.~Rabadan and F.~Zamora, \emph{{Inflationary scenarios from
  branes at angles}},
  \href{http://dx.doi.org/10.1088/1126-6708/2002/01/036}{\emph{JHEP} {\bf 01}
  (2002) 036}, [\href{http://arxiv.org/abs/hep-th/0112147}{{\tt
  hep-th/0112147}}].

\bibitem{Dvali:2001fw}
G.~R. Dvali, Q.~Shafi and S.~Solganik, \emph{{D-brane inflation}},  in
  \emph{{4th European Meeting From the Planck Scale to the Electroweak Scale
  (Planck 2001) La Londe les Maures, Toulon, France, May 11-16, 2001}}, 2001.
\newblock \href{http://arxiv.org/abs/hep-th/0105203}{{\tt hep-th/0105203}}.

\bibitem{Kachru:2003sx}
S.~Kachru, R.~Kallosh, A.~D. Linde, J.~M. Maldacena, L.~P. McAllister and S.~P.
  Trivedi, \emph{{Towards inflation in string theory}},
  \href{http://dx.doi.org/10.1088/1475-7516/2003/10/013}{\emph{JCAP} {\bf 0310}
  (2003) 013}, [\href{http://arxiv.org/abs/hep-th/0308055}{{\tt
  hep-th/0308055}}].

\bibitem{Kallosh:2013yoa}
R.~Kallosh, A.~Linde and D.~Roest, \emph{{Superconformal Inflationary
  $\alpha$-Attractors}},
  \href{http://dx.doi.org/10.1007/JHEP11(2013)198}{\emph{JHEP} {\bf 11} (2013)
  198}, [\href{http://arxiv.org/abs/1311.0472}{{\tt 1311.0472}}].

\bibitem{Ferrara:2013rsa}
S.~Ferrara, R.~Kallosh, A.~Linde and M.~Porrati, \emph{{Minimal Supergravity
  Models of Inflation}},
  \href{http://dx.doi.org/10.1103/PhysRevD.88.085038}{\emph{Phys. Rev.} {\bf
  D88} (2013) 085038}, [\href{http://arxiv.org/abs/1307.7696}{{\tt
  1307.7696}}].

\bibitem{Wands:2007bd}
D.~Wands, \emph{{Multiple field inflation}},
  \href{http://dx.doi.org/10.1007/978-3-540-74353-8_8}{\emph{Lect. Notes Phys.}
  {\bf 738} (2008) 275--304},
  [\href{http://arxiv.org/abs/astro-ph/0702187}{{\tt astro-ph/0702187}}].

\bibitem{ArmendarizPicon:1999rj}
C.~Armendariz-Picon, T.~Damour and V.~F. Mukhanov, \emph{{k - inflation}},
  \href{http://dx.doi.org/10.1016/S0370-2693(99)00603-6}{\emph{Phys. Lett.}
  {\bf B458} (1999) 209--218}, [\href{http://arxiv.org/abs/hep-th/9904075}{{\tt
  hep-th/9904075}}].

\bibitem{Garriga:1999vw}
J.~Garriga and V.~F. Mukhanov, \emph{{Perturbations in k-inflation}},
  \href{http://dx.doi.org/10.1016/S0370-2693(99)00602-4}{\emph{Phys. Lett.}
  {\bf B458} (1999) 219--225}, [\href{http://arxiv.org/abs/hep-th/9904176}{{\tt
  hep-th/9904176}}].

\bibitem{Salopek:1988qh}
D.~S. Salopek, J.~R. Bond and J.~M. Bardeen, \emph{{Designing Density
  Fluctuation Spectra in Inflation}},
  \href{http://dx.doi.org/10.1103/PhysRevD.40.1753}{\emph{Phys. Rev.} {\bf D40}
  (1989) 1753}.

\bibitem{Futamase:1987ua}
T.~Futamase and K.-i. Maeda, \emph{{Chaotic Inflationary Scenario in Models
  Having Nonminimal Coupling With Curvature}},
  \href{http://dx.doi.org/10.1103/PhysRevD.39.399}{\emph{Phys. Rev.} {\bf D39}
  (1989) 399--404}.

\bibitem{Fakir:1990eg}
R.~Fakir and W.~G. Unruh, \emph{{Improvement on cosmological chaotic inflation
  through nonminimal coupling}},
  \href{http://dx.doi.org/10.1103/PhysRevD.41.1783}{\emph{Phys. Rev.} {\bf D41}
  (1990) 1783--1791}.

\bibitem{Kaiser:1994vs}
D.~I. Kaiser, \emph{{Primordial spectral indices from generalized Einstein
  theories}}, \href{http://dx.doi.org/10.1103/PhysRevD.52.4295}{\emph{Phys.
  Rev.} {\bf D52} (1995) 4295--4306},
  [\href{http://arxiv.org/abs/astro-ph/9408044}{{\tt astro-ph/9408044}}].

\bibitem{Komatsu:1999mt}
E.~Komatsu and T.~Futamase, \emph{{Complete constraints on a nonminimally
  coupled chaotic inflationary scenario from the cosmic microwave background}},
  \href{http://dx.doi.org/10.1103/PhysRevD.59.064029}{\emph{Phys. Rev.} {\bf
  D59} (1999) 064029}, [\href{http://arxiv.org/abs/astro-ph/9901127}{{\tt
  astro-ph/9901127}}].

\bibitem{Barvinsky:2008ia}
A.~O. Barvinsky, A.~{\relax Yu}. Kamenshchik and A.~A. Starobinsky,
  \emph{{Inflation scenario via the Standard Model Higgs boson and LHC}},
  \href{http://dx.doi.org/10.1088/1475-7516/2008/11/021}{\emph{JCAP} {\bf 0811}
  (2008) 021}, [\href{http://arxiv.org/abs/0809.2104}{{\tt 0809.2104}}].

\bibitem{Barvinsky:2008cya}
A.~O. Barvinsky, A.~{\relax Yu}. Kamenshchik and A.~A. Starobinsky,
  \emph{{Inflation in the Standard Model with a strong non-minimal curvature
  coupling and the Higgs boson mass}},  in \emph{{Proceedings, 15th
  International Seminar on High Energy Physics (Quarks 2008): Sergiev Posad,
  Russia. May 23-29, 2008}}, 2008.

\bibitem{Bezrukov:2009db}
F.~Bezrukov and M.~Shaposhnikov, \emph{{Standard Model Higgs boson mass from
  inflation: Two loop analysis}},
  \href{http://dx.doi.org/10.1088/1126-6708/2009/07/089}{\emph{JHEP} {\bf 07}
  (2009) 089}, [\href{http://arxiv.org/abs/0904.1537}{{\tt 0904.1537}}].

\bibitem{Hertzberg:2010dc}
M.~P. Hertzberg, \emph{{On Inflation with Non-minimal Coupling}},
  \href{http://dx.doi.org/10.1007/JHEP11(2010)023}{\emph{JHEP} {\bf 11} (2010)
  023}, [\href{http://arxiv.org/abs/1002.2995}{{\tt 1002.2995}}].

\bibitem{Okada:2010jf}
N.~Okada, M.~U. Rehman and Q.~Shafi, \emph{{Tensor to Scalar Ratio in
  Non-Minimal $\phi^4$ Inflation}},
  \href{http://dx.doi.org/10.1103/PhysRevD.82.043502}{\emph{Phys. Rev.} {\bf
  D82} (2010) 043502}, [\href{http://arxiv.org/abs/1005.5161}{{\tt
  1005.5161}}].

\bibitem{Linde:2011nh}
A.~Linde, M.~Noorbala and A.~Westphal, \emph{{Observational consequences of
  chaotic inflation with nonminimal coupling to gravity}},
  \href{http://dx.doi.org/10.1088/1475-7516/2011/03/013}{\emph{JCAP} {\bf 1103}
  (2011) 013}, [\href{http://arxiv.org/abs/1101.2652}{{\tt 1101.2652}}].

\bibitem{Horndeski:1974wa}
G.~W. Horndeski, \emph{{Second-order scalar-tensor field equations in a
  four-dimensional space}},
  \href{http://dx.doi.org/10.1007/BF01807638}{\emph{Int. J. Theor. Phys.} {\bf
  10} (1974) 363--384}.

\bibitem{Nicolis:2008in}
A.~Nicolis, R.~Rattazzi and E.~Trincherini, \emph{{The Galileon as a local
  modification of gravity}},
  \href{http://dx.doi.org/10.1103/PhysRevD.79.064036}{\emph{Phys. Rev.} {\bf
  D79} (2009) 064036}, [\href{http://arxiv.org/abs/0811.2197}{{\tt
  0811.2197}}].

\bibitem{Deffayet:2009mn}
C.~Deffayet, S.~Deser and G.~Esposito-Farese, \emph{{Generalized Galileons: All
  scalar models whose curved background extensions maintain second-order field
  equations and stress-tensors}},
  \href{http://dx.doi.org/10.1103/PhysRevD.80.064015}{\emph{Phys. Rev.} {\bf
  D80} (2009) 064015}, [\href{http://arxiv.org/abs/0906.1967}{{\tt
  0906.1967}}].

\bibitem{Deffayet:2009wt}
C.~Deffayet, G.~Esposito-Farese and A.~Vikman, \emph{{Covariant Galileon}},
  \href{http://dx.doi.org/10.1103/PhysRevD.79.084003}{\emph{Phys. Rev.} {\bf
  D79} (2009) 084003}, [\href{http://arxiv.org/abs/0901.1314}{{\tt
  0901.1314}}].

\bibitem{Gleyzes:2014dya}
J.~Gleyzes, D.~Langlois, F.~Piazza and F.~Vernizzi, \emph{{Healthy theories
  beyond Horndeski}},
  \href{http://dx.doi.org/10.1103/PhysRevLett.114.211101}{\emph{Phys. Rev.
  Lett.} {\bf 114} (2015) 211101}, [\href{http://arxiv.org/abs/1404.6495}{{\tt
  1404.6495}}].

\bibitem{Langlois:2015cwa}
D.~Langlois and K.~Noui, \emph{{Degenerate higher derivative theories beyond
  Horndeski: evading the Ostrogradski instability}},
  \href{http://dx.doi.org/10.1088/1475-7516/2016/02/034}{\emph{JCAP} {\bf 1602}
  (2016) 034}, [\href{http://arxiv.org/abs/1510.06930}{{\tt 1510.06930}}].

\bibitem{Ezquiaga:2016nqo}
J.~M. Ezquiaga, J.~Garc\'ia-Bellido and M.~Zumalac\'arregui, \emph{{Towards the
  most general scalar-tensor theories of gravity: a unified approach in the
  language of differential forms}},
  \href{http://dx.doi.org/10.1103/PhysRevD.94.024005}{\emph{Phys. Rev.} {\bf
  D94} (2016) 024005}, [\href{http://arxiv.org/abs/1603.01269}{{\tt
  1603.01269}}].

\bibitem{Motohashi:2016ftl}
H.~Motohashi, K.~Noui, T.~Suyama, M.~Yamaguchi and D.~Langlois, \emph{{Healthy
  degenerate theories with higher derivatives}},
  \href{http://dx.doi.org/10.1088/1475-7516/2016/07/033}{\emph{JCAP} {\bf 1607}
  (2016) 033}, [\href{http://arxiv.org/abs/1603.09355}{{\tt 1603.09355}}].

\bibitem{Crisostomi:2016czh}
M.~Crisostomi, K.~Koyama and G.~Tasinato, \emph{{Extended Scalar-Tensor
  Theories of Gravity}},
  \href{http://dx.doi.org/10.1088/1475-7516/2016/04/044}{\emph{JCAP} {\bf 1604}
  (2016) 044}, [\href{http://arxiv.org/abs/1602.03119}{{\tt 1602.03119}}].

\bibitem{BenAchour:2016fzp}
J.~Ben~Achour, M.~Crisostomi, K.~Koyama, D.~Langlois, K.~Noui and G.~Tasinato,
  \emph{{Degenerate higher order scalar-tensor theories beyond Horndeski up to
  cubic order}}, \href{http://dx.doi.org/10.1007/JHEP12(2016)100}{\emph{JHEP}
  {\bf 12} (2016) 100}, [\href{http://arxiv.org/abs/1608.08135}{{\tt
  1608.08135}}].

\bibitem{Achour:2016rkg}
J.~Ben~Achour, D.~Langlois and K.~Noui, \emph{{Degenerate higher order
  scalar-tensor theories beyond Horndeski and disformal transformations}},
  \href{http://dx.doi.org/10.1103/PhysRevD.93.124005}{\emph{Phys. Rev.} {\bf
  D93} (2016) 124005}, [\href{http://arxiv.org/abs/1602.08398}{{\tt
  1602.08398}}].

\bibitem{Motohashi:2017eya}
H.~Motohashi, T.~Suyama and M.~Yamaguchi, \emph{{Ghost-free theory with
  third-order time derivatives}},  \href{http://arxiv.org/abs/1711.08125}{{\tt
  1711.08125}}.

\bibitem{STAROBINSKY198099}
A.~Starobinsky, \emph{A new type of isotropic cosmological models without
  singularity},
  \href{http://dx.doi.org/https://doi.org/10.1016/0370-2693(80)90670-X}{\emph{Physics
  Letters B} {\bf 91} (1980) 99 -- 102}.

\bibitem{Parker:2009uva}
L.~E. Parker and D.~Toms, \emph{{Quantum Field Theory in Curved Spacetime}}.
\newblock Cambridge Monographs on Mathematical Physics. Cambridge University
  Press, 2009,
  \href{http://dx.doi.org/10.1017/CBO9780511813924}{10.1017/CBO9780511813924}.

\bibitem{Whitt:1984pd}
B.~Whitt, \emph{{Fourth Order Gravity as General Relativity Plus Matter}},
  \href{http://dx.doi.org/10.1016/0370-2693(84)90332-0}{\emph{Phys. Lett.} {\bf
  145B} (1984) 176--178}.

\bibitem{Coule:1987wt}
D.~H. Coule and M.~B. Mijic, \emph{{Quantum Fluctuations and Eternal Inflation
  in the $r^{2}$ Model}},
  \href{http://dx.doi.org/10.1142/S0217751X88000266}{\emph{Int. J. Mod. Phys.}
  {\bf A3} (1988) 617--629}.

\bibitem{Barrow:1988xh}
J.~D. Barrow and S.~Cotsakis, \emph{{Inflation and the Conformal Structure of
  Higher Order Gravity Theories}},
  \href{http://dx.doi.org/10.1016/0370-2693(88)90110-4}{\emph{Phys. Lett.} {\bf
  B214} (1988) 515--518}.

\bibitem{Maeda:1988ab}
K.-i. Maeda, \emph{{Towards the Einstein-Hilbert Action via Conformal
  Transformation}},
  \href{http://dx.doi.org/10.1103/PhysRevD.39.3159}{\emph{Phys. Rev.} {\bf D39}
  (1989) 3159}.

\bibitem{Burrage:2010cu}
C.~Burrage, C.~de~Rham, D.~Seery and A.~J. Tolley, \emph{{Galileon inflation}},
  \href{http://dx.doi.org/10.1088/1475-7516/2011/01/014}{\emph{JCAP} {\bf 1101}
  (2011) 014}, [\href{http://arxiv.org/abs/1009.2497}{{\tt 1009.2497}}].

\bibitem{Kamada:2010qe}
K.~Kamada, T.~Kobayashi, M.~Yamaguchi and J.~Yokoyama, \emph{{Higgs
  G-inflation}},
  \href{http://dx.doi.org/10.1103/PhysRevD.83.083515}{\emph{Phys. Rev.} {\bf
  D83} (2011) 083515}, [\href{http://arxiv.org/abs/1012.4238}{{\tt
  1012.4238}}].

\bibitem{Ohashi:2012wf}
J.~Ohashi and S.~Tsujikawa, \emph{{Potential-driven Galileon inflation}},
  \href{http://dx.doi.org/10.1088/1475-7516/2012/10/035}{\emph{JCAP} {\bf 1210}
  (2012) 035}, [\href{http://arxiv.org/abs/1207.4879}{{\tt 1207.4879}}].

\bibitem{Kamada:2013bia}
K.~Kamada, T.~Kobayashi, T.~Kunimitsu, M.~Yamaguchi and J.~Yokoyama,
  \emph{{Graceful exit from Higgs $G$ inflation}},
  \href{http://dx.doi.org/10.1103/PhysRevD.88.123518}{\emph{Phys. Rev.} {\bf
  D88} (2013) 123518}, [\href{http://arxiv.org/abs/1309.7410}{{\tt
  1309.7410}}].

\bibitem{Arnowitt:1962hi}
R.~L. Arnowitt, S.~Deser and C.~W. Misner, \emph{{The Dynamics of general
  relativity}}, \href{http://dx.doi.org/10.1007/s10714-008-0661-1}{\emph{Gen.
  Rel. Grav.} {\bf 40} (2008) 1997--2027},
  [\href{http://arxiv.org/abs/gr-qc/0405109}{{\tt gr-qc/0405109}}].

\bibitem{Malik:2009aa}
K.~A. Malik and D.~Wands, \emph{Cosmological perturbations}, {\emph{Phys.Rept.}
  {\bf 475} (2009) 1--51}, [\href{http://arxiv.org/abs/0809.4944}{{\tt
  0809.4944}}].

\bibitem{Maldacena:2002vr}
J.~M. Maldacena, \emph{{Non-Gaussian features of primordial fluctuations in
  single field inflationary models}},
  \href{http://dx.doi.org/10.1088/1126-6708/2003/05/013}{\emph{JHEP} {\bf 05}
  (2003) 013}, [\href{http://arxiv.org/abs/astro-ph/0210603}{{\tt
  astro-ph/0210603}}].

\bibitem{Dvorkin:2009ne}
C.~Dvorkin and W.~Hu, \emph{{Generalized Slow Roll for Large Power Spectrum
  Features}}, \href{http://dx.doi.org/10.1103/PhysRevD.81.023518}{\emph{Phys.
  Rev.} {\bf D81} (2010) 023518}, [\href{http://arxiv.org/abs/0910.2237}{{\tt
  0910.2237}}].

\bibitem{Adshead:2011bw}
P.~Adshead, W.~Hu, C.~Dvorkin and H.~V. Peiris, \emph{{Fast Computation of
  Bispectrum Features with Generalized Slow Roll}},
  \href{http://dx.doi.org/10.1103/PhysRevD.84.043519}{\emph{Phys. Rev.} {\bf
  D84} (2011) 043519}, [\href{http://arxiv.org/abs/1102.3435}{{\tt
  1102.3435}}].

\bibitem{Adshead:2011jq}
P.~Adshead, C.~Dvorkin, W.~Hu and E.~A. Lim, \emph{{Non-Gaussianity from Step
  Features in the Inflationary Potential}},
  \href{http://dx.doi.org/10.1103/PhysRevD.85.023531}{\emph{Phys. Rev.} {\bf
  D85} (2012) 023531}, [\href{http://arxiv.org/abs/1110.3050}{{\tt
  1110.3050}}].

\bibitem{Hu:2011vr}
W.~Hu, \emph{{Generalized Slow Roll for Non-Canonical Kinetic Terms}},
  \href{http://dx.doi.org/10.1103/PhysRevD.84.027303}{\emph{Phys. Rev.} {\bf
  D84} (2011) 027303}, [\href{http://arxiv.org/abs/1104.4500}{{\tt
  1104.4500}}].

\bibitem{Miranda:2012rm}
V.~Miranda, W.~Hu and P.~Adshead, \emph{{Warp Features in DBI Inflation}},
  \href{http://dx.doi.org/10.1103/PhysRevD.86.063529}{\emph{Phys. Rev.} {\bf
  D86} (2012) 063529}, [\href{http://arxiv.org/abs/1207.2186}{{\tt
  1207.2186}}].

\bibitem{Miranda:2013wxa}
V.-c. Miranda and W.~Hu, \emph{{Inflationary Steps in the Planck Data}},
  \href{http://dx.doi.org/10.1103/PhysRevD.89.083529}{\emph{Phys. Rev.} {\bf
  D89} (2014) 083529}, [\href{http://arxiv.org/abs/1312.0946}{{\tt
  1312.0946}}].

\bibitem{Adshead:2013aa}
P.~Adshead, W.~Hu and V.~Miranda, \emph{Bispectrum in single-field inflation
  beyond slow-roll},  \href{http://arxiv.org/abs/1303.7004}{{\tt 1303.7004}}.

\bibitem{Hu:2014hoa}
W.~Hu, \emph{{Generalized slow roll for tensor fluctuations}},
  \href{http://dx.doi.org/10.1103/PhysRevD.89.123503}{\emph{Phys. Rev.} {\bf
  D89} (2014) 123503}, [\href{http://arxiv.org/abs/1405.2020}{{\tt
  1405.2020}}].

\bibitem{Miranda:2014wga}
V.-c. Miranda, W.~Hu and P.~Adshead, \emph{{Steps to Reconcile Inflationary
  Tensor and Scalar Spectra}},
  \href{http://dx.doi.org/10.1103/PhysRevD.89.101302}{\emph{Phys. Rev.} {\bf
  D89} (2014) 101302}, [\href{http://arxiv.org/abs/1403.5231}{{\tt
  1403.5231}}].

\bibitem{Obied:2018qdr}
G.~Obied, C.~Dvorkin, C.~Heinrich, W.~Hu and V.~Miranda, \emph{{Inflationary
  versus reionization features from $Planck$ 2015 data}},
  \href{http://dx.doi.org/10.1103/PhysRevD.98.043518}{\emph{Phys. Rev.} {\bf
  D98} (2018) 043518}, [\href{http://arxiv.org/abs/1803.01858}{{\tt
  1803.01858}}].

\bibitem{Motohashi:2015hpa}
H.~Motohashi and W.~Hu, \emph{{Running from Features: Optimized Evaluation of
  Inflationary Power Spectra}},
  \href{http://dx.doi.org/10.1103/PhysRevD.92.043501}{\emph{Phys. Rev.} {\bf
  D92} (2015) 043501}, [\href{http://arxiv.org/abs/1503.04810}{{\tt
  1503.04810}}].

\bibitem{Motohashi:2017gqb}
H.~Motohashi and W.~Hu, \emph{{Generalized Slow Roll in the Unified Effective
  Field Theory of Inflation}},
  \href{http://dx.doi.org/10.1103/PhysRevD.96.023502}{\emph{Phys. Rev.} {\bf
  D96} (2017) 023502}, [\href{http://arxiv.org/abs/1704.01128}{{\tt
  1704.01128}}].

\bibitem{Cabass:2016aa}
G.~Cabass, E.~D. Valentino, A.~Melchiorri, E.~Pajer and J.~Silk, \emph{Running
  the running}, {\emph{Phys. Rev. D} {\bf 94} (2016) 023523},
  [\href{http://arxiv.org/abs/1605.00209}{{\tt 1605.00209}}].

\bibitem{Bruck:2016aa}
C.~van~de Bruck and C.~Longden, \emph{Running of the running and entropy
  perturbations during inflation}, {\emph{Phys. Rev. D} {\bf 94} (2016)
  021301}, [\href{http://arxiv.org/abs/1606.02176}{{\tt 1606.02176}}].

\bibitem{Array:2015xqh}
{\scshape BICEP2, Keck Array} collaboration, P.~A.~R. Ade et~al.,
  \emph{{Improved Constraints on Cosmology and Foregrounds from BICEP2 and Keck
  Array Cosmic Microwave Background Data with Inclusion of 95 GHz Band}},
  \href{http://dx.doi.org/10.1103/PhysRevLett.116.031302}{\emph{Phys. Rev.
  Lett.} {\bf 116} (2016) 031302}, [\href{http://arxiv.org/abs/1510.09217}{{\tt
  1510.09217}}].

\bibitem{Finelli:2016cyd}
{\scshape CORE} collaboration, F.~Finelli et~al., \emph{{Exploring cosmic
  origins with CORE: Inflation}},
  \href{http://dx.doi.org/10.1088/1475-7516/2018/04/016}{\emph{JCAP} {\bf 1804}
  (2018) 016}, [\href{http://arxiv.org/abs/1612.08270}{{\tt 1612.08270}}].

\bibitem{Remazeilles:2017szm}
{\scshape CORE} collaboration, M.~Remazeilles et~al., \emph{{Exploring cosmic
  origins with CORE: $B$-mode component separation}},
  \href{http://dx.doi.org/10.1088/1475-7516/2018/04/023}{\emph{JCAP} {\bf 1804}
  (2018) 023}, [\href{http://arxiv.org/abs/1704.04501}{{\tt 1704.04501}}].

\bibitem{Delabrouille:2017rct}
{\scshape CORE} collaboration, J.~Delabrouille et~al., \emph{{Exploring cosmic
  origins with CORE: Survey requirements and mission design}},
  \href{http://dx.doi.org/10.1088/1475-7516/2018/04/014}{\emph{JCAP} {\bf 1804}
  (2018) 014}, [\href{http://arxiv.org/abs/1706.04516}{{\tt 1706.04516}}].

\bibitem{2015}
J.~Aasi, B.~P. Abbott, R.~Abbott, T.~Abbott, M.~R. Abernathy, K.~Ackley et~al.,
  \emph{Advanced ligo},
  \href{http://dx.doi.org/10.1088/0264-9381/32/7/074001}{\emph{Classical and
  Quantum Gravity} {\bf 32} (Mar, 2015) 074001}.

\bibitem{:2016ab}
\emph{Gw150914: The advanced ligo detectors in the era of first discoveries},
  \href{http://dx.doi.org/10.1103/PhysRevLett.116.131103}{\emph{Physical Review
  Letters} {\bf 116} (2016) }.

\bibitem{Amaro-Seoane:2017aa}
P.~Amaro-Seoane, H.~Audley, S.~Babak, J.~Baker, E.~Barausse, P.~Bender et~al.,
  \emph{Laser interferometer space antenna},
  \href{http://arxiv.org/abs/1702.00786}{{\tt 1702.00786}}.

\bibitem{Hobbs_2010}
G.~Hobbs, A.~Archibald, Z.~Arzoumanian, D.~Backer, M.~Bailes, N.~D.~R. Bhat
  et~al., \emph{The international pulsar timing array project: using pulsars as
  a gravitational wave detector},
  \href{http://dx.doi.org/10.1088/0264-9381/27/8/084013}{\emph{Classical and
  Quantum Gravity} {\bf 27} (Apr, 2010) 084013}.

\bibitem{KRAMER2004993}
M.~Kramer, D.~Backer, J.~Cordes, T.~Lazio, B.~Stappers and S.~Johnston,
  \emph{Strong-field tests of gravity using pulsars and black holes},
  \href{http://dx.doi.org/https://doi.org/10.1016/j.newar.2004.09.020}{\emph{New
  Astronomy Reviews} {\bf 48} (2004) 993 -- 1002}.

\bibitem{Abbott:2016blz}
{\scshape LIGO Scientific, Virgo} collaboration, B.~P. Abbott et~al.,
  \emph{{Observation of Gravitational Waves from a Binary Black Hole Merger}},
  \href{http://dx.doi.org/10.1103/PhysRevLett.116.061102}{\emph{Phys. Rev.
  Lett.} {\bf 116} (2016) 061102}, [\href{http://arxiv.org/abs/1602.03837}{{\tt
  1602.03837}}].

\bibitem{Abbott:2016nmj}
{\scshape LIGO Scientific, Virgo} collaboration, B.~P. Abbott et~al.,
  \emph{{GW151226: Observation of Gravitational Waves from a 22-Solar-Mass
  Binary Black Hole Coalescence}},
  \href{http://dx.doi.org/10.1103/PhysRevLett.116.241103}{\emph{Phys. Rev.
  Lett.} {\bf 116} (2016) 241103}, [\href{http://arxiv.org/abs/1606.04855}{{\tt
  1606.04855}}].

\bibitem{TheLIGOScientific:2016pea}
{\scshape LIGO Scientific, Virgo} collaboration, B.~P. Abbott et~al.,
  \emph{{Binary Black Hole Mergers in the first Advanced LIGO Observing Run}},
  \href{http://dx.doi.org/10.1103/PhysRevX.6.041015,
  10.1103/PhysRevX.8.039903}{\emph{Phys. Rev.} {\bf X6} (2016) 041015},
  [\href{http://arxiv.org/abs/1606.04856}{{\tt 1606.04856}}].

\bibitem{GBM:2017lvd}
{\scshape LIGO Scientific, Virgo, Fermi GBM, INTEGRAL, IceCube, AstroSat
  Cadmium Zinc Telluride Imager Team, IPN, Insight-Hxmt, ANTARES, Swift, AGILE
  Team, 1M2H Team, Dark Energy Camera GW-EM, DES, DLT40, GRAWITA, Fermi-LAT,
  ATCA, ASKAP, Las Cumbres Observatory Group, OzGrav, DWF (Deeper Wider Faster
  Program), AST3, CAASTRO, VINROUGE, MASTER, J-GEM, GROWTH, JAGWAR,
  CaltechNRAO, TTU-NRAO, NuSTAR, Pan-STARRS, MAXI Team, TZAC Consortium, KU,
  Nordic Optical Telescope, ePESSTO, GROND, Texas Tech University, SALT Group,
  TOROS, BOOTES, MWA, CALET, IKI-GW Follow-up, H.E.S.S., LOFAR, LWA, HAWC,
  Pierre Auger, ALMA, Euro VLBI Team, Pi of Sky, Chandra Team at McGill
  University, DFN, ATLAS Telescopes, High Time Resolution Universe Survey,
  RIMAS, RATIR, SKA South Africa/MeerKAT} collaboration, B.~P. Abbott et~al.,
  \emph{{Multi-messenger Observations of a Binary Neutron Star Merger}},
  \href{http://dx.doi.org/10.3847/2041-8213/aa91c9}{\emph{Astrophys. J.} {\bf
  848} (2017) L12}, [\href{http://arxiv.org/abs/1710.05833}{{\tt 1710.05833}}].

\bibitem{TheLIGOScientific:2017qsa}
{\scshape LIGO Scientific, Virgo} collaboration, B.~P. Abbott et~al.,
  \emph{{GW170817: Observation of Gravitational Waves from a Binary Neutron
  Star Inspiral}},
  \href{http://dx.doi.org/10.1103/PhysRevLett.119.161101}{\emph{Phys. Rev.
  Lett.} {\bf 119} (2017) 161101}, [\href{http://arxiv.org/abs/1710.05832}{{\tt
  1710.05832}}].

\bibitem{Abbott:2017oio}
{\scshape LIGO Scientific, Virgo} collaboration, B.~P. Abbott et~al.,
  \emph{{GW170814: A Three-Detector Observation of Gravitational Waves from a
  Binary Black Hole Coalescence}},
  \href{http://dx.doi.org/10.1103/PhysRevLett.119.141101}{\emph{Phys. Rev.
  Lett.} {\bf 119} (2017) 141101}, [\href{http://arxiv.org/abs/1709.09660}{{\tt
  1709.09660}}].

\bibitem{Abbott:2017gyy}
{\scshape LIGO Scientific, Virgo} collaboration, B.~P. Abbott et~al.,
  \emph{{GW170608: Observation of a 19-solar-mass Binary Black Hole
  Coalescence}},
  \href{http://dx.doi.org/10.3847/2041-8213/aa9f0c}{\emph{Astrophys. J.} {\bf
  851} (2017) L35}, [\href{http://arxiv.org/abs/1711.05578}{{\tt 1711.05578}}].

\bibitem{Abbott:2017vtc}
{\scshape LIGO Scientific, VIRGO} collaboration, B.~P. Abbott et~al.,
  \emph{{GW170104: Observation of a 50-Solar-Mass Binary Black Hole Coalescence
  at Redshift 0.2}}, \href{http://dx.doi.org/10.1103/PhysRevLett.118.221101,
  10.1103/PhysRevLett.121.129901}{\emph{Phys. Rev. Lett.} {\bf 118} (2017)
  221101}, [\href{http://arxiv.org/abs/1706.01812}{{\tt 1706.01812}}].

\bibitem{Monitor:2017mdv}
{\scshape LIGO Scientific, Virgo, Fermi-GBM, INTEGRAL} collaboration, B.~P.
  Abbott et~al., \emph{{Gravitational Waves and Gamma-rays from a Binary
  Neutron Star Merger: GW170817 and GRB 170817A}},
  \href{http://dx.doi.org/10.3847/2041-8213/aa920c}{\emph{Astrophys. J.} {\bf
  848} (2017) L13}, [\href{http://arxiv.org/abs/1710.05834}{{\tt 1710.05834}}].

\bibitem{LIGOScientific:2018mvr}
{\scshape LIGO Scientific, Virgo} collaboration, B.~P. Abbott et~al.,
  \emph{{GWTC-1: A Gravitational-Wave Transient Catalog of Compact Binary
  Mergers Observed by LIGO and Virgo during the First and Second Observing
  Runs}},  \href{http://arxiv.org/abs/1811.12907}{{\tt 1811.12907}}.

\bibitem{Moore_2014}
C.~J. Moore, R.~H. Cole and C.~P.~L. Berry, \emph{Gravitational-wave
  sensitivity curves},
  \href{http://dx.doi.org/10.1088/0264-9381/32/1/015014}{\emph{Classical and
  Quantum Gravity} {\bf 32} (Dec, 2014) 015014}.

\bibitem{Caprini:2018mtu}
C.~Caprini and D.~G. Figueroa, \emph{{Cosmological Backgrounds of Gravitational
  Waves}}, \href{http://dx.doi.org/10.1088/1361-6382/aac608}{\emph{Class.
  Quant. Grav.} {\bf 35} (2018) 163001},
  [\href{http://arxiv.org/abs/1801.04268}{{\tt 1801.04268}}].

\bibitem{Cook:2011hg}
J.~L. Cook and L.~Sorbo, \emph{{Particle production during inflation and
  gravitational waves detectable by ground-based interferometers}},
  \href{http://dx.doi.org/10.1103/PhysRevD.86.069901,
  10.1103/PhysRevD.85.023534}{\emph{Phys. Rev.} {\bf D85} (2012) 023534},
  [\href{http://arxiv.org/abs/1109.0022}{{\tt 1109.0022}}].

\bibitem{Senatore:2011sp}
L.~Senatore, E.~Silverstein and M.~Zaldarriaga, \emph{{New Sources of
  Gravitational Waves during Inflation}},
  \href{http://dx.doi.org/10.1088/1475-7516/2014/08/016}{\emph{JCAP} {\bf 1408}
  (2014) 016}, [\href{http://arxiv.org/abs/1109.0542}{{\tt 1109.0542}}].

\bibitem{Mylova:2018yap}
M.~Mylova, O.~zsoy, S.~Parameswaran, G.~Tasinato and I.~Zavala, \emph{{A new
  mechanism to enhance primordial tensor fluctuations in single field
  inflation}},  \href{http://arxiv.org/abs/1808.10475}{{\tt 1808.10475}}.

\bibitem{Mukhanov:2013tua}
V.~Mukhanov, \emph{{Quantum Cosmological Perturbations: Predictions and
  Observations}},
  \href{http://dx.doi.org/10.1140/epjc/s10052-013-2486-7}{\emph{Eur. Phys. J.}
  {\bf C73} (2013) 2486}, [\href{http://arxiv.org/abs/1303.3925}{{\tt
  1303.3925}}].

\bibitem{Barranco:2014ira}
L.~Barranco, L.~Boubekeur and O.~Mena, \emph{{A model-independent fit to Planck
  and BICEP2 data}},
  \href{http://dx.doi.org/10.1103/PhysRevD.90.063007}{\emph{Phys. Rev.} {\bf
  D90} (2014) 063007}, [\href{http://arxiv.org/abs/1405.7188}{{\tt
  1405.7188}}].

\bibitem{Collaboration:2011aa}
C.~Collaboration, C.~Armitage-Caplan, M.~Avillez, D.~Barbosa, A.~Banday,
  N.~Bartolo et~al., \emph{Core (cosmic origins explorer) a white paper},
  \href{http://arxiv.org/abs/1102.2181}{{\tt 1102.2181}}.

\bibitem{Draper:2011aa}
P.~Draper, P.~Meade, M.~Reece and D.~Shih, \emph{{Implications of a 125 GeV
  Higgs for the MSSM and Low-Scale SUSY Breaking}},
  \href{http://dx.doi.org/10.1103/PhysRevD.85.095007}{\emph{Phys. Rev.} {\bf
  D85} (2012) 095007}, [\href{http://arxiv.org/abs/1112.3068}{{\tt
  1112.3068}}].

\bibitem{Heisenberg:2018aa}
L.~Heisenberg, \emph{A systematic approach to generalisations of general
  relativity and their cosmological implications},
  \href{http://arxiv.org/abs/1807.01725}{{\tt 1807.01725}}.

\bibitem{deRham:2014zqa}
C.~de~Rham, \emph{{Massive Gravity}},
  \href{http://dx.doi.org/10.12942/lrr-2014-7}{\emph{Living Rev. Rel.} {\bf 17}
  (2014) 7}, [\href{http://arxiv.org/abs/1401.4173}{{\tt 1401.4173}}].

\bibitem{deRham:2010tw}
C.~de~Rham, G.~Gabadadze, L.~Heisenberg and D.~Pirtskhalava, \emph{{Cosmic
  Acceleration and the Helicity-0 Graviton}},
  \href{http://dx.doi.org/10.1103/PhysRevD.83.103516}{\emph{Phys. Rev.} {\bf
  D83} (2011) 103516}, [\href{http://arxiv.org/abs/1010.1780}{{\tt
  1010.1780}}].

\bibitem{Motohashi:2015aa}
H.~Motohashi, \emph{Third order equations of motion and the ostrogradsky
  instability},
  \href{http://dx.doi.org/10.1103/PhysRevD.91.085009}{\emph{Physical Review D}
  {\bf 91} (2015) }.

\bibitem{Dvali:2000hr}
G.~R. Dvali, G.~Gabadadze and M.~Porrati, \emph{{4-D gravity on a brane in 5-D
  Minkowski space}},
  \href{http://dx.doi.org/10.1016/S0370-2693(00)00669-9}{\emph{Phys. Lett.}
  {\bf B485} (2000) 208--214}, [\href{http://arxiv.org/abs/hep-th/0005016}{{\tt
  hep-th/0005016}}].

\bibitem{Langlois:2015skt}
D.~Langlois and K.~Noui, \emph{{Hamiltonian analysis of higher derivative
  scalar-tensor theories}},
  \href{http://dx.doi.org/10.1088/1475-7516/2016/07/016}{\emph{JCAP} {\bf 1607}
  (2016) 016}, [\href{http://arxiv.org/abs/1512.06820}{{\tt 1512.06820}}].

\bibitem{Deffayet:2010zh}
C.~Deffayet, S.~Deser and G.~Esposito-Farese, \emph{{Arbitrary $p$-form
  Galileons}}, \href{http://dx.doi.org/10.1103/PhysRevD.82.061501}{\emph{Phys.
  Rev.} {\bf D82} (2010) 061501}, [\href{http://arxiv.org/abs/1007.5278}{{\tt
  1007.5278}}].

\bibitem{Deffayet:2013tca}
C.~Deffayet, A.~E. Gmrkolu, S.~Mukohyama and Y.~Wang, \emph{{A no-go theorem
  for generalized vector Galileons on flat spacetime}},
  \href{http://dx.doi.org/10.1007/JHEP04(2014)082}{\emph{JHEP} {\bf 04} (2014)
  082}, [\href{http://arxiv.org/abs/1312.6690}{{\tt 1312.6690}}].

\bibitem{Deffayet:2016von}
C.~Deffayet, S.~Mukohyama and V.~Sivanesan, \emph{{On p-form theories with
  gauge invariant second order field equations}},
  \href{http://dx.doi.org/10.1103/PhysRevD.93.085027}{\emph{Phys. Rev.} {\bf
  D93} (2016) 085027}, [\href{http://arxiv.org/abs/1601.01287}{{\tt
  1601.01287}}].

\bibitem{Deffayet:2017eqq}
C.~Deffayet, S.~Garcia-Saenz, S.~Mukohyama and V.~Sivanesan, \emph{{Classifying
  Galileon $p$-form theories}},
  \href{http://dx.doi.org/10.1103/PhysRevD.96.045014}{\emph{Phys. Rev.} {\bf
  D96} (2017) 045014}, [\href{http://arxiv.org/abs/1704.02980}{{\tt
  1704.02980}}].

\bibitem{doi:10.1063/1.522837}
G.~W. Horndeski, \emph{Conservation of charge and the einstein--maxwell field
  equations}, \href{http://dx.doi.org/10.1063/1.522837}{\emph{Journal of
  Mathematical Physics} {\bf 17} (1976) 1980--1987},
  [\href{http://arxiv.org/abs/https://doi.org/10.1063/1.522837}{{\tt
  https://doi.org/10.1063/1.522837}}].

\bibitem{Barrow:2012ay}
J.~D. Barrow, M.~Thorsrud and K.~Yamamoto, \emph{{Cosmologies in Horndeski's
  second-order vector-tensor theory}},
  \href{http://dx.doi.org/10.1007/JHEP02(2013)146}{\emph{JHEP} {\bf 02} (2013)
  146}, [\href{http://arxiv.org/abs/1211.5403}{{\tt 1211.5403}}].

\bibitem{Jimenez:2013qsa}
J.~Beltran~Jimenez, R.~Durrer, L.~Heisenberg and M.~Thorsrud, \emph{{Stability
  of Horndeski vector-tensor interactions}},
  \href{http://dx.doi.org/10.1088/1475-7516/2013/10/064}{\emph{JCAP} {\bf 1310}
  (2013) 064}, [\href{http://arxiv.org/abs/1308.1867}{{\tt 1308.1867}}].

\bibitem{Heisenberg:2014rta}
L.~Heisenberg, \emph{{Generalization of the Proca Action}},
  \href{http://dx.doi.org/10.1088/1475-7516/2014/05/015}{\emph{JCAP} {\bf 1405}
  (2014) 015}, [\href{http://arxiv.org/abs/1402.7026}{{\tt 1402.7026}}].

\bibitem{Allys:2015sht}
E.~Allys, P.~Peter and Y.~Rodriguez, \emph{{Generalized Proca action for an
  Abelian vector field}},
  \href{http://dx.doi.org/10.1088/1475-7516/2016/02/004}{\emph{JCAP} {\bf 1602}
  (2016) 004}, [\href{http://arxiv.org/abs/1511.03101}{{\tt 1511.03101}}].

\bibitem{Jimenez:2016isa}
J.~Beltran~Jimenez and L.~Heisenberg, \emph{{Derivative self-interactions for a
  massive vector field}},
  \href{http://dx.doi.org/10.1016/j.physletb.2016.04.017}{\emph{Phys. Lett.}
  {\bf B757} (2016) 405--411}, [\href{http://arxiv.org/abs/1602.03410}{{\tt
  1602.03410}}].

\bibitem{Tasinato:2014eka}
G.~Tasinato, \emph{{Cosmic Acceleration from Abelian Symmetry Breaking}},
  \href{http://dx.doi.org/10.1007/JHEP04(2014)067}{\emph{JHEP} {\bf 04} (2014)
  067}, [\href{http://arxiv.org/abs/1402.6450}{{\tt 1402.6450}}].

\bibitem{Tasinato:2014mia}
G.~Tasinato, \emph{{A small cosmological constant from Abelian symmetry
  breaking}},
  \href{http://dx.doi.org/10.1088/0264-9381/31/22/225004}{\emph{Class. Quant.
  Grav.} {\bf 31} (2014) 225004}, [\href{http://arxiv.org/abs/1404.4883}{{\tt
  1404.4883}}].

\bibitem{DeFelice:2016uil}
A.~De~Felice, L.~Heisenberg, R.~Kase, S.~Mukohyama, S.~Tsujikawa and Y.-l.
  Zhang, \emph{{Effective gravitational couplings for cosmological
  perturbations in generalized Proca theories}},
  \href{http://dx.doi.org/10.1103/PhysRevD.94.044024}{\emph{Phys. Rev.} {\bf
  D94} (2016) 044024}, [\href{http://arxiv.org/abs/1605.05066}{{\tt
  1605.05066}}].

\bibitem{DeFelice:2016yws}
A.~De~Felice, L.~Heisenberg, R.~Kase, S.~Mukohyama, S.~Tsujikawa and Y.-l.
  Zhang, \emph{{Cosmology in generalized Proca theories}},
  \href{http://dx.doi.org/10.1088/1475-7516/2016/06/048}{\emph{JCAP} {\bf 1606}
  (2016) 048}, [\href{http://arxiv.org/abs/1603.05806}{{\tt 1603.05806}}].

\bibitem{deFelice:2017paw}
A.~de~Felice, L.~Heisenberg and S.~Tsujikawa, \emph{{Observational constraints
  on generalized Proca theories}},
  \href{http://dx.doi.org/10.1103/PhysRevD.95.123540}{\emph{Phys. Rev.} {\bf
  D95} (2017) 123540}, [\href{http://arxiv.org/abs/1703.09573}{{\tt
  1703.09573}}].

\bibitem{Cisterna:2016nwq}
A.~Cisterna, M.~Hassaine, J.~Oliva and M.~Rinaldi, \emph{{Static and rotating
  solutions for Vector-Galileon theories}},
  \href{http://dx.doi.org/10.1103/PhysRevD.94.104039}{\emph{Phys. Rev.} {\bf
  D94} (2016) 104039}, [\href{http://arxiv.org/abs/1609.03430}{{\tt
  1609.03430}}].

\bibitem{Minamitsuji:2016ydr}
M.~Minamitsuji, \emph{{Solutions in the generalized Proca theory with the
  nonminimal coupling to the Einstein tensor}},
  \href{http://dx.doi.org/10.1103/PhysRevD.94.084039}{\emph{Phys. Rev.} {\bf
  D94} (2016) 084039}, [\href{http://arxiv.org/abs/1607.06278}{{\tt
  1607.06278}}].

\bibitem{DeFelice:2016cri}
A.~De~Felice, L.~Heisenberg, R.~Kase, S.~Tsujikawa, Y.-l. Zhang and G.-B. Zhao,
  \emph{{Screening fifth forces in generalized Proca theories}},
  \href{http://dx.doi.org/10.1103/PhysRevD.93.104016}{\emph{Phys. Rev.} {\bf
  D93} (2016) 104016}, [\href{http://arxiv.org/abs/1602.00371}{{\tt
  1602.00371}}].

\bibitem{Fan:2016jnz}
Z.-Y. Fan, \emph{{Black holes with vector hair}},
  \href{http://dx.doi.org/10.1007/JHEP09(2016)039}{\emph{JHEP} {\bf 09} (2016)
  039}, [\href{http://arxiv.org/abs/1606.00684}{{\tt 1606.00684}}].

\bibitem{Chagoya:2016aar}
J.~Chagoya, G.~Niz and G.~Tasinato, \emph{{Black Holes and Abelian Symmetry
  Breaking}},
  \href{http://dx.doi.org/10.1088/0264-9381/33/17/175007}{\emph{Class. Quant.
  Grav.} {\bf 33} (2016) 175007}, [\href{http://arxiv.org/abs/1602.08697}{{\tt
  1602.08697}}].

\bibitem{Heisenberg:2017xda}
L.~Heisenberg, R.~Kase, M.~Minamitsuji and S.~Tsujikawa, \emph{{Hairy
  black-hole solutions in generalized Proca theories}},
  \href{http://dx.doi.org/10.1103/PhysRevD.96.084049}{\emph{Phys. Rev.} {\bf
  D96} (2017) 084049}, [\href{http://arxiv.org/abs/1705.09662}{{\tt
  1705.09662}}].

\bibitem{Heisenberg:2017hwb}
L.~Heisenberg, R.~Kase, M.~Minamitsuji and S.~Tsujikawa, \emph{{Black holes in
  vector-tensor theories}},
  \href{http://dx.doi.org/10.1088/1475-7516/2017/08/024}{\emph{JCAP} {\bf 1708}
  (2017) 024}, [\href{http://arxiv.org/abs/1706.05115}{{\tt 1706.05115}}].

\bibitem{Chagoya:2017fyl}
J.~Chagoya, G.~Niz and G.~Tasinato, \emph{{Black Holes and Neutron Stars in
  Vector Galileons}},
  \href{http://dx.doi.org/10.1088/1361-6382/aa7c01}{\emph{Class. Quant. Grav.}
  {\bf 34} (2017) 165002}, [\href{http://arxiv.org/abs/1703.09555}{{\tt
  1703.09555}}].

\bibitem{Heisenberg:2016eld}
L.~Heisenberg, R.~Kase and S.~Tsujikawa, \emph{{Beyond generalized Proca
  theories}},
  \href{http://dx.doi.org/10.1016/j.physletb.2016.07.052}{\emph{Phys. Lett.}
  {\bf B760} (2016) 617--626}, [\href{http://arxiv.org/abs/1605.05565}{{\tt
  1605.05565}}].

\bibitem{Heisenberg:2018acv}
L.~Heisenberg, \emph{{Scalar-Vector-Tensor Gravity Theories}},
  \href{http://dx.doi.org/10.1088/1475-7516/2018/10/054}{\emph{JCAP} {\bf 1810}
  (2018) 054}, [\href{http://arxiv.org/abs/1801.01523}{{\tt 1801.01523}}].

\bibitem{Kase:2018nwt}
R.~Kase and S.~Tsujikawa, \emph{{Dark energy in scalar-vector-tensor
  theories}},
  \href{http://dx.doi.org/10.1088/1475-7516/2018/11/024}{\emph{JCAP} {\bf 1811}
  (2018) 024}, [\href{http://arxiv.org/abs/1805.11919}{{\tt 1805.11919}}].

\bibitem{Heisenberg:2018vti}
L.~Heisenberg and S.~Tsujikawa, \emph{{Hairy black hole solutions in $U(1)$
  gauge-invariant scalar-vector-tensor theories}},
  \href{http://dx.doi.org/10.1016/j.physletb.2018.03.059}{\emph{Phys. Lett.}
  {\bf B780} (2018) 638--646}, [\href{http://arxiv.org/abs/1802.07035}{{\tt
  1802.07035}}].

\bibitem{Kobayashi:2011nu}
T.~Kobayashi, M.~Yamaguchi and J.~Yokoyama, \emph{{Generalized G-inflation:
  Inflation with the most general second-order field equations}},
  \href{http://dx.doi.org/10.1143/PTP.126.511}{\emph{Prog. Theor. Phys.} {\bf
  126} (2011) 511--529}, [\href{http://arxiv.org/abs/1105.5723}{{\tt
  1105.5723}}].

\bibitem{Einhorn:2009bh}
M.~B. Einhorn and D.~R.~T. Jones, \emph{{Inflation with Non-minimal
  Gravitational Couplings in Supergravity}},
  \href{http://dx.doi.org/10.1007/JHEP03(2010)026}{\emph{JHEP} {\bf 03} (2010)
  026}, [\href{http://arxiv.org/abs/0912.2718}{{\tt 0912.2718}}].

\bibitem{Kallosh:2010ug}
R.~Kallosh and A.~Linde, \emph{{New models of chaotic inflation in
  supergravity}},
  \href{http://dx.doi.org/10.1088/1475-7516/2010/11/011}{\emph{JCAP} {\bf 1011}
  (2010) 011}, [\href{http://arxiv.org/abs/1008.3375}{{\tt 1008.3375}}].

\bibitem{Ferrara:2010yw}
S.~Ferrara, R.~Kallosh, A.~Linde, A.~Marrani and A.~Van~Proeyen, \emph{{Jordan
  Frame Supergravity and Inflation in NMSSM}},
  \href{http://dx.doi.org/10.1103/PhysRevD.82.045003}{\emph{Phys. Rev.} {\bf
  D82} (2010) 045003}, [\href{http://arxiv.org/abs/1004.0712}{{\tt
  1004.0712}}].

\bibitem{Lee:2010hj}
H.~M. Lee, \emph{{Chaotic inflation in Jordan frame supergravity}},
  \href{http://dx.doi.org/10.1088/1475-7516/2010/08/003}{\emph{JCAP} {\bf 1008}
  (2010) 003}, [\href{http://arxiv.org/abs/1005.2735}{{\tt 1005.2735}}].

\bibitem{Ferrara:2010in}
S.~Ferrara, R.~Kallosh, A.~Linde, A.~Marrani and A.~Van~Proeyen,
  \emph{{Superconformal Symmetry, NMSSM, and Inflation}},
  \href{http://dx.doi.org/10.1103/PhysRevD.83.025008}{\emph{Phys. Rev.} {\bf
  D83} (2011) 025008}, [\href{http://arxiv.org/abs/1008.2942}{{\tt
  1008.2942}}].

\bibitem{Kallosh:2010xz}
R.~Kallosh, A.~Linde and T.~Rube, \emph{{General inflaton potentials in
  supergravity}},
  \href{http://dx.doi.org/10.1103/PhysRevD.83.043507}{\emph{Phys. Rev.} {\bf
  D83} (2011) 043507}, [\href{http://arxiv.org/abs/1011.5945}{{\tt
  1011.5945}}].

\bibitem{Kaiser:2013sna}
D.~I. Kaiser and E.~I. Sfakianakis, \emph{{Multifield Inflation after Planck:
  The Case for Nonminimal Couplings}},
  \href{http://dx.doi.org/10.1103/PhysRevLett.112.011302}{\emph{Phys. Rev.
  Lett.} {\bf 112} (2014) 011302}, [\href{http://arxiv.org/abs/1304.0363}{{\tt
  1304.0363}}].

\bibitem{Chiba:2014sva}
T.~Chiba and K.~Kohri, \emph{{Consistency Relations for Large Field Inflation:
  Non-minimal Coupling}},
  \href{http://dx.doi.org/10.1093/ptep/ptv007}{\emph{PTEP} {\bf 2015} (2015)
  023E01}, [\href{http://arxiv.org/abs/1411.7104}{{\tt 1411.7104}}].

\bibitem{Pallis:2014cda}
C.~Pallis and Q.~Shafi, \emph{{Gravity Waves From Non-Minimal Quadratic
  Inflation}},
  \href{http://dx.doi.org/10.1088/1475-7516/2015/03/023}{\emph{JCAP} {\bf 1503}
  (2015) 023}, [\href{http://arxiv.org/abs/1412.3757}{{\tt 1412.3757}}].

\bibitem{Tenkanen:2017jih}
T.~Tenkanen, \emph{{Resurrecting Quadratic Inflation with a non-minimal
  coupling to gravity}},
  \href{http://dx.doi.org/10.1088/1475-7516/2017/12/001}{\emph{JCAP} {\bf 1712}
  (2017) 001}, [\href{http://arxiv.org/abs/1710.02758}{{\tt 1710.02758}}].

\bibitem{Kobayashi:2010cm}
T.~Kobayashi, M.~Yamaguchi and J.~Yokoyama, \emph{{G-inflation: Inflation
  driven by the Galileon field}},
  \href{http://dx.doi.org/10.1103/PhysRevLett.105.231302}{\emph{Phys. Rev.
  Lett.} {\bf 105} (2010) 231302}, [\href{http://arxiv.org/abs/1008.0603}{{\tt
  1008.0603}}].

\bibitem{DeFelice:2011zh}
A.~De~Felice and S.~Tsujikawa, \emph{{Primordial non-Gaussianities in general
  modified gravitational models of inflation}},
  \href{http://dx.doi.org/10.1088/1475-7516/2011/04/029}{\emph{JCAP} {\bf 1104}
  (2011) 029}, [\href{http://arxiv.org/abs/1103.1172}{{\tt 1103.1172}}].

\bibitem{BazrafshanMoghaddam:2016tdk}
H.~Bazrafshan~Moghaddam, R.~Brandenberger and J.~Yokoyama, \emph{{Note on
  Reheating in G-inflation}},
  \href{http://dx.doi.org/10.1103/PhysRevD.95.063529}{\emph{Phys. Rev.} {\bf
  D95} (2017) 063529}, [\href{http://arxiv.org/abs/1612.00998}{{\tt
  1612.00998}}].

\bibitem{Heisenberg:2018mxx}
L.~Heisenberg, R.~Kase and S.~Tsujikawa, \emph{{Cosmology in
  scalar-vector-tensor theories}},
  \href{http://dx.doi.org/10.1103/PhysRevD.98.024038}{\emph{Phys. Rev.} {\bf
  D98} (2018) 024038}, [\href{http://arxiv.org/abs/1805.01066}{{\tt
  1805.01066}}].

\bibitem{Adams:2001vc}
J.~A. Adams, B.~Cresswell and R.~Easther, \emph{{Inflationary perturbations
  from a potential with a step}},
  \href{http://dx.doi.org/10.1103/PhysRevD.64.123514}{\emph{Phys. Rev.} {\bf
  D64} (2001) 123514}, [\href{http://arxiv.org/abs/astro-ph/0102236}{{\tt
  astro-ph/0102236}}].

\bibitem{Covi:2006ci}
L.~Covi, J.~Hamann, A.~Melchiorri, A.~Slosar and I.~Sorbera, \emph{{Inflation
  and WMAP three year data: Features have a Future!}},
  \href{http://dx.doi.org/10.1103/PhysRevD.74.083509}{\emph{Phys. Rev.} {\bf
  D74} (2006) 083509}, [\href{http://arxiv.org/abs/astro-ph/0606452}{{\tt
  astro-ph/0606452}}].

\bibitem{Hamann:2007pa}
J.~Hamann, L.~Covi, A.~Melchiorri and A.~Slosar, \emph{{New Constraints on
  Oscillations in the Primordial Spectrum of Inflationary Perturbations}},
  \href{http://dx.doi.org/10.1103/PhysRevD.76.023503}{\emph{Phys. Rev.} {\bf
  D76} (2007) 023503}, [\href{http://arxiv.org/abs/astro-ph/0701380}{{\tt
  astro-ph/0701380}}].

\bibitem{Pahud:2008ae}
C.~Pahud, M.~Kamionkowski and A.~R. Liddle, \emph{{Oscillations in the inflaton
  potential?}}, \href{http://dx.doi.org/10.1103/PhysRevD.79.083503}{\emph{Phys.
  Rev.} {\bf D79} (2009) 083503}, [\href{http://arxiv.org/abs/0807.0322}{{\tt
  0807.0322}}].

\bibitem{Joy:2008qd}
M.~Joy, A.~Shafieloo, V.~Sahni and A.~A. Starobinsky, \emph{{Is a step in the
  primordial spectral index favored by CMB data ?}},
  \href{http://dx.doi.org/10.1088/1475-7516/2009/06/028}{\emph{JCAP} {\bf 0906}
  (2009) 028}, [\href{http://arxiv.org/abs/0807.3334}{{\tt 0807.3334}}].

\bibitem{Mortonson:2009qv}
M.~J. Mortonson, C.~Dvorkin, H.~V. Peiris and W.~Hu, \emph{{CMB polarization
  features from inflation versus reionization}},
  \href{http://dx.doi.org/10.1103/PhysRevD.79.103519}{\emph{Phys. Rev.} {\bf
  D79} (2009) 103519}, [\href{http://arxiv.org/abs/0903.4920}{{\tt
  0903.4920}}].

\bibitem{Stewart:2001cd}
E.~D. Stewart, \emph{{The Spectrum of density perturbations produced during
  inflation to leading order in a general slow roll approximation}},
  \href{http://dx.doi.org/10.1103/PhysRevD.65.103508}{\emph{Phys.Rev.} {\bf
  D65} (2002) 103508}, [\href{http://arxiv.org/abs/astro-ph/0110322}{{\tt
  astro-ph/0110322}}].

\bibitem{Kase:2014cwa}
R.~Kase and S.~Tsujikawa, \emph{{Effective field theory approach to modified
  gravity including Horndeski theory and Ho{\v r}ava--Lifshitz gravity}},
  \href{http://dx.doi.org/10.1142/S0218271814430081}{\emph{Int. J. Mod. Phys.}
  {\bf D23} (2014) 1443008}, [\href{http://arxiv.org/abs/1409.1984}{{\tt
  1409.1984}}].

\bibitem{Dvorkin:2011ui}
C.~Dvorkin and W.~Hu, \emph{{Complete WMAP Constraints on Bandlimited
  Inflationary Features}},
  \href{http://dx.doi.org/10.1103/PhysRevD.84.063515}{\emph{Phys. Rev.} {\bf
  D84} (2011) 063515}, [\href{http://arxiv.org/abs/1106.4016}{{\tt
  1106.4016}}].

\bibitem{Obied:2017tpd}
G.~Obied, C.~Dvorkin, C.~Heinrich, W.~Hu and V.~Miranda, \emph{{Inflationary
  Features and Shifts in Cosmological Parameters from Planck 2015 Data}},
  \href{http://dx.doi.org/10.1103/PhysRevD.96.083526}{\emph{Phys. Rev.} {\bf
  D96} (2017) 083526}, [\href{http://arxiv.org/abs/1706.09412}{{\tt
  1706.09412}}].

\bibitem{Miranda:2015cea}
V.~Miranda, W.~Hu, C.~He and H.~Motohashi, \emph{{Nonlinear Excitations in
  Inflationary Power Spectra}},
  \href{http://dx.doi.org/10.1103/PhysRevD.93.023504}{\emph{Phys. Rev.} {\bf
  D93} (2016) 023504}, [\href{http://arxiv.org/abs/1510.07580}{{\tt
  1510.07580}}].

\bibitem{Motohashi:2017kbs}
H.~Motohashi and W.~Hu, \emph{{Primordial Black Holes and Slow-Roll
  Violation}}, \href{http://dx.doi.org/10.1103/PhysRevD.96.063503}{\emph{Phys.
  Rev.} {\bf D96} (2017) 063503}, [\href{http://arxiv.org/abs/1706.06784}{{\tt
  1706.06784}}].

\bibitem{Adshead:2012aa}
P.~Adshead and W.~Hu, \emph{Fast computation of first-order feature-bispectrum
  corrections}, {\emph{Phys. Rev. D} {\bf 85} (2012) 103531},
  [\href{http://arxiv.org/abs/1203.0012}{{\tt 1203.0012}}].

\bibitem{Adshead:2013ab}
P.~Adshead, W.~Hu and V.~Miranda, \emph{Bispectrum in single-field inflation
  beyond slow-roll}, {\emph{Phys. Rev. D88 023507 (2013)} (03, 2013) },
  [\href{http://arxiv.org/abs/1303.7004}{{\tt 1303.7004}}].

\bibitem{Passaglia:2018aa}
S.~Passaglia and W.~Hu, \emph{Scalar bispectrum beyond slow-roll in the unified
  eft of inflation}, {\emph{Phys. Rev. D} {\bf 98} (2018) 023526},
  [\href{http://arxiv.org/abs/1804.07741}{{\tt 1804.07741}}].

\bibitem{Bardeen:1980kt}
J.~M. Bardeen, \emph{{Gauge Invariant Cosmological Perturbations}},
  \href{http://dx.doi.org/10.1103/PhysRevD.22.1882}{\emph{Phys. Rev.} {\bf D22}
  (1980) 1882--1905}.

\end{thebibliography}\endgroup

\newpage
\cleardoublepage
\thispagestyle{empty}

$$ $$\\[8cm]
\begin{center} $\star\qquad\star\qquad\star$ \end{center}

\end{document}